\documentclass[12pt]{article}
\usepackage{amsfonts,amsthm,amsmath,amssymb,upgreek}
\usepackage{hyperref}
\usepackage[paper=letterpaper,margin=1in]{geometry}
\usepackage{graphicx}
\usepackage{units}
\usepackage{color}
\usepackage{xcolor}
\usepackage[export]{adjustbox}
\usepackage{float}
\usepackage [english]{babel}
\usepackage{empheq}
\usepackage [autostyle, english = american]{csquotes}
\MakeOuterQuote{"}
\numberwithin{equation}{section}

\def\Tr{{\rm Tr }}
\def\hat{\widehat}
\newcommand{\be}{\begin{equation}}
\newcommand{\ee}{\end{equation}}

\def\disk{\text{Disk}}
\def\cyl{\text{Cyl}}

\begin{document}
\thispagestyle{empty}

\vspace*{.5cm}
\begin{center}

{\bf {\LARGE Comments on wormholes and factorization}\\
\vspace{1cm}}

\begin{center}

 {\bf Phil Saad$^1$, Stephen H. Shenker$^2$, and Shunyu Yao$^2$}\\
  \bigskip \rm
  
\bigskip

$^1$School of Natural Sciences,\\
Institute for Advanced Study, Princeton, NJ 08540\\

\bigskip

$^2$Stanford Institute for Theoretical Physics,\\Stanford University, Stanford, CA 94305 \\

\rm
  \end{center}

\vspace{1.2cm}
{\bf Abstract}
\end{center}

\begin{quotation}
\noindent

In AdS/CFT partition functions of decoupled copies of the CFT factorize. In bulk computations of such quantities contributions from spacetime wormholes which link separate asymptotic boundaries threaten to spoil this property, leading to a ``factorization puzzle.''  Certain simple models like JT gravity have wormholes, but bulk computations in them correspond to averages over an ensemble of boundary systems. These averages need not factorize. We can formulate a toy version of the factorization puzzle in such models by focusing on a specific member of the ensemble where partition functions will again factorize.

As Coleman and Giddings-Strominger pointed out in the 1980s,  fixed members of ensembles are described in the bulk  by ``$\alpha$-states" in a many-universe Hilbert space.   In this paper we analyze in detail the bulk mechanism for factorization in such  $\alpha$-states  in the topological model introduced by Marolf and Maxfield (the ``MM model") and in JT gravity.   
In these models geometric calculations in $\alpha$ states  are poorly controlled. We circumvent this complication by working in {\it approximate} $\alpha$ states where bulk calculations just involve the simplest topologies: disks and cylinders.

One of our main results is an effective description of the factorization mechanism. In this effective description the many-universe contributions from the full  $\alpha$ state are replaced by a small number of effective boundaries. Our motivation in constructing this effective description, and more generally in studying these simple ensemble models, is that the lessons learned might have wider applicability.   In fact the effective description lines up with a recent discussion of the SYK model with fixed couplings \cite{Saad:2021rcu}. We conclude with some discussion about the possible applicability of this effective model in more general contexts. 

\end{quotation}

%
%


\setcounter{page}{0}
\setcounter{tocdepth}{2}
\setcounter{footnote}{0}
\newpage

\tableofcontents

\pagebreak

\section{Introduction}\label{SectionIntroduction}
\subsection{Overview}

Gauge/gravity duality currently provides the only known complete definition of quantum gravity.   In this framework bulk quantum gravity, in certain situations,  has a dual description as a  well defined non-gravitational boundary quantum system.   Many parts of the dictionary between bulk and boundary quantities have been established, but our understanding of the bulk is still incomplete.   

There are certain sharp tensions between our current knowledge of the bulk and the well-defined boundary description whose resolution would certainly deepen our understanding of  quantum gravity.   The one we will focus on in this paper is the puzzle of factorization, first described in this context in \cite{Witten:1999xp,Maldacena:2004rf}.

Consider two decoupled boundary CFTs, called L and R, with corresponding partition functions $Z_L$ and $Z_R$.   The boundary description immediately shows that the partition function of the combined system $Z_{LR}$ factorizes -- that is, it is equal to $Z_L Z_R$.  But a bulk calculation seems  to include gravitational configurations connecting the two boundaries,  spacetime wormholes are the paradigmatic example.   Naively such configurations spoil factorization.    The problem is to understand how a complete bulk description resolves this apparent contradiction.

\begin{figure}[H]
\centering
\includegraphics[scale=0.35]{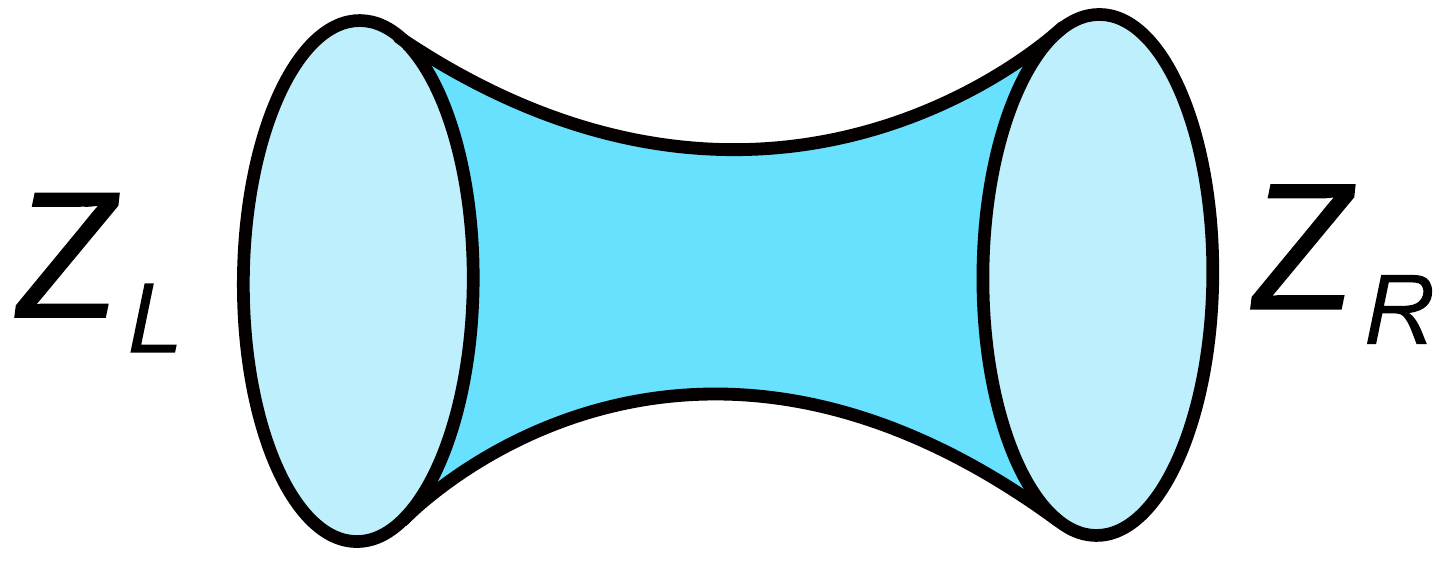}
\caption{\small Here we picture a spacetime wormhole which contributes to $Z_L Z_R$. }
\end{figure}

This question has become more pressing recently  because spacetime wormholes have been argued to be useful in explaining a number of nonperturbative properties of quantum black holes in the context gauge/gravity duality.  A first example is  the ``eternal traversable wormhole" \cite{Maldacena:2018lmt};  Another example  is the ``ramp" region in the long-time behavior of the spectral form factor \cite{Saad:2018bqo,Saad:2019lba} and of  correlation functions \cite{Saad:2019pqd} that is a signature of random matrix statistics in the quantum chaotic boundary system. Another example is the replica wormhole calculation of  the Page curve \cite{Almheiri:2019qdq,Penington:2019kki} and of squared matrix elements \cite{Stanford:2020wkf} in models of evaporating black holes.  Some of these situations, in particular the spectral form factor and squared matrix elements, are described by decoupled boundary systems and so the wormhole explanation gives rise to a factorization puzzle.

The most complete calculations of these quantities have been done in systems like the SYK model \cite{Sachdev:1992fk,KitaevTalks,Kitaev:2017awl} and its low energy limit, JT gravity \cite{Jackiw:1984je,Teitelboim:1983ux,Jensen:2016pah,Maldacena:2016upp,Engelsoy:2016xyb}. These systems are dual to \textit{ensembles} of boundary quantum systems \cite{Saad:2019lba,Stanford:2019vob}. In the SYK model one averages over an ensemble of Hamiltonians with different values of the couplings between the fermions.  In JT gravity one averages over Hamiltonian matrices drawn from a certain random matrix ensemble. Then instead of computing genuine products of partition functions, one considers \textit{averages} of products of partition functions.  Denoting the average over the ensemble by $\langle \cdot \rangle$, the average over L and R systems $\langle Z_L Z_R \rangle$ in general will \textit{not} factorize into the product $\langle Z_L\rangle \langle Z_R\rangle $.  In the the bulk these averaged products typically receive contributions from wormholes, which accurately calculate the deviation from factorization.  There is no puzzle here.\footnote{There has been a substantial amount of additional recent work  on wormholes and ensembles. Examples include \cite{Okuyama:2019xbv,Pollack:2020gfa,Afkhami-Jeddi:2020ezh,Maloney:2020nni,Belin:2020hea,Cotler:2020ugk,Bousso:2020kmy,Altland:2020ccq,Anous:2020lka,Chen:2020tes,Liu:2020jsv,Langhoff:2020jqa,Marolf:2021kjc,Casali:2021ewu,Meruliya:2021utr,Eberhardt:2021jvj,Godet:2021cdl,Verlinde:2021jwu,Mukhametzhanov:2021nea,Johnson:2021zuo,Marolf:2020rpm,Giddings:2020yes}.}

The  connection  between wormholes and ensembles  is actually an old one, going back to the ideas of Coleman \cite{Coleman:1988cy} and Giddings-Strominger \cite{Giddings:1988cx} in the 1980s.  These ideas have recently been reformulated in the AdS/CFT context, with important extensions, by Marolf and Maxfield \cite{Marolf:2020xie} in work we will discuss extensively below.

  But the paradigmatic examples of gauge/gravity duality consist of  bulk gravitational systems dual to  \textit{single} boundary theories, not ensembles.   Bulk spacetime wormhole configurations joining decoupled boundaries may well exist in these systems \cite{Maldacena:2004rf,ArkaniHamed:2007js,Cotler:2020ugk,Cotler:2020lxj,Marolf:2021kjc,Mahajan:2021maz,Cotler:2021cqa}.
 How could this be compatible with factorization?

 Perhaps the wormholes are actually absent, or don't contribute  for some reason.  But on general grounds, especially the universality of random matrix behavior in such quantum chaotic systems, we expect that decoupled non-averaged systems display certain features reminiscent of the wormhole effects.  An  example we will focus on is the ramp in the spectral form factor, the product of analytically continued partition functions $Z_L(\beta +iT) Z_R(\beta +iT)$.   This product must, of necessity, factorize.  How can this be compatible with the wormhole signal?

For a chaotic non-averaged system, including the boundary theories in standard examples of gauge/gravity duality, one expects, again based on random matrix universality, that the spectral form factor is a noisy function for sufficiently late times, including the ramp region \cite{1997PhRvL..78.2280P}. This noise is typically the same order as  the signal.  For systems described by ensembles, like SYK and JT gravity, one expects  similar noisy behavior for a fixed draw from the ensemble. In suitably rich ensembles, like those in SYK and JT gravity, this noise can largely be removed by ensemble averaging; the resulting averaged spectral form factor has a smooth ramp rather than an erratic one, giving the signal well described by wormholes.      These large noisy fluctuations can be quantified by studying ensemble averaged statistics like the variance, higher moments, and time autocorrelation functions.  Such statistics can also be computed in the bulk using wormholes.\footnote{This is the case as long as we restrict attention to finite order moments, and times short compared to the plateau time.}

So while in systems like SYK and JT gravity, the wormhole does not give a good approximation for a fixed draw from the ensemble, the wormhole has a clear role in computing the ensemble averages. However, in non-averaged systems  the wormhole cannot be isolated this way, and due to the large fluctuations it does not give a uniformly good approximation to the answer.  So one might expect that wormholes are not relevant here.

But random matrix universality also indicates that the autocorrelation time of the noise in the spectral form factor is short, of order the inverse size of the energy window considered \cite{1997PhRvL..78.2280P,1999JPhA...32.6903H},\footnote{This autocorrelation time will be derived in the JT system in appendix \ref{autoJTappend}.} and so performing a time average along the ramp leads to a smooth curve which matches the wormhole result.   So it seems appealing to entertain the possibility that wormholes play some kind of role in the bulk description of non-averaged systems.\footnote{Of course this might not be the case -- these features might have a completely different origin.}
  Note that this time averaging correlates the L and R observables so there is no factorization puzzle.\footnote{The Renyi entropies computed by replica wormholes \cite{Penington:2019kki,Almheiri:2019qdq} are other examples of correlated observables which do not raise a factorization issue, and where noise is suppressed.}

 The goal of this paper is to explore, compare and contrast some scenarios for how wormholes and factorization can coexist in non-averaged gauge/gravity duals.   We currently do not have enough control in any  standard holographic system to address these subtle questions.   So we must of necessity return to  simple models, in particular those  described by ensembles.  Here, as mentioned above,  we can study a version of the factorization puzzle by looking at  just one element of the ensemble. Computations for a single ensemble element will include contributions from wormholes, but nevertheless still manage to produce a factorized answer.  Our hope, and at this point it is only a hope, is that the mechanisms we examine might give some insight into how standard non-averaged holographic systems behave.
 
Specifically,  we will examine  the Marolf-Maxfield (MM) model \cite{Marolf:2020xie} where a fixed element of the ensemble is referred to as an $\alpha$ state \cite{Coleman:1988cy,Giddings:1988cx}).  We will then turn to JT gravity where a single element of the ensemble corresponds to a particular random matrix drawn from the appropriate distribution\cite{Blommaert:2019wfy}. To set some context we will begin by discussing a simple analogy -- the periodic orbit theory of semiclassical quantum chaos \cite{Haake:1315494};

 We will also make some brief remarks comparing these models to another system, the  SYK model with a single instance of the random fermion couplings.  We  have recently presented an analysis  of this system in another paper, written jointly with Douglas Stanford \cite{Saad:2021rcu}. 
 
 In each model we will endeavor to study the simplest situation where the factorization puzzle arises.   In  JT gravity we will only study the ramp region of the spectral form factor and ignore the plateau.  This amounts to ignoring the discreteness of the random matrix eigenvalue spectrum, and treating it as a continuous density.\footnote{See \cite{Blommaert:2021gha} for progress in controlling effects related to the discreteness of the spectrum.} This allows us to focus on the simplest topologies which we will call the disk (the Euclidean black hole) and the cylinder (the wormhole).   Similarly in the MM model we will ignore the discreteness of the spectrum of $\alpha$ states and consider sharply peaked but smooth approximate  $\alpha$ states.  Again only disks and cylinders will contribute.

 \subsection{Summary of the paper}
 
 We now give a brief summary of this rather lengthy paper and its main results.

 In Section 2, \hyperref[periodicorbits]{\textbf{Periodic orbits}}. we briefly review the computation of the spectral form factor in the periodic orbit theory of semiclassical quantum chaos. In these systems, one uses the Gutzwiller trace formula to express the spectral form factor as a double-sum over periodic orbits, one for each copy of the system. The off-diagonal summands have erratic phases, but the diagonal terms are smooth. As a result, averaging removes the off-diagonal terms, with the diagonal terms giving the resulting ``ramp''. Making an analogy between this diagonal orbit sum and the wormhole, we are led to ask what is the analog of the off-diagonal structure in gravity.
 
 In Section 3. \hyperref[SectionAlphaStates]{\textbf{The disk-and-cylinder approximation}}, we study a toy model of gravity with wormholes, which we call the Coleman-Giddings-Strominger (CGS) model. In this model, the analogs of partition functions are computed by sums over surfaces with topologies limited to disks and cylinders.\footnote{A similar model has recently been discussed in this context in \cite{Godet:2021cdl}.} We view the simplest version of this model as an approximation to the MM model, and a more complicated variant of this model as an approximation to JT gravity. We leave the justification of this approximation to sections 4 and 5. The general goal of this section is to describe the simple picture of factorization which applies in the disk-and-cylinder approximation, and introduce an effective description  based on this simple structure.
 
 The main points of Section 3 are as follows:
 
 \begin{itemize}
 \item Using the framework of Marolf and Maxfield \cite{Marolf:2020xie}, building on \cite{Coleman:1988cy,Giddings:1988cx},  we view the path integral for this model as computing correlation functions of boundary operators $\hat{Z}_I$ in a state $|\psi\rangle$ of closed universes. Eigenstates $|\psi_\alpha\rangle$ of the boundary operators correspond to fixed members of an ensemble of theories. Then our interest is in studying correlation functions in a eigenstate, or ``$\alpha$ state'' \cite{Giddings:1988cx,Coleman:1988cy,Marolf:2020xie}.
 
 \item While in general an $\alpha$ state $|\psi_\alpha\rangle$ is a superposition of states with any number of closed universes, correlation functions of $n$ boundary operators depend solely on the components of a state $|\psi^2_\alpha\rangle$ (which is simply related to $|\psi_\alpha\rangle$) with up to $n$ universes. So rather than depending on the whole many-universe structure of the closed-universe wavefunction, an $n$-point function only depends on an $n$-universe state.
 
 \item In an $\alpha$ state, the $n$-universe components of $|\psi^2_\alpha\rangle$ needed to compute an $n$-point function are simply determined by the one-universe component. This follows from the fact that an $n$-point function in an $\alpha$ state factorizes into a product of one-point functions. In many ways this one-universe component provides a simpler description of a fixed member of the ensemble than the many-universe $\alpha$ state.
 
\item In this model, studying the factorization problem amounts to comparing a two-point function in an $\alpha$ state with the product of one-point functions. In our formalism, this involves comparing the two-universe component of $|\psi^2_\alpha\rangle$ with the product of one-universe components. Schematically,
\begin{align}
\hspace{-22pt}\langle \psi_\alpha | \hat{Z}_I \hat{Z}_J |\psi_\alpha\rangle &\supset \text{Cylinder} + \text{Two-universe component}
\cr
\hspace{-22pt}\langle \psi_\alpha | \hat{Z}_I|\psi_\alpha\rangle \langle \psi_\alpha| \hat{Z}_J |\psi_\alpha\rangle &\supset \big(\text{One-universe component}\big) \hspace{-2pt} \cdot \hspace{-2pt} \big(\text{One-universe component}\big)
\end{align}
In appropriate cases, we can use the analogy with periodic orbit theory and view the cylinder as a diagonal sum in some basis of one-universe states. Expressed in that basis, the one-universe component of $|\psi^2_\alpha\rangle$ is a sum over one-universe states with random coefficients, and the product of one-point functions $\langle \psi_\alpha | \hat{Z}_I|\psi_\alpha\rangle\langle \psi_\alpha| \hat{Z}_J |\psi_\alpha\rangle$ is a double sum. Then the two-universe component of $|\psi^2_\alpha\rangle$, which is added to the "diagonal" cylinder, behaves like an off-diagonal sum. This off-diagonal character has a geometric origin in the many-universe description, resulting from a geometric "exclusion effect."

\item We can capture the behavior of the CGS model in an $\alpha$ state with an effective model, in which the many-universe $\alpha$ state is traded for a random "$\Psi$" boundary condition. This $\Psi$ boundary condition is essentially an effective description of the one-universe component of $|\psi^2_\alpha\rangle$.  "Broken cylinders" with one $\hat{Z}_I$ boundary and one $\Psi$ boundary summarize the full many-universe structure of the $\alpha$ state. As we explain in the \hyperref[discussion]{\textbf{Discussion}}, these broken cylinders are closely related to the "half-wormholes" found in \cite{Saad:2021rcu}.

 Upon averaging, pairs of broken cylinders are glued together along their $\Psi$ boundaries to form the usual cylinders. While this effective description naturally captures some aspects of the full theory, it does not have an "exclusion effect" to enforce the off-diagonal character of pairs of $\Psi$ boundaries. Instead one must implement this by hand with an ad-hoc "exclusion rule."

\begin{figure}[H]
\centering
\includegraphics[scale=0.3]{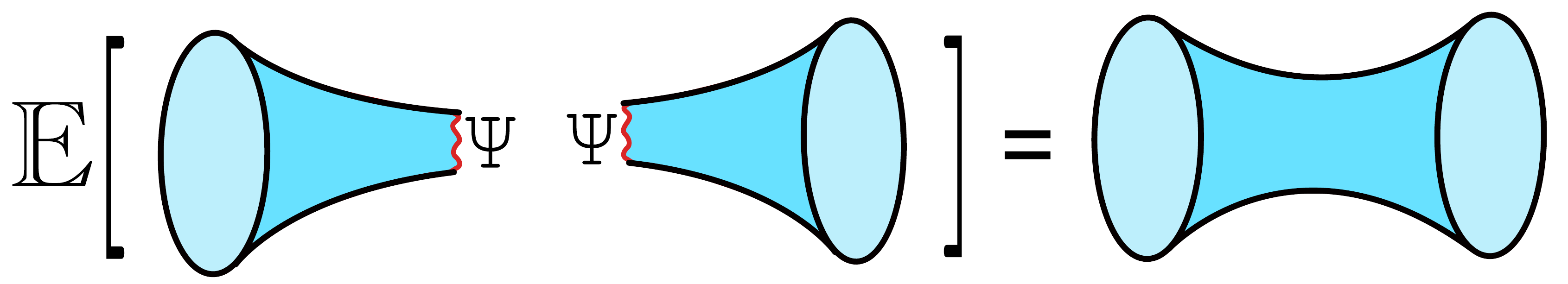}
\caption{\small Averaging glues pairs of broken cylinders together along their $\Psi$ boundaries to produce full cylinders.}
\end{figure}

 \end{itemize}

The rest of the paper is devoted to applying the simple picture found in the CGS model to the MM model and JT gravity.

 In Section 4, \hyperref[MMmodel]{\textbf{Approximate $\alpha$ states in the MM model}}, we study the MM model. We show that the disk-and-cylinder approximation of the MM model, which is equivalent to the CGS model from the previous section, is valid for appropriate correlation functions in \textit{approximate} $\alpha$ states. Establishing the validity of this approximation then allows us to apply the ideas of Section 3 to this model.

In Section 5 \hyperref[JTgravity]{\textbf{JT gravity}}, we study JT gravity. The main idea of the section is that the disk-and-cylinder approximation is valid for appropriate correlation functions in approximate $\alpha$ states, and thus the ideas from Section 3 may also be applied to JT gravity. However, we postpone the justification of this approximation to the end of the section, and instead begin by assuming the validity of the disk-and-cylinder approximation and exploring the results. In particular, we focus on the effective description of JT gravity with $\Psi$ boundaries, which is especially vivid in this context. In this model, the broken cylinders are naturally constructed by bisecting the cylinder along a circular geodesic, and giving this new boundary a random $\Psi$ boundary condition. This boundary condition can be described by a random function $\Psi(b)$, where $b$ is the length of the boundary. In this effective description, we see how both ensemble-averaging and time-averaging can glue the $\Psi(b)$ boundaries together to form wormholes.
 
We end Section 5 by studying approximate $\alpha$ states in JT gravity. Exact $\alpha$ states in JT gravity correspond to fixing a discrete set of energy levels in the boundary theory, which may be described in the bulk using "Eigenbranes" \cite{Blommaert:2019wfy,Blommaert:2020seb}. However, the physics on timescales less than the plateau time\footnote{For simple observables. Correlation functions with exponentially many operator insertions may be sensitive to the discreteness of the boundary spectrum.} is not sensitive to the discreteness of the boundary spectrum. In the averaged theory, physics on these timescales can be described in the bulk with wormholes, and in the boundary as Gaussian fluctuations of a \textit{coarse-grained}, continuous approximation to the boundary spectrum. Then to construct approximate $\alpha$ states, rather than following an approach which directly mirrors the approach in Section 4, which would involve \textit{approximately} fixing a discrete set of energy levels, we instead approximately fix a coarse-grained, continuous version of the density of states. In these approximate $\alpha$ states, we argue that the disk-and-cylinder approximation is valid at least until a timescale determined by the  error specified in the approximation. This error is quantified by the small variance in the approximate $\alpha$ state. Specifying an exponentially small variance, this timescale can be made exponentially long in the entropy, $\sim e^{ c S}$, with $c<1$.

We begin the \hyperref[discussion]{\textbf{Discussion}} by comparing our findings in this paper with recent results about the SYK model published in a joint paper with Douglas Stanford \cite{Saad:2021rcu}. In that work we analyzed the SYK model for a fixed choice of boundary couplings, which is analogous to studying gravity in an exact $\alpha$ state. As we briefly explain in this section, the results from that work are reminiscent of our findings in this paper. In the remainder of the Discussion, we raise several questions and give brief comments on each.

 \section{Periodic orbits}\label{periodicorbits}
 Here we briefly review an idea, originally  due to Berry \cite{berry1985semiclassical},  that explains the ramp in the spectral form factor for few-body quantum chaotic systems in the semiclassical approximation.\footnote{For a detailed review see \cite{Haake:1315494}.}  It suggests a picture of how wormholes and factorization can be reconciled \cite{Saad:2019lba}.  This picture gives a reference example that we will compare and contrast to other model systems in subsequent sections.
 
   Consider a chaotic classical mechanical system, like a billiard ball moving on a suitably shaped billiard table.   In a semiclassical limit we can write a Minkowski time partition function $ Z(it) = {\rm Tr} [e^{-iHt}]$ using a suitable formulation of the Feynman path integral called the Gutzwiller trace formula.\footnote{Strictly speaking this trace should be restricted to an energy window as in the quantity $Y(it)$ defined subsequently.} Schematically we have 
 \be
Z(iT) =  {\rm Tr} [e^{-iHT} ]=  \sum_{a} D_a e^{iS_a}
 \ee
 where $a$ runs over classical periodic orbits, $S_a$ is the classical action of the orbit,  and $D_a$ is the one loop fluctuation determinant. The spectral form factor is given by  the product of partition functions
 \be\label{doublesum}
 Z_L(iT)Z_R(-iT) =  \sum_{ab} D_a D_b e^{i(S_a - S_b)}
 \ee
 In this non-averaged system this quantity manifestly factorizes, as reflected in the double sum.
 
 But this double sum simplifies if we do various kinds of averaging.   For example we could average over a window of times $T$, or we could average over an ensemble of related chaotic Hamiltonians.    The time averaged case does not factorize because the observable is correlated, the ensemble case doesn't because the Hamiltonians are.

 At large $T$ the actions $S_a$ are large and the phases rapidly oscillate.  For large, but not too large, times the only terms in \eqref{doublesum} that survive averaging are the "diagonal" ones $a=b$, up to a relative time translation.  There are exponentially many such orbits, compensated for by exponentially small one-loop factors $D_a$.  The relative time translation mode gives a factor of $T$, explaining the linear behavior of the ramp in the spectral form factor.   This is Berry's  "diagonal approximation" \cite{berry1985semiclassical}.  It is clear that the passage from the double sum to the diagonal single sum destroys factorization, as expected in an averaged system.

 All of the above is a "boundary" quantum mechanical  analysis, but it is appealing to consider the diagonal sum as an analog of the bulk wormhole configuration, where the L and R factors are linked by the diagonal correlation pattern.   This analogy suggests a way to restore the factorization in a non-averaged system.    We keep the wormhole but need to add back in the bulk analog of the  off-diagonal terms that restore the manifestly factorizing double sum in \eqref{doublesum}.

 Note that in this analogy the wormhole is an {\it emergent }concept, it plays no role in the fundamental description of the system.  Note also that just two copies of the boundary Hilbert space are involved here, and hence just two sums.   To study correlators with more $Z$'s, like $\langle (Z_L(iT)Z_R(-iT))^k \rangle$, $2k$ sums would be required.

\section{The disk-and-cylinder approximation}\label{SectionAlphaStates}

\subsection{Introduction to section}

 In this section we begin our discussion of simple bulk models that display a factorization problem.  We focus here on a very interesting model introduced by Marolf and Maxfield (the MM model) \cite{Marolf:2020xie}  that illustrates many of the basic issues.  The model is a reformulation and extension of the old ideas of Coleman  \cite{Coleman:1988cy} and Giddings-Strominger \cite{Giddings:1988cx} to the AdS/CFT context. The only degree of freedom in this model is the topology of the bulk manifold and contains arbitrarily many one dimensional asymptotic AdS type boundaries connected by 2-manifolds of arbitrary genus.    We will focus on an  approximation that amounts to limiting the 2-manifold topologies to disks and cylinders (spacetime wormholes).   This simpler model is close to the one studied by Coleman and Giddings-Strominger and we refer to it as the CGS model.  The MM and CGS models are reviewed in \hyperref[SectionMMCGS]{\textbf{The MM and CGS model}}.

These models describe states in a "many-universe" Hilbert space which is reviewed in \hyperref[statesofcloseduniverses]{\textbf{States of closed universes}}.  There are certain states in this Hilbert space called $\alpha$ states in which partition functions $Z$ assume a definite value -- eigenstates of the ${\hat Z}$ operator.  Expectation values of products of $Z$'s manifestly factorize in such states.   The analysis of such exact $\alpha$ states in the full MM model is subtle, so to simplify our analysis we will study {\it approximate} $\alpha$ states in which expectation values of $Z$ products approximately factorize, so there still is a sharp factorization puzzle.  This amounts to focusing on the simple disk and cylinder topologies of the CGS model.  Factorization in the CGS model and its "multi-species" extension is discussed in \hyperref[factorizationcgs]{\textbf{Factorization in the CGS model}} and \hyperref[cgsspeciesalphastates]{\textbf{Factorization in the CGS model with species}}.   We postpone the analysis of the errors in approximating the full MM model in approximate $\alpha$ states by the CGS model until \hyperref[MMmodel]{\textbf{The MM model}} and show that some of the subtle properties of the MM model, null states and discrete $\alpha$ spectrum, do not play a role in resolving the approximate factorization puzzle. The analysis of errors in approximating JT gravity by the CGS model with species is postponed to \hyperref[JTgravity]{\textbf{JT gravity}}.

 We should emphasize that in these models the spacetime wormhole connecting different boundaries is a {\it fundamental} part of the description of the system.  This is in contrast to the periodic orbit picture discussed in Section \ref{periodicorbits} where the wormhole is {\it emergent}.  Nonetheless in an (approximate) $\alpha$ state factorization is (approximately) restored.  Roughly speaking in an (approximate) $\alpha$ state additional contributions  corresponding to the "off-diagonal" terms in the periodic orbit story are present, allowing the entire result to be written in the factorized double sum form.   We make this analogy more precise by describing these contributions in terms of a random single-universe wavefunction.

We end with \hyperref[effectivemodel]{\textbf{An effective description with random boundaries}}, in which we introduce an effective description of the CGS model in an $\alpha$ state. In this effective description, the contributions from the many-universe $\alpha$ state are replaced with contributions from a few "effective" boundaries,  with a random boundary condition. An unusual feature of this effective description is that we must modify the conventional sum over geometries to include an "exclusion rule", which instructs us to remove either the cylinders connecting partition function boundaries, or the corresponding "diagonal" contribution of pairs of cylinders with effective boundaries. This exclusion rule has a geometric origin in the full theory, but must be put in by hand in the effective model.

\subsection{The MM and CGS model}\label{SectionMMCGS}

To set the context we first review the elementary parts of the MM model introduced in \cite{Marolf:2020xie}. This model is defined by a Euclidean path integral which computes quantities $Z_n$. $Z_n$ is computed by summing over (oriented) two-dimensional spacetimes $M$ with $n$ circular boundaries. The surface $M$ may have any number of disconnected components. 

\begin{figure}[H]
\centering
\includegraphics[scale=0.3]{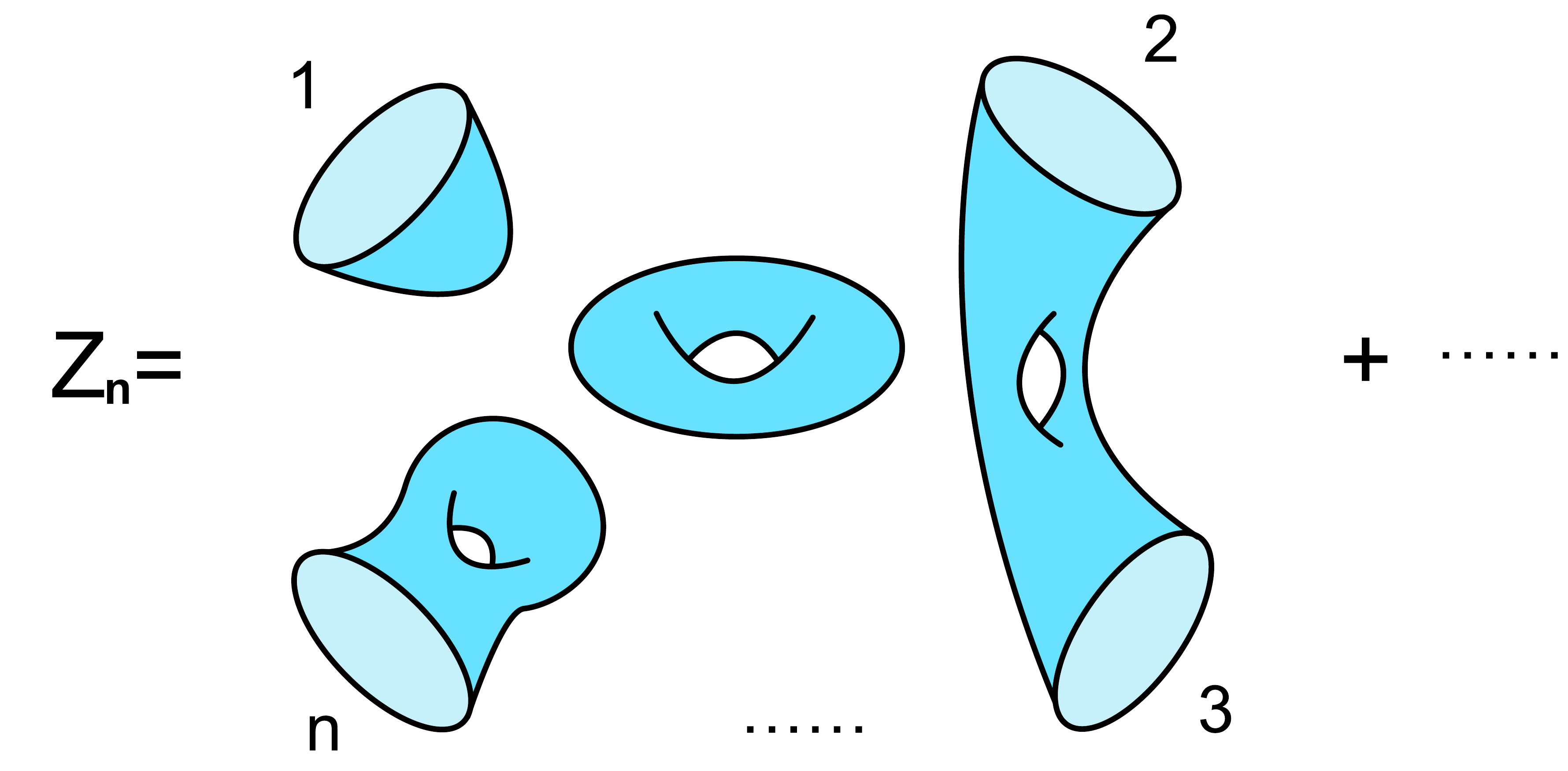}
\caption{\small Here we show an example contribution to $Z_n$ (for $n\geq 4)$. Each spacetime contributing to $Z_n$ has $n$ circular boundaries, and may have any number of connected components. The lower set of dots represent the $n-4$ boundaries not explicitly drawn.}
\end{figure}

The path integral for $Z_n$ is analogous to the AdS gravity path integral which computes a product of $n$ partition functions.
The $n$ circular boundaries are analogous to the asymptotically AdS boundaries.  However, in this model the action depends solely on the topology of the spacetime, so the ``partition functions'' $Z_n$ computed in this model do not depend on a temperature.

The Euclidean action in this model is simply proportional to the Euler character of the spacetime $M$, $I(M) = - S_0\chi(M)$. We think of $S_0$ as a large number analogous to the black hole entropy.\footnote{Note that our conventions differ from those used in \cite{Marolf:2020xie} by a shift of one unit of $\chi$.   The choice used in \cite{Marolf:2020xie} gives integer values for ``partition functions'' appropriate to a boundary Hilbert space trace.   We choose to use the geometrically more natural normalization.  None of the lessons we draw are affected by this.} Then the path integral reduces to a sum over topologies, weighed by the action $\exp[-I(M)]$ and a measure factor $\mu(M)$,
\be
Z_n = \sum_{\text{Surfaces }M} \mu(M) e^{- I(M)}.
\ee
The measure factor is defined to treat disconnected spacetimes without boundary as indistinguishable. If $M$ has $m_g$ connected components of genus $g$ and no boundary,
\be
\mu(M) = \prod_{g} \frac{1}{m_g !}.
\ee
The measure factor $\mu(M)$ will not play a large role in our discussion. $\mu(M)$ is not equal to one only for spacetimes without boundaries; the contributions of these spacetimes factor out of all quantities we compute and are normalized away.

For practice, we will briefly describe the computations of $Z_0$, $Z_1$, and $Z_2$. 

To compute $Z_0$ we sum over all oriented surfaces with no boundaries. We can view this sum as a sum over the numbers $m_g$ of connected components with genus $g$. The Euler character of a surface $M$ with $m_g$ connected components of genus $g$ is simply the sum of the the Euler characters of the connected components
\be
\chi(M) = \sum_{g=0}^\infty m_g (2-2g). 
\ee
Then,
\be
Z_0 = \prod_{g=0}^\infty \bigg[\sum_{m_g=0}^\infty \frac{1}{m_g !} e^{ m_g S_0 (2-2 g)}\bigg] = \prod_{g=0}^\infty \exp\bigg[ e^{S_0(2-2g)} \bigg] = \exp\bigg[ \frac{e^{2 S_0}}{1- e^{-2 S_0}}\bigg].
\ee
\begin{figure}[H]
\centering
\includegraphics[scale=0.3]{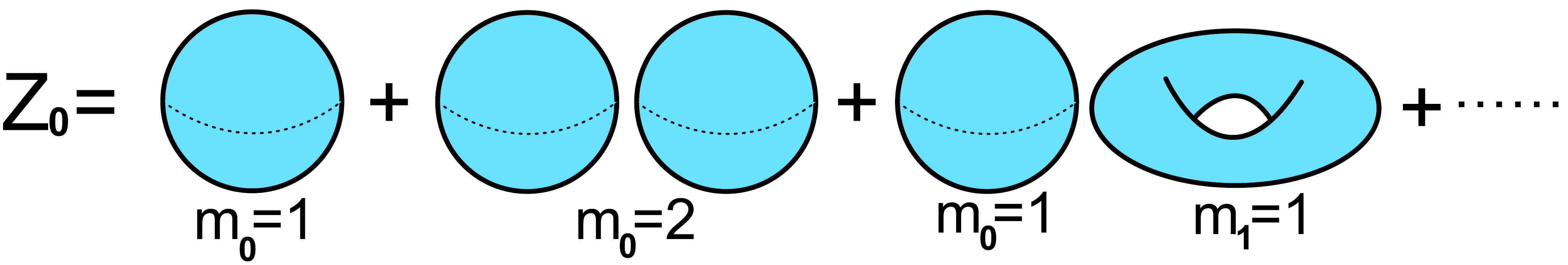}
\caption{\small Here we illustrate some examples of contributions to $Z_0$, labeled by the number of connected components for each genus, $m_g$.}
\end{figure}
 
 To compute $Z_1$ we sum over surfaces with one circular boundary. We can view this sum as a sum over the numbers $m_g$ of connected components with no boundary and genus $g$, and over the genus $g_1$ of the connected component with a boundary. 
 
Because the Euler character is a sum of the Euler characters of the connected components, the path integral factorizes into the product of the sum over surfaces with no boundary $Z_0$, and the sum over connected surfaces with one boundary.
\be
Z_1 = Z_0 \times \sum_{g_1=0}^\infty e^{ S_0 (2-2g_1-1)}. 
\ee
 The second factor in the above formula is the sum over connected surfaces with one boundary, which are classified by their genus $g_1$. The Euler character of a connected surface with $k$ boundaries and genus $g_1$ is $\chi = 2-2g_1-n$; in the present case we have $n=1$.

Focusing on this factor, and defining a normalized path integral $\tilde{Z}_1$,
\be\label{eq:OneZPathIntegral}
\tilde{Z}_1\equiv \frac{Z_1}{Z_0} = e^{S_0} + e^{-S_0} + \dots = \frac{e^{S_0}}{1 - e^{-2 S_0}}.
\ee
\begin{figure}[H]
\centering
\includegraphics[scale=0.3]{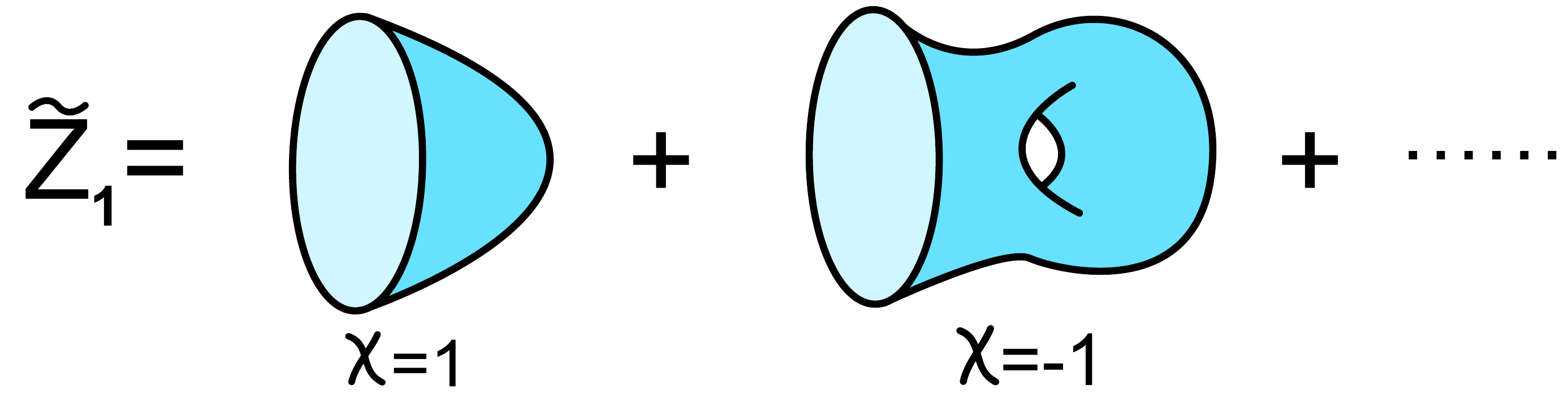}
\caption{\small Here we have pictured the leading contributions to $\tilde{Z}_1$, which is given by a sum over spacetimes with one connected component and a single circular boundary. We have labeled the contributions by their Euler character.}
\end{figure}

Finally, we compute $Z_2$, the sum over surfaces with two circular boundaries. Again, as in the computation of $Z_1$, the contributions of connected components with no boundaries factorizes out, leaving a sum over connected surfaces with boundaries. This will happen for any $Z_n$.

In the case of two boundaries, the components with boundaries may have one or two boundaries. The sum over these surfaces can be split into the sum over surfaces with two connected components with genus $g_1$ and $g_2$, each with one boundary, and the sum over surfaces with one connected component with genus $g_{12}$ and two boundaries.
\be\label{eq:2ZPathIntegral}
\tilde{Z}_2\equiv \frac{Z_2}{Z_0} = \bigg(\sum_{g_1=0}^\infty e^{S_0(2-2g_1 - 1)} \bigg)\bigg(\sum_{g_2=0}^\infty e^{S_0(2-2g_2 - 1)} \bigg)+ \sum_{g_{12}=0}^\infty e^{S_0 (2-2g_{12}-2)}.
\ee
\begin{figure}[H]
\centering
\includegraphics[scale=0.3]{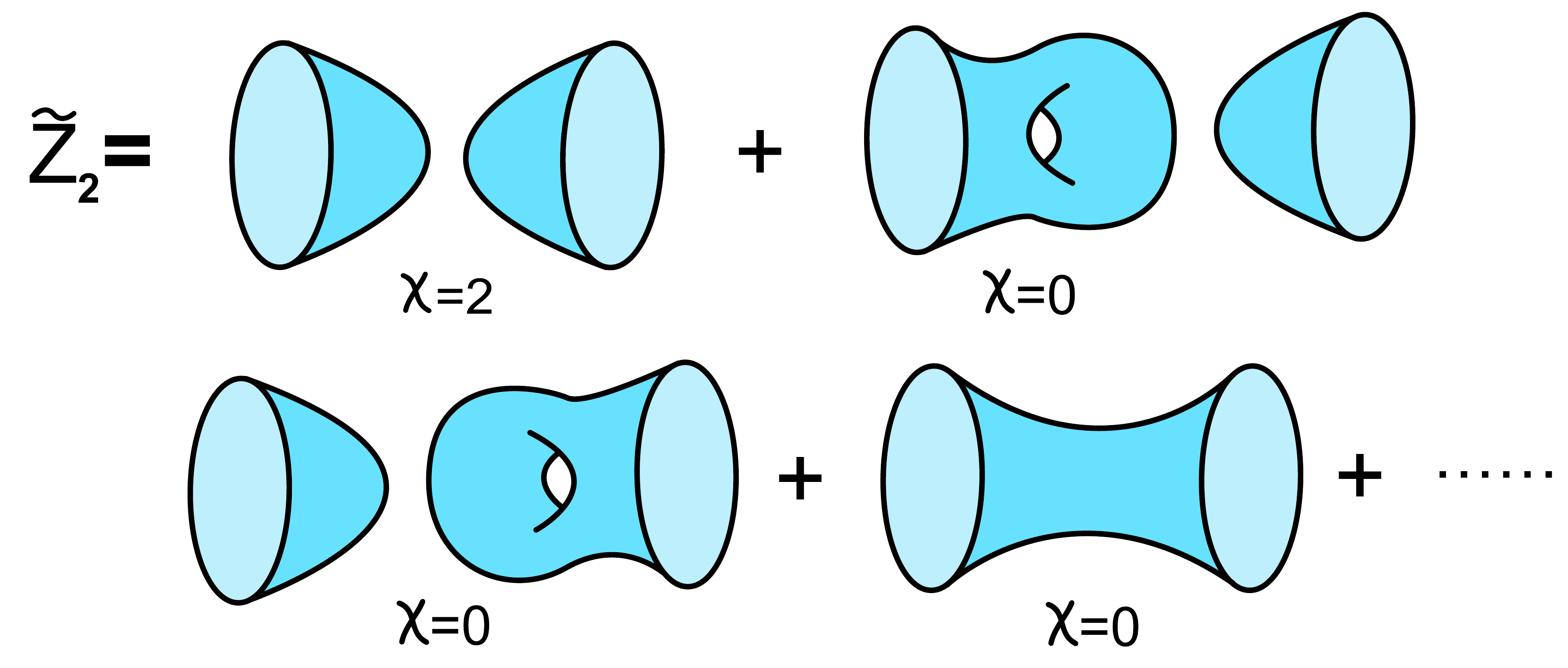}
\caption{\small Here we have pictured the leading contributions to $\tilde{Z}_2$, which is given by a sum over spacetimes with two circular boundaries, including only spacetimes where each connected component has at least one of these two boundaries. The cylinder contribution, with $\chi=0$, is an example of a spacetime wormhole, connecting two separate boundaries.}
\end{figure}

At large $S_0$, complicated topologies provide small contributions to the $Z_n$ for $n\ll e^{S_0}$. For example, we look at the calculation of $\tilde{Z}_2 - \tilde{Z}_1^2$. This quantity diagnoses the failure of factorization of the ``partition functions'' in this model, which will be of central interest to us.

$\tilde{Z}_2 - \tilde{Z}_1^2$ is given by a sum over surfaces with one connected component with two circular boundaries. The cylinder gives a contribution of one, while the cylinder with a handle added gives a contribution of $e^{-2S_0}$.
\be
\tilde{Z}_2 - \tilde{Z}_1^2= 1 + e^{-2S_0} +\dots
\ee
\begin{figure}[H]
\centering
\includegraphics[scale=0.3]{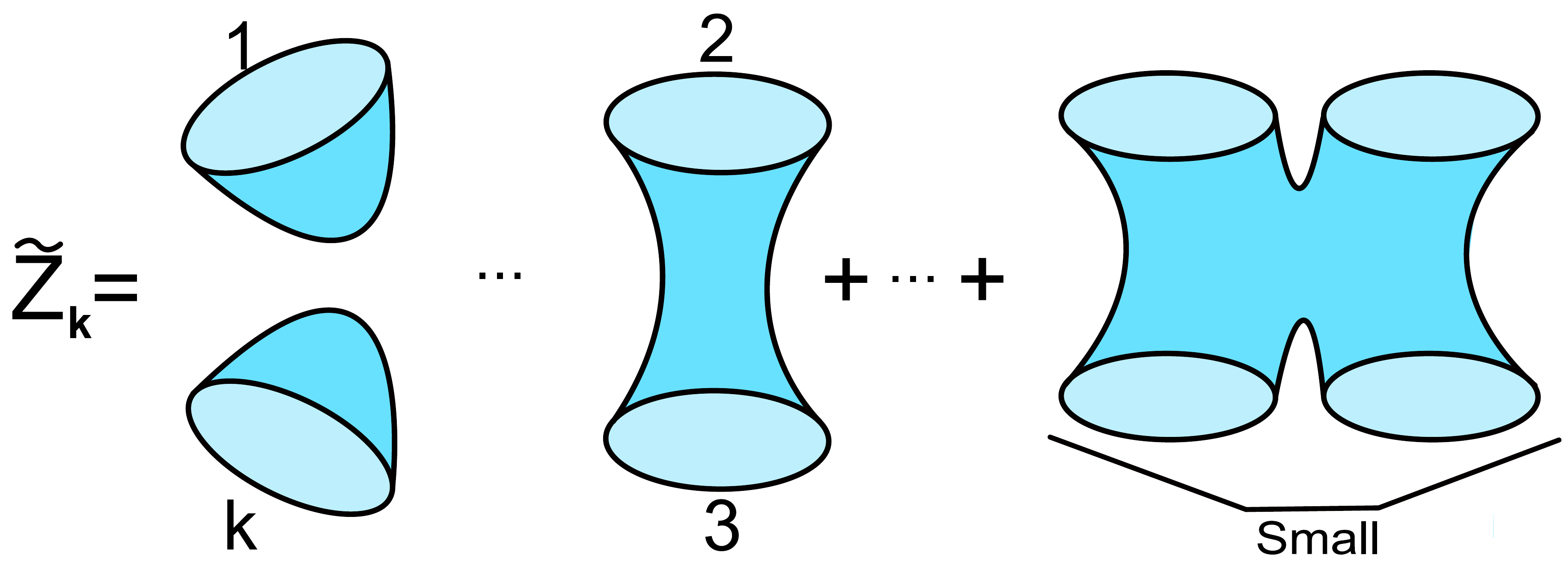}
\caption{\small The leading fully connected contribution to $\tilde{Z}_k$, which includes only contributions from spacetimes with a single connected component, is given by the $k$-holed sphere. This contribution has Euler character $-k+2$, so the connected correlators for $k>2$ are exponentially small, $\sim e^{-(k-2)S_0}$.}
\end{figure}

More general ``connected correlators'' such as $\tilde{Z}_{4,c}\equiv \tilde{Z}_4 - 2 \tilde{Z}_2^2- 2 \tilde{Z}_2 \tilde{Z}_1^2 - \tilde{Z_1}^4$, will be small when $S_0$ is large. For $k\ll e^{S_0}$, $\tilde{Z}_{k,c}$, the `connected k-boundary correlator', will be given to leading order by the contribution of the spacetime with the topology of a sphere with $k$ holes. This has an Euler character of $\chi = 2-k$, so
\be
\tilde{Z}_{k,c}\approx e^{-(k-2) S_0},\hspace{20pt}k\ll e^{S_0}.
\ee

Neglecting these exponentially small terms, the $\tilde{Z}_{k}$ behave approximately like moments of a Gaussian distributed random variable. This interpretation will be made precise in the later sections. 

Motivated by this Gaussian behavior, we define a modification of the MM model, which we dub the Coleman-Giddings-Strominger (CGS) model. This model is obtained from the MM model by restricting the sum over topologies to only include spacetimes whose connected components have the topology of the disk or the cylinder.

In this CGS model the $\tilde{Z}_n$ behave exactly like the moments of a Gaussian random variable with mean $\tilde{Z}_1$ and variance $\cyl$,
\begin{align}
\tilde{Z}_1 &= \disk = e^{S_0}
\cr
\tilde{Z}_2 & = \disk^2 + \cyl = e^{2S_0}+1
\cr
\tilde{Z}_{2,c} & = \cyl = 1
\cr
\tilde{Z}_{k,c} &= 0, \hspace{20pt} k>2.
\end{align}
Here the subscript ``c'' denotes the connected contribution to $\tilde{Z}_k$.

The CGS model lacks some of the interesting features of the MM model, which are linked to the contributions from higher-genus surfaces. However,  this model has its own factorization problem, since $\tilde{Z}_{2,c}=\cyl \neq 0$. We will focus on this issue. 

 In Section \ref{MMmodel} we will describe in more detail how and when the MM model can be reduced to the CGS model in computations related to factorization. However, for now we will focus solely on the CGS model.

\subsection{States of Closed Universes}\label{statesofcloseduniverses}
Marolf and Maxfield \cite{Marolf:2020xie} also introduced a Hilbert space structure associated to spacetimes with boundaries, interpreting quantities like the $Z_k$ as amplitudes in a quantum theory of closed universes.  
The Gaussian behavior of the $\tilde{Z}_n$ at large $S_0$ discussed in the previous section can be interpreted in this way as expectation values of a quantum harmonic oscillator position operator $\hat{Z}$ in a certain state of closed universes.  In this section we briefly review the discussion in \cite{Marolf:2020xie} of the quantum mechanics of closed universes. The main points that we will explain are:

\begin{itemize}
\item The Hilbert space of closed universes is spanned by the states $\hat{Z}^k |NB\rangle$. The No-Boundary state, $|NB\rangle$,  is a state of no closed universes. $\hat{Z}$ is a Hermitian operator which creates and destroys closed universes with the spatial topology of a circle.
\item The inner product between states $\hat{Z}^k |NB\rangle$, $\hat{Z}^{k'}|NB\rangle$ is defined to be equal to the path integral $\tilde{Z}_{k+k'}$.\footnote{Normalizing the No-Boundary state correponds to dividing out by the sum over spacetimes without boundaries, $Z_0$.}
\item The operators $\hat{Z}$ are observables. By expressing a general state $|\psi\rangle$ as a superposition of the eigenvectors $|Z_\alpha \rangle$ of $\hat{Z}$, $|\psi\rangle = \sum_\alpha \psi_\alpha |Z_\alpha \rangle$, we can relate correlation functions of the operator $\hat{Z}$ to moments of a variable $Z$ in a classical probability distribution. The wavefunction $\psi_\alpha$ determines the classical probability distribution $|\psi_\alpha|^2$.

\item The path integrals $\tilde{Z}_n$ compute expectation values of $\hat{Z}$ in the No-Boundary state, which has a Gaussian wavefunction. The variance of the corresponding probability distribution is given by the cylinder contribution to the path integral.

\item Computing correlation functions of ``observables,''   that is $\hat{Z}$s,  in more general states involves path integrals with two classes of circular boundaries - boundaries associated with the observables, and boundaries associated with the state of the closed universes.

\end{itemize}

\subsubsection{The Hilbert space of closed universes: Z basis}\label{zbasis}
To describe the Hilbert space of closed universes, we start by introducing the ``No-Boundary'' state $|NB\rangle$.\footnote{The No-Boundary state is sometimes referred to as the Hartle-Hawking state.} We can think of $|NB\rangle$ as the state of no closed universes. There are complications in defining an appropriate notion of closed universe number, since closed universes can  nucleate, annihilate, join and split. However, we  will use a certain notion of closed universe number appropriate in the CGS model and the ``perturbative'' limit of the MM model, where we ignore joining and splitting effects. The No-Boundary state is the zero closed universe state with this definition of closed universe number.

We will discuss the rest of the Hilbert space in three ways. First we describe an overcomplete basis of states which is closely related to the path integral description of the model; the ``Z basis''. Then we will discuss an orthonormal basis which is appropriate for describing the CGS model and some limits of the MM model, the ``Number" or ``N" basis. Finally we describe the eigenbasis of the observable $\hat{Z}$, the ``$\alpha$ basis''. This basis naturally connects the physics of closed universes to ensemble averaging. 

To obtain the Z basis, we act on the No-Boundary state $|NB\rangle$ with the Hermitian operator $\hat{Z}$ raised to a non-negative integer power $k$. We may define the action of this operator, and the states $\hat{Z}^k|NB\rangle \equiv |Z^k\rangle$, by giving a formula for their inner product, $\langle NB| \hat{Z}^{k'} \hat{Z}^k |NB\rangle$. The path integral gives a natural definition of these inner products.\be
\langle NB|\hat{Z}^{k'} \hat{Z}^k |NB\rangle = \tilde{Z}_{k+k'} ~.
\ee

\begin{figure}[H]
\centering
\includegraphics[scale=0.25]{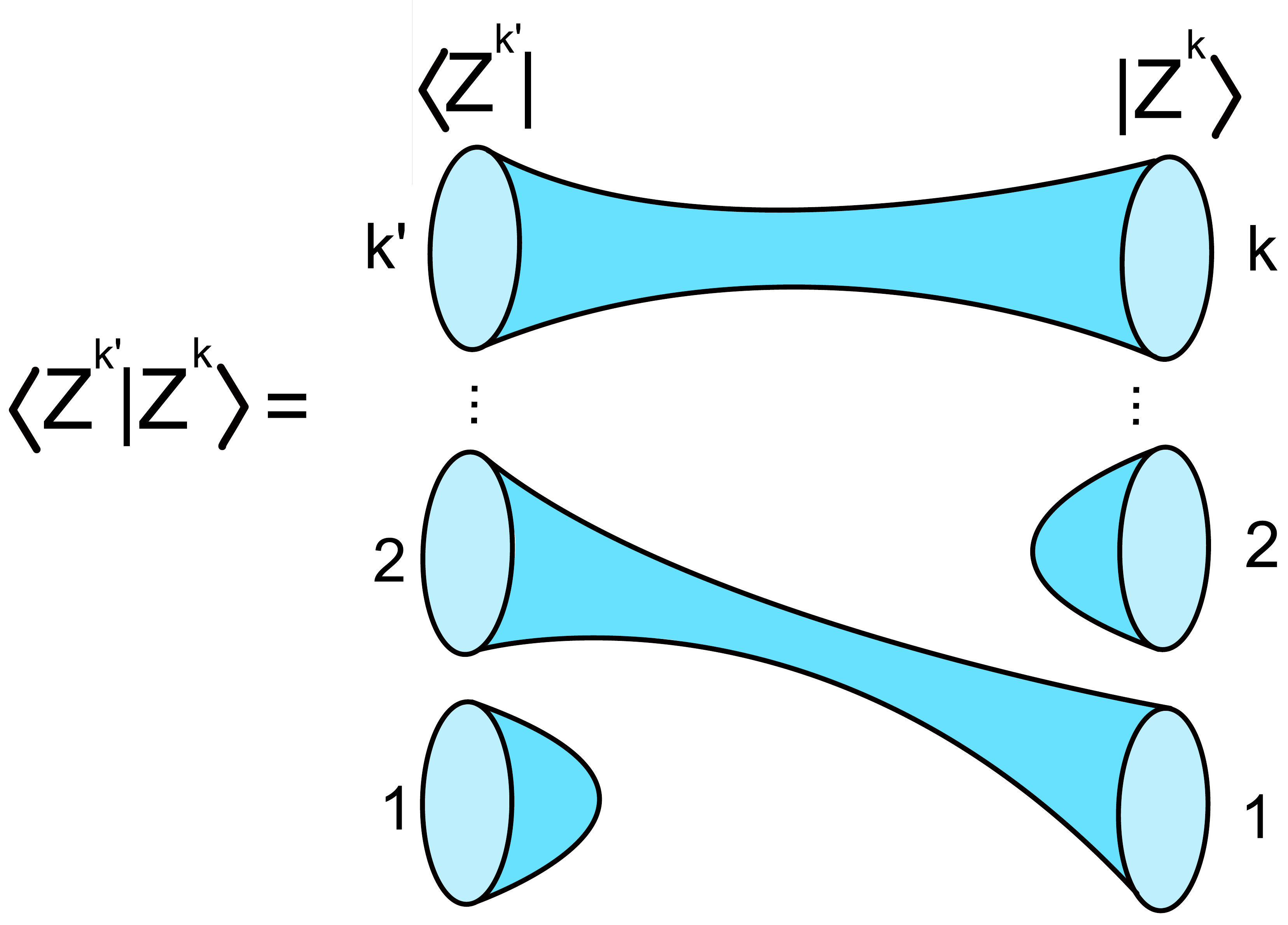}
\caption{\small Here we depict an example contribution to the overlap between states $|Z^k\rangle$ and $|Z^{k'}\rangle$.}
\end{figure}

With this definition, we can compute general overlaps of normalized states $|\psi\rangle$, $|\psi'\rangle$ and matrix elements of products of the operator $\hat{Z}$ in these states by expressing the states as superpositions of the $\hat{Z}^k |NB\rangle$ to write these quantities as sums over path integrals $\tilde{Z}_k$,\footnote{In the MM model, nonperturbative effects in the topological expansion for $\tilde{Z}_k$ render this inner product positive semidefinite; the proper Hilbert space is then obtained by modding out by null states \cite{Marolf:2020xie}. As a consequence, the $Z$ basis expression for a state $|\psi\rangle$ is non-unique. Subtleties related to null states will not be important in this section, but we will discuss them briefly in Sections \ref{MMmodel} and \ref{JTgravity}.}
\begin{align}
\langle \psi' | \hat{Z}^n |\psi\rangle &= \bigg(\sum_{k'} (\psi'_{k'})^* \langle NB|\hat{Z}^{k'}\bigg)  \hat{Z}^n\bigg(\sum_{k} \psi_{k} \hat{Z}^k |NB\rangle\bigg)
\cr
&= \sum_{k,k'} (\psi'_{k'})^* \psi_k  \;\; \langle NB| \hat{Z}^{k'+k+n} |NB\rangle 
\cr
&=  \sum_{k,k'} (\psi'_{k'})^* \psi_k  \; \tilde{Z}_{k+k'+n}.
\end{align}

\begin{figure}[H]
\centering
\includegraphics[scale=0.25]{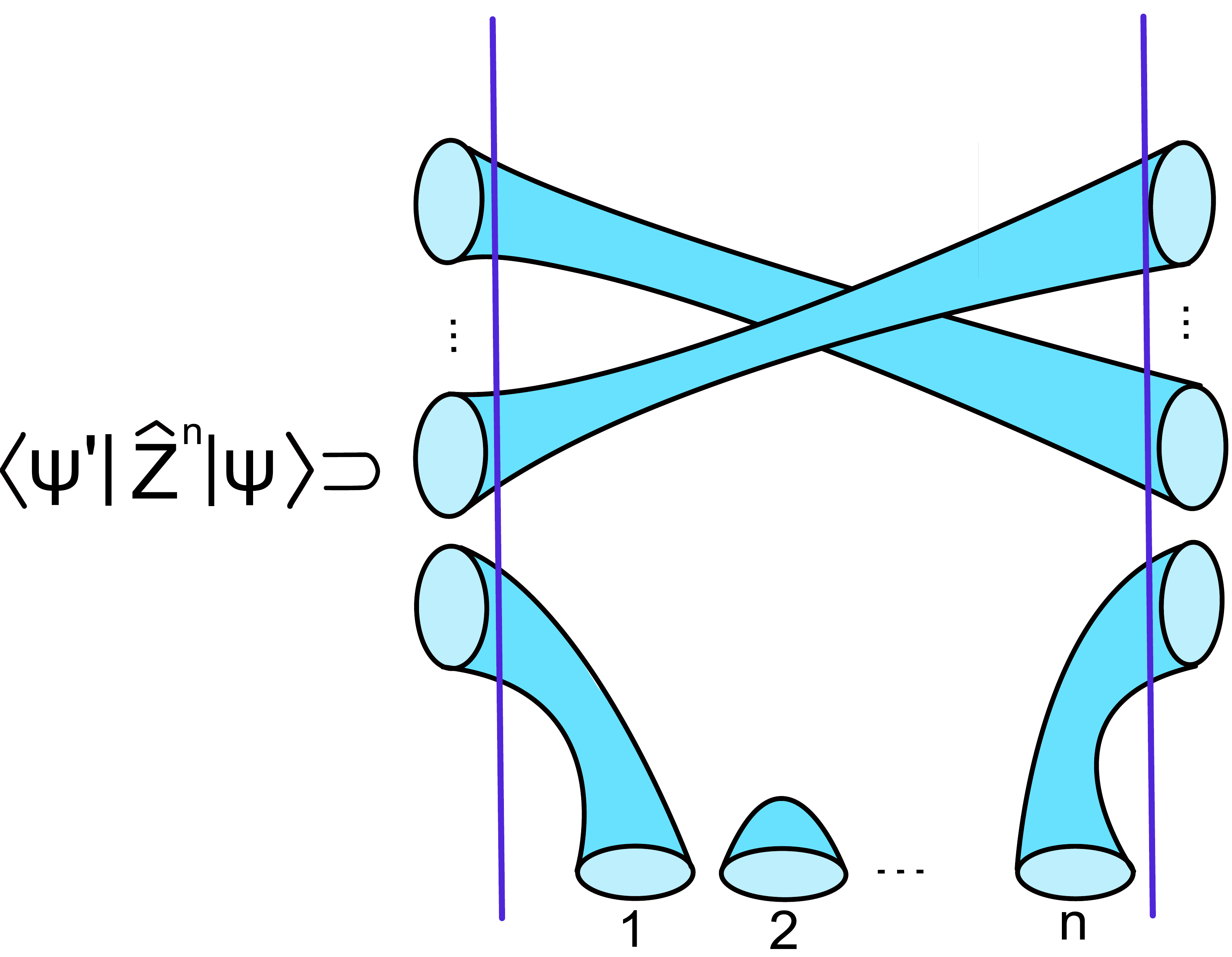}
\caption{\small In representing the path integral computing these inner products between normalized states, we draw diagrams like the above example. Normalized states $|\psi\rangle$ and $\langle \psi'|$ are denoted with  boundaries behind vertical purple lines, while operator insertions in correlation functions are denoted by boundaries inserted between the bra and ket.}
\end{figure}
 It is clear that in this inner product the operator $\hat{Z}$ that is Hermitian. We will often denote expectation values $\langle \psi | \hat{O} | \psi \rangle$ by $\langle \hat{O} \rangle_{\psi}$~.

\subsubsection{The Hilbert space of closed universes: N basis}\label{Nbasis}
To describe the next basis of interest, the Number or N basis, we begin by discussing the state of a single closed universe.

In the CGS or MM models, the states of closed universes are rather simple. First we consider a single closed universe.\footnote{When we refer to a single  closed universe we are referring to a closed {\it one dimensional} manifold, topologically a circle.} There are no degrees of freedom characterizing a single closed universe, so there is only one state associated to a single closed universe. We can try to create a state with a single closed universe by acting with the operator $\hat{Z}$ on the No-Boundary state. However, we can see that $\hat{Z}|NB\rangle\equiv |Z\rangle$ might not be a good definition of a single-universe state, as the overlap with the No-Boundary state is nonzero. In the CGS model, the overlap between the normalized states is\footnote{Here, and in the remainder of this paper, $|NB\rangle$ represents the normalized No-Boundary state.}
\be
\frac{\langle NB | Z\rangle}{\sqrt{\langle Z|Z\rangle}} = \frac{\disk}{\sqrt{\disk^2+\cyl}} = 1-\mathcal{O}(e^{-2S_0}).
\ee
The closed universe created by $\hat{Z}$ has a large amplitude to disappear via the disk diagram. 

\begin{figure}[H]
\centering
\includegraphics[scale=0.4]{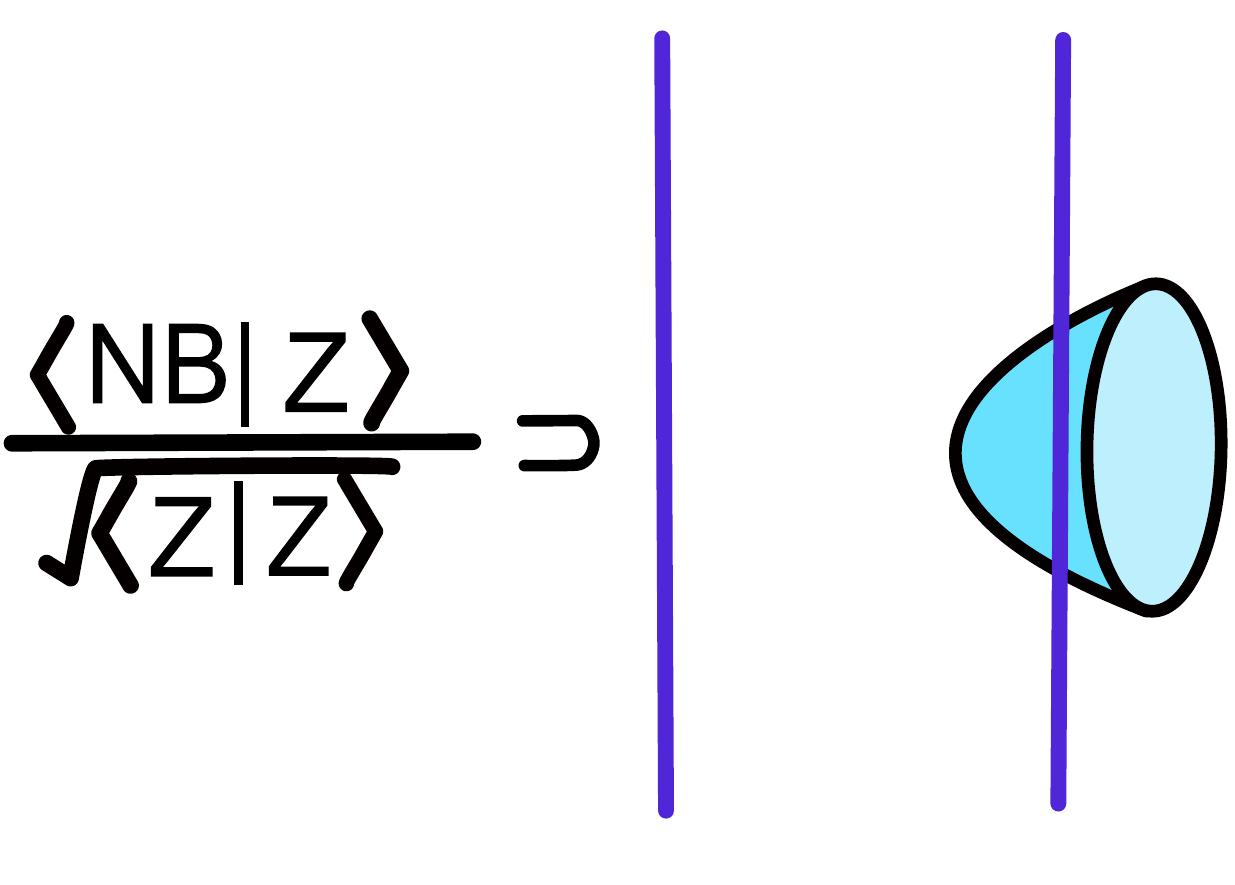}
\caption{\small Here we have pictured the disk contribution to the overlap between $|Z\rangle $ and the No-Boundary state. The purple lines denote that the bra and ket in the figure are normalized. }
\end{figure}

The disk acts like a tadpole in field theory. In that context we subtract the tadpole to define a field which creates a one-particle state.

Here, the one-closed universe state is naturally defined as
\be
|1\rangle \propto (\hat{Z}-\disk) |NB\rangle.
\ee

\begin{figure}[H]
\centering
\includegraphics[scale=0.3]{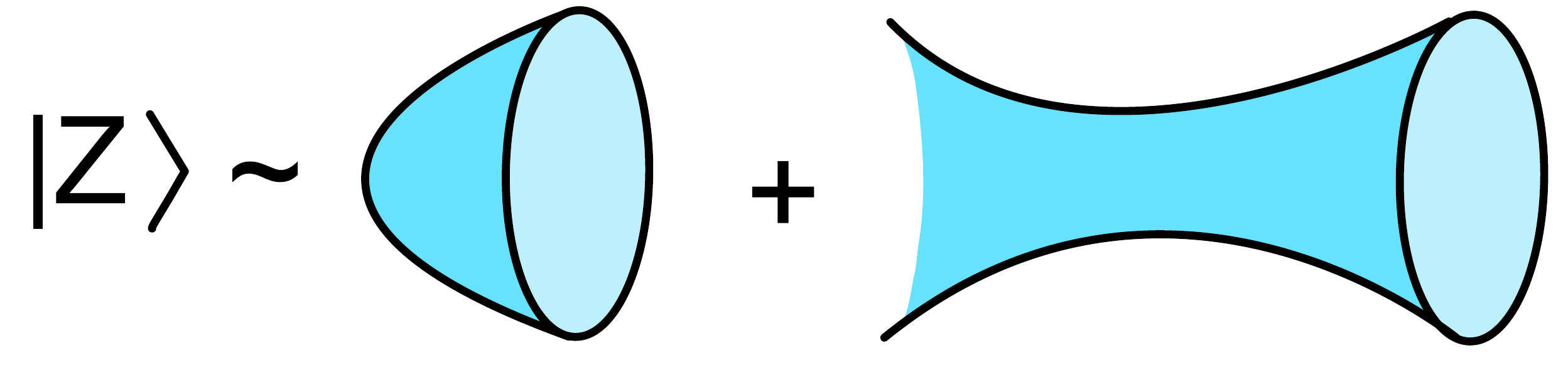}
\caption{\small We can represent the state $|Z\rangle$ as disk plus a single universe state, as shown above. The disk describes the amplitude for the closed universe created by $\hat{Z}$ to annihilate, and thus gives the overlap between $|Z\rangle$ and the No-Boundary state. The cylinder describes the amplitude for the closed universe created by $\hat{Z}$ to end up in the state $|1\rangle$.}
\end{figure}

The operator $\hat{Z}-\disk$ can create and destroy single closed universes, so it is natural to express it as a sum of creation and annihilation operators,
\begin{align}\label{eq:Zcreationannihilation}
\hat{Z}-\disk &= a+ a^\dagger
\cr
[a,a^\dagger]&=\cyl=1
\cr
a|NB\rangle&=0
\cr
a^\dagger |NB\rangle &= |1\rangle.
\end{align}
In general, we define the (unnormalized!) $n$-(closed) universe state as $|n\rangle= (a^\dagger)^n|NB\rangle$. These span the Hilbert space of a harmonic oscillator in the orthonormal  N basis, and we see that $\hat{Z}-\disk$ is just the conventional position operator.

It is instructive to use creation and annihilation operators to describe the state $\hat{Z}^2|NB\rangle$. By normal ordering the operator $\hat{Z}^2$, we find
\begin{align}\label{eq:2Znormalorder}
\hat{Z}^2|NB\rangle &= \big(\disk^2 + 2\; \disk (a+ a^\dagger) + (a+ a^\dagger)^2\big)|NB\rangle
\cr
&= \big(\disk^2 + 2\; \disk (a+ a^\dagger) +a^2 + (a^\dagger)^2+ 2 a^\dagger a + \cyl \big)|NB\rangle
\cr
& = (\disk^2 +\cyl) |NB\rangle + 2 \; \disk |1\rangle + |2\rangle.
\end{align}

\begin{figure}[H]
\centering
\includegraphics[scale=0.3]{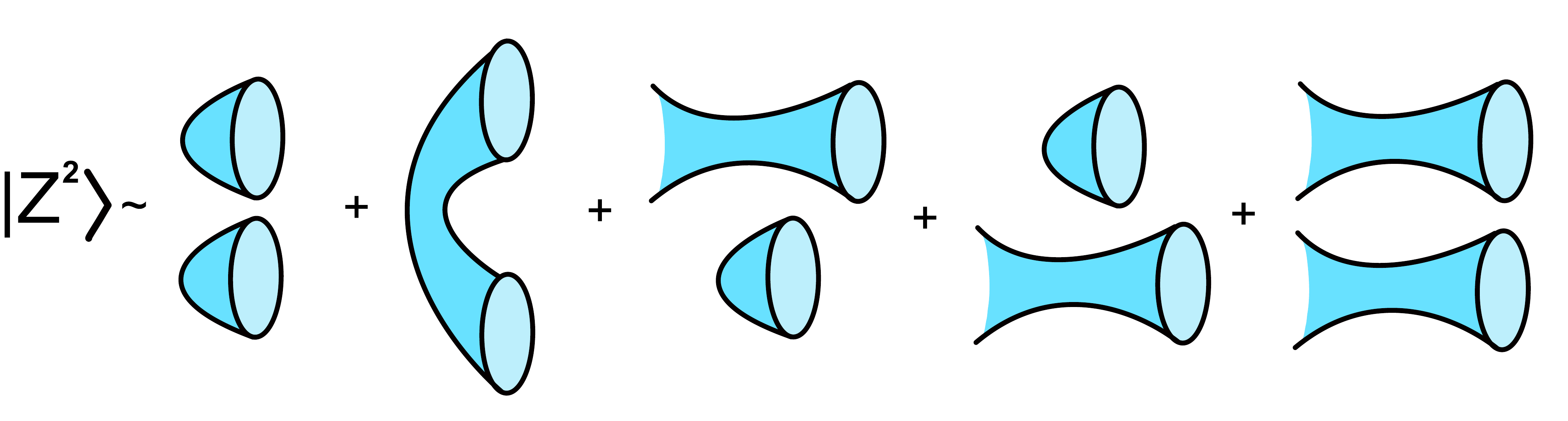}
\caption{\small Here we have depicted the state $|Z^2\rangle$ in the N basis. The first two terms, with two disks and a cylinder, describe the component with zero closed universes; the disks describe the amplitude for the closed universes created by $\hat{Z}$ to separately annihilate, and the cylinder describes the amplitude for them to annihilate with each other. The remaining terms describe the components with one or two universes, in which either one or zero of the closed universes are annihilated via the disk spacetime.}
\end{figure}

Here we can see pictorially why the normal ordering is related to the cylinder topology.

The N basis is particularly useful because states of different closed universe number are orthogonal\footnote{The normalization of these states is chosen to simplify later formulas.}
\be
\langle n |m\rangle = \cyl^n\; n! \;\delta_{nm} = n! \;\delta_{nm} .
\ee
This is in contrast with the states $\hat{Z}^n|NB\rangle$ or even $(\hat{Z}-\disk)^n|NB\rangle$ which are highly non-orthogonal due to tadpoles (the disk) and normal ordering (the cylinder). 

In the CGS model, the number eigenstates form an orthogonal basis of the Hilbert space, and the Hilbert space is identical to that of a harmonic oscillator. However the MM model  includes surfaces of arbitrary topology which give nonzero overlaps between states of different universe number. For example, the three holed sphere gives an overlap between the one-universe and two-universe states. Furthermore, nonperturbative effects in the topological expansion drastically change the structure of the Hilbert space  -- there are now a large number of null states. Marolf and Maxfield demonstrated a dramatic consequence of such nonperturbative effects.  They change the spectrum of the  $\hat{Z}$ operator from a continuous one -- the values of the harmonic oscillator position -- to a discrete one in the full model.

In Section \ref{MMmodel} we will return to the full MM model and address  these issues.   We will find that while these effects play a role in the computations of correlators in exact $\hat{Z}$ eigenstates, there are many \textit{approximate} eigenstates for which they are unimportant. In these states, the CGS model serves as a good approximation to the full MM model. The situation will be similar in JT gravity, discussed in Section \ref{JTgravity}. There, a suitable generalization of the CGS model serves as a good approximation in computations in certain approximate eigenstates, for which spacetimes with topologies other than the disk and cylinder can be ignored.

We now turn to a formal discussion of such eigenstates.

\subsubsection{The Hilbert space of closed universes: $\alpha$ basis}\label{alphabasis}

We now introduce the $\alpha$ basis, the eigenbasis of the observable $\hat{Z}$. We label the eigenstates of $\hat{Z}$ as $|\alpha\rangle$, with real eigenvalues $Z_\alpha$,
\be
\hat{Z}|\alpha\rangle = Z_\alpha |\alpha\rangle.
\ee
$\hat{Z}$ may or may not have a continuous spectrum, with delta function normalized eigenstates: In the CGS model $\hat{Z}$ has a continuous spectrum; while in the MM model $\hat{Z}$ has a discrete spectrum.

It is interesting to consider the expression for expectation values of powers of $\hat{Z}$ in a general state $|\psi\rangle$ when expressed in the orthonormal $|\alpha\rangle$ basis.

\begin{align}\label{eq:alphaprob}
\langle \hat{Z}^n \rangle_\psi \equiv \langle \psi| \hat{Z}^n |\psi\rangle &=  \sum_{\alpha,\beta}\psi^*_\beta \psi_\alpha \langle \beta| \hat{Z}^n |\alpha\rangle
\cr
&= \sum_{\alpha,\beta}\psi^*_\beta \psi_\alpha \langle \beta |\alpha\rangle Z_\alpha^n
\cr
& = \sum_{\alpha} |\psi_\alpha|^2 Z_\alpha^n.
\end{align}
Here we have assumed that the spectrum of $\hat{Z}$ is discrete, though the generalization to the case with a continuous spectrum is clear. 

The last line in equation \eqref{eq:alphaprob}  allows us to describe quantum expectation values as  expectation values in a  classical ensemble of systems with probability distribution given by $P^{(\psi)}(Z_\alpha) = |\psi_\alpha|^2$.

We can summarize the relationship between the closed universe states and ensemble averages as follows:\footnote{The relationship between the physical states $|\psi\rangle$ and distributions $P^{(\psi)}(Z_\alpha)$ is not one-to-one; states which differ only by the phases of their components $\psi_\alpha$ map to the same distribution. In general, correlation functions of physically relevant operators are invariant under $|\psi\rangle\rightarrow e^{i \theta(\hat{Z})}|\psi\rangle$. }
\begin{align}
\text{Eigenvectors $|\alpha\rangle$ of $\hat{Z}$} \;&\longleftrightarrow \; \text{Elements $\alpha$ of an ensemble of dual boundary theories.}
\cr
\text{States $|\psi\rangle$ of closed universes} \; &\longleftrightarrow \; \text{Probability distributions $P^{(\psi)}(Z_\alpha) = |\psi_\alpha|^2$ over the ensemble.}
\cr
\text{Eigenvalues $Z_\alpha$ of $\hat{Z}$} \; & \longleftrightarrow \; \text{Partition function $Z_\alpha$ of the member $\alpha$ of the ensemble.}
\cr
\text{Correlation functions $\langle \hat{Z}^n \rangle_\psi$} \; & \longleftrightarrow \; \text{Ensemble averages $\langle Z^n\rangle_{P} = \sum_\alpha P^{(\psi)}(Z_\alpha) \;Z_\alpha^n$}.
\end{align} 

In particular, we can consider expectation values in the No-Boundary state $|NB\rangle$. These describe a probability distribution
\be
P^{(NB)}(Z_\alpha)  = |\langle \alpha|NB\rangle|^2.
\ee
In the CGS model, the spectrum of $\hat{Z}$ is continuous, and we can just label the eigenvalues by a continuous real parameter  $z$. Then the probability distribution corresponding to the No-Boundary state is Gaussian; $|NB\rangle$ is just the (shifted) harmonic oscillator ground state:
\be
P^{(NB)}(z) dz \propto \exp\bigg[-\frac{(z-\disk)^2}{2 \;\cyl}\bigg]dz.
\ee

In more general theories, we expect that the No-Boundary state can  often be well approximated by a Gaussian state. This is true, for example, in the MM model when $S_0$ is large, and is a good approximation for many observables in JT gravity at large entropy. The variance of the Gaussian wavefunction is described by spacetimes with the topology of a cylinder which connects two asymptotic boundaries.

To connect to AdS/CFT we interpret $\langle \hat{Z} \rangle_{NB}$ as the  partition function with one boundary computed using the Euclidean black hole spacetime. In examples such as the MM and CGS  models and JT gravity the apparent lack of factorization of $\langle \hat{Z} \hat{Z} \rangle_{NB}$ is resolved by interpreting this expectation value as an ensemble average of quantum mechanical boundary systems. Such ensemble averages need not factorize since the averaging over parameters in the boundary Hamiltonian correlates the two $Z$'s. The failure of factorization, measured by the connected correlator $\langle \hat{Z}^2\rangle_{NB,c}$, is the nonzero variance of the partition function in this ensemble.
 
Of course this immediately suggests a way to calculate a product of partition functions which \textit{does} factorize; don't compute in the No-Boundary state, but in an $\alpha$ state $|\alpha\rangle$. The probability distribution for this state has no variance, and correlation functions clearly factorize.
\be
\langle \alpha| \hat{Z} \hat{Z} | \alpha \rangle = Z_{\alpha} Z_{\alpha} .
\ee

In examples like the MM model and JT gravity, there are many $\alpha$ states $|\alpha\rangle$, and thus many members of an ensemble of boundary duals to focus on. This mechanism for factorization seems incompatible with many conventional examples of AdS/CFT, where there are no hints of an ensemble of boundary duals, and thus no sign of the existence of multiple distinct $\alpha$ states in the bulk dual \cite{Marolf:2020xie,McNamara:2020uza}. However, one might hope that by studying this mechanism for factorization one might learn lessons that are useful for understanding factorization in conventional AdS/CFT.

\subsubsection{Reformulating correlation functions as overlaps}\label{correlationfunctionsoverlaps}

In the CGS and MM models, a general state $|\psi\rangle$ of the closed universes can be expressed in terms of a function\footnote{In the full MM model, different choices of $\psi(\hat{Z})$ may create states which may differ only by a null state.} $\psi(\hat{Z})$
 of the boundary operator $\hat{Z}$ acting on the No-Boundary state,
\be
|\psi\rangle = \psi(\hat{Z}) |NB\rangle.
\ee
We will always choose $\psi(\hat{Z})$ so that $|\psi\rangle$ is normalized.
Consider the $n$-point function
\begin{align}
\langle \psi | \hat{Z}^n|\psi\rangle &= \langle NB| \psi(\hat{Z})^* \hat{Z}^n \psi(\hat{Z})|NB\rangle
\cr
&= \langle NB | \hat{Z}^n \;\; |\psi(\hat{Z})|^2 |NB\rangle.
\end{align}
Defining the (unnormalized) state $|\psi^2\rangle \equiv |\psi(\hat{Z})|^2|NB\rangle\rightarrow \psi(\hat{Z})^2|NB\rangle$,\footnote{Note that the state $|\psi^2\rangle$, as well as correlation functions computed with it, only depend on the modulus of the function $\psi(\hat{Z})$. We can then take $\psi(\hat{Z})$ to be real without loss of generality.} we can view the $n$-point function as an overlap of two states
\be
\langle \hat{Z}^n\rangle_\psi = \langle Z^n| \psi^2\rangle.
\ee
Essentially, $|\psi^2\rangle$ combines the closed universes from the bra and the ket into one state. This reformulation will play an important role in our discussion of factorization.\footnote{Had we considered a correlation function of operators which do not commute with $\hat{Z}$, we would not have been able to express the correlation function as an overlap with $|\psi^2\rangle$. However, there is no clear gravity interpretation of such correlation functions so we will not consider them.}${}^,$

\begin{figure}[H]
\centering
\includegraphics[scale=0.3]{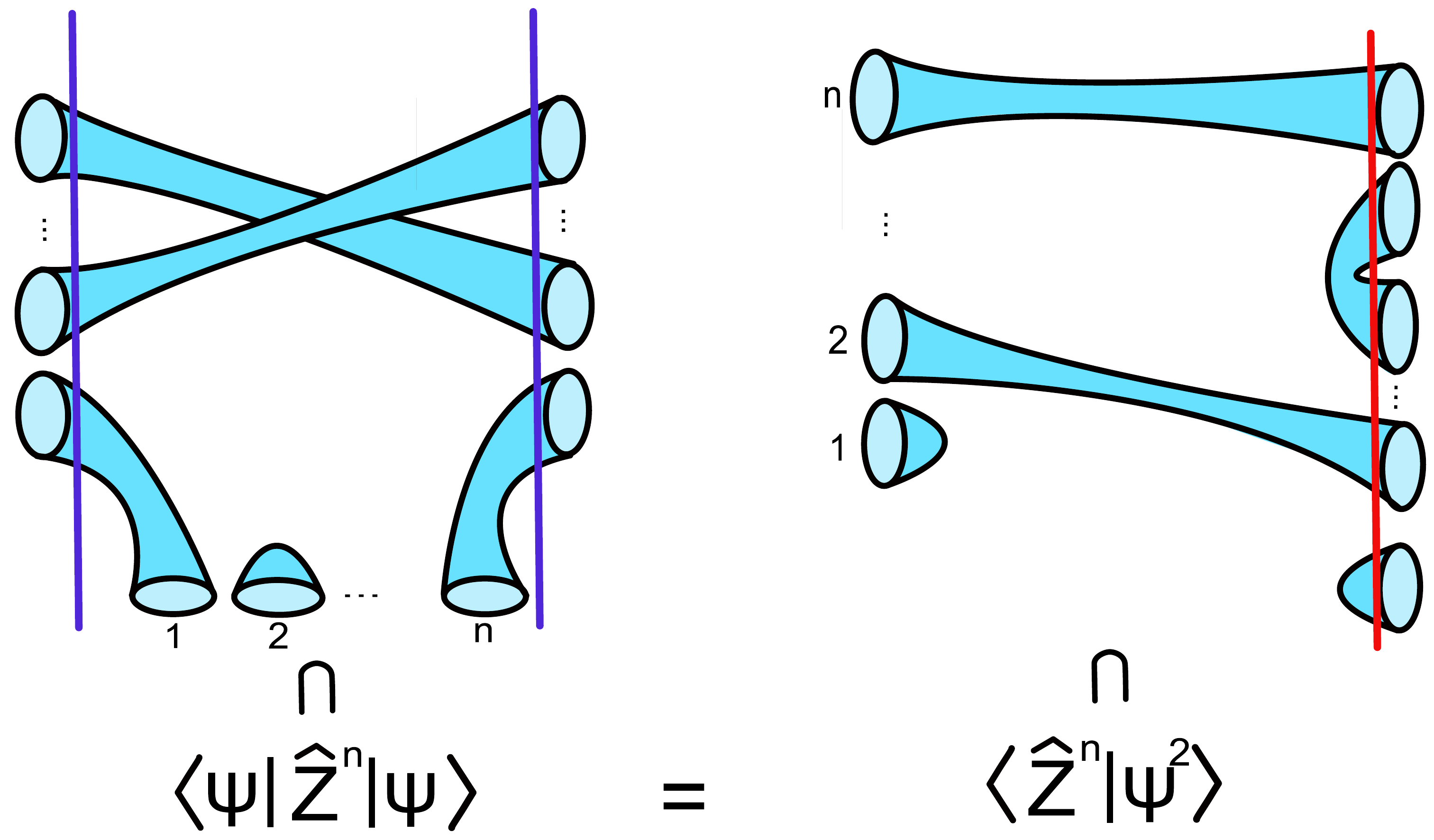}
\caption{\small In this figure, we show how to view an expectation value as an overlap of two states. The operator insertion $\hat{Z}^n$ in the correlation function on the LHS is then viewed as creating a state $|Z^n\rangle$ on the RHS. The red line on the RHS, associated with the ket, denotes that the state is not normalized, but is instead a state $|\psi^2\rangle$ for $|\psi\rangle$ normalized. Then the RHS depicts an overlap between two unnormalized states.}
\end{figure}

Viewing correlation functions as overlaps with the state $|\psi^2\rangle$ is particularly useful when states with different numbers of closed universes are orthogonal, as is the case in the CGS model. The state $|Z^n\rangle$ is a superposition of number eigenstates $|k\rangle$ with a maximum of $k=n$ universes. For example, in \eqref{eq:2Znormalorder} we found that $|Z^2\rangle$ has components with up to two universes.

Then, by inserting a complete set of states in the N-basis, the overlap $\langle Z^n|\psi^2\rangle$ \textit{only} involves the number basis components of $|\psi^2\rangle$ with up to $n$ universes
\be\label{kpointgeneral}
\langle Z^n|\psi^2\rangle = \sum_{k=0}^n \frac{1}{\cyl^k k!}\langle Z^n |k\rangle\langle k|\psi^2\rangle.
\ee
The factors of $1/\cyl$ come from the normalization of the resolution of the identity. $\hat{1}=\sum_{k=0}^\infty \frac{1}{\langle k|k\rangle} |k\rangle \langle k|$, where $\langle k|k\rangle = \cyl^k k!$.
For the case $n=0$, we are computing the norm of $|\psi\rangle$, which is equal to the overlap of $|\psi^2\rangle$ with the No-Boundary state. This is equal to the norm of $|\psi\rangle$, which is set to one.

The one-point function of $\hat{Z}$ involves only one nontrivial component of $|\psi^2\rangle$, the one-universe component
\be\label{eq:onepointfunction}
\langle \hat{Z}\rangle_\psi = \langle Z|\psi^2\rangle = \langle Z|NB\rangle + \frac{1}{\cyl}\langle Z|1\rangle \langle 1|\psi^2\rangle.
\ee
Physically, the factor of $1/\cyl$ from the resolution of the identity tells us that when we glue two cylinders together, we multiply their weights and divide by $\cyl$ so that normalized cylinders are glued together to make normalized cylinders: $\frac{1}{\cyl} \cyl \times \cyl = \cyl$.

\begin{figure}[H]
\centering
\includegraphics[scale=0.25]{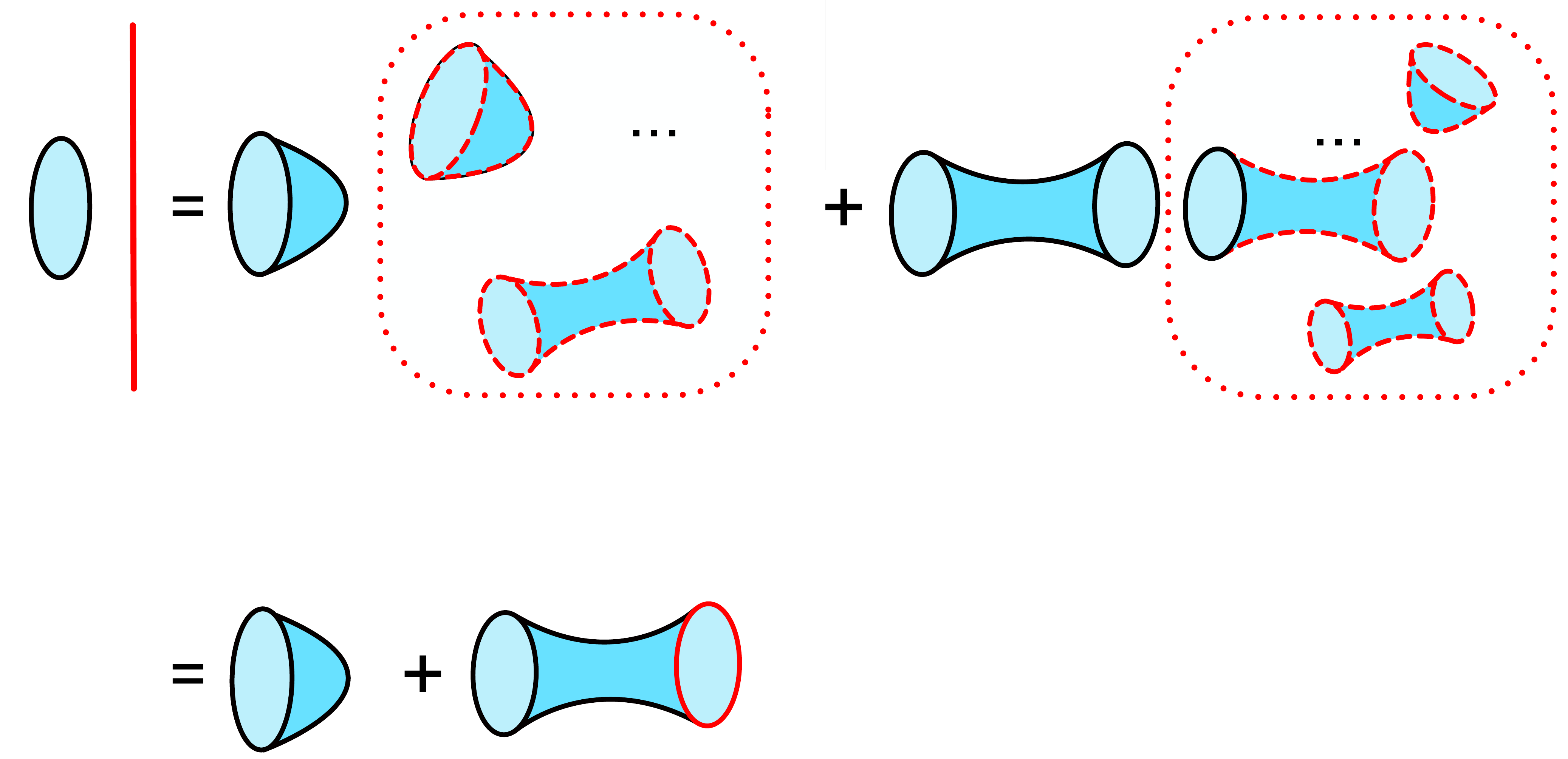}
\caption{\small Here we have depicted the computation of the one-point function $\langle \hat{Z}\rangle_\psi$, viewed as the overlap $\langle Z|\psi^2\rangle$. On the LHS we see a boundary corresponding to the bra $\langle Z|$, and a red line denoting the ket $|\psi^2\rangle$. In the first term on the RHS of the first line, we picture the contribution where the boundary for the state $\langle Z|$ is capped off with a disk. This contribution also includes a sum over spacetimes ending on the $|\psi^2\rangle$ boundaries, which are denoted by dashed red lines. The entire contribution, which gives the overlap $\langle NB|\psi^2\rangle=1$, is circled with a red dotted line. In the second term, we picture the contribution for which the boundary for the state $\langle Z|$ connects to the $|\psi^2\rangle$ boundaries. In the top line we have depicted this contribution along the lines of \eqref{eq:onepointfunction}, as a product of the amplitude $\langle Z| 1 \rangle$ and the one-universe component $\langle 1|\psi^2\rangle$. The one-universe component is computed by the spacetimes within the dotted red line. In the second line we view this contribution more directly as an overlap with the one-universe component of $|\psi^2\rangle$. The red circular line denotes a boundary corresponding to the unnormalized state $\langle 1|\psi^2\rangle |1\rangle$, rather than the normalized state $|1\rangle$.}\label{OnePointFunction}
\end{figure}

Pictorially, we represent this equation in Figure \ref{OnePointFunction}.

The full, many-universe state $|\psi^2\rangle$ is prepared by including any number of $\hat{Z}$ boundaries in the path integral. However, the one-point function only involves the zero and one-universe components. 

The overlap $\langle Z|NB\rangle$ is equal to the one-point function of $\hat{Z}$ in the No-Boundary state, given by the disk. We can think of this as the contribution from nucleating a single closed universe from the closed universe vacuum.

The overlap $\langle Z|1\rangle=1$ is the amplitude between a single closed universe, in the normalized state $|1\rangle$, and the unnormalized state $|Z\rangle$. In the computation of the one-point function, we must multiply this amplitude by the one-universe component of $|\psi^2\rangle$.

The amplitude $\langle Z|1\rangle$ is computed by the cylinder topology, while the one-universe component $\langle 1|\psi^2\rangle$ is computed by summing over many disconnected spacetimes with boundaries created by the operator $\psi(\hat{Z})^2$. The product $\frac{1}{\cyl}\langle Z|1\rangle\langle 1|\psi^2\rangle$ then represents the contribution to the overlap $\langle Z |\psi^2\rangle$ from the $\hat{Z}$ boundary absorbing a single closed universe created by $\psi(\hat{Z})^2$. Pictorially, we represent this with the cylinder in Figure \ref{OnePointFunction}, with one boundary corresponding to the state $|Z\rangle$, and the other red boundary corresponding to the one-universe component of $|\psi^2\rangle$. 

\begin{figure}[H]
\centering
\includegraphics[scale=0.35]{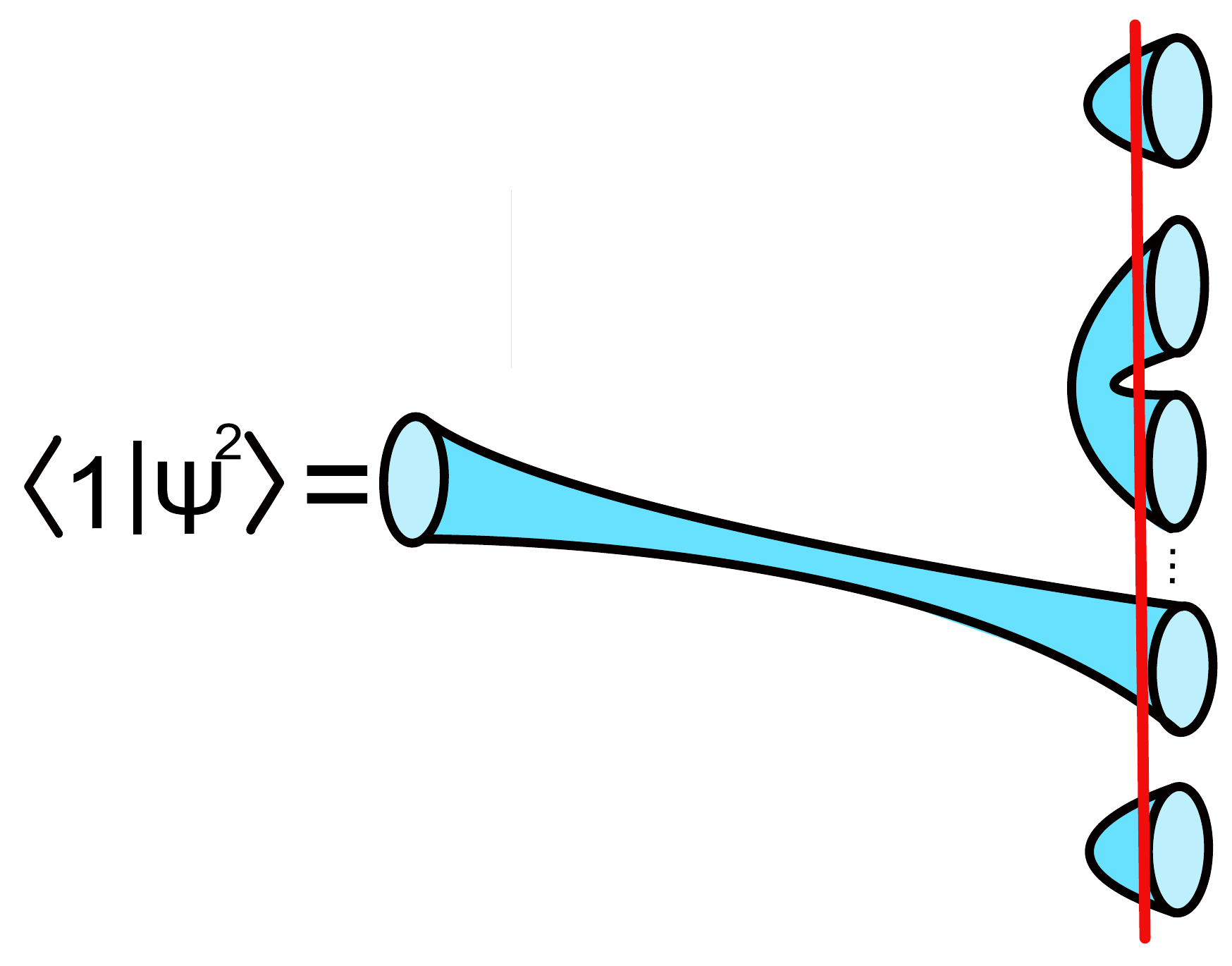}
\caption{\small Here we have pictured an example contribution to the one-universe component of $|\psi^2\rangle$.}
\label{oneuniversecomponentfig}
\end{figure}

In the CGS model, there are no contributions to the overlap $\langle Z|\psi^2\rangle$ in which the $\hat{Z}$ boundary absorbs two or more closed universes created by $\psi(\hat{Z})^2$. These contributions would involve spacetimes with connected components with three or more boundaries, such as the three-holed sphere, and these spacetimes are not present in the CGS model.

\begin{figure}[H]
\centering
\includegraphics[scale=0.4]{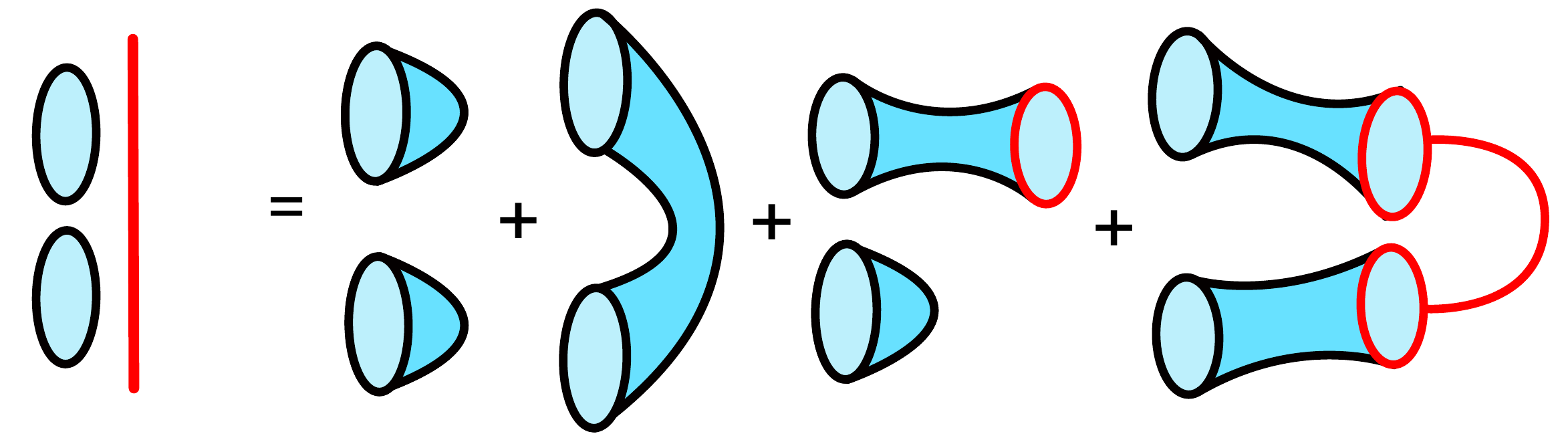}
\caption{\small Here we have pictured the contributions to the RHS of \eqref{eq:twopointfunction}. The pair of disks and the cylinder correspond to $\langle Z^2|NB\rangle$, the disk and cylinder with a red circular boundary corresponds to $\frac{1}{\cyl}\langle Z^2\rangle|1\rangle \langle 1|\psi^2\rangle$, and the cylinders with linked red boundaries corresponds to $\frac{1}{2\; \cyl^2}\langle Z^2|2\rangle\langle 2|\psi^2\rangle$. The linked boundaries represent the two-universe component $\langle 2|\psi^2\rangle$; this is distinct from a pair of unlinked red circular boundaries, which would denote two copies of the one-universe component $\langle 1|\psi^2\rangle$.}
\label{TwoPointFunction}
\end{figure}

The two-point function of $\hat{Z}$ involves only one additional component of the many-universe state $|\psi^2\rangle$, the two-universe component. 
\be\label{eq:twopointfunction}
\langle \hat{Z}^2\rangle_\psi = \langle Z^2|\psi^2\rangle = \langle Z^2|NB\rangle + \frac{1}{\cyl}\langle Z^2|1\rangle \langle 1|\psi^2\rangle + \frac{1}{2 \;\cyl^2}\langle Z^2|2\rangle \langle 2|\psi^2\rangle.
\ee
Pictorially, we represent this in Figure \ref{TwoPointFunction}

Some of the contributions to the two-point function are simply products of contributions to the one-point function. However, the two-point function involves the cylinder connecting the two $\hat{Z}$ boundaries (included in the overlap $\langle Z^2|NB\rangle$), as well as the amplitude $\frac{1}{2\; \cyl^2} \langle Z^2|2\rangle=1$ for the two $\hat{Z}$ boundaries to each absorb two closed universes in a normalized state, multiplied by the two-universe component $\langle 2|\psi^2\rangle$. In Figure \ref{TwoPointFunction}, we have represented this second contribution with two cylinders with linked red boundaries. The pair of linked red boundaries represents the two-universe state $\langle 2|\psi^2\rangle$.  Here the link differentiates this state from two copies of the one-universe state $\langle 1|\psi^2\rangle$. As with the one-universe component, this two-universe component is computed by summing over many disconnected spacetimes with boundaries created by the operator $\psi(\hat{Z})^2$.

\subsection{Factorization in the CGS model}\label{factorizationcgs}

In the special case that $|\psi\rangle$ is an eigenstate of $\hat{Z}$ (an $\alpha$ state) with eigenvalue $z$, correlation functions should factorize. In particular, the two-point function should be equal to the square of the one-point function,
\be
\langle \hat{Z}^2\rangle_{z} = \langle \hat{Z}\rangle_{z}^2 = z^2.
\ee
The subscript $z$ denotes the $\alpha$ state $|z\rangle \equiv |\psi_{z}\rangle$.

When viewing the correlation functions as overlaps of the form $\langle Z^n|\psi^2_{z}\rangle$, factorization implies a particularly simple relationship between the components $\langle k |\psi^2_{z}\rangle$ with different universe number. This simple relationship will be the basis of an effective description we describe in Section \ref{effectivemodel}.

Terms in the two-point function only involving the disk topology manifestly factorize into contributions from the square of the one-point function.   So to isolate the nontrivial contributions we consider correlation functions of $\hat{Z}-\disk$ and compare our expressions for
\be\label{eq:onepointfunctionalpha}
\langle \hat{Z}-\disk \rangle_{z}^2 = \langle 1|\psi_{z}^2\rangle^2.
\ee
and
\be\label{eq:twopointfunctionalpha}
\langle (\hat{Z}-\disk))^2\rangle_{z} = \cyl+ \langle 2|\psi_{z}^2\rangle.
\ee
in terms of the components of $|\psi^2\rangle$, using equations \eqref{eq:onepointfunction} and \eqref{eq:twopointfunction} with $|\psi\rangle = |\psi_{z}\rangle$. The fact that $\langle \hat{Z}\rangle_{z} = z$ allows us to identify $\langle 1 |\psi_{z}^2\rangle = z-\disk$. Since $z$ is Gaussian distributed in the No-Boundary state/ensemble, the one-universe component $\langle 1 |\psi_{z}^2\rangle$ is as well (but here with zero mean). Denoting averaging $z$ over the No-Boundary ensemble with $\mathbb{E}[\dots]_{NB}$,
\be\label{averagingoneuniverse}
\mathbb{E}\big[\langle 1|\psi_{z}^2\rangle \big]_{NB}=0,\hspace{20pt} \mathbb{E}\big[ \big(\langle 1|\psi_{z}^2\rangle \big)^2\big]_{NB}=\cyl=1.
\ee
Equation \eqref{eq:onepointfunctionalpha} tells us that the one-universe component $\langle 1|\psi^2_{z}\rangle$ contains all the information about the $\alpha$ state. We can see this indirectly by using the fact that all other correlation functions in the $\alpha$ state must factorize into products of this one-point function, which is determined by $\langle 1|\psi^2_{z}\rangle$, or more directly by noting that the identification $\langle 1 |\psi_{z}^2\rangle = z-\disk$ fixes the $\alpha$ state $|z\rangle$ in terms of $\langle1|\psi^2_{z}\rangle$. We can then view $\langle 1 |\psi_{z}^2\rangle$ as the parameter describing the set of eigenstates, or equivalently the random variable describing the ensemble of boundary theories; rather than describing the theory in an $\alpha$ state in terms of the many-universe $\alpha$ state $|\psi_{z}\rangle$, we can simply use the random one-universe wavefunction $\langle 1|\psi^2_{z}\rangle$. This reduction in the number of universes/boundaries, as well as the simple Gaussian statistics of this one-universe component, are the key simplifications in our description of factorization.  

As the one-universe component $\langle 1|\psi^2_{z}\rangle$ contains all the information about the $\alpha$ state, the remaining components must be redundant variables. Using \eqref{kpointgeneral} to generalize \eqref{eq:twopointfunctionalpha}, an $n$-point function can be expressed in terms of the components $\langle k|\psi^2_{z}\rangle$ with $k\leq n$. On the other hand, we can write the $k$-point function as a product of one-point functions, which only depend on the one-universe component. Factorization of the $k$-point function then implies identities that relate the $k$-universe components to the one-universe component.

For example, the equality between the square of the one-point function \eqref{eq:onepointfunctionalpha} and the two-point function \eqref{eq:twopointfunctionalpha} implies that the two-universe component is equal to the square of the one-universe component,  minus the cylinder,
\be\label{eq:twouniverserelation}
\langle 2|\psi_{z}^2\rangle=\langle 1|\psi_{z}^2\rangle^2 - \cyl.
\ee

\begin{figure}[H]
\centering
\includegraphics[scale=0.5]{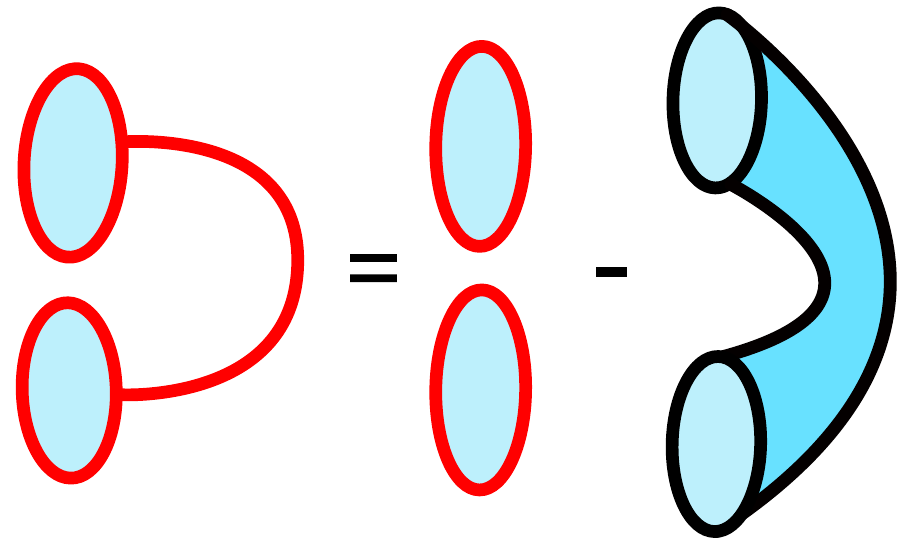}
\caption{\small The two-universe wavefunction $\langle 2|\psi^2_{z}\rangle$ is related to the one-universe wavefunction $\langle 1|\psi^2_{z}\rangle$ through this identity. This identity ensures that the two-point function in an $\alpha$ state factorizes.}
\end{figure}

We briefly note that using \eqref{averagingoneuniverse}, we see that the two-universe component averages to zero, $\mathbb{E}\big[\langle 2|\psi^2_{z}\rangle \big]=0$.

Now we take a moment to compare this computation to the computation of the spectral form factor in  periodic orbit theory \eqref{doublesum}. The rough analogy we will make is to think of the one-universe component $\langle 1|\psi_{z}^2\rangle$ as the sum over orbits describing $Z(iT)$ or $Z(-iT)$, and the cylinder as the \textit{diagonal} part of the double-sum over orbits. This second comparison is motivated by the fact that the cylinder and diagonal sum both describe the ensemble average. 

The two-universe component $\langle 2|\psi_{z}^2\rangle$ then plays the role of the \textit{off-diagonal} part of the double-sum over orbits; this is what we need to add to the cylinder (analogous to the diagonal sum) that computes the averaged answer  in order to find an answer that factorizes.

Of course, this analogy is very coarse, especially since in this simple model the one-universe and two-universe components of $|\psi_{z}^2\rangle$ are not given by any natural sum. However, in more complicated models with a richer closed universe Hilbert space, such as the models we consider in the remainder of this paper, this analogy will be more appropriate.

So far we have just considered the factorization of the two-point function into the square of the one-point function. As these correlation functions depend only on the one and two-universe components of $|\psi^2_{z}\rangle$, we found no constraints on the higher universe number components. To find the analogous relationship between the $n$-universe component $\langle n|\psi^2_{z}\rangle$ and the one-universe component, we demand that the $n$-point function factorize. This tells us that that the $n$-universe component is equal to the $n$th power of the one-universe component, minus corrections from replacing pairs of the one-universe component with the cylinder. For example, in the case $n=3$,
\be\label{eq:threeuniverserelation}
\langle 3|\psi_{z}^2\rangle = \langle 1|\psi_{z}^2\rangle^3 - 3 \;\cyl \; \langle 1|\psi_{z}^2\rangle.
\ee

\begin{figure}[H]
\centering
\includegraphics[scale=0.5]{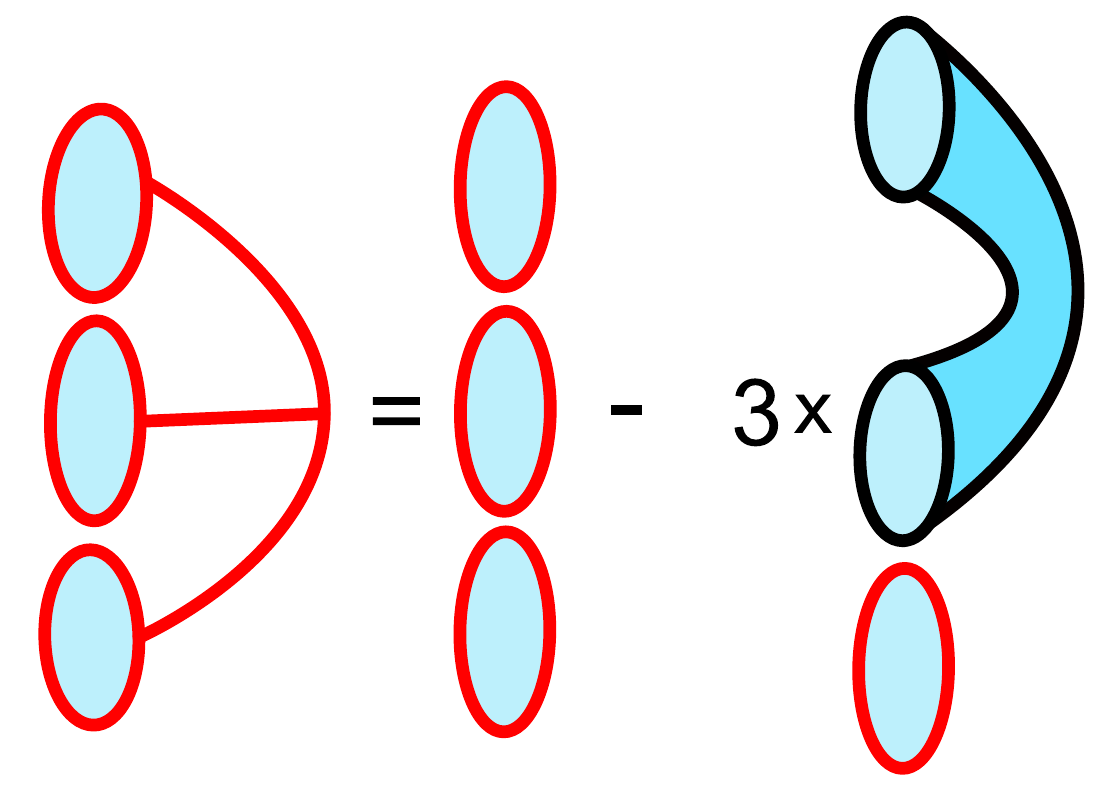}
\caption{\small The three-universe wavefunction $\langle 3|\psi^2_{z}\rangle$, depicted by the three linked circles, is related the the one-universe wavefunction $\langle 1|\psi^2_{z}\rangle$ through this identity, which we may derive by demanding that the three-point function in an $\alpha$ state factorizes. Similar identities describing the $n$-universe components of $|\psi^2_{z}\rangle$ can be derived by demanding that the $n$-point function factorizes.}
\end{figure}

Through these relations we can see that the full many-universe $\alpha$ state can be simply described in terms of a one-universe state, which has simple Gaussian statistics. Though we can think of this wavefunction as computed by a sum involving spacetimes with many boundaries and many connected components, along the lines of Figure \ref{oneuniversecomponentfig}, we may also instead think of independently defining this wavefunction through its statistics \eqref{averagingoneuniverse}, without recourse to the many-universe description, and defining the remaining components of $|\psi^2_{z}\rangle$ through these recursion relations. We will exploit this point of view later, in Section \ref{effectivemodel}.

\subsubsection{Explicit calculation}\label{explicitcalculationcgs}

It is useful to check that the relations \eqref{eq:twouniverserelation} and \eqref{eq:threeuniverserelation} for $|\psi_{z}^2\rangle$ are obeyed using the explicit expression for the eigenstate $|z\rangle$,
\be
|z\rangle \propto \delta(\hat{Z}-z) |NB\rangle.
\ee
We then identify $\psi_{z}(\hat{Z}) \propto \delta(\hat{Z}-z)$. 

This expression is difficult to work with directly; instead we express the delta function as an integral,

\be\label{deltafunctionalphacgs}
\psi_{z}(\hat{Z}) \propto \int_{-\infty}^\infty dp\; e^{  i p (\hat{Z}-z)}.
\ee
This expression leads to an expression for the eigenstate $|z\rangle$ in the ``spacetime D-brane'' basis introduced in \cite{Marolf:2020xie}. The state $|p\rangle = e^{i p \hat{Z}} |NB\rangle$ is thought of as a spacetime D-brane with a parameter $p$, since the computation of correlation functions in these states mirrors Polchinski's perturbative computation of string theory amplitudes in the presence of a D-brane \cite{Polchinski:1994fq}.\footnote{These are close analogs  of the D-branes considered in \cite{Saad:2019lba} which are also created by the exponential of a closed universe operator.  In that case the operator being exponentiated is the Laplace transform of the partition function. In the closely related minimal string theory, versions of these states describe D-branes of FZZT type \cite{Aganagic:2003qj,Maldacena:2004sn}.} In the CGS model, we can view the brane state $|p\rangle$ as a coherent state of closed universes. Using the expression $\hat{Z}=\disk+ a + a^\dagger$ and normal ordering the exponential $e^{i p \hat{Z}}$ using the BCH formula we find
\be
e^{i p \hat{Z}} |NB\rangle = e^{i p \disk} e^{- \cyl \frac{p^2}{2}} e^{i p a^\dagger} |NB\rangle.
\ee

In order to verify that the $n$-universe components of $|\psi_{z}^2\rangle$ for $|\psi\rangle = |z\rangle$ obey the relations \eqref{eq:twouniverserelation} and \eqref{eq:threeuniverserelation}, we must normal order the operator $\psi_{z}(\hat{Z})^2\propto \delta(\hat{Z}-z)^2$. To do this, it is helpful to regulate the delta function integral by adding a small Gaussian convergence factor,
\be
\psi_{z,\Delta}(\hat{Z}) \propto \int_{-\infty}^\infty dp \; e^{-\Delta^2 p^2 + i p (\hat{Z}-z)}.
\ee

This operator creates a sharp Gaussian state $|z,\Delta\rangle$, peaked around $z$ with a small width proportional to $\Delta$.
\be\label{eq:psideltaexpression}
|z\rangle_\Delta \propto \int_{-\infty}^\infty dp\; e^{-\Delta^2 p^2+  i p (\hat{Z}-z)} |NB\rangle \propto e^{-\frac{(\hat{Z}-z)}{4\Delta^2} } |NB\rangle.
\ee
We compute $\psi_{z,\Delta}(\hat{Z})^2$ by expressing the product of two gaussian integrals over $p$ and $p'$ as the product of two gaussian integrals in the sum and difference variables $p_+=p+p'$ and $p_-=p-p'$.
\be\label{psisquaredcgs}
\psi_{z,\Delta}(\hat{Z})^2 \propto \int_{-\infty}^\infty dp dp' e^{-\Delta^2 (p^2+p'^2)+  i (p+p') (\hat{Z}-z)} = \bigg[ \int_{-\infty}^\infty  dp_-  e^{-\frac{\Delta^2 p_-^2}{2}}\bigg] \int_{-\infty}^\infty dp_+ e^{-\frac{\Delta^2 p_+^2}{2}+  i p_+ (\hat{Z}-z)}.
\ee
The integral over $p_-$ doesn't depend on $\hat{Z}$, it simply contributes a normalization factor. $\psi_{z,\Delta}(\hat{Z})^2$ is also a superposition of exponential brane operators $e^{i p_+ \hat{Z}}$, so we can simply normal order using the BCH formula. Normal ordering introduces a Gaussian term $e^{-\cyl\frac{p_+^2}{2}}$ so we can safely take the limit $\Delta\rightarrow 0$ to find
\be
|\psi_{z}^2\rangle \propto \int_{-\infty}^\infty dp_+\; e^{-\cyl\frac{p_+^2}{2}+ i p+ (\disk- z) } e^{i p_+ a^\dagger} |NB\rangle.
\ee
Expanding the exponential $e^{ip_+ a^\dagger}$, we find
\be\label{eq:explicitalphastate}
|\psi_{z}^2\rangle \propto \sum_{k=0}^\infty \int_{-\infty}^\infty dp_+ \; e^{-\cyl\frac{p_+^2}{2}+ i p_+ (\disk-z)  } \frac{(i p_+)^k}{k!} |k\rangle.
\ee
The constant of proportionality is fixed by the condition that $\langle NB|\psi_{z}^2\rangle=1$.

To take the $n$-universe component we use the overlap $\langle n|k\rangle = \cyl^n n! \delta_{nk}$. Then the $n$-universe component is given by the $n$'th moment of $i p_+ \cyl$ in this Gaussian. We can rescale our variables $p_+\rightarrow p_+'= p_+\cyl$ so that the Gaussian has a center $ip_+'= z-\disk$ and variance $\cyl$. 

The one-universe component is given by the center of the Gaussian.
\be
\langle 1|\psi_{z}^2\rangle = z-\disk,
\ee
and the two-universe component is proportional to the two-point function of $i p_+'$,
\begin{align}
\langle 2|\psi_{z}^2\rangle &= (z-\disk)^2-\cyl 
\cr
 &= \langle 1|\psi_{z}^2\rangle^2 -\cyl.
\end{align}
The general pattern describing the $n$-universe components is precisely described by the Gaussian integral \eqref{eq:explicitalphastate}, with Wick contractions between the $n$ factors of $p_+'$ contributing the cylinder corrections in \eqref{eq:twouniverserelation} and \eqref{eq:threeuniverserelation}.

As an aside, we will mention that the use of the delta function (or regularized delta function) to describe the eigenstate $|z\rangle$ is related to how one would naturally compute correlation functions from the ensemble-average point of view discussed in Section \ref{alphabasis}. In this point of view, we think about correlation functions in a state $|\psi\rangle$ as ensemble averages over the eigenvalue $z'$ of $\hat{Z}$ with a distribution $P^{(\psi)}(z')$
\be
\langle \hat{Z}^n\rangle_\psi = \int dz'\; P^{(\psi)}(z') \; z'^n.
\ee
Choosing $|\psi\rangle$ to be an eigenstate means that we choose $P^{(\psi)}(z')$ to focus on a single element $z'=z$ of the ensemble; in other words we choose $P^{(\psi)}(z')= \delta(z-z')$.   If we multiply and divide this distribution by the No-Boundary distribution, then correlation functions in the eigenstate $|z\rangle$ can be described as correlation functions in the No-Boundary state with an operator proportional to $\delta(\hat{Z}-z)$ inserted. Comparing with our prescription for computing a correlation function in the state $|\psi\rangle$ by inserting the operator $\psi(\hat{Z})^2$ in the No-Boundary state, we can identify $\psi_{z}(\hat{Z})^2\propto \delta(\hat{Z}-z)$. To compare this with our earlier identification of $\psi_{z}(\hat{Z}) \propto \delta(\hat{Z}-z)$, it is useful to use the regulated integral expression for the delta function; the square of the delta function is itself a delta function, up to an (infinite) normalization.

\subsection{The CGS model with ``species''}\label{cgsspecies}

To summarize the results of the previous section: in the CGS model, $n$-point correlation functions in a state $|\psi\rangle$ are simple functions of the $k$-universe components of the state $|\psi^2\rangle$ with $k \le n$. The contributions of the $k$-universe components describe the absorption of $k$ closed universes by the $n$ $\hat{Z}$ operators in the correlation function. In the case that $|\psi\rangle$ is an eigenstate $|z\rangle$, an $n$-point correlation function factorizes into the $n$'th power of the one-point function; after subtracting off the disks, this is simply the $n$'th power of the one-universe component $\langle 1|\psi^2\rangle$. In order for this to happen, the various contributions to the $n$-point function from cylinders and $k$-universe components $\langle k|\psi^2\rangle$ with $k\geq 2$ must sum up to powers of the one-universe component $\langle 1|\psi^2\rangle$. The basic way in which this happens is illustrated by the two-point function; the two-universe component plus the cylinder is equal to the square of the one-universe component. This is roughly analogous to the periodic orbit story from Section \ref{periodicorbits}, with the cylinder playing the role of a diagonal sum over orbits, and the two-universe component playing the role of the off-diagonal sum.

The CGS model is very simple; there is only a single one-universe state, $|1\rangle$, created by the operator $\hat{Z}$. This model will serve as an approximation to the MM model, as we will demonstrate in Section \ref{MMmodel}. However, in order to approximate more complicated theories of gravity we must generalize the CGS model so that a single closed universe can have many possible states. The appropriate generalizations of the relations \eqref{eq:twouniverserelation} and \eqref{eq:threeuniverserelation} describing the components of $|\psi^2\rangle$ for $|\psi\rangle$ an eigenstate sharpens the analogy with the periodic orbit story from Section \ref{periodicorbits}. Because this generalization has many internal states of a single closed universe, we refer to this model as the ``CGS model with species''.

To motivate this generalization, we begin by discussing some more general aspects of the quantum mechanics of closed universes with a theory with a negative cosmological constant, following \cite{Marolf:2020xie}. To construct the Hilbert space of closed universes, we mirror the procedure used in Section \ref{statesofcloseduniverses}, with the addition of a larger set of closed universe operators $\hat{Z}\rightarrow \hat{Z}[\mathcal{J}]$. $\hat{Z}[\mathcal{J}]$ is an operator which creates an asymptotically AdS boundary in the path integral (which we can think of as periodic in complexified time), with boundary conditions labeled by $\mathcal{J}$.\footnote{It is not necessary that these operators create states on \textit{asymptotic} boundaries. Later in Section \ref{JTgravity} we will work with states of finite size closed universes.}$^,$\footnote{In three or more dimensions, we could also consider operators which create closed universes with more general spatial topologies.} $\mathcal{J}$ may represent a fixed metric, as in the special case of the partition function $\hat{Z}[\mathcal{J}]\rightarrow \hat{Z}(\beta)$. More generally $\mathcal{J}$ denotes the boundary conditions of bulk fields $\phi$ (including the metric) at an asymptotically AdS boundary.

Following the procedure reviewed in Section \ref{statesofcloseduniverses}, we define the operators $\hat{Z}[\mathcal{J}]$ by their matrix elements in the No-Boundary state, and we define these matrix elements via the gravity path integral with the corresponding boundary conditions:
\be
\langle NB|\hat{Z}[\mathcal{J}_1]\dots \hat{Z}[\mathcal{J}_n]|NB\rangle = \int_{\phi\rightarrow \mathcal{J}} D\phi \; e^{-I[\phi]}.
\ee
We can then roughly define the Hilbert space as the span of states created by acting on the No-Boundary state with the operators $\hat{Z}[\mathcal{J}]$.\footnote{See \cite{Marolf:2020xie} for a more precise and detailed discussion.}

This definition of the matrix elements makes clear that the $\hat{Z}[\mathcal{J}]$ commute. Hermitian conjugation of the $\hat{Z}[\mathcal{J}]$ is defined by $\hat{Z}[\mathcal{J}]^\dagger =\hat{Z}[\mathcal{J}^*]$, with $\mathcal{J}^*$ the CPT conjugation of the boundary conditions $\mathcal{J}$. For simplicity, in the remainder of this section we will restrict our attention to Hermitian operators.\footnote{The operators $\hat{Z}(\beta+iT)$, and related fixed-energy operators we will consider in later sections are not Hermitian, but we can take their real and imaginary parts to form Hermitian operators. Then the formulas from this section can be applied directly.}

In this more general setting, $\alpha$ states now diagonalize all of the $\hat{Z}[\mathcal{J}]$ for different sources $\mathcal{J}$,
\be
\hat{Z}[\mathcal{J}]|\alpha\rangle= Z_\alpha [\mathcal{J}]|\alpha\rangle.
\ee
 Expectation values of products of operators $\hat{Z}(\beta)$ in a state $|\psi\rangle$ are then equal to ensemble averages of products of partition functions with a distribution described by the components of $|\psi\rangle$, $P^{(\psi)}_\alpha = |\langle \alpha| \psi\rangle |^2$,
\be
\langle \hat{Z}[\mathcal{J}_1]\dots \hat{Z}[\mathcal{J}_n]\rangle_\psi = \sum_\alpha P^{(\psi)}_\alpha \; Z_\alpha [\mathcal{J}_1] \dots Z_\alpha[\mathcal{J}_1] .
\ee

In the case that we take $\hat{Z}[\mathcal{J}]$ to be a partition function $\hat{Z}(\beta)$, the eigenvalues $Z_\alpha(\beta)$ describe the partition function of an element of a boundary ensemble. As will be the case when we study JT gravity in Section \ref{JTgravity}, we can then think of elements of the boundary ensemble as being labeled by the set of energy eigenvalues of the boundary Hamiltonian.

\subsubsection{Introduction to the CGS model with ``species''}\label{cgsspeciesintro}

We now define a generalization of the CGS model as a theory in which a set of boundary operators $\hat{Z}_I$, which can be thought of as linear superpositions of operators $\hat{Z}[\mathcal{J}]$, are independent Gaussian variables in the No-Boundary state.
\begin{align}\label{eq:gaussiancorrelatorspecies}
\langle \hat{Z}_I \rangle_{NB} &= \disk_I
\cr
\langle \hat{Z}_I \hat{Z}_J\rangle_{NB,c} &= \cyl\; \delta_{IJ} = \delta_{IJ}
\cr
\langle \hat{Z}_{I_1} \dots \hat{Z}_{I_n} \rangle_{NB,c} &=0,\hspace{20pt} k>2.
\end{align}
In the remainder of this section, we will not explicitly write factors of $\cyl$, which is set equal to one here.

In general the operators $\hat{Z}[\mathcal{J}]$ may have a more complicated inner product, $\langle \hat{Z}[\mathcal{J}]\hat{Z}[\mathcal{J}']\rangle_{NB,c} = \cyl(\mathcal{J},\mathcal{J}')$, which is computed by the path integral with two boundaries. For example, partition function operators $\hat{Z}(\beta)$ in JT gravity have a two-point function which is not diagonal in $\beta$. In that case we can simply diagonalize the correlation matrix $\cyl(\mathcal{J},\mathcal{J}')$ and rescale the operators to find independent, unit-variance variables $\hat{Z}_I$. For simplicity we will assume that there are finitely many independent Gaussian variables $\hat{Z}_I$.

\begin{figure}[H]
\centering
\includegraphics[scale=0.3]{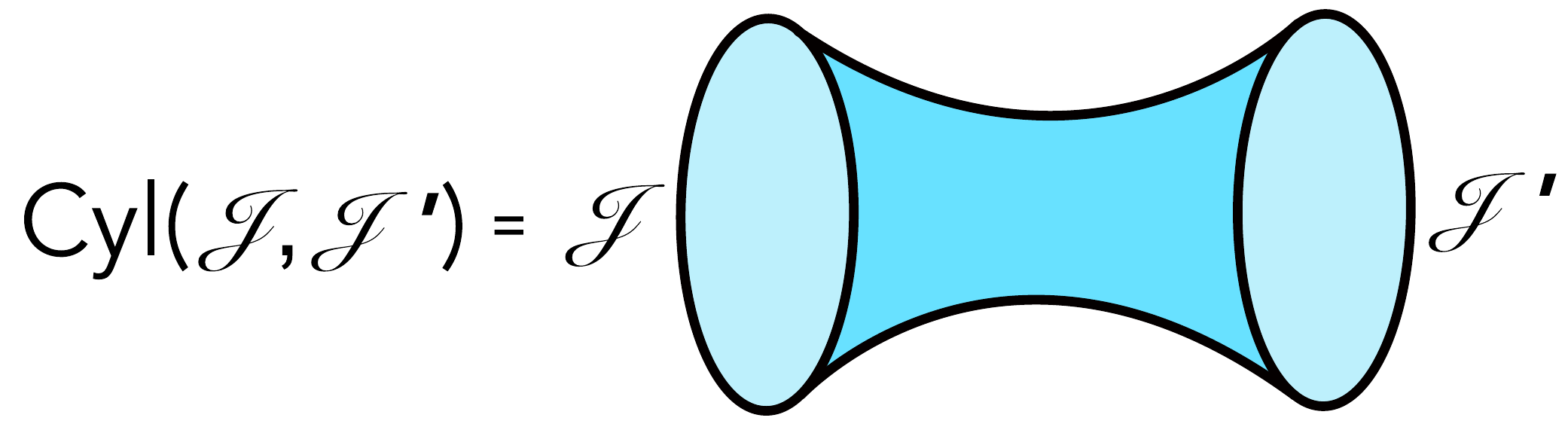}
\caption{\small The connected two-point function $\langle \hat{Z}[\mathcal{J}]\hat{Z}[\mathcal{J}']\rangle_{NB,c} = \cyl(\mathcal{J},\mathcal{J}')$ is computed by the path integral over cylindrical spacetimes with $\mathcal{J}$ boundary conditions on one boundary and $\mathcal{J}'$ boundary conditions on the other.}
\end{figure}

In our applications we will use this version of the CGS model to approximate JT gravity. In that case the natural operators $\hat{Z}[\mathcal{J}]$ are analytically continued partition functions $\hat{Z}(\beta+i T)$ (or their fixed energy counterparts $\hat{Y}_{E} (T)$, to be defined in Section \ref{JTgravity}). However, it is possible that this version of the CGS model may be relevant for describing higher dimensional theories of gravity as well, in cases where higher dimensional spacetime wormholes give approximately Gaussian correlators of operators $\hat{Z}[\mathcal{J}]$ in the No-Boundary state.\footnote{For example, in higher dimensional theories of gravity, the double cone \cite{Saad:2018bqo}, when it dominates, describes Gaussian correlation functions of $\hat{Y}_{E}(T)$.   Other calculations in higher dimensions include \cite{Cotler:2020ugk,Belin:2020hea,Marolf:2021kjc,Mahajan:2021maz}.}

\subsubsection{$n$-universe states in the CGS model with species}\label{cgsspeciesstates}

The Gaussian correlators \eqref{eq:gaussiancorrelatorspecies} prompt us to think of the operators $\hat{Z}_I$ as the shifted position operators of many independent harmonic oscillators. The $\hat{Z}_I$ can be expressed in terms of creation and annihilation operators
\be
\hat{Z}_I = \disk_I + a^\dagger_I + a_I.
\ee
The $a^\dagger_I$ and $a_I$ satisfy
\begin{align}
[a_I, a^\dagger_J] &= \delta_{IJ}
\cr
a_I |NB\rangle &=0.
\end{align}
The $a^\dagger_I$ then create orthonormal one-universe states,
\be
|I\rangle \equiv a^\dagger_I |NB\rangle ,\hspace{20pt} \langle I|J\rangle = \delta_{IJ},
\ee
which form a basis of the one-universe Hilbert space. Note that
\be\label{eq:Zstatedef}
|Z_I\rangle \equiv \hat{Z}_I |NB\rangle= \disk_I |NB\rangle + |I\rangle.
\ee
 The full closed universe Hilbert space is spanned by $n$-universe states labeled by the indices $I_1\dots I_n$ up to reordering,
\be
|\{I_1\dots I_n\}\rangle = a^\dagger_{I_1} \dots a^\dagger_{I_n}|NB\rangle.
\ee
With a slight abuse of terminology, we will also refer to this basis as the ``N basis''.
The states $|\{I_1\dots I_n\}\rangle$ with different sets of species $\{I_1\dots I_n\}$ are orthogonal,
\be
\langle \{J_1\dots J_m\}| \{I_1\dots I_n\} \rangle = C_{ \{I_1\dots I_n\}} \delta_{ \{J_1\dots J_m\}, \{I_1\dots I_n\}},
\ee
where $C_{ \{I_1\dots I_n\}}$ can be determined from normal ordering, and is equal to a product of factorials of the multiplicites of each species appearing in the set $\{I_1\dots I_n\}$.

A general state $|\psi\rangle$ can be described by its components $\psi_{\{I_1\dots I_n\}}$,
\be
|\psi\rangle = \psi_0 |NB\rangle +\sum_{n=1}^\infty \sum_{\{I_1\dots I_n\}}\frac{\psi_{\{I_1\dots I_n\}}}{C_{ \{I_1\dots I_n\}}}|\{I_1\dots I_n\} \rangle .
\ee
The components $\psi_{\{I_1\dots I_n\}} = \langle \{I_1\dots I_n\}|\psi\rangle$ can be thought of as a symmetric tensor.

\subsubsection{Correlation functions}\label{cgsspeciescorrelators}

Now we will discuss the computations of the correlators $\langle \hat{Z}_{I_1}\dots \hat{Z}_{I_n}\rangle_\psi$ in a general state $|\psi\rangle$ before focusing on the case that $|\psi\rangle$ is a joint eigenstate of all of the $\hat{Z}_I$.

We again approach this by viewing correlation functions in a state $|\psi\rangle = \psi(\hat{Z}_I)|NB\rangle$ as an overlap between the state $|\psi^2\rangle =  \psi(\hat{Z}_I)^2|NB\rangle$\footnote{As in the single-species version of the CGS model, we can take $\psi(\hat{Z}_I)$ to be real.} and a state $|Z_{I_1} \dots Z_{I_n}\rangle = \hat{Z}_{I_1}\dots \hat{Z}_{I_n}|NB\rangle$ created by the operators inserted in the correlation function. We then insert a complete set of states in the basis $|\{I_1\dots I_k\}\rangle$. The simplification that we take advantage of is that the states $|Z_{I_1} \dots Z_{I_n}\rangle$ contain only up to $n$ universes. Denoting the projector onto the $k$-universe subspace as 
\be\label{projectordef}
\hat{P}_k= \sum_{\{J_1\dots J_k\}} \frac{1}{C_{\{J_1\dots J_k\}}}| \{J_1\dots J_k\}\rangle\langle \{J_1\dots J_k\}|,
\ee
we may express a correlation function as a sum over overlaps in the N basis,
\begin{align}
\langle Z_{I_1} \dots& Z_{I_n} |\psi^2\rangle = \langle \hat{Z}_{I_1} \dots \hat{Z}_{I_n} |NB\rangle+ \sum_{k=1}^\infty  \langle Z_{I_1}\dots Z_{I_n}|\hat{P}_k |\psi^2\rangle
\cr
&= \langle Z_{I_1} \dots Z_{I_n} |NB\rangle+ \sum_{k=1}^n \sum_{\{J_1\dots J_k\}} \frac{\langle Z_{I_1}\dots Z_{I_n}| \{J_1\dots J_k\}\rangle}{C_{\{J_1\dots J_k\}}}\langle \{J_1\dots J_k\}|\psi^2\rangle.
\end{align}
Here we used the fact that if $|\psi\rangle$ is normalized, $\langle NB|\psi^2\rangle=1$. $\langle Z_{I_1} \dots Z_{I_n} |NB\rangle$ is equal to the correlation function in the No-Boundary state, and is a sum of products of $\disk_{I}$ and the cylinder $\delta_{IJ}$. 

Now we focus on the one and two-point functions. Using \eqref{eq:Zstatedef}, the one-point function is simply equal to the disk plus the cylinder, with one boundary of the cylinder in the state $\hat{P}_1 |\psi^2\rangle$, where $\hat{P}_1$ is the projector onto the one-universe sector of the Hilbert space. Using the orthonormality of the states $|I\rangle, \; |J\rangle$, this cylinder is just equal to the component $\langle I|\psi^2\rangle$,
\begin{align}\label{eq:onepointfunctionspeciesnormalized}
\langle Z_I\rangle_\psi &= \disk_I + \sum_J \langle Z_I |J\rangle\langle J|\psi^2\rangle
\cr
&=\disk_I + \langle I|\psi^2\rangle.
\end{align}

Subtracting off the disks for simplicity, as in \eqref{eq:Zstatedef}, the two-point function of operators $\hat{Z}_I$ and $\hat{Z}_J$ is a sum of the cylinder between the two operators, which is equal to $\delta_{IJ}$, and a pair of cylinders with two boundaries in the two-universe state $\hat{P}_2|\psi^2\rangle$,
\begin{align}\label{eq:twopointfunctionspeciesnormalized}
\langle (\hat{Z}_I-\disk_I)(\hat{Z}_J-\disk_J) \rangle_\psi &= \delta_{IJ} +\sum_{\{KL\}}\frac{ \langle Z_I Z_J | \{K L\}\rangle}{C_{\{KL\}}}\langle \{K L\} |\psi^2\rangle
\cr
&= \delta_{IJ} + \langle \{I J\} |\psi^2\rangle.
\end{align}
Here we used the fact that $\langle Z_I Z_J | \{K L\}\rangle = C_{\{KL\}} \delta_{\{KL\},\{IJ\}}$.

In the applications considered in this paper, most correlation functions of interest are correlation functions of non-independent operators $\hat{Z}[\mathcal{J}]$, such as the analytically continued partition function $\hat{Z}(\beta+iT)$. These operators can be expressed as linear superpositions of the independent operators $\hat{Z}_I$. The relation between these two operators is captured by the overlap $\langle Z[\mathcal{J}]| Z_I\rangle$ or $\langle Z[\mathcal{J}] |I\rangle$. This second overlap subtracts off the disk contributions and is computed by the cylinder with a $Z[\mathcal{J}]$ boundary and a boundary in the state $|I\rangle$.

Subtracting off the disks, we can then express a correlator of the more general $Z[\mathcal{J}]$ by ``gluing'' the $\langle Z[\mathcal{J}] |I\rangle$ cylinders to the correlators of disk-subtracted orthonormal $\hat{Z}_I$.

For the one-point function of $\hat{Z}[\mathcal{J}]$, the one-universe wavefunction $\langle I |\psi^2\rangle$ is glued to the cylinder $\langle Z[\mathcal{J}]|I\rangle$,
\be\label{eq:onepointfunctionspecies}
\langle \hat{Z}[\mathcal{J}]\rangle_\psi = \disk(\mathcal{J}) + \sum_I \langle Z[\mathcal{J}]|I\rangle \langle I |\psi^2\rangle.
\ee
Pictorally, we may represent this in a way similar to Figure \ref{OnePointFunction}. In this case the red circular boundary represents the one-universe wavefunction $\hat{P}_1 |\psi^2\rangle$, which contains single universes of many species.

The disk-subtracted two-point function of more general operators $\hat{Z}[\mathcal{J}]$, $\hat{Z}[\mathcal{J}']$, is obtained by gluing the cylinders $\langle Z[\mathcal{J}]|I\rangle$ to \eqref{eq:twopointfunctionspeciesnormalized},
\be\label{twopointfunctionspeciesnew}
\langle (\hat{Z}[\mathcal{J}]-\disk(\mathcal{J}))(\hat{Z}[\mathcal{J}']-\disk(\mathcal{J}')) \rangle_\psi = \cyl(\mathcal{J},\mathcal{J}') + \sum_{IJ} \langle Z[\mathcal{J}]|I\rangle \langle Z[\mathcal{J}']|J\rangle\langle \{IJ\} |\psi^2\rangle.
\ee
In the above equation the cylinder $\cyl(\mathcal{J},\mathcal{J}')$ comes from gluing two $\langle Z[\mathcal{J}]|I\rangle$ cylinders to the $\langle I|J\rangle = \delta_{IJ}$ cylinder,
\begin{align}\label{gluingequation}
\sum_{IJ} \langle Z[\mathcal{J}] |I\rangle \langle Z[\mathcal{J}'] |J\rangle \delta_{IJ} &= \sum_I \langle Z[\mathcal{J}] |I\rangle\langle I |Z[\mathcal{J}']\rangle
\cr
&=\langle \hat{Z}[\mathcal{J}]\hat{Z}[\mathcal{J}']\rangle_{NB,c}\equiv \cyl(\mathcal{J},\mathcal{J}').
\end{align}
In the first line we used the fact that the $\hat{Z}[\mathcal{J}]$ are Hermitian, and we can choose the overlaps $\langle Z[\mathcal{J}]|I\rangle$ to be real, and in the second line we used that the states $|Z[\mathcal{J}]\rangle$ are orthogonal to the states with $k\geq 2$ universes, so that, subtracting off contributions from disks by considering the connected correlator, the projector $\hat{P}_1 = \sum_I |I\rangle\langle I|$ can be replaced with the identity.

\begin{figure}[H]
\centering
\includegraphics[scale=0.3]{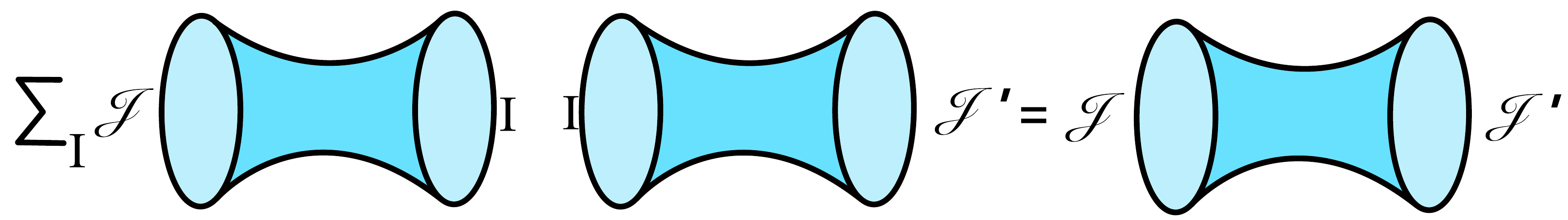}
\caption{\small We can split the path integal over cylinders with $\hat{Z}[\mathcal{J}]$ and $\hat{Z}[\mathcal{J}]$ boundaries by inserting a complete set of intermediate one-universe states.
   }
\end{figure}

\subsubsection{Factorization in the CGS model with species}\label{cgsspeciesalphastates}

In order to study factorization we now consider the case in which $|\psi\rangle = |\psi_{\{z_I\}}\rangle$ is an eigenstate of the $\hat{Z}_I$,
\be
\hat{Z}_I|\psi_{\{z_I\}}\rangle =z_I |\psi_{\{z_I\}}\rangle.
\ee
The eigenvalues $z_I$ can be thought of as random variables labeling the ensemble of boundary theories.   In the probability distribution provided by the No-Boundary state these variables are independently Gaussian with unit variance and a mean given by the disk. 

Using equation \eqref{eq:onepointfunctionspeciesnormalized}, we can identify $\langle I |\psi^2_{\{z_I\}}\rangle = z_I- \disk_I$, so the one-universe components $\langle I |\psi^2_{\{z_I\}}\rangle$ can equivalently be thought of as the random Gaussian variables characterizing the ensemble. These variables have mean zero, and are independently distributed with unit variance. Denoting averaging the $z_I$ with the No-Boundary distribution by $\mathbb{E}[\dots]_{NB}$,
\be\label{averagingspeciesdef}
\mathbb{E}\big[\langle I |\psi^2_{\{z_I\}}\rangle \big]_{NB}=0,\hspace{20pt} \mathbb{E}\big[\langle I |\psi^2_{\{z_I\}}\rangle \langle J |\psi^2_{\{z_I\}}\rangle \big]_{NB} = \delta_{IJ}.
\ee

 A single draw from this ensemble is the one-universe wavefunction $\hat{P}_1|\psi^2_{\{z_I\}}\rangle$ for a specific choice of $\{z_I\}$. As in the single-species version of the CGS model, this one-universe wavefunction contains all of the information about the full many-universe $\alpha$ state. The remaining components are related to the one-universe component by identities which generalize \eqref{eq:twouniverserelation} and \eqref{eq:threeuniverserelation}, which we will derive by demanding that correlation functions factorize.

To understand these identities intuitively, it will be useful to make an analogy between the expressions for correlation functions in this model as sums over the components of $|\psi^2_{\{z_I\}}\rangle$, and the expressions from Section \ref{periodicorbits} for products of $Z(i T)$ in periodic orbit theory as sums over orbits.

Equation \eqref{eq:onepointfunctionspecies} expresses the one point function of an operator $\hat{Z}[\mathcal{J}]$ in an $\alpha$ state as a sum over single closed universe states $I$. Typically, for an $\alpha$ state there will be many nonzero terms in this sum. We will make an analogy between this sum and the sum over orbits describing $Z(iT)$ in Section \ref{periodicorbits},
\be
\sum_a D_a e^{i S_a} \hspace{10pt} \sim \hspace{10pt} \sum_I \langle Z[\mathcal{J}]|I\rangle \langle I |\psi^2_{\{z_I\}}\rangle.
\ee
We should note that in this analogy we are not literally identifying the sum over closed universe states $I$ with an orbit sum for some system describing the CGS model. Instead we are trying to make a formal analogy between two sums over many random variables, the random phases $e^{iS_a}$ and the random compontents $\langle I |\psi^2_{\{z_I\}}\rangle$.\footnote{ We refer to Section \ref{JTeffectivemodel} for some attempts to make a more dynamical analogy.} We also emphasize that the sum $\sum_I$ over closed universe states is \textit{not} equivalent to the sum over energies $E_n$ computing $Z(iT)$ in the boundary theory.\footnote{For example, in the application to JT gravity in Section \ref{JTgravity}, we will see that the ramp in the spectral form factor is given by a diagonal sum in the $|I\rangle$ basis, while the ramp is \textit{not} given by a diagonal sum over boundary energies.}

In the periodic orbit story, the spectral form factor $Z(iT) Z(iT)^*$ is given by a double sum over orbits, which is decomposed into the diagonal and off-diagonal parts. The diagonal sum gives the ensemble average of the spectral form factor. In the gravity computation, this average is given by the cylinder topology, so we will think of the diagonal sum over orbits as analogous to the cylinder,
\be\label{oneuniverseperiodicorbits}
\sum_{a} |D_a|^2 \hspace{10pt} \sim \hspace{10pt} \cyl(\mathcal{J},\mathcal{J}').
\ee
Since we are making an analogy between the orbit sum and the sum over closed universes $I$, we should find that the diagonal part of the double sum over $I,J$ describing the product of one-point functions of operators $\hat{Z}[\mathcal{J}],\; \hat{Z}[\mathcal{J}']$ to also be approximately equal to the cylinder. In the case that the dimension $d_1$ of the single closed universe Hilbert space is large, then for typical operators $\hat{Z}[\mathcal{J}]$ and a typical $\alpha$ state this is indeed the case,
\begin{align}\label{eq:diagequalscyl}
\sum_{I=1}^{d_1} \langle Z[\mathcal{J}]|I\rangle \langle Z[\mathcal{J}']|I\rangle\; \langle I |\psi^2_{\{z_I\}}\rangle^2&=\sum_{I=1}^{d_1} \langle Z[\mathcal{J}]|I\rangle \langle I |Z[\mathcal{J}']\rangle\bigg(1 +\mathcal{O}\bigg(\sqrt{\frac{1}{d_1}}\bigg) \bigg)
\cr
&= \cyl(\mathcal{J},\mathcal{J}') +\mathcal{O}\bigg(\sqrt{\frac{1}{d_1}}\bigg).
\end{align}
In the first equality we used the reality of the matrix elements and the results of Appendix \ref{diagselfaveraging}, assuming that the $\hat{Z}[\mathcal{J}]|I\rangle$ are ``smooth''. This will be true for many natural observables which do not diagonalize the cylinder, including the observables we will study in JT gravity in Section \ref{JTgravity}. In these cases, essentially what happens is that the fluctuations of the random $\langle I |\psi^2_{\{z_I\}}\rangle^2$ around their average value add incoherently, so for large $d_1$ we can replace them by the average $\langle I |\psi^2_{\{z_I\}}\rangle^2 \rightarrow 1$ for typical choices of $\{z_I\}$.

\begin{figure}[H]
\centering
\includegraphics[scale=0.35]{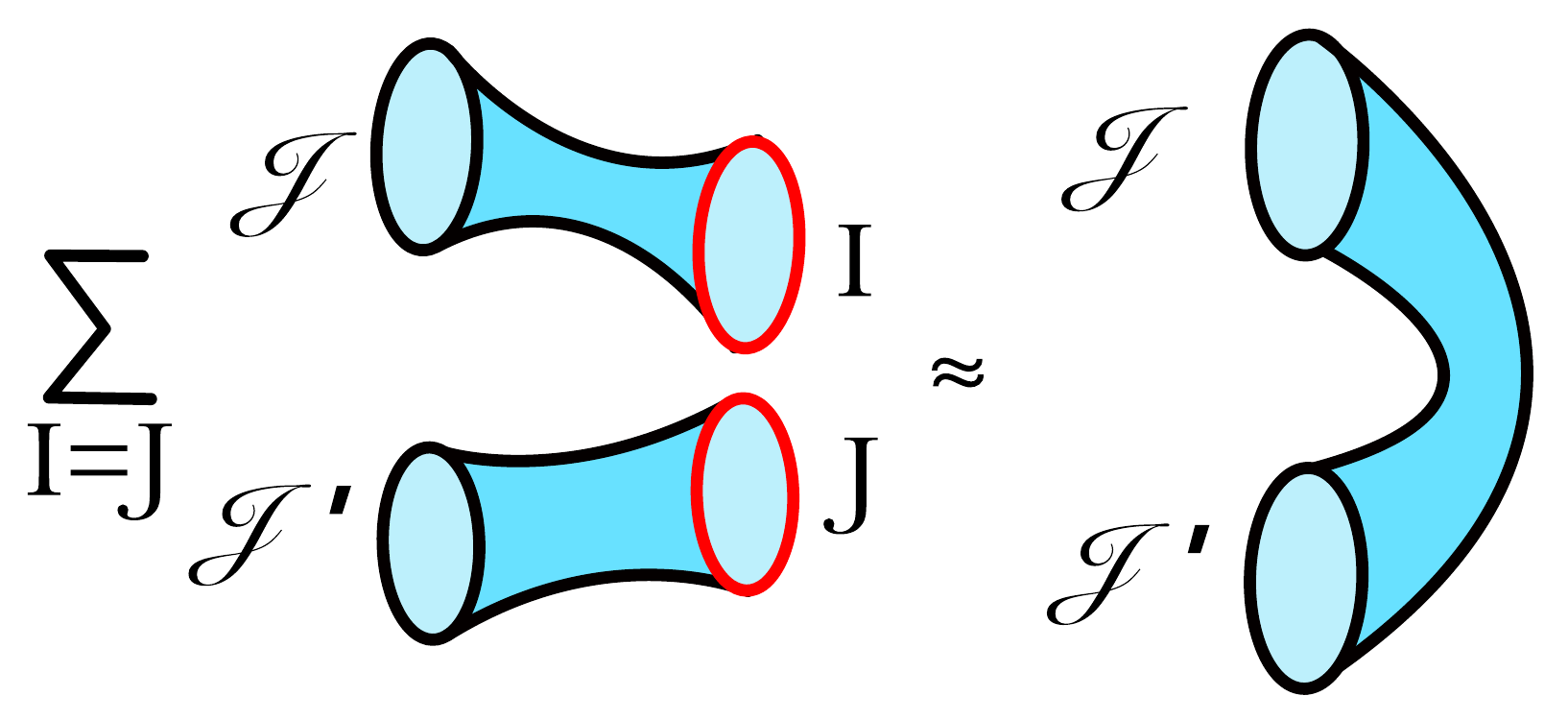}
\caption{\small For appropriate choices of operators $\hat{Z}[\mathcal{J}]$, $\hat{Z}[\mathcal{J}']$, the diagonal part of the product of cylinders $\langle Z[\mathcal{J}]|\hat{P}_1 |\psi^2_{\{z_I\}}\rangle\langle Z[\mathcal{J}']|\hat{P}_1 |\psi^2_{\{z_I\}}\rangle$ is approximately equal to the cylinder $\cyl(\mathcal{J},\mathcal{J}')$ at large $d_1$ for typical choices of $\{z_I\}$.
}
\end{figure}
We will refer to this as an approximate ``diagonal = cylinder'' identity. This further sharpens our analogy between the orbit sum and the sum over $I$; ensemble averages can be computed by restricting to diagonal sums. Though this ``diagonal = cylinder'' identity only holds in certain cases, the diagonal and off-diagonal structure exhibited in these cases serves as a useful heuristic for understanding the identities relating the components of $|\psi^2_{\{z_I\}}\rangle$. We will use these terms to describe our formulas, keeping quotation marks around the words "diagonal" and "off-diagonal" to remind the reader of our heuristic use of these terms.

With this analogy  in mind we examine the two-point function in an eigenstate, and demand that it factorize into the product of one-point functions. First we will start with the two-point function of the $\hat{Z}_I$ with the disks subtracted \eqref{eq:twopointfunctionspeciesnormalized}.
\be\label{eq:twopointfunctionspeciesnormalized2}
\langle (\hat{Z}_I-\disk_I)(\hat{Z}_J-\disk_J)\rangle_{\{z_I\}} = \delta_{IJ}+\langle \{I J\}|\psi^2_{\{z_I\}}\rangle .
\ee

Using \eqref{eq:onepointfunctionspeciesnormalized} and demanding that this two-point function factorizes, $\langle (\hat{Z}_I-\disk_I)(\hat{Z}_J-\disk_J)\rangle_{\{z_I\}} = \langle (\hat{Z}_I-\disk_I)\rangle_{\{z_I\}}\langle(\hat{Z}_J-\disk_J)\rangle_{\{z_I\}}$, we see that the one-universe components and the two-universe components of $|\psi^2_{\{z_I\}}\rangle$ must be related by
\be\label{eq:twouniversefactorizespecies}
\boxed{\langle \{I J\}|\psi^2_{\{z_I\}}\rangle = \langle I |\psi^2_{\{z_I\}}\rangle \langle J |\psi^2_{\{z_I\}}\rangle -\delta_{IJ}.}
\ee
This equation is the many-species generalization of equation \eqref{eq:twouniverserelation}, containing the essential requirement for factorization. However, in this case it is relationship between symmetric matrices with indices $I$ and $J$, with the matrix $\delta_{IJ}$ interpreted as the cylinder $\langle I|J\rangle$.

Looking at the off-diagonal components, $I\neq J$, we find
\be
\langle \{I J\}|\psi^2_{\{z_I\}}\rangle = \langle I |\psi^2_{\{z_I\}}\rangle \langle J |\psi^2_{\{z_I\}}\rangle ,\hspace{20pt} I\neq J.
\ee
This equation is a nontrivial relation between components of $|\psi^2_{\{z_I\}}\rangle$ with different universe number; the off-diagonal components of the two-universe wavefunction \textit{factorize} into products of the one-universe component. While we arrived at this relation in a simple way by demanding that the two-point functions factorize, we will see in Section \ref{secdeltafunctionspecies} and Appendix \ref{appendixexplicitcalculationspecies} that this relation is complicated in terms of the geometric representation of the states $|\psi^2_{\{z_I\}}\rangle$.

While the off-diagonal components of $\langle \{I J\}|\psi^2_{\{z_I\}}\rangle$ factorize, the fact that the cylinder is diagonal in the $I,J$ indices picks out the diagonal, $I=J$, components of $\langle \{I J\}|\psi^2_{\{z_I\}}\rangle$ as special. These components are related by
\be
\langle \{I I\}|\psi^2_{\{z_I\}}\rangle = \langle I |\psi^2_{\{z_I\}}\rangle^2 -1.
\ee
The factor of one subtracted on the RHS corresponds to the cylinder $1/\langle 1|1\rangle$. Using \eqref{averagingspeciesdef}, the average of the RHS is zero, indicating that on average, the diagonal part of the product of one-universe components compensates for the cylinder so that the two-universe component is off-diagonal on average. The fluctuations of $\langle \{I I\}|\psi^2_{\{z_I\}}\rangle$ around their average value of zero are nonzero. However, the fluctuations are independent for different $I$. In applying this formula to correlators of operators $\hat{Z}[\mathcal{J}]$, we will find sums over these diagonal components, which then add incoherently. Using the results in Appendix \ref{diagselfaveraging}, we will see that in appropriate correlators, we will be able to ignore the diagonal components and treat them as if they were zero. In this situation, the "diagonal=cylinder" identity allows us to heuristically view the two-universe components as purely "off-diagonal". 

This fits with our analogy to periodic orbits; in order for the two-point function to factorize, we must add an "off-diagonal" sum to the (diagonal) cylinder, with the off-diagonal summands being simply a product of summands for the one-point functions.

To see this in more detail, we apply equation \eqref{eq:twouniversefactorizespecies} to the two-point functions of more general operators $\hat{Z}[\mathcal{J}]$. Gluing the cylinders $\langle Z[\mathcal{J}]|I\rangle$ to \eqref{eq:twouniversefactorizespecies}, we find
\begin{align}\label{eq:twouniversecomparisonspecies}
\sum_{I,J=1}^{d_1} \langle Z[\mathcal{J}]|I\rangle\langle Z[\mathcal{J}'] |J \rangle \;\langle \{IJ\} |\psi^2_{\{z_I\}}\rangle &=
\cr
 \bigg( \sum_{I=1}^{d_1} \langle Z[\mathcal{J}]| I \rangle\langle I |\psi^2_{\{z_I\}}\rangle \bigg) \bigg( \sum_{J=1}^{d_1} \langle Z[\mathcal{J}']| J \rangle\langle J |\psi^2_{\{z_I\}}\rangle \bigg) &- \cyl(\mathcal{J},\mathcal{J}').
\end{align}

\begin{figure}[H]
\centering
\includegraphics[scale=0.35]{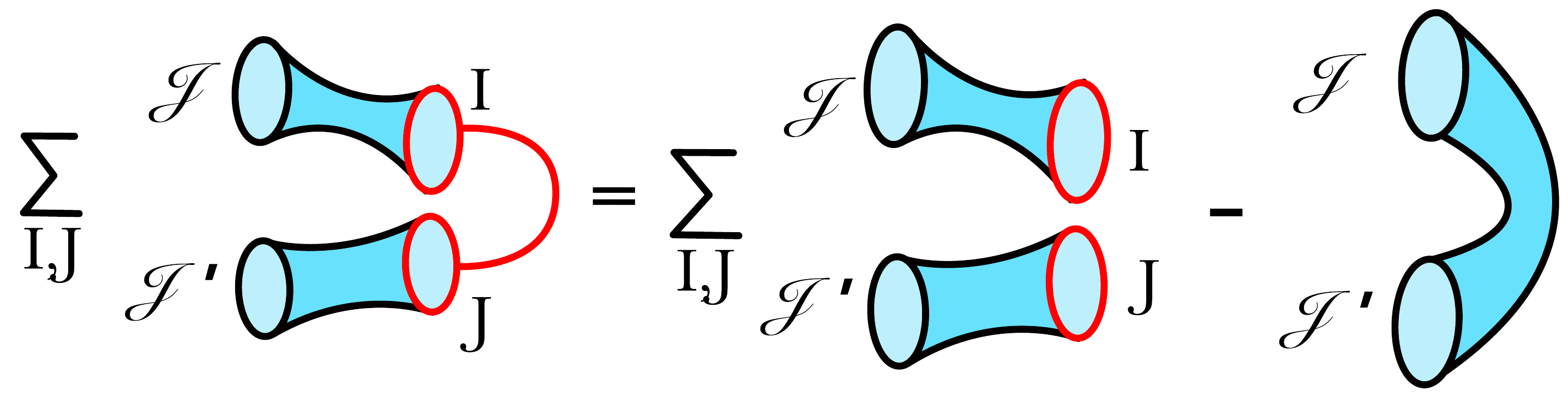}
\caption{\small The identity \eqref{eq:twouniversefactorizespecies} implies the pictured relationship between contributions to two-point functions and one-point functions in an $\alpha$ state. The linked red boundaries denote the two-universe component of $|\psi^2_{\{z_I\}}\rangle$, while the unlinked red boundaries denote the one-universe component.}
\label{figuretwouniversefactor}
\end{figure}
This equation is precise, independent of $d_1$. However, at large $d_1$ we can use the ``diagonal=cylinder'' identity to connect more sharply with the periodic orbits story and allow us to simplify the description of factorization.

The (disk subtracted) two-point function is the sum of the cylinder $\cyl(\mathcal{J},\mathcal{J}')$ and the contribution of the two-universe component $\sum_{IJ} \langle Z[\mathcal{J}]|I\rangle\langle Z[\mathcal{J}'] |J \rangle \;\langle \{IJ\} |\psi^2_{\{z_I\}}\rangle$. This second contribution is a double-sum over closed universe states $I$ and $J$. Since we are identifying the diagonal part of the orbit sum with the cylinder, we should identify the off-diagonal part of the sum with the contribution of the two-universe component.

Along with \eqref{eq:diagequalscyl}, \eqref{eq:twouniversecomparisonspecies} tells us that this identification is sound. The diagonal part of the double-sum in \eqref{eq:twouniversecomparisonspecies} is small for large $d_1$, so the double-sum is essentially off-diagonal.
\begin{figure}[H]
\centering
\includegraphics[scale=0.38]{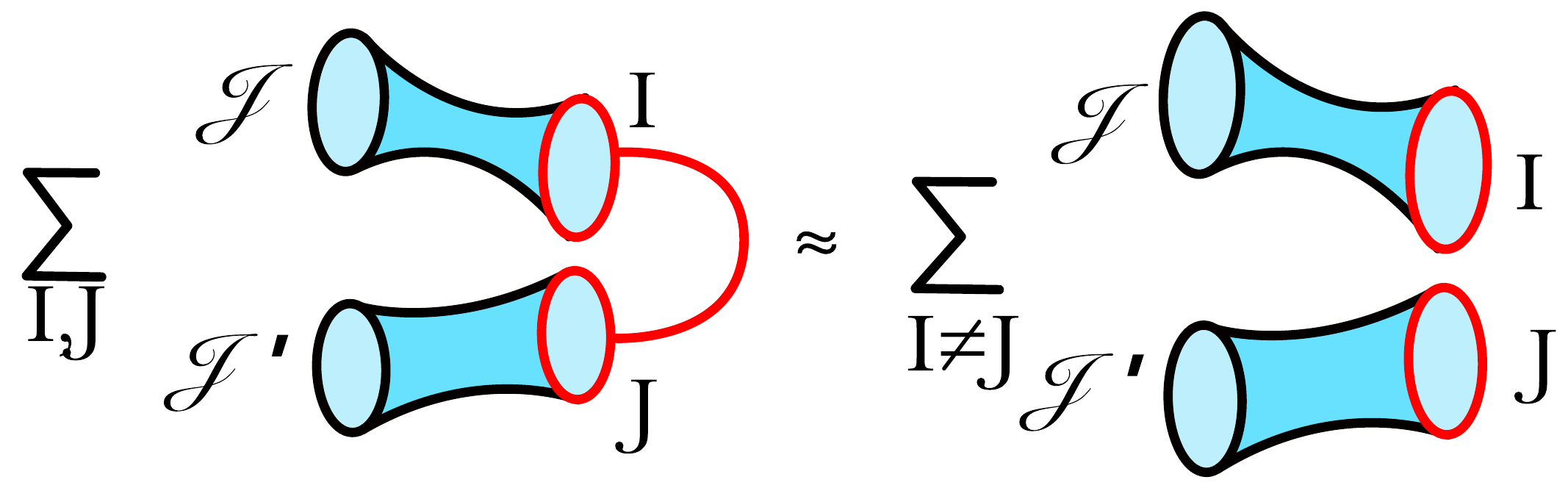}
\caption{\small When the "diagonal=cylinder" identity applies, the identity relating contributions to the two-point function and one-point function simplies; the contribution on the LHS to the two-point function is approximately equal to the off-diagonal part of the contribution of the $\alpha$ state to the product of one-point functions. In the computation of the two-point function, the "off-diagonal" LHS is added to the "diagonal" cylinder, yielding a factorizing double-sum.}
\label{figureoffdiag}
\end{figure}

To summarize, we can see the factorization of the two-point function by denoting $(\text{1-universe})_I \equiv \langle\hat{Z}[\mathcal{J}]|I\rangle\langle I|\psi^2_{\{z_I\}}\rangle$ and $(\text{2-universe})_{IJ} \equiv \langle\hat{Z}[\mathcal{J}]\hat{Z}[\mathcal{J'}]|I\rangle\langle \{IJ\}|\psi^2_{\{z_I\}}\rangle$, and schematically writing
\begin{align}
\langle (\hat{Z}-\disk)^2 \rangle_{\{z_I\}}& = \hspace{35pt}\cyl\hspace{39pt}+ \hspace{10pt}\sum_{IJ} (\text{2-universe})_{IJ}
\cr
&\approx \sum_{I=J}  (\text{1-universe})_I^2 \hspace{10pt}+\hspace{10pt} \sum_{I\neq J}   (\text{1-universe})_I  (\text{1-universe})_J
\cr
&=\sum_{IJ}  (\text{1-universe})_I  (\text{1-universe})_J
\cr
&= \bigg(\sum_I (\text{1-universe})_I  \bigg)^2 = \big(\langle \hat{Z}-\disk\rangle_{\{z_I\}} \big)^2.
\end{align}
We emphasize that though at large $d_1$ the calculation approximately has this ``diagonal + off-diagonal'' structure, giving a simple picture of factorization, \eqref{eq:twouniversefactorizespecies} allows the two-point function to factorize even at small $d_1$.

The ``off-diagonal'' structure of the two-universe wavefunction $\langle \{IJ\}|\psi^2_{\{z_I\}}\rangle$ described by \eqref{eq:twouniversefactorizespecies} has a geometrical origin. To see this, compare the computations of the product of one-universe components $\langle I|\psi^2_{\{z_I\}}\rangle \langle J|\psi^2_{\{z_I\}}\rangle$ and a two-universe wavefunction $\langle \{I J\}|\psi^2_{\{z_I\}}\rangle$. 

$\langle I|\psi^2_{\{z_I\}}\rangle$ is computed by the sum over spacetimes with a boundary in the state $|I\rangle$ and boundaries in the state $|\psi^2_{\{z_I\}}\rangle$. The $|I\rangle$ boundary can connect to just one of the $|\psi^2_{\{z_I\}}\rangle$ boundaries at a time with a cylinder, and the component $|I\rangle$ is equal to a sum over the $|\psi^2_{\{z_I\}}\rangle$ boundaries the $|I\rangle$ boundary connects to. The product of components $\langle I|\psi^2_{\{z_I\}}\rangle \langle J|\psi^2_{\{z_I\}}\rangle$ can then be thought of as a double-sum over the $|\psi^2_{\{z_I\}}\rangle$ boundaries that the $|I\rangle$ and $|J\rangle$ boundaries connect to.

\begin{figure}[]
\centering
\includegraphics[scale=0.39]{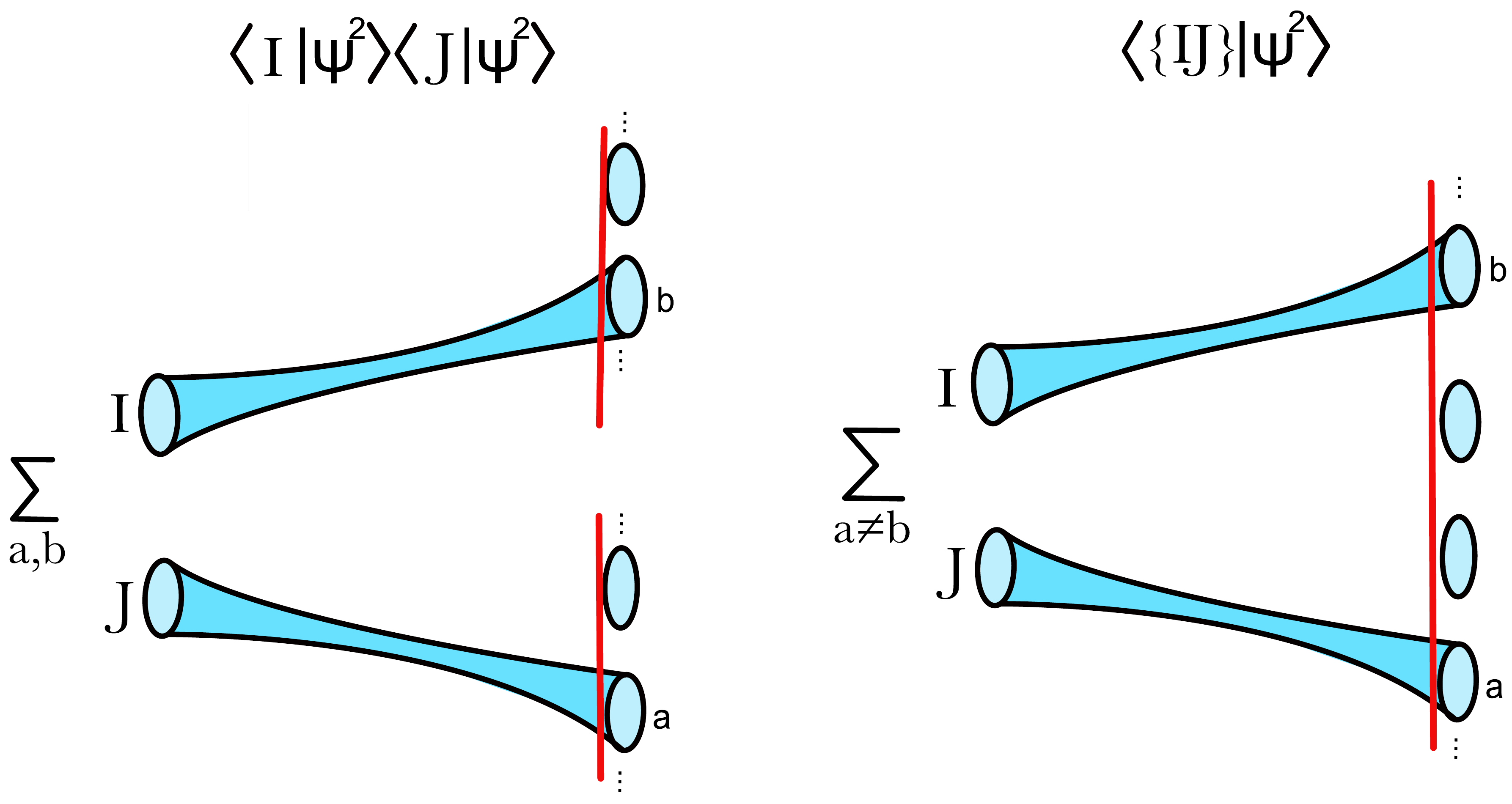} 
\caption{\small Here we compare the computation of the product of one-universe components $\langle I|\psi^2\rangle \langle J|\psi^2\rangle$ with the two-universe component $\langle \{IJ\}|\psi^2\rangle$. In the computation of the one-universe component, the $|I\rangle$ boundary connects to a boundary in $|\psi^2\rangle$  labeled by $a$ with a cylinder and the $|J\rangle$ boundary connects to $b$. For fixed $a,b$, we sum over the spacetimes connecting to the remaining boundaries in $|\psi^2\rangle$. Then we sum over the choices $a,b$. In the computation of the two-universe component, we similarly sum over spacetimes for which the $|I\rangle$ and $|J\rangle$ boundaries connect to boundaries in $|\psi^2\rangle$, labeled by $a,b$ with cylinders. However, in this case the $|\psi^2\rangle$ boundaries are shared, and there are no contributions from $a=b$; these "diagonal" terms are "excluded". In an $\alpha$ state $|\psi_{\{z_I\}}\rangle$, the state is tuned so that this exclusion results in the diagonal/off-diagonal structure in $I,J$, \eqref{eq:twouniversefactorizespecies}.
}
\label{exclusioneffect}
\end{figure}

The two-universe component $\langle \{I J\} |\psi^2_{\{z_I\}}\rangle $ is given by a similar computation, where we sum over spacetimes with an $|I\rangle$ boundary and a $|J\rangle$ boundary, but now these boundaries ``share'' the $|\psi^2_{\{z_I\}}\rangle$ boundaries. When we sum over the $|\psi^2_{\{z_I\}}\rangle$ boundaries that $|I\rangle$ and $|J\rangle$ connect to with cylinders, we do not include the \textit{diagonal} term in the sum, since the $|I\rangle$ and $|J\rangle$ boundaries cannot connect to the same boundary with a cylinder simultaneously.\footnote{If we had included more complicated topologies, as in the MM model,  there could be such a connection, but it would involve a more complicated topology like the three-holed sphere. This would be  suppressed because of its Euler character.}

We will refer to this phenomenon as the ``exclusion effect'' and is illustrated in Figure \ref{exclusioneffect}.

This geometrical picture is rough, and does not explain other aspects of the relation between the one-universe and two-universe components of $|\psi^2_{\{z_I\}}\rangle$. For example, it is clear geometrically why the two-universe component is generally not equal to the product of one-universe components. However, the fact that the two-universe component factorizes precisely into the product of one-universe components is a result of a fine-tuning in the many-universe state.\footnote{This property is more clearly understood using the explicit description of $|\psi^2_{\{z_I\}}\rangle$ as a regulated delta-function in \ref{secdeltafunctionspecies}.} However, identifying the geometric origin of the difference between the linked and unlinked pair of red boundaries in \ref{figuretwouniversefactor} and \ref{figureoffdiag} (the two-universe component of $|\psi^2_{\{z_I\}}\rangle$ and the product of one-universe components) will be important in our effective description of factorization in Section \ref{effectivemodel}. In Appendix \ref{appendixsharing}, we will also use this exclusion effect in a more general context to emphasize the importance of the "sharing" of the $\alpha$ state boundaries in allowing correlation functions to factorize.

\subsubsection{Factorization of higher-point correlators}
As an aside, we briefly discuss the factorization of higher point correlators.

To generalize equation \eqref{eq:twouniversefactorizespecies} to higher point correlators, it is useful to first express this equation in a basis-independent way. Viewing the multi-universe states $|\{I_1\dots I_n\}\rangle$ as (symmetrized) tensor products of single-universe states, and introducing the maximally  entangled state of two universes $|MAX\rangle = \sum_I |I\rangle\otimes |I\rangle$, we can express \eqref{eq:twouniversefactorizespecies} in a basis independent way as
\be\label{twouniversefactorizebasisindep}
\hat{P}_2|\psi^2_{\{z_I\}}\rangle = (\hat{P}_1 |\psi^2_{\{z_I\}}\rangle)^{\otimes 2} - |MAX\rangle.
\ee
The projection operators onto $n$-universe states $\hat{P}_n$ were defined in equation \eqref{projectordef}.
Geometrically, the state $|MAX\rangle$ corresponds to the state of two universe produced by the cylinder spacetime, and $\langle Z[\mathcal{J}] Z[\mathcal{J}']|MAX\rangle = \cyl(\mathcal{J},\mathcal{J}')$. 

We can think of the lack of factorization of correlators in the No-Boundary state as follows: The cylinder corresponds to the pair creation of two closed universes in the maximally entangled state, and the contribution of the cylinder to correlation functions describes the absorption of these entangled universes by the operators in the correlator. The entanglement between different closed universes then spoils factorization.

In order for correlators to factorize, the closed universes absorbed by the operators in the correlator must not be entangled. We can think{} of this many-universe state as the sum of an $n$-universe component of $|\psi^2_{\{z_I\}}\rangle$ and the state of pairwise-entangled universes created by cylinders. Then the $n$-universe component of $|\psi^2_{\{z_I\}}\rangle$ must be a product state, $(\hat{P}_1|\psi^2_{\{z_I\}}\rangle)^{\otimes n}$, minus corrections which cancel the pairwise-entangled cylinder states, somewhat as in \eqref{eq:threeuniverserelation}. For example, 
\be\label{threeuniversefactorizebasisindep}
\hat{P}_3 |\psi^2_{\{z_I\}}\rangle = (\hat{P}_1|\psi^2_{\{z_I\}}\rangle)^{\otimes 3} - {3\choose 2}\mathcal{S}\big[ (\hat{P}_1 |\psi^2_{\{z_I\}}\rangle) \otimes |MAX\rangle\big].
\ee
where the operator $\mathcal{S}[\dots]$ symmetrizes its argument. This can be derived by demanding that the three-point function, which probes the three-universe component of the many-universe state, factorizes.

\subsubsection{Explicit representation of the $\alpha$ state}\label{secdeltafunctionspecies}

We now briefly discuss the computation of correlators using the explicit (regularized) delta function representation of an $\alpha$-state, generalizing the results of Section \ref{eq:twouniversefactorizespecies} to the many-species case. 

We can represent $\psi(\hat{Z}_I)$ for an approximate eigenstate of the $\hat{Z}_I$ with eigenvalues $z_{I}$ as a product of regularized delta functions,
\be
\psi_{\{z_I\},\Delta}(\hat{Z}_I) \propto \int_{-\infty}^\infty dp_I\; e^{-\Delta^2\sum_I p_I^2 +  i \sum_I p_I (\hat{Z}_I- z_I)}.
\ee
We have chosen the same width for each of the Gaussian regularizations of the delta functions. Squaring this to obtain $\psi_{\{z_I\},\Delta}(\hat{Z}_I)^2$, we can change variables to the sum and difference variables of the $p_I$ and $p_I'$ and integrate out the difference variables to find an expression for $\psi_{\{z_I\},\Delta}(\hat{Z}_I)^2$ as a product of Gaussians integrals in the sum variables. This is just many copies of the single delta function expression \eqref{eq:explicitalphastate} for $\psi_{z}(\hat{Z})^2$ in the one-species model. 

Now we can normal order the exponential $e^{i \sum_I \hat{Z}_I}$ to find an expression for the alpha state in the N-basis $|I_1\dots I_k\rangle$,
\be \label{psi2spec}
\psi_{\{z_I\},\Delta}(\hat{Z}_I)^2|NB\rangle \propto \int_{-\infty}^\infty dp_{+I} e^{ \frac{\Delta+1}{2} \sum_I p_{+I}^2 + i \sum_I p_{+I}(\disk_I - z_I) } \sum_{k=0}^\infty \sum_{\{I_1\dots I_k\}} \frac{(i p_{+I_1})\dots (i p_{+I_k})}{C_{\{I_1\dots I_k\}}} |\{I_1\dots I_k\}\rangle.
\ee
We can safely take the limit $\Delta\rightarrow 0$, so that the Gaussian integrals have a width determined by the cylinder, which is equal to one.

The components $\langle \{I_1\dots I_k\}|\psi^2_{\{z_I\}}\rangle$ are then given by moments of a multi-dimensional Gaussian integral. The Wick contractions describing{} these moments exactly match the pattern \eqref{twouniversefactorizebasisindep},\eqref{threeuniversefactorizebasisindep}, which we expect to describe these components, with the identification $\langle I|\psi^2_{\{z_I\}}\rangle = z_I-\disk_I$.

In Appendix \ref{appendixexplicitcalculationspecies}, we discuss the explicit computation of the factorization of the two-universe components into products of one-universe components in more detail.

\subsection{An effective description with random boundaries}\label{effectivemodel}

In the CGS model, as well as in the MM model and JT gravity, an $\alpha$ state is a complicated many-universe state. The full description of factorization in these states depends crucially on the many-universe structure, with identities such as \eqref{eq:twouniversefactorizespecies} resulting from the "exclusion effect" in a fine-tuned many-universe state. On the other hand, if we take the identities such as \eqref{eq:twouniversefactorizespecies} as given, the many-universe structure of the theory does not play a large role; correlation functions can be described simply in terms of the one-universe state $\hat{P}_1|\psi^2_\alpha\rangle$. The components of this one-universe state have simple statistics,  they are independent Gaussian variables, so the ensemble of boundary theories is also more simply described in terms of this one-universe state.

Furthermore, these relations often simplify when applied to correlators at large dimension $d_1$ of the one-universe Hilbert space, when the $n$-universe components in a non-orthormal basis described by operators $\hat{Z}[\mathcal{J}]$ have an ``off-diagonal'' structure. In this case, the description of correlators in terms of the one-universe wavefunction $\hat{P}_1 |\psi^2_\alpha\rangle$ becomes especially simple, resembling the periodic orbits story. The cylinder behaves like a ``diagonal'' sum, the two-universe component behaves like an ``off-diagonal'' sum, and they add to a factorizing double-sum.

In contrast to the periodic orbits story, though,  the (diagonal) cylinder is a \textit{fundamental} part of the model. We add the two-universe component to the cylinder, which is then required to have an ``off-diagonal'' structure, as pictured in Figure \ref{figureoffdiag}. The ``exclusion effect'' underlying this relation has a fundamentally many-universe explanation. It is the essential way in which the many-universe character of the theory enters into the calculation of correlators in an $\alpha$ state.

In this section we introduce an effective model meant to describe the CGS model in an $\alpha$ state, but not in a more general state of the closed universes. This effective description takes advantage of this simple description of the CGS model in an $\alpha$ state in terms of one-universe states $\hat{P}_1|\psi^2_\alpha\rangle$. In the effective model, we replace the contributions from the random many-universe $\alpha$ state with contributions from a small number of random "$\Psi$" boundaries; to compute a product of $n$ partition functions, we need only $n$ $\Psi$ boundaries.

For a given computation, there are different choices of effective description which give the same result. These choices correspond in the full theory to the choice of computing an $n$-point function using a manifestly factorized description or not, $\langle \hat{Z}[\mathcal{J}]^n\rangle_\alpha = \langle \hat{Z}[\mathcal{J}]\rangle_\alpha^n $. 

In one version of the effective model, we only allow "unlinked" $\Psi$ boundaries in the path integral. Each unlinked $\Psi$ boundary comes with a random boundary condition, fixed to be the same for each boundary. A single $\Psi$ boundary replaces the contribution of one-universe states $\hat{P}_1|\psi^2_\alpha\rangle$. For example, the contribution of one $\Psi$ boundary to a partition function is a "broken cylinder"- a cylinder with a $\hat{Z}[\mathcal{J}]$ boundary, and a $\Psi$ boundary. This broken cylinder replaces the contribution of the one-universe components $\langle I|\psi^2_{\{z_I\}}\rangle$, \eqref{oneuniverseperiodicorbits}, in the full model. In this version of the effective model, we disallow, or "exclude", cylinders that connect partition function boundaries. Then this version of the effective model corresponds to the manifestly factorized computation of correlation functions in the full theory.

In the other version of the effective model, we allow cylinders connecting partition functions, but introduce "linked" $\Psi$ boundaries rather than unlinked boundaries. As in the other version of the effective model, we sum over spacetimes with different numbers of circular $\Psi$ boundaries; however, linked $\Psi$ boundaries come with a different boundary condition. Rather than giving each $\Psi$ boundary the same random boundary condition, the boundary condition is "shared" between the linked $\Psi$ boundaries. Roughly, the linked $\Psi$ boundary condition is like a product of unlinked boundary conditions, with the "diagonal" contributions "excluded". These linked $\Psi$ boundaries then replace the contributions of the $n$-universe components $\hat{P}_n|\psi^2\rangle$, and this version of the effective description corresponds to the not-manifestly-factorized computation of correlators in the full theory.

This distinction between linked and unlinked $\Psi$ boundaries is a consequence of abandoning the many-universe description of the full theory. In the effective model, we have a choice between using unlinked $\Psi$ boundaries, but excluding the full cylinders, or using linked $\Psi$ boundaries with the diagonal components excluded. This ad-hoc "exclusion rule", which instructs us to exclude either the cylinder, or the "diagonal" piece of the linked $\Psi$ boundaries (related by a "diagonal=cylinder identity"), is put in by hand in this effective model, but has a natural geometrical origin in the full theory. For example, the "off-diagonal" structure of the two-universe components of $|\psi^2_\alpha\rangle$ results from the geometric exclusion preventing two cylinders from ending on the same $\alpha$ state boundary.

This effective description will turn out to be useful for describing JT gravity, discussed in Section \ref{JTgravity}. The effective description of JT gravity in an $\alpha$ state has similar advantages and disadvantages to the effective description of the CGS model; in particular the ``exclusion rule'' must be explained by appealing to the full many-universe description of JT gravity. However, JT gravity could possibly serve as a laboratory for exploring some generalizations in which the random $\Psi$ boundaries are replaced by pseudorandom boundary conditions, with the pseudorandomness resulting from chaotic bulk dynamics.

This effective description is similar to the description of the SYK model for a fixed choice of couplings. However, in the SYK model the distinction between different versions of the effective model, with linked and unlinked $\Psi$ boundaries, is explained in a simple, but nongeometrical way in the full theory, as discussed in \cite{Saad:2021rcu}.  We briefly review these ideas in the \hyperref[discussion]{\textbf{Discussion}}.

\subsubsection{Definition of the effective model}

Now we describe this effective description in more detail. This effective description is meant to capture the behavior of correlation functions of the CGS model in an $\alpha$ state, but without using the apparatus of the many-closed-universe quantum mechanics. Since we will no longer be making use of many-universe states, we will no longer think about the observables we are computing as correlation functions. In this effective description, partition functions always factorize, so we can think of them as numbers,\footnote{We could also think of the partition functions computed in the effective description as correlation functions in the No-Boundary state of the bulk theory with $\Psi$ boundaries. We will comment on this perspective in the Appendix \ref{appendixsharing}.}
\be
\langle \hat{Z}[\mathcal{J}_1] \dots  \hat{Z}[\mathcal{J}_n]\rangle_\alpha \longrightarrow Z^{(\Psi)}[\mathcal{J}_1] \dots  Z^{(\Psi)}[\mathcal{J}_n].
\ee
Here $Z^{(\Psi)}[\mathcal{J}]$ is a fixed number. The symbol $\Psi$ denotes the choice of the element of the ensemble of boundary theories to which the $\alpha$ state corresponds, so we should identify $Z^{(\Psi)}[\mathcal{J}] = \langle \hat{Z}[\mathcal{J}]\rangle_\alpha$. We are using the symbol $\Psi$ instead of $\alpha$ to label members of the ensemble in order to emphasize that we do not think of $\Psi$ as described by some many-universe $\alpha$ state.

Now we give a prescription for computing the observables $Z^{(\Psi)}[\mathcal{J}_1]\dots Z^{(\Psi)}[\mathcal{J}_n]$ using this effective description. First, we discuss the version of the effective model with no cylinders, and only unlinked $\Psi$ boundaries.

\begin{figure}[H]
\centering
\includegraphics[scale=0.4]{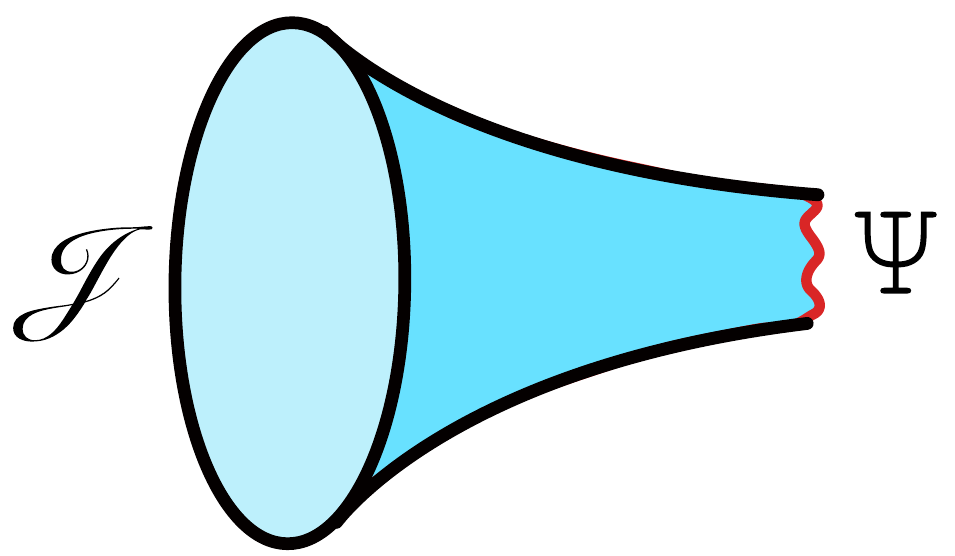} 
\caption{\small Here we have pictured the contribution to $Z^{(\Psi)}[\mathcal{J}]$ from one $\Psi$ boundary, denoted by the wiggly red boundary. This ``broken cylinder'' replaces the contribution of the one-universe components of $|\psi^2_\alpha\rangle$ in the full model, such as the cylinder with the red circular boundary in Figure \ref{OnePointFunction}.
}
\label{sumpsiboundaries}
\end{figure}

$Z^{(\Psi)}[\mathcal{J}]$ is computed by a path integral over spacetimes with each connected component has a boundary with a $\mathcal{J}$ boundary condition; we do not include spacetimes disconnected from these boundaries. In the sum over geometries, we also freely include any number of ``$\Psi$ boundaries''.\footnote{The rules for including $\Psi$ boundaries in the path integral are similar to the rules for including EOW brane loops in a modification of the MM model \cite{Marolf:2020xie}.} Restricting our attention to the case that the full theory is a two-dimensional theory of gravity, these are circular boundaries with an identical random boundary condition labeled by $\Psi$,  as pictured in Figure \ref{sumpsiboundaries}.\footnote{In this effective model we relax the earlier condition that we consider states of closed universes only on asymptotic boundaries, so that we may think of the $\Psi$ boundary as living on a ``bulk'' circular boundary.}. We refer to such configurations as ``broken cylinders.''\footnote{As we explain in the \hyperref[discussion]{\textbf{Discussion}}, these broken cylinders are closely related to the "half-wormholes" found in \cite{Saad:2021rcu}.}

Computations in this effective model then resemble computations of correlators in the full theory, with the $\Psi$ boundaries replacing the boundaries on which the components of $|\psi^2_\alpha\rangle$ lived. However, physically the $\Psi$ boundaries and the boundaries introduced by the state $|\psi^2_\alpha\rangle$ have very different properties. Most importantly, in the full CGS model, different $\hat{Z}[\mathcal{J}]$ boundaries ``share'' the same $|\psi^2_\alpha\rangle$ boundaries, and as a result, the components of $|\psi^2_\alpha\rangle$ with more than one universe do not factorize. The exclusion effect, illustrated in Figure \ref{exclusioneffect}, relies crucially on this fact.

In this version of the effective model, partition functions do \textit{not} "share" $\Psi$ boundaries. For example, the computation of two copies of a partition function $Z^{(\Psi)}[\mathcal{J}]Z^{(\Psi)}[\mathcal{J}']$ includes a contribution from two $\Psi$ boundaries, with each partition function connecting to a $\Psi$ boundary with a "broken cylinder". The analogous contribution in the full model involves the two-universe component of $|\psi^2_\alpha\rangle$, with each partition function connecting to one of the two universes in $\hat{P}_2|\psi^2_\alpha\rangle$ with a cylinder. But because of the exclusion effect in the full theory, this two-universe component had the cylinder subtracted out. In this effective model, the $\Psi$ boundaries contribute separately. Because pairs of $\Psi$ boundaries do not have an exclusion effect, we must disallow the cylinders connecting partition function boundaries \textit{by hand}. We must also disallow disconnected spacetimes with just $\Psi$ boundaries. Then in this version of the effective model, a product of $n$ partition functions manifestly factorizes,
\be
Z^{(\Psi)}[\mathcal{J}_1] \dots  Z^{(\Psi)}[\mathcal{J}_n] = \big(\disk(\mathcal{J}_1)+ \text{Broken cylinder})\dots \big(\disk(\mathcal{J}_n)+ \text{Broken cylinder}).
\ee
Note that for the computation of $n$ partition functions, we only need to include up to $n$ $\Psi$ boundaries. This is in contrast to the full model.  There, even though we only needed the components of $|\psi^2_\alpha\rangle$ with up to $n$ universes, each of these components is computed by a sum over spacetimes with arbitrarily many boundaries. This dramatic reduction in the number of boundaries is a simplifying feature of this effective model.

In order to match the effective model with the full CGS model we need a dictionary between the contributions of $\Psi$ boundaries and the components of $|\psi^2_\alpha\rangle$. To do this, it is convenient to describe the $\Psi$ boundary condition in terms of a vector $|\Psi\rangle$ which lives in a Hilbert space isomorphic to the one-closed-universe Hilbert space. We may do this because $\Psi$ and a one-universe state $|\Psi\rangle$ both correspond to boundary conditions for bulk fields on a circular spatial slice.

To match with the full theory in an $\alpha$ state $|\psi_\alpha\rangle$, we take $|\Psi \rangle = \hat{P}_1|\psi_\alpha^2\rangle$.\footnote{We emphasize that in the effective description, we should not think of $|\Psi\rangle$ as coming from a many-universe $\alpha$ state.} Using the symbol $\mathbb{E}[\dots]_\Psi$ to denote averaging over the $\Psi$ boundary conditions, the components of $|\Psi\rangle$ obey\footnote{Here $\langle I|\Psi\rangle$ is given by the cylinder with a $\hat{Z}_I$ boundary and a $\Psi$ boundary.}
\be\label{eq:psistatistics}
 \mathbb{E}[\langle I|\Psi\rangle]=0, \hspace{20pt} \mathbb{E}[\langle I|\Psi\rangle \langle J|\Psi\rangle]=\delta_{IJ}.
\ee
$Z^{(\Psi)}[\mathcal{J}]$ is computed by a sum over two topologies, the disk and the cylinder with a $\Psi$ boundary. By using the description of the $\Psi$ boundary condition as a random vector $|\Psi\rangle$, we can express the one-$\Psi$ boundary contribution to $Z^{(\Psi)}[\mathcal{J}]$, the "broken cylinder", as an overlap $\langle Z[\mathcal{J}] | \Psi\rangle$, computed using the GGS theory without $\Psi$ boundaries.
\be\label{eq:effectivemodelz}
Z^{(\Psi)}[\mathcal{J}] = \disk(\mathcal{J}) + \langle Z[\mathcal{J}] | \Psi\rangle.
\ee

\begin{figure}[H]
\centering
\includegraphics[scale=0.3]{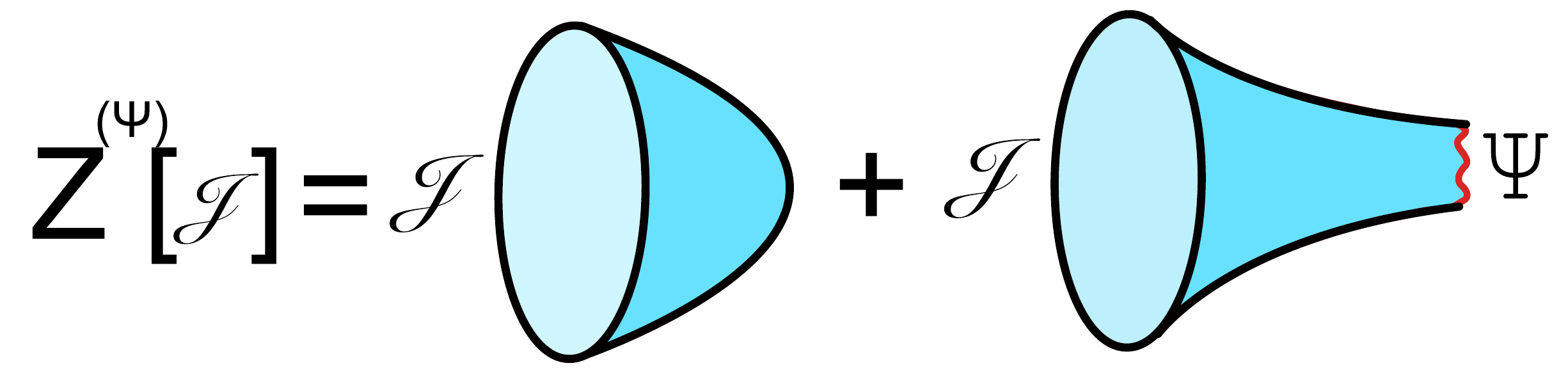} 
\caption{\small $Z^{(\Psi)}[\mathcal{J}]$ is given by a sum over spacetimes with zero or one $\Psi$ boundaries.}
\end{figure}
We can see that the broken cylinder directly replaces the contribution of the one-universe state $\hat{P}_1|\psi^2_\alpha \rangle$ to the correlator in the full CGS model.

\begin{figure}[H]
\centering
\includegraphics[scale=0.3]{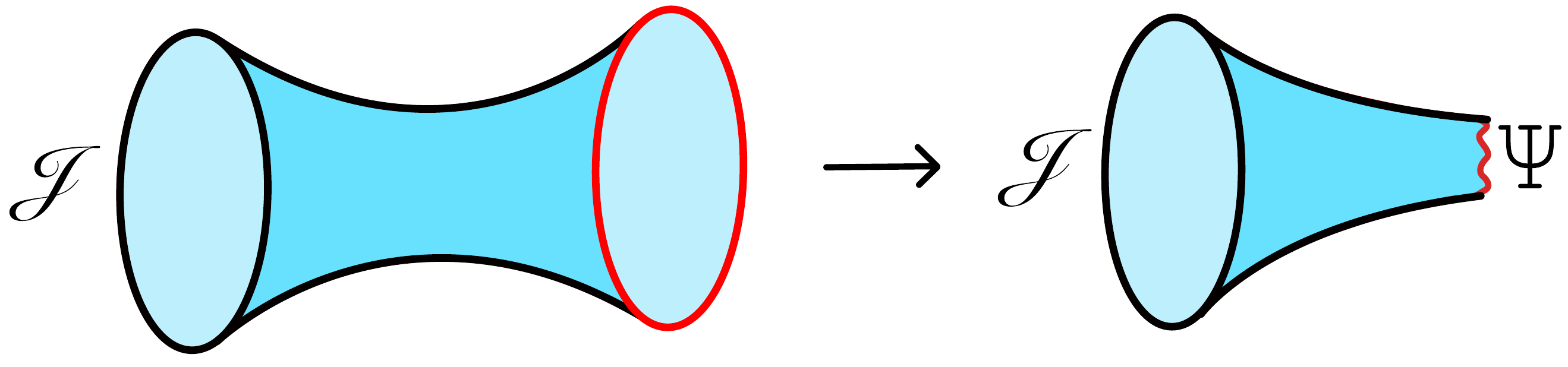} 
\caption{\small The "broken cylinder" with a $\Psi$ boundary replaces the cylinder with a circular red boundary pictured in Figure \ref{OnePointFunction}.}
\end{figure}

We can check that upon ensemble averaging, this effective model matches with the full theory. Using \eqref{eq:psistatistics},
\begin{align}
\mathbb{E}\big[ Z^{(\Psi)}[\mathcal{J}]Z^{(\Psi)}[\mathcal{J}]\big]_\Psi &= \mathbb{E}\big[ \big(\disk(\mathcal{J}) +  \langle Z[\mathcal{J}] | \Psi\rangle\big)\big(\disk(\mathcal{J}') +  \langle Z[\mathcal{J}'] | \Psi\rangle\big)\big]_\Psi
\cr & = \disk(\mathcal{J})\disk(\mathcal{J}') + \sum_{IJ} \langle Z[\mathcal{J}] |I\rangle\langle Z[\mathcal{J}'] |J\rangle\; \mathbb{E}\big[\langle I|\Psi\rangle \langle J|\Psi\rangle\big]_\Psi
\cr
&= \disk(\mathcal{J})\disk(\mathcal{J}')+ \sum_I \langle Z[\mathcal{J}] |I\rangle\langle Z[\mathcal{J}'] |I\rangle
\cr
&= \disk(\mathcal{J})\disk(\mathcal{J}')+ \cyl(\mathcal{J},\mathcal{J}').
\end{align}
Essentially, averaging glues $\Psi$ boundaries together, so that broken cylinders are glued together to make a full cylinder.

\begin{figure}[H]
\centering
\includegraphics[scale=0.3]{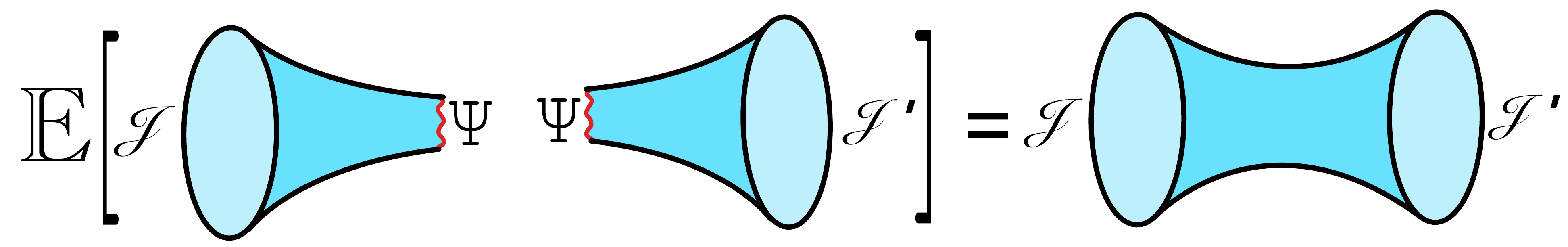} 
\caption{\small Averaging over the random $\Psi$ boundary condition glues together the broken cylinders to form a full cylinder.}
\end{figure}

To illustrate the other version of the effective model, with linked $\Psi$ boundaries, we consider the computation of a product of two partition functions. In the full theory, this either corresponds to the product of one-point functions $\langle\hat{Z}[\mathcal{J}]\rangle_\alpha\langle\hat{Z}[\mathcal{J}']\rangle_\alpha$ or the two-point function $\langle\hat{Z}[\mathcal{J}]\hat{Z}[\mathcal{J}']\rangle_\alpha$. These quantities are numerically equal, but the computations are somewhat different. The first version of the effective model, which we have just described, corresponds to the computation $\langle\hat{Z}[\mathcal{J}]\rangle_\alpha\langle\hat{Z}[\mathcal{J}']\rangle_\alpha$. There are no contributions from cylinders connecting the partition function, and the contributions from connecting to the the $\alpha$ state or $\Psi$ boundaries manifestly factorize. 

Computing the product $Z^{(\Psi)}[\mathcal{J}]Z^{(\Psi)}[\mathcal{J}']$ using the second version of the effective model corresponds to the computation of the two-point function $\langle\hat{Z}[\mathcal{J}]\hat{Z}[\mathcal{J}']\rangle_\alpha$. In this version of the effective model, we allow cylinders which connect the partition function boundaries, but the $\Psi$ boundary condition is modified. The product $Z^{(\Psi)}[\mathcal{J}]Z^{(\Psi)}[\mathcal{J}']$ receives a contribution from two $\Psi$ boundaries, with each partition function connecting to a $\Psi$ boundary with a broken cylinder, but in this case the pair of $\Psi$ boundaries are given a "linked" boundary condition. Rather than giving each $\Psi$ boundary the same boundary condition, corresponding to the state $|\Psi\rangle$, the linked $\Psi$ boundaries have a "shared" boundary condition, which can be described in terms of an entangled state for the two boundaries, $|\Psi^{(2)}\rangle$. This state replaces the two-universe component $\hat{P}_2|\psi^2_\alpha\rangle$ in the full theory, and thus has the "off-diagonal" structure 
\be\label{linkedpsiboundaries}
\langle \{I, J\}|\Psi^{(2)}\rangle= \langle I |\Psi\rangle \langle J|\Psi\rangle - \delta_{IJ} \hspace{10pt}\longrightarrow \hspace{10pt} \begin{cases} \langle I|\Psi\rangle \langle J|\Psi\rangle,&\hspace{20pt} I\neq J,
\cr
0 &\hspace{20pt} I=J.
\end{cases}
\ee
The last part of this equation applies for appropriate observables $Z[\mathcal{J}]$, for which the diagonal pieces can be ignored using the results of Appendix \ref{diagselfaveraging}. If we restrict our attention to these observables, we may simply define $|\Psi^{(2)}\rangle$ in the effective description using the last part of this equation.

\begin{figure}[H]
\centering
\includegraphics[scale=0.35]{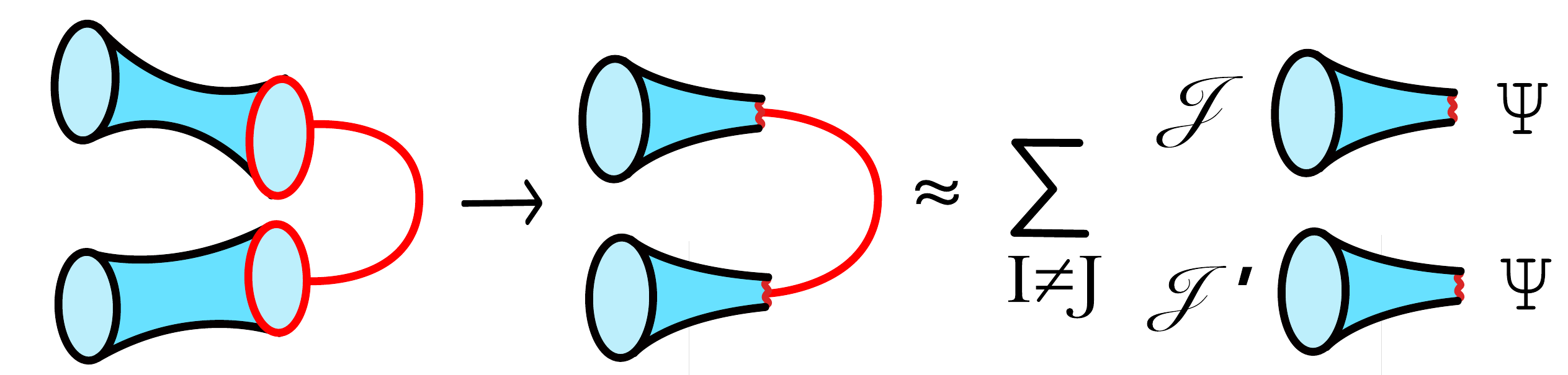} 
\caption{\small Here we illustrate the consequences of equation \eqref{linkedpsiboundaries} in the computation of $Z^{(\Psi)}[\mathcal{J}]Z^{(\Psi)}[\mathcal{J}']$ in the effective model. The contribution of the two-universe component in the full model, pictured on the left, is traded for a contribution from linked $\Psi$ boundaries described by the state $|\Psi^{(2)}\rangle$. For appropriate observables, we can then use \eqref{linkedpsiboundaries} to express the contribution as an off-diagonal sum with unlinked $\Psi$ boundaries. }
\label{figlinkedpsi}
\end{figure}

In the computation of an $n$-point function, using this version of the effective model, we must also consider linked $\Psi$ boundary conditions for up to $n$ boundaries. These linked boundary conditions are similarly described by a product of the unlinked boundary conditions, with the diagonal parts "excluded" pairwise.

This "off-diagonal" structure of the linked boundary condition, enforced by the "exclusion rule", was put in by hand in this effective model. In the full theory, this rule originates from the geometric exclusion effect. Later in this paper, in Section \ref{JTeffectivemodel}, we will find that this effective model describes JT gravity in a certain approximation. In this case as well the exclusion rule must be explained using the full many-universe description of the theory. 

The advantages of this effective model relative to the full description will be particularly apparent in this application.  For example, the effective description gives a vivid picture of how both ensemble averaging and time averaging (in the context of computing the spectral form factor) glue together pairs of broken cylinders along their $\Psi$ boundaries to form full cylinders.

\section{Approximate $\alpha$ states in the MM model}\label{MMmodel}

In the previous section we studied a truncation of the MM model, the CGS model, which involves only the disk and cylinder topologies. For correlators of a small (fewer than $\sim e^{S_0}$) number of $\hat{Z}$s in the No-Boundary state, the GGS model serves as a good approximation to the MM model, since the contributions of other topologies are suppressed by powers of $e^{-S_0}$. However, typical $\alpha$ states contain arbitrarily many closed universes, and as a result surfaces with many boundaries dominate the computation of correlators in an $\alpha$ state.

Does the simple picture of factorization we discussed in the CGS model have any relevance to the full MM model? In this section we will discuss how in  certain \textit{approximate} $\alpha$ states the CGS model still serves as a good approximation to the full MM model. These approximate $\alpha$ states can be chosen to have a small but nonzero width, corresponding to a small variance in the ensemble, so that correlators of $\hat{Z}$ factorize up to (exponentially) small errors. Similar approximate $\alpha$ states will allow us to study an approximate version of the factorization problem in JT gravity.

We begin this section by reviewing some aspects of the Hilbert space of the MM model and then  introducing the approximate $\alpha$ states. We then discuss the limits of validity of the CGS approximation in these approximate $\alpha$ states. 

There are two main sources of error in the CGS approximation to the MM model. The first involves corrections from surfaces with more than two boundaries, such as the three-holed sphere. In the CGS approximation, the No-Boundary state is approximated by a Gaussian.  In the full MM model these surfaces describe the deviation from Gaussianity. For correlators in approximate $\alpha$ states peaked far from the peak of the No-Boundary state, where these non-Gaussianities are large, $n$-holed spheres start to become important, with the three-holed sphere providing the leading correction.

The second source of error we discuss involves effects from the discreteness of the spectrum of $\hat{Z}$ in the full MM model. These effects are related to the existence of ``null states'', which are only visible nonperturbatively in the topological expansion. If we choose our approximate $\alpha$ states to be have Gaussian wavefunctions in the $\hat{Z}$ eigenbasis, with a small width $\Delta$ and mean $z$, correlation functions are insensitive to this discreteness if $\Delta \gg e^{-S_0}$ so that the wavefunction is spread over many of the exponentially finely spaced discrete eigenvalues. For example, the correlators will be smooth functions of $z$. However, if we take the width of the wavefunction to be less than the spacing between eigenvalues, correlators will oscillate sharply with $z$, peaked where $z$ is equal to one of the discrete $\hat{Z}$ eigenvalues, and will be small in between.

The width $\Delta$ of the approximate $\alpha$ state quantifies the failure of factorization. In order for the approximate $\alpha$ states to provide a good (approximate) model for factorization within the CGS approximation, we must take $\Delta$ to be small, but not so small that effects from the discreteness of the spectrum are important. In this section we find that we may still take $\Delta$ to be exponentially small, $\Delta \sim \sqrt{S}_0 e^{-S_0}$, and stay within the CGS approximation.

\subsection{Review of the MM model} 

We begin this subsection by reviewing some facts about the spectrum of $\hat{Z}$, and the wavefunction of the No-Boundary state in the $\hat{Z}$ eigenbasis, following \cite{Marolf:2020xie}. These facts are conveniently captured by the computation of the generating function $\langle \exp[u \hat{Z}]\rangle_{NB}$. To compute this, we use the cumulant expansion,
\be\label{eq:cumulant}
\langle \exp[u \hat{Z}]\rangle_{NB} = \exp\big[\sum_{n=1}^\infty \frac{u^n}{n!} \langle \hat{Z}^n\rangle_{NB,c}\big],
\ee
with the $n=0$ term of the sum removed so that the No-Boundary state is normalized.

The connected correlators $\langle \hat{Z}^n\rangle_{NB,c}$ are given by a sum over connected surfaces with $n$ circular boundaries, which are $n$-holed spheres with $g$ handles added. Summing over the $g$ we find\footnote{We remind the reader that our normalizations differ from \cite{Marolf:2020xie}.  We take surfaces to be weighted by $e^{S_0 \chi}$, so for example the disk has weight $e^{S_0}$.}
\be
\langle\hat{Z}^n\rangle_{NB,c} = e^{(2-n)S_0} \big(1 + e^{-2S_0} +\dots\big) = e^{-n S_0} \frac{e^{2 S_0}}{1-e^{-2 S_0}}.
\ee
Inputting this into equation \eqref{eq:cumulant}, we find that the sum over $n$ exponentiates. Labeling the combination $\frac{e^{2 S_0}}{1-e^{-2 S_0}} \equiv \lambda$, 
\be\label{resumcumulant}
\langle \exp[u \hat{Z}]\rangle_{NB} = e^{-\lambda}\exp\big[\lambda e^{u e^{-S_0}}\big].
\ee
Expanding the outermost exponential, we find
\be
\langle \exp[u \hat{Z}]\rangle_{NB} = \sum_{k=0}^\infty e^{-\lambda} \frac{\lambda^k}{k!} \; e^{u e^{-S_0} k}.
\ee
This is precisely the expression for the generating function in the $\hat{Z}$ eigenbasis. With the eigenvalues $z_k$ of $\hat{Z}$ labeled by a nonnegative integer $k$,
\be
\langle \exp[u \hat{Z}]\rangle_{NB} = \sum_{k=0}^\infty |\langle z_k| NB\rangle|^2 e^{u z_k}.
\ee
We then identify
\be
z_k = e^{-S_0} k ,\hspace{20pt} |\langle z_k| NB\rangle|^2 = e^{-\lambda} \frac{\lambda^k}{k!}.
\ee
The discrete eigenvalues of $\hat{Z}$ are spaced by $e^{-S_0}$, and the No-Boundary state gives a Poisson distribution over the $z_k$. The mean of this distribution is $\langle \hat{Z}\rangle_{NB} = e^{-S_0}\lambda$, and the variance is $\langle \hat{Z}^2\rangle_{NB,c} = e^{-2 S_0}\lambda$.

\begin{figure}[H]
\vspace{-1em}
\centering
\includegraphics[scale=0.55]{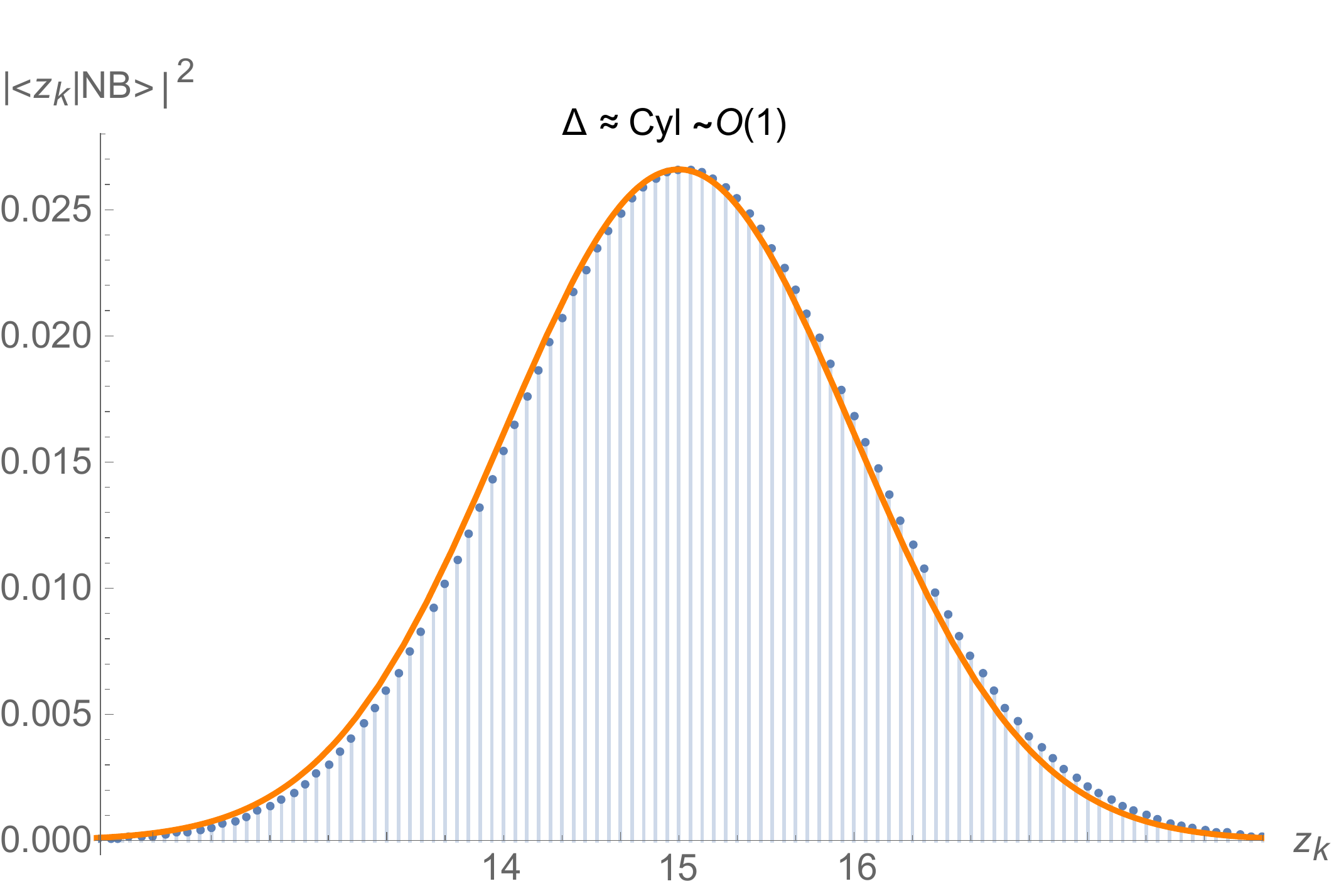} 
\caption{In this picture we illustrate the difference between Poisson (discrete blue points) and Gaussian (continuous orange line) distributions. We take  $e^{S_0}\approx 15$ to accentuate the difference. In typical gravitational systems, $e^{S_0}$ is much larger and the difference between distributions much smaller.}
\end{figure}

We take a moment to note that these features of the spectrum do not depend very sensitively on the surfaces with handles. To ignore the contributions of these surfaces, we can simply take $\lambda\rightarrow e^{2 S_0}$. Then the spacing of the eigenvalues is unchanged, and the mean and variance of of the No-Boundary wavefunction become $\langle \hat{Z}\rangle_{NB} \rightarrow e^{S_0}=\disk$, and $\langle \hat{Z}^2\rangle_{NB,c}\rightarrow 1=\cyl$. These shifts in the mean and variance are by factors of relative order $e^{-2S_0}$; the shift in the mean is exponentially smaller than the spacing between eigenvalues. When we study correlators in approximate $\alpha$ states, the errors introduced by ignoring the surfaces with handles and taking $\lambda\rightarrow e^{2 S_0}$ will be smaller than the errors coming from surfaces with no handles but more than two boundaries, so we will safely be able to ignore them.\footnote{An exception is contributions from spacetimes with both disks and surfaces with handles. The large weight $e^{S_0}$ of the disk can cancel the suppression of the handles. However, as the disks manifestly factorize, by focusing on connected quantities we can safely ensure that the contributions of surfaces with handles are small. In any case, the surfaces with handles do not play an important role in our discussion, so we will simply ignore them in this section. Their contributions can simply be reintroduced by shifting the contributions $\disk$, $\cyl$, etc. appearing in our formulas by exponentially small amounts, for example by taking $\disk \rightarrow \disk + \text{Handle-disk}+\dots = \disk (1 + \mathcal{O}(e^{-2 S_0}))$ .}

At large $S_0$, and near the peak of the distribution, we can ignore the contributions of all surfaces except for the disk and cylinder. In this limit the No-Boundary Poisson distribution can be approximated as a Gaussian
\be
|\langle z_k| NB\rangle|^2 \approx \frac{1}{\sqrt{ 2\pi \cyl}}\exp\bigg[ - \frac{( z_k- \disk)^2}{2 \cyl} \bigg], \hspace{20pt} |z_k-\disk| \ll e^{S_0}.
\ee
We can then associate the CGS approximation with this Gaussian region of the No-Boundary wavefunction.

\subsection{Approximate $\alpha$ states in the MM model}

Now we introduce an approximate $\alpha$ state $|z,\Delta\rangle$, by acting on the No-Boundary state with an approximate projector onto $\hat{Z}=z$,
\be\label{eq:defapproxalphastate}
|z,\Delta\rangle \propto \int_{-\infty}^\infty dp \;e^{-\Delta^2 p^2 + i p (\hat{Z}-z)} |NB\rangle = e^{-\frac{(\hat{Z}-z)^2}{4\Delta^2}} |NB\rangle.
\ee
This expression is identical to the expression \eqref{eq:psideltaexpression} for the regularized delta-function expression for an $\alpha$ state in the CGS model. In order to find an expression for $|\psi^2_{z,\Delta}\rangle$ with $|\psi_{z,\Delta}\rangle\equiv |z,\Delta\rangle$, we can identify $\psi_{z,\Delta}(\hat{Z})$ as the approximate projector $ e^{-\frac{(\hat{Z}-z)^2}{4\Delta^2}} $ and square this. As in the CGS model, in order to do computations it is useful to then express this squared projector as a Gaussian integral,
\be\label{eq:psisquaredmm}
\big(\psi_{z,\Delta}(\hat{Z})\big)^2 \propto \int_{-\infty}^\infty dp\; e^{-\frac{\Delta^2}{2}p^2 + i p(\hat{Z}-z)}.
\ee

The expression \eqref{eq:defapproxalphastate} for $|z,\Delta\rangle$ hides many important features which are essential to understanding the role of the disk-and-cylinder CGS approximation in these states. In order to understand this state in more detail we first express it in the $\alpha$ basis. Inserting a complete set of states $|z_k\rangle$ we find
\begin{align}\label{eq:alphabasisapproxalphastate}
|z,\Delta\rangle& \propto \sum_{k=0}^\infty e^{-\frac{(\hat{Z}-z)^2}{4\Delta^2}}|z_k\rangle\langle z_k |NB\rangle
\cr
& \propto \sum_{k=0}^\infty \sqrt{\frac{\lambda^k}{k!}}e^{-\frac{(z_k-z)^2}{4\Delta^2}} |z_k\rangle.
\end{align}
The wavefunction in this basis is then the product of the Gaussian profile of the approximate projector $e^{-\frac{(z_k-z)^2}{4\Delta^2}}$ and the Poisson distribution of the No-Boundary state. However, for sufficiently small $\Delta$, the Poisson distribution is essentially constant over the width $\sim \Delta$ of this projector, and we can approximate the state $|z,\Delta\rangle$ with just a Gaussian wavefunction. We will comment more on the non-Gaussianities shortly. For now we simply approximate
\be\label{eq:approxapproxalphastate}
|z,\Delta\rangle \approx \mathcal{N} \sum_{k=0}^\infty e^{-\frac{(z_k-z)^2}{4\Delta^2}} |z_k\rangle,
\ee
with $\mathcal{N}$ a normalization factor. Of course, we could have \textit{defined} the approximate $\alpha$ state $|z,\Delta\rangle$ with this Gaussian expression. However, our definition \eqref{eq:defapproxalphastate} is more useful for connecting to the results of the previous section on the CGS model, as the expressions \eqref{eq:psisquaredmm} and \eqref{psisquaredcgs} are identical and simple to calculate with.

 For $\Delta\rightarrow 0$ we can see that \eqref{eq:approxapproxalphastate} approaches an exact alpha state if $z=z_k$ for one of the discrete $z_k$. If $z$ is not equal to one of the $z_k$, then this approaches the zero vector. 

For the purposes of this section, we will take $\Delta \ll 1$, so that the variance of $\hat{Z}$ in this state is much smaller than in the No-Boundary state, so correlators approximately factorize. However, as we have discussed, we should take $\Delta$ large enough so that $|z,\Delta\rangle$ is spread over many discrete eigenvalues. Taking $\Delta = C e^{-S_0}$, the state is spread over $\sim C$ eigenvalues, with $C\gg 1$.

\begin{figure}[H]
\vspace{-1em}
\centering
\includegraphics[scale=0.55]{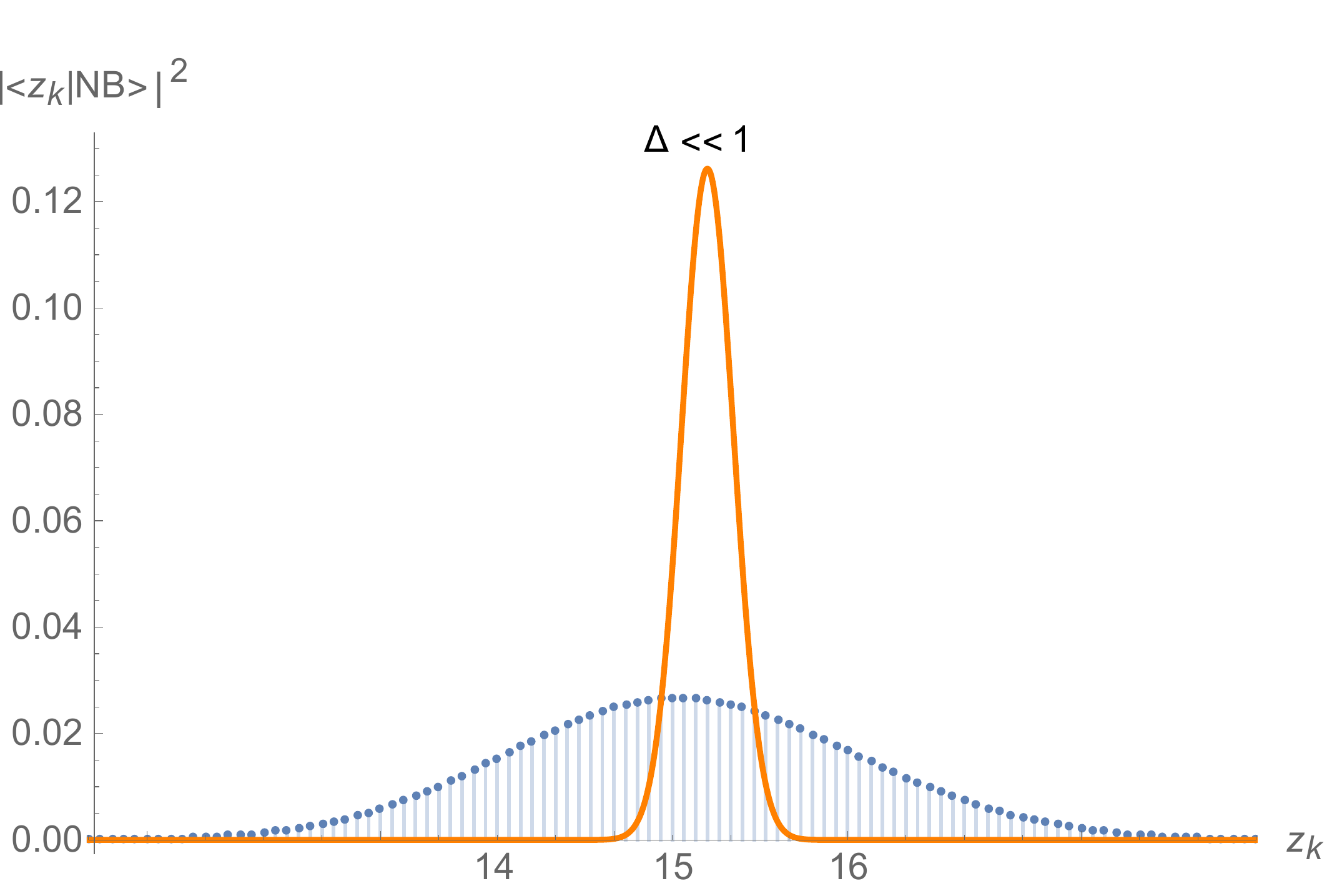} 
\caption{Here we show the probability distribution for an approximate $\alpha$ state, compared to the Poisson No-Boundary distribution. The approximate $\alpha$ state is sharply peaked around $z_k= z$, with $|z-\disk|\ll 1$ so that the approximate $\alpha$ state approximates a \textit{typical} $\alpha$ state with large probability in the No-Boundary ensemble. The width of this state is exponentially smaller than the No-Boundary state, but is still much larger than the exponentially small spacing between the $z_k$.}
\end{figure}

\subsection{ CGS approximation errors in an approximate $\alpha$ state}

Suppose we compute  correlators in the approximate $\alpha$ state $|z,\Delta\rangle$.  To what extent can we apply the CGS approximation and sum only over disk and cylinder spacetimes? For simplicity, we will focus only on correlators with fewer than exponentially many $\hat{Z}$s.\footnote{Correlators of exponentially many $\hat{Z}$s will be sensitive to the tails of the wavefunction, and we also expect that the disk-and-cylinder CGS approximation breaks down. See \cite{Marolf:2020xie} for some related discussion of the behavior of high moments of $\hat{Z}$ in the No-Boundary state.}

In the introduction to this section, we alluded to two distinct sources of error in the CGS approximation. First, perturbative corrections in the topological expansion from connected surfaces with more than two boundaries (such as the three-holed sphere) can become important if the center $z$ of the approximate $\alpha$ state is too far from the center of the No-Boundary state, $z=\disk$. This is the case because these surfaces account for the non-Gaussianities in the No-Boundary wavefunction, which enters into our expression \eqref{eq:alphabasisapproxalphastate} for $|z,\Delta\rangle$. 

For $|z-\disk|$ and $\Delta$ sufficiently small, we can approximate the No-Boundary Poisson distribution near $z$ as Gaussian or constant, and these corrections will be unimportant. However, if we take $|z-\disk|$ or $\Delta$ large enough, these non-Gaussianities will become important and the disk-and-cylinder CGS approximation will fail. 

The second source of error is due to non-perturbative effects which account for the discreteness of the spectrum of $\hat{Z}$. For $\Delta$ sufficiently large, $|z,\Delta\rangle$ is spread over many discrete eigenvalues and these corrections are small. For $\Delta \sim e^{-S_0}$, this discreteness, visible nonperturbatively in the topological expansion, becomes important and the CGS approximation fails. 

In this section we will find that the CGS approximation for correlation function computations in states $|z,\Delta\rangle$ has exponentially small errors $\sim e^{-S_0}$, if $z$ is \textit{typical} in the No-Boundary distribution ($|\disk-z|\sim 1$) and $\Delta \sim \sqrt{S_0} e^{-S_0}$. In this case the errors in factorization, described by the variance, are exponentially small. Then up to exponentially small errors, the CGS model serves as a good approximation to the full MM model in appropriate $\alpha$ states and the picture of factorization and the effective description developed in the previous section can be applied.\footnote{We note that the perturbative source of errors is specificially a result of our choice \eqref{eq:alphabasisapproxalphastate} of $|z,\Delta\rangle$ as a non-Gaussian state, as opposed to the Gaussian approximate $\alpha$ state \eqref{eq:approxapproxalphastate}. On the other hand, the nonperturbative errors from the discreteness of the spectrum should be an issue for any smooth approximation to an $\alpha$ state.}

\subsubsection{Errors in the CGS approximation from non-Gaussianities} \label{sec:MM3holesphere}

Consider the formula for an $n$-point correlation function in the state $|z,\Delta\rangle$ using the representation \eqref{eq:psisquaredmm} of $\psi_{z,\Delta}(\hat{Z})$,
\be\label{eq:correlatormm}
\langle \hat{Z}^n\rangle_{z,\Delta} = \langle \hat{Z}^n \; \big(\psi_{z,\Delta}(\hat{Z})\big)^2\rangle_{NB} \propto \int_{-\infty}^\infty dp \; e^{-\frac{\Delta^2}{2}p^2 -i p z} \langle \hat{Z}^n e^{i p \hat{Z}}\rangle_{NB}.
\ee
For simplicity, we first focus on the case $n=0$, then describe the case $n\neq 0$.

We can compute the expectation value $\langle e^{i p \hat{Z}}\rangle_{NB}$ using the cumulant expansion \eqref{eq:cumulant}.\footnote{We note again that we are ignoring the contributions from surfaces with handles, which may be simply included by rescaling $\disk$, $\cyl$, $\text{3-holed sphere}$, etc. by exponentially small amounts.}
\begin{align}\label{eq:cumulant2}
\langle e^{i p \hat{Z}}\rangle_{NB} &= \exp\bigg[ \sum_{k=1}^\infty \frac{(ip)^k}{k!} \langle \hat{Z}^k\rangle_{NB,c} \bigg]
\cr
& = \exp\bigg[ i p \;\disk + \frac{(ip)^2}{2!} \cyl + \frac{(ip)^3}{3!} (\text{3-holed sphere})+\dots\bigg]
\cr
& = \exp\bigg[ i p \;e^{S_0}- \frac{p^2}{2}  - i \frac{p^3}{6} e^{-S_0}+\dots\bigg].
\end{align}
Keeping only disks and cylinders, we would find a Gaussian function
\be
\langle e^{i p \hat{Z}}\rangle_{NB} \approx e^{i p\; \disk -\frac{\cyl}{2} p^2}.
\ee
If $|p|$ is sufficiently small, this is a good approximation, since the leading correction in the cumulant expansion, from the three-holed sphere, is of order $p^3 e^{-S_0}$. Using our formula \eqref{eq:correlatormm} for the correlator, it is natural to treat the integral over $p$ as a perturbed Gaussian integral, with the center and width of the Gaussian determined by the Gaussian in \eqref{eq:correlatormm} as well as the disk and cylinder terms in \eqref{eq:cumulant2}, with the higher-order terms in \eqref{eq:cumulant2} treated as perturbations. In the next subsection, we will justify focusing on this small $p$ region by arguing that the contributions from additional saddle points at large $p$ can be made small. For now, we just study the perturbative expansion around this Gaussian integral.\footnote{It is important to note, as we discussed earlier in the section, that surfaces with handles will not be important in this discussion. This is because the power of $p$ associated with a surface is equal to the number of boundaries. Adding a handle to a surface does not increase the associated power of $p$, but it does exponentially suppress its contribution.}

Separating out the the disk and cylinder contributions from \eqref{eq:cumulant2} in \eqref{eq:correlatormm}, we find
\be\label{correlatorperturb}
\langle 1\rangle_{z,\Delta} \propto \int_{-\infty}^\infty dp \; \exp\bigg[-\frac{\Delta^2+\cyl}{2} p^2 + i p(\disk-z) + \sum_{k=3}^\infty \frac{(ip)^k}{k!} e^{(2-k)S_0}\bigg].
\ee
For now we assume that we can safely take $\Delta\ll \cyl$. This will be justified in the next subsection. Then the Gaussian part of the integral over $p$ has a center at $p=i\frac{\disk-z}{\cyl}$ with a variance $1/\cyl$. Then we shift $p\rightarrow p' = p- i\frac{\disk-z}{\cyl}$. Implimenting this shift on the $k\geq 3$ terms gives $p'$-independent terms, as well as lower order polynomials in $p'$ with coefficients proportional to powers of $\frac{\disk-z}{\cyl}$.\footnote{Note that the cylinder contribution to the Gaussian term in $p$ keeps $p\sim 1$, even for $\Delta$ small.}

In order for the disk-and-cylinder approximation to be valid, we should be able to treat these terms as perturbations to the Gaussian disk-and-cylinder approximation to the integral. This is certainly true for \textit{typical} choices of $z$, drawn from the No-Boundary distribution; $|\disk-z|$ is typically of order one, so that the $k\geq 3$ terms in \eqref{correlatorperturb}, after shifting $p\rightarrow p'$, are polynomials in $p'$ with coefficients that are exponentially small. The Gaussian part of the integral has a variance of $1/\cyl= 1$, so that these terms are small.

We can also see that these perturbative errors are small even for some \textit{atypical} choices of $z$. As $|\disk-z|$ is made exponentially large $\sim e^{c S_0}$, the $k=3$ term, corresponding to the three-holed sphere, gives the leading correction. For $|\disk-z|\ll e^{\frac{S_0}{3}}$, the corrections from the three-holed sphere are small but not necessarily exponentially small.

Now we discuss the errors in correlation functions with $n\geq 1$. From the formula \eqref{eq:correlatormm}, we see that we must evaluate
\be
\langle \hat{Z}^n e^{i p \hat{Z}}\rangle_{NB}.
\ee
As mentioned in Section \ref{explicitcalculationcgs}, we can think of the operator $e^{i p \hat{Z}}$ as creating a spacetime D-brane, and $\langle \hat{Z}^n e^{i p \hat{Z}}\rangle_{NB}$ represents an $n$-point amplitude in the presence of this D-brane. Each of the $n$ ``probe'' $\hat{Z}$s can connect to each other, and they can connect to the brane any number of times. We then sum over surfaces with connected components involving some probe $\hat{Z}$ boundaries, as well as any number of brane boundaries. Each brane boundary comes with a factor of $i p$, and the brane boundaries are treated as distinguishable so they also come with symmetry factors. Altogether, the sum over spacetimes with probe $\hat{Z}$ gives a series in $p$, where each power of $p$ is accompanied by a power of $e^{-S_0}$ because each boundary decreases the Euler character of the connected manifold by one. We label this series as $F^{(n)}( p) = \sum_{k=0}^\infty C^{(n)}_k (i p e^{-S_0})^k$. The $k=0$ term in this sum corresponds to the No-Boundary correlation function $C^{(n)}_0=\langle \hat{Z}^n\rangle_{NB}$.

In addition to summing over spacetimes with probe $\hat{Z}$ boundaries, we include all of the spacetimes which only have brane boundaries. These sum up to the $n=0$ correlation function, $\langle e^{i p\hat{Z}} \rangle_{NB}$.

Combining these, we find
\be\label{summingboundaries}
\langle \hat{Z}^n e^{i p \hat{Z}}\rangle_{NB}= F^{(n)}( p) \langle e^{i p\hat{Z}} \rangle_{NB}.
\ee

To compute a correlation function in the state $|z,\Delta\rangle$, we then insert this into the formula \eqref{eq:correlatormm} and perform the integral over $p$. Now we must argue that if $|z-\disk|\ll e^{\frac{S_0}{3}}$, the contributions of surfaces with more than two boundaries can be treated as perturbations to the Gaussian CGS result. Since we have already argued that such contributions to $\langle e^{i p\hat{Z}} \rangle_{NB}$ are small, we must now simply argue that the contributions to $F^{(n)}( p)$ from these surfaces are also small. 

This is simple, as we have alread seen that the Gaussian $p$ integral is peaked around $|p|\sim |\disk - z|$. Since the contribution of each ``brane'' boundary in $F^{(n)}( p)$ comes with a factor of $ p e^{-S_0}$ relative to the sphere (with weight $e^{2 S_0}$), the brane boundaries are exponentially suppressed if $|\disk-z|\ll e^{\frac{S_0}{3}}$. Cylinder spacetimes with a single brane boundary are then not suppressed, giving contributions which scale like $|\disk-z|$, but spacetimes with more than two brane boundaries, such as the three-holed sphere, give exponentially small contributions. Spacetimes with no brane boundaries make up the $k=0$ term in $F^{(n)}( p) = \sum_{k=0}^\infty C^{(n)}_k (i p e^{-S_0})^k$. This term is equal to the No-Boundary correlation function $C^{(n)}_0=\langle \hat{Z}^n\rangle_{NB}$, to which surfaces with more than two boundaries give small corrections.

Furthermore, if $z$ is chosen to be \textit{typical} in the No-Boundary distribution, so that $|\disk-z|\sim 1$, then the errors from the $k$-holed spheres are exponentially small in these correlation functions.

\subsubsection{Errors in the CGS approximation from discreteness}
Consider the formula \eqref{eq:alphabasisapproxalphastate} for the state $|z,\Delta\rangle$ in the discrete $\alpha$ basis,
\be\label{eq:alphabasisalphastate2}
|z,\Delta\rangle \propto \sum_{k=0}^\infty e^{-\frac{(z-z_k)^2}{4 \Delta^2}} \sqrt{\frac{\lambda^k}{k!}} |z_k\rangle.
\ee
In order to understand the errors in the disk-and-cylinder approximation related to the discreteness of the spectrum, it is useful to understand a particular derivation of our orginal expression \eqref{eq:defapproxalphastate} for $|z,\Delta\rangle$ in the form $\psi_{z,\Delta}(\hat{Z})|NB\rangle$ starting from this $\alpha$ basis expression. 

We make use of the ``spacetime D-brane'' expression for the $\alpha$ state $|z_k\rangle$ \cite{Marolf:2020xie}, the discrete version of the delta-function expression for an $\alpha$ state \eqref{deltafunctionalphacgs}
\be
|z_k\rangle \propto \sqrt{\frac{k!}{\lambda^k}} \int_{-\pi e^{S_0}}^{\pi e^{S_0}} dp\; e^{ i p (\hat{Z}- k e^{-S_0})} |NB\rangle.
\ee
Note that in this expression the $p$ integral is over a finite interval, as this integral represents a Kronecker delta.

Inserting this expression into our definition \eqref{eq:alphabasisalphastate2} of $|z,\Delta\rangle$, we find
\be\label{eq:ksumexpression}
|z,\Delta\rangle \propto \sum_{k=0}^\infty  \exp\bigg[ - \frac{(z-z_k)^2}{4 \Delta^2}\bigg] \int_{-\pi e^{S_0}}^{\pi e^{S_0}} dp\; e^{ i p (\hat{Z}- k e^{-S_0})} |NB\rangle.
\ee
Note that the Poisson No-Boundary wavefunction $\propto \sqrt{\frac{\lambda^k}{k!}}$ has cancelled out.

We can freely extend the sum over $k$ to negative integer $k$. This is because the state $\int_{-\pi e^{S_0}}^{\pi e^{S_0}} dp\; e^{ i p (\hat{Z}- k e^{-S_0})} |NB\rangle$ with $k<0$ is equal to the zero vector, as the integral projects onto $\hat{Z} =z_k= k e^{-S_0}$, but there are no such states in the theory for $k<0$. We then may use the Poisson summation formula to replace the sum over $k$ with a sum over integers $\ell$,
\be\label{eq:poissonsum}
|z,\Delta\rangle \propto \int_{-\pi e^{S_0}}^{\pi e^{S_0}} dp\; \sum_{\ell=-\infty}^\infty e^{-\Delta^2 (p+ 2\pi \ell e^{S_0})^2 + i (p+ 2\pi \ell e^{S_0}) (\hat{Z}-z)} |NB\rangle.
\ee
By shifting $p\rightarrow p- 2\pi \ell e^{S_0}$, we can combine all of the terms in the sum over $\ell$ into a single $p$ integral over the whole real line. This leaves us with the expression \eqref{eq:defapproxalphastate} for the state $|z,\Delta\rangle$,
\be\label{eq:approxalphadefagain}
|z,\Delta\rangle \propto \int_{-\infty}^\infty dp \; e^{-\Delta^2 p^2 + i p (\hat{Z}-z)}|NB\rangle.
\ee

Now we return to the beginning of this derivation and consider what happens if we ignore the discreteness of the spectrum. We can do this by replacing the sum over $k$ in \eqref{eq:alphabasisalphastate2}, or perhaps more appropriately \eqref{eq:ksumexpression}, with an integral over $k$. This corresponds to keeping only the $\ell=0$ term in the Poisson sum \eqref{eq:poissonsum}, so that the $p$ integral is over a finite interval, $-\pi e^{S_0} <p<\pi e^{S_0}$.

We can then associate the effects from the discreteness of the spectrum with the $\ell\neq 0 $ terms in \eqref{eq:poissonsum}, and therefore with the large $p$ region, $|p|> \pi e^{S_0}$, of the integral \eqref{eq:approxalphadefagain}. In the previous subsection, we focused on the perturbative expanions around $p \propto |\disk-z|$, ignoring the contributions from this large $|p|$ region of the integral. We will see that the corrections from this region correspond to additional saddle points, correcting the result from the saddle point studied in that section.

At sufficiently large $\Delta$, we expect that the errors from ignoring this discretness are small. We can get a rough estimate of their size by looking at the $\ell = \pm 1$ terms in the Poisson sum \eqref{eq:poissonsum}. The $p$ integral is over a region where $|p|\sim e^{S_0}$, so the Gaussian term in the integral gives a suppression of order $\exp\big[-  c \Delta^2 e^{2S_0}\big]$. This suggests that if we take $\Delta \sim \sqrt{S_0} e^{-S_0}$, these errors will be exponentially suppressed, competing with the perturbative errors discussed in the previous section.

This estimate is a bit rough. We can get a more accurate estimation of the error, which connects more clearly to the disk-and-cylinder CGS approximation, by studying the formula \eqref{eq:correlatormm} for correlation functions in the state $|z,\Delta\rangle$. To isolate the effects related to the discreteness of the spectrum, we focus on the behavior of the large $|p|$ region of the integral. 

Again, we first focus on the calculation of the $n=0$ case of \eqref{eq:correlatormm}, the normalization of the state $|z,\Delta\rangle$, before discussing correlation functions with $n>0$. 

At large $|p|\sim e^{S_0}$, we cannot truncate the cumulant expansion for $\langle e^{i p \hat{Z}}\rangle_{NB}$, since the $n$-holed spheres give a contribution of order $|p|^n e^{(2-n)S_0}$, which is large. Instead, we do the full sum over $n$-holed spheres in the cumulant expansion, only ignoring surfaces with handles. As we have seen from equation \eqref{resumcumulant}, these contributions exponentiate and we find
\be\label{eq:nonperturbperiodic}
\langle e^{i p \hat{Z}}\rangle_{NB} \approx e^{-e^{2 S_0}} \exp\bigg[e^{2 S_0} e^{i p e^{-S_0}}\bigg].
\ee
This function is \textit{periodic} in $p$, with a period $\Delta p = 2\pi e^{S_0}$. Inserting this expression into the formula \eqref{eq:correlatormm}, we find
\be\label{eq:normalizationnonperturbative}
\langle 1\rangle_{z,\Delta} \approx \mathcal{N} \int_{-\infty}^\infty dp\; e^{- \frac{\Delta^2}{2} p^2 - i p z + e^{2S_0} e^{i p e^{-S_0}}},
\ee
with $\mathcal{N}$ a normalization factor. For $\Delta$ exponentially small, the Gaussian term does not suppress the region of the integral with $|p|\sim e^{S_0}$ and the integral is sensitive to the periodicity of the function \eqref{eq:nonperturbperiodic}. It is clear that for small $\Delta$, the periodicity leads to saddle points in the integral for $|p|\approx 2\pi e^{S_0}$, similar to the saddle point at $p=0$ which leads to the perturbative answer. These new saddle points have actions of order $\Delta^2 e^{2S_0}$, so they give exponentially suppressed contributions if $\Delta \sim \sqrt{S_0} e^{-S_0}$. This matches our expectation from our earlier analysis.

In the disk-and-cylinder CGS approximation, the periodicity of \eqref{eq:nonperturbperiodic} is not visible, and the large $|p|$ region is suppressed even for small $\Delta \ll e^{-S_0}$. This approximation is then only valid when the errors from the large $|p|$ region of the integral, associated with the discreteness of the spectrum, are small. Fortunately, this still allows us to choose $\Delta$ to be exponentially small, albeit with a factor of $\sqrt{S_0}$ as well, and remain within the regime where the CGS approximation is valid. Note that this means that the approximate $\alpha$ state is spread over $\sim \sqrt{S_0}$ discrete eigenvalues.

Now we briefly discuss the correlation functions with $n>1$. For simplicity, we focus on the one-point function. We first compute
\begin{align}
\langle \hat{Z} e^{i p \hat{Z}}\rangle_{NB} &= \frac{\partial}{\partial (ip)} \langle e^{i p \hat{Z}}\rangle_{NB} 
\cr 
&= e^{S_0} e^{i p e^{-S_0}}\langle e^{i p \hat{Z}}\rangle_{NB}.
\end{align}
This gives a nonperturbative expression for the function $F^{(1)}(p)$ in \eqref{summingboundaries}.

Inserting this into \eqref{eq:correlatormm}, we find
\be
\langle \hat{Z} \rangle_{z,\Delta} \approx \mathcal{N} \int_{-\infty}^\infty dp\; e^{- \frac{\Delta^2}{2} p^2 - i p (z- e^{-S_0}) + e^{2S_0} e^{i p e^{-S_0}}},
\ee
This integral is simply proportional to the integral \eqref{eq:normalizationnonperturbative} up to an exponentially small shift in $z$, so the analysis of the nonperturbative corrections is the same. 

For higher point functions, we insert
\be
\langle \hat{Z}^n e^{i p \hat{Z}}\rangle_{NB} = \frac{\partial^n}{\partial (ip)^n} \langle e^{i p \hat{Z}}\rangle_{NB} .
\ee 
into the formula \eqref{eq:correlatormm}. These are simply sums of powers of $\exp[ i p e^{-S_0}]$ times $\langle e^{i p \hat{Z}}\rangle_{NB} $, so the behavior of the $p$ integrals are similar.\footnote{At large $n$, these terms can exponentiate and change the behavior of the integral. See \cite{Marolf:2020xie} for a related calculation of high moments of $\hat{Z}$ in the No-Boundary state.}

We briefly note that it should be possible to alternatively understand these effects by expanding out the exponential $ e^{i p \hat{Z}}$ in the expression $\langle \hat{Z}^n\rangle_{z,\Delta} \propto \int_{-\infty}^\infty dp e^{-\frac{\Delta^2}{2} p^2 - i p z } \langle \hat{Z}^n e^{i p \hat{Z}}\rangle_{NB}$, evaluating the resulting Gaussian integrals over $p$ term by term, and examining the large-order behavior of this expansion.

\section{JT gravity}\label{JTgravity}

In this section we study JT gravity in approximate $\alpha$ states and connect to the ideas discussed in the previous sections. In particular, following the ideas of Section \ref{MMmodel}, we demonstrate that in JT gravity the disk-and-cylinder CGS approximation is valid for appropriate correlation functions in a class of approximate $\alpha$ states. The reduction to the CGS model, and its effective description, essentially follows directly from this. However, some of the rather formal aspects of Section \ref{SectionAlphaStates} come to life in this model, in a rather concrete way. In particular, a version of the effective description of JT gravity provides a particularly vivid realization of the effective model described in Section \ref{effectivemodel}.

In Section \ref{jtreview}, we give a brief review of some relevant aspects JT gravity, and introduce some notation. In this section we define the operators $\hat{Y}_E(T)$, which serve as as convenient choices of boundary operators $\hat{Z}[\mathcal{J}]$ in JT gravity. $\hat{Y}_E(T)$ is essentially a microcanonical version of the analytically continued partition function $\hat{Z}(\beta + i T)$.\footnote{Note that here $T$ corresponds to a time, not a temperature.}

In Section \ref{JTeffectivemodel}, we study a truncation of JT gravity  to disks and cylinder topologies. Formally this is just an example of the CGS model with species studied in Section \ref{cgsspecies}, but here the geometrical interpretation is clearer.

The main points that we cover in Section \ref{JTeffectivemodel} are
\begin{itemize}
\item We discuss some of the results of Section \ref{SectionAlphaStates} in a natural basis, the ``$b$'' basis. States in the $b$ basis describe closed universes on circular geodesic spatial slices with fixed and \textit{finite} size $b$. This is in a sense a ``bulk'' basis, rather than a "boundary" basis associated with operators $\hat{Z}[\mathcal{J}]$.

\item In the effective model of JT gravity, we trade away the many-universe description for random $\Psi$ boundaries. In JT gravity, these are circular geodesic boundaries in the bulk with finite sizes. A random function $\Psi(b)$ describes the boundary condition. Averaging glues together the random $\Psi(b)$ boundaries to make wormholes.

\item The effective model is useful for illustrating some expected properties of JT gravity. In particular, we use the effective model to demonstrate that appropriate time-averaging can approximately average out the noise in the spectral form factor, by approximately gluing together the random $\Psi(b)$ boundaries.

\item One can consider models where the $\Psi(b)$ boundaries are replaced with \textit{dynamical} boundaries. An obvious choice is to replace the $\Psi(b)$ with dynamical end-of-the-world (EOW) branes. If these EOW branes have appropriate chaotic internal dynamics, they may produce a pseudorandom approximation to the random $\Psi(b)$ function. We briefly highlight some challenges for reproducing the effective model with these EOW brane models.

\end{itemize}

Section \ref{jtapproximatealpha} is devoted to justifying the picture developed in Section \ref{JTeffectivemodel}, by verifying that the disk-and-cylinder CGS approximation is accurate in certain approximate $\alpha$ states. We give evidence for a bound on the validity of the disk-and-cylinder approximation in certain approximate $\alpha$ states, for which the value of $\hat{Y}_E(T)$ for some range of time with some tolerated error. Our bound allows for the time to be exponentially long, with the error exponentially small. Our argument for this bound is based on assuming certain behaviors of nonperturbative contributions to correlation functions of the operators $\hat{Y}_E(T)$, which we discuss in Appendix \ref{appendixbound}.\footnote{Here, and throughout this paper, we refer to effects that are nonperturbative in $e^{-S}$, where $S$ is the entropy, as nonperturbative. These effects are "doubly-nonperturbative" in the entropy, \cite{Cotler:2016fpe,Saad:2019lba}.}

The points we cover in \ref{jtapproximatealpha} are as follows:
\begin{itemize}
\item We explain how the random function $\Psi(b)$ in the effective description is related to the exact boundary ensemble, parametrized by the eigenvalues of a random matrix. The disk-and-cylinder approximation describes Gaussian fluctuations of the coarse-grained continuous approximation to the discrete boundary spectrum. Choosing a fixed function $\Psi(b)$ corresponds to fixing a coarse-grained density of states.

\item The observables we focus on, the $\hat{Y}_E(T)$, are insensitive to fine details of the energy spectrum for times that are not too long. We expect that these observables cannot differentiate between an exact $\alpha$ state, corresponding to fixed discrete boundary energy levels (or equivalently, a fixed set of ``Eigenbranes'' \cite{Blommaert:2019wfy,Blommaert:2020seb} in the bulk theory), and an approximate $\alpha$ state with only coarse-grained features of the density fixed. These coarse-grained features are described by a random draw of a continuous density of states, or equivalently a function $\Psi(b)$.\footnote{The discreteness of the boundary spectrum can also be related to the existence of null states in the bulk. In our approximate $\alpha$ states, null states are unimportant.}$^,$\footnote{We should point out a recent paper by Blommaert and Kruthoff \cite{Blommaert:2021gha}. In this paper, the authors explore another method of constructing approximate $\alpha$ states, which may be more appropriate for incorporating nonperturbative effects related to the discreteness of the spectrum.}

\item To argue more quantitatively for this expectation, we consider approximate $\alpha$ states in which the observables $\hat{Y}_E(T)$ are approximately fixed at evenly spaced times, up to a maximum time $T_{max}$. This roughly corresponds to fixing many Fourier modes of the continuous density of states up to a frequency $\sim T_{max}$. For $T_{max}$ not too large we argue, based on some conjectured behavior for the behavior of certain nonperturbative contributions which we discuss in Appendix \ref{appendixmoments}, that the variance of $\hat{Y}_E(T)$ in these states can be made exponentially small in the entropy while remaining in the regime of validity of the disk-and-cylinder approximation. Then the disk-and-cylinder computations discussed in the first part of this section are sufficient to explain approximate factorization in JT gravity for exponentially long timescales. We give a brief overview of this argument in the main text, with a more detailed discussion in Appendix \ref{appendixbound}.

\end{itemize}

\subsection{Review of JT gravity}\label{jtreview}

Before moving on, we take a moment to review some relevant aspects of JT gravity and introduce some notation.\footnote{For more background see for example \cite{Jensen:2016pah,Maldacena:2016upp,engelsoy2016investigation,Sarosi:2017ykf,Saad:2019lba}.} In pure JT gravity, we can compute partition functions $Z(\beta_1 \dots \beta_n)$ using the Euclidean path integral, summing over spacetimes with $n$ asymptotic boundaries with renormalized lengths $\beta_1 \dots \beta_n$, with the renormalized value of the dilaton fixed and equal along each of these boundaries. The action has a term proportional to the Euler character $\chi$ of the spacetime, $I_{JT}\supset - S_0 \chi$, so that handles and boundaries descrease the path integral weight of a spacetime by powers of $e^{\chi S_0}$. $S_0$ is the zero-temperature entropy, which we take to be large.

We can view these multi-boundary partition functions as correlation functions of partition function operators $\hat{Z}(\beta)$ in the No-Boundary state: $Z(\beta_1\dots \beta_n) \rightarrow \langle \hat{Z}(\beta_1)\dots \hat{Z}(\beta_n)\rangle_{NB}$. Including contributions from surfaces of any genus, the results of \cite{Saad:2019lba} demonstrate that these correlation functions are equal to ensemble-averages of genuine quantum mechanical partition functions $\Tr[ e^{-\beta H}]$, where the Hamiltonian $H$ is a (double-scaled) random matrix drawn from a certain distribution.\footnote{For simplicity, in this paper we focus on the version of JT gravity in which we only include contributions from orientable spacetimes. This theory is described by an ensemble of random Hamiltonians with GUE statistics. Different versions of JT gravity and JT supergravity are related to different random matrix ensembles \cite{Stanford:2019vob}. Versions of JT gravity with a modified dilaton potential have also been related to ensembles of random Hamiltonians \cite{Maxfield:2020ale,Witten:2020wvy,Turiaci:2020fjj,Gao:2021uro}.} To leading order, the average density of states $\langle \hat{\rho}(E)\rangle_{NB}$ can be obtained via an inverse Laplace transform of the disk contribution to the partition function $\langle \hat{Z}(\beta) \rangle_{NB}$. Fluctuations around this average are exponentially small in the entropy.

We can then identify a, fixed choice of Hamiltonian $H$ drawn from this ensemble, or more precisely a fixed set of eigenvalues $\{E_i\}$ of this matrix, with an $\alpha$ state, which may be represented in terms of "Eigenbranes" \cite{Blommaert:2019wfy}.\footnote{In this paper, we restrict our attention to operators such as $\hat{Z}(\beta)$, or superpositions of the $\hat{Z}(\beta)$, which depend only on the eigenvalues of $H$. See \cite{Blommaert:2020seb} for a discussion of observables which depend on the eigenvectors.} Each Hamiltonian in the ensemble has a discrete spectrum.  In JT gravity this discreteness does not follow in a simple way from the genus expansion.\footnote{In a microcanonical ensemble, the discreteness can be understood as due to doubly nonperturbative (in the entropy) D-brane effects. In the canonical ensemble these effects may well be due to a suitably resummed genus expansion \cite{Okuyama:2019xbv,2001math......1201O}.}

In both the MM model and JT gravity, a large part of our focus is on controlling nonperturbative effects related to a discrete spectrum. In the MM model, this was the discrete spectrum of the operator $\hat{Z}$, while in JT gravity this is the discrete spectrum of $H$. We will often view the boundary operators $\hat{Y}_E(T)$ (to be defined momentarily) at many different times as analogous to many copies of $\hat{Z}$. However, in JT gravity these boundary operators themselves do \textit{not} have a discrete spectrum, and the relationship between the nonperturbative effects we study and this discreteness is somewhat more subtle.

We now define the operators $\hat{Y}_E(T)$ and describe their behavior in the No-Boundary state. 
\be\label{JTYdef}
\hat{Y}_E(T)\equiv \frac{\Delta E}{2\pi i} \int_{\epsilon+ i \mathbb{R}} d\beta \; e^{\beta E + \frac{1}{2} \beta^2 \Delta E^2} \;\hat{Z}(\beta +i T).
\ee
$\hat{Y}_E(T)$ can roughly be thought of as a microcanonical version of the analytically continued partition function $\hat{Z}(\beta + i T)$, with the energy fixed in a Gaussian window centered on energy $E$ with width $ \Delta E$. For simplicity we consider only correlation functions of operators with the same energy window $\Delta E$, so we do not need to include an explicit $\Delta E$ label on these operators.\footnote{We focus on the case where $\Delta E$ is of order $S(E)^0$, so that there are $\sim e^{S(E)}$ discrete energy eigenvalues within an energy window.}

In the No-Boundary state, the one-point function of $\hat{Y}_E(T)$ is dominated by the disk topology .\footnote{Until exponentially long times.} The disk has Euler character $\chi=1$, so the topological term in the JT gravity action gives this contribution a weight that is exponentially large in the entropy. But this one-point function decays in time, so at sufficiently late times it is nonetheless small. If the energy window has a small overlap with the edge of the spectrum at $E=0$, then this one-point function decays as a Gaussian, with a width proportional to $1/\Delta E$. We will focus on late times, so we can approximate this one-point function by zero.

The two-point function $\langle \hat{Y}_E(T)^\dagger \hat{Y}_{E'}(T')\rangle_{NB}$ encodes the variance of $\hat{Y}_E(T)$ (and implicitly the density of states), as well as the autocorrelation of $\hat{Y}_E(T)$ for different times and energies. For times less than the "plateau time", $T_{plateau}= 2\pi \rho_0(E)$ with $\rho_0(E)$ the disk contribution to $\langle \hat{\rho}(E)\rangle_{NB}$, the leading connected contribution to the two-point function is given by the cylinder topology. The cylinder has Euler character $\chi=0$, so the topological term in the JT gravity action gives this contribution a weight of order one in the entropy. Unlike the disk, this cylinder does not always decay; for fixed $T-T'$, the cylinder grows linearly in $T+T'$. So the cylinder can dominate at late times. Higher genus contributions to the two-point function are suppressed due to their larger Euler character, and also turn out to decay in time, so they can be ignored.\footnote{If the energy window is chosen to have a large overlap with the edge of the spectrum, these higher-genus contributions can be important at late times \cite{Okuyama:2019xbv,2001math......1201O}.} At the plateau time, nonperturbative corrections become important, and the cylinder no longer dominates. In this paper, we restrict our attention to times $T<T_{plateau}$, so that the cylinder dominates this two-point function.

In gravity, we can view this two-point function as an overlap of two states. Defining $|Y_E(T)\rangle \equiv \hat{Y}_E(T)|NB\rangle$, we find $\langle \hat{Y}_E(T)^\dagger \hat{Y}_{E'}(T')\rangle_{NB} = \langle Y_E(T) | Y_{E'}(T')\rangle$. Inserting a resolution of the identity into this overalap in the form of a sum over projection operators onto states with fixed universe number, the disconnected, two disk contribution to this overlap comes from the contribution of the zero-universe projector $\langle Y_E(T) |NB\rangle\langle NB| Y_{E'}(T')\rangle$, and the cylinder corresponds to the contribution of the one-universe projector.

\begin{figure}[H]
\centering
\includegraphics[scale=0.35]{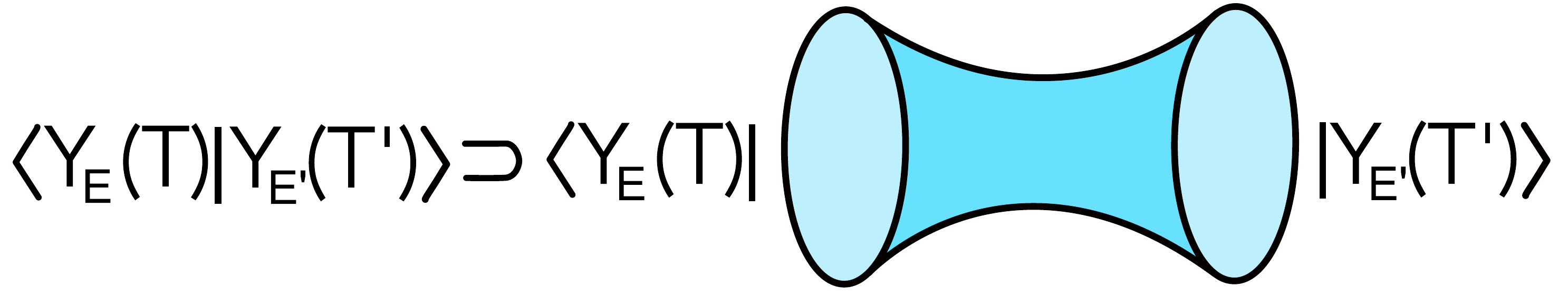}
\caption{\small The cylinder gives the inner product between the one-universe projections of the states $|Y_E(T)\rangle$.}
\end{figure}

In Appendix \ref{autoJTappend} we study the cylinder contribution to the two-point function and give some explicit formulas. Here we briefly review some of the key features. For simplicity, we focus on energy windows which have small overlap with the edge of the spectrum.

For $E=E'$ and $T=T'$, the two-point function $\langle |\hat{Y}_E(T)|^2\rangle_{NB}$ is the averaged (microcanonical) spectral form factor. As expected from random matrix universality, this function has a linear ramp, $\langle |\hat{Y}_E(T)|^2\rangle_{NB}\propto T$, and is independent of the energy $E$. In the gravity computation, this linear ramp results from relative time translations between the two asymptotic boundaries of the cylinder. For $E=E'$, but $T\neq T'$, the two-point function decays as a Gaussian in $|T-T'|$, with a width $\sim 1/\Delta E$. For $E\neq E'$ but $T=T'$, it decays as a Gaussian in $|E-E'|$, with a width $\sim \Delta E$.

The connected contributions to higher point correlation functions of $\hat{Y}_E(T)$ are computed by surfaces with more than two boundaries. and are thus exponentially suppressed in $S_0$.\footnote{In fact, we expect that for correlation functions of $\hat{Y}_E(T)$, these surfaces are suppressed by powers of $e^{-S(E)}$.}  For energy windows which have small overlap with the edge of the spectrum, these contributions decay rapidly in time. Then higher point functions, such as $\langle |\hat{Y}_E(T)|^{2k}\rangle_{NB}$, are dominated by cylinders, with small connected contributions.\footnote{For $k\sim e^{2 S(E)}$ we expect large contributions from nonperturbative effects. These are discussed in Appendix \ref{appendixmoments}.}

Altogether, we see that the $\hat{Y}_E(T)$ for sufficiently widely separated times and energies behave as approximately independent Gaussian variables in the No-Boundary state, with average zero and variance described by the cylinder. In an $\alpha$ state corresponding to a fixed set of boundary energy levels, the spectral form factor oscillates erratically about its average ramp. The size of these oscillations is described the variance of the spectral form factor, which is computed by pairs of cylinders. As in Section \ref{periodicorbits}, these pairings tell us that the typical size of these fluctuations are of order the size of the signal itself.\footnote{For an illustration of this see Figure 20 in \cite{Cotler:2016fpe}.} In a typical $\alpha$ state, this noise is a large correction, of size comparable to the cylinder. Further, we can see that the noise is correlated on a timescale of order $1/\Delta E$. We return to this issue in the next section.

From here, we could analyze JT gravity following the steps from Section \ref{SectionAlphaStates}, using these independent Gaussian variables to construct an (approximately) orthonormal basis for the closed universe Hilbert space, and studying the wavefunctions of $|\psi^2_\alpha\rangle$ in this basis. However, we will adopt a different strategy here, focusing on a basis which is more appropriate for connecting to the effective description.

\subsection{JT in the disk-and-cylinder approximation and an effective model}\label{JTeffectivemodel}

In this section, we describe JT gravity truncated to include only contributions from the disk and cylinder topologies. This truncated version of JT gravity reproduces many aspects of full JT gravity.  For example, correlation functions of $\hat{Y}_E(T)$ in the No-Boundary state for times less than the plateau time will be unchanged by this truncation, up to exponentially small errors. However, for times after the plateau time, and for insertions of many copies of $\hat{Y}_E(T)$, these theories will differ.

 We will show that the disk-and-cylinder version of JT gravity is described by a Gaussian ensemble of theories with \textit{continuous} densities of states, rather than the discrete densities of the full theory. Then $\alpha$ states in this model are described by continuous functions, corresponding to a fixed continuous density of states, rather than a set of discrete energy levels.

As we will see in Section \ref{jtapproximatealpha}, this disk-and-cylinder approximation to JT gravity will be valid not just in the No-Boundary state, but in appropriate approximate $\alpha$ states, with the continuous density describing this truncated theory corresponding to a coarse-grained version of the discrete spectrum in the full theory. However, for now we simply study this disk-and-cylinder model on its own, with the main goal of describing the effective description of this model in an $\alpha$ state.

Our first step is to introduce a basis for the closed universe Hilbert space of this disk-and-cylinder version of JT gravity - the ``$b$'' basis. This basis is a generalization of the $N$ basis in Section \ref{cgsspecies}, with sectors of fixed numbers of closed universes. In that context, the single-universe states were naturally thought of as being defined on spatial slices at an asymptotic boundary, with infinite spatial size. In contrast, the one-universe states in the $b$ basis we use here are defined on geodesic spatial slices with a fixed and finite length $b\geq 0$.\footnote{This basis is a closed universe analog of the $\ell$ basis introduced by \cite{Yang:2018gdb}. Our strategy for computing the inner product mirrors the strategy in that work.}

We denote the one-universe states with fixed length $b$ as $|b\rangle$, and the states of $n$ universes with lengths $b_1\dots b_n$ as $|b_1\dots b_n\rangle$. 
As always, closed universes are identical so $|b_1\dots b_n\rangle$ is symmetric in $b_1\dots b_n$.

To compute the inner product between these states, it is useful to take an indirect approach. We can infer the inner product $\langle b|b'\rangle$ from the inner product of states $\hat{P}_1 |Y_E(T)\rangle$, which is computed by the cylinder with $\hat{Y}_E(T)$ boundaries.  We then write the projector in the $b$ basis, $\hat{P}_1 = \int_0^\infty db \int_0^\infty db' (\langle b|b'\rangle)^{-1} |b\rangle\langle b'|$, with $(\langle b|b'\rangle)^{-1}$ the inverse function of the inner product $\langle b|b'\rangle$. Then
\be\label{innerproductindirect}
\langle Y_E(T)|\hat{P}_1 |Y_{E'}(T')\rangle = \int_0^\infty db \int_0^\infty db' (\langle b|b'\rangle)^{-1}  \; \langle Y_E(T)|b\rangle\langle b'|Y_{E'}(T')\rangle.
\ee
The overlap $\langle b |Y_E(T)\rangle$ is computed by the path integral over cylindrical geometries with a $|b\rangle$ boundary condition and a $| Y_E(T)\rangle$ boundary condition. The $|b\rangle$ boundary condition corresponds to fixing one circular boundary to be a geodesic with length $b$. In this case, we think of the $| Y_E(T)\rangle$ boundary condition as corresponding to fixing the renormalized length of an asymptotic boundary to $\beta+i T$, then integrating over $\beta$ with the weight \eqref{JTYdef}. 

\begin{figure}[H]
\centering
\includegraphics[scale=0.4]{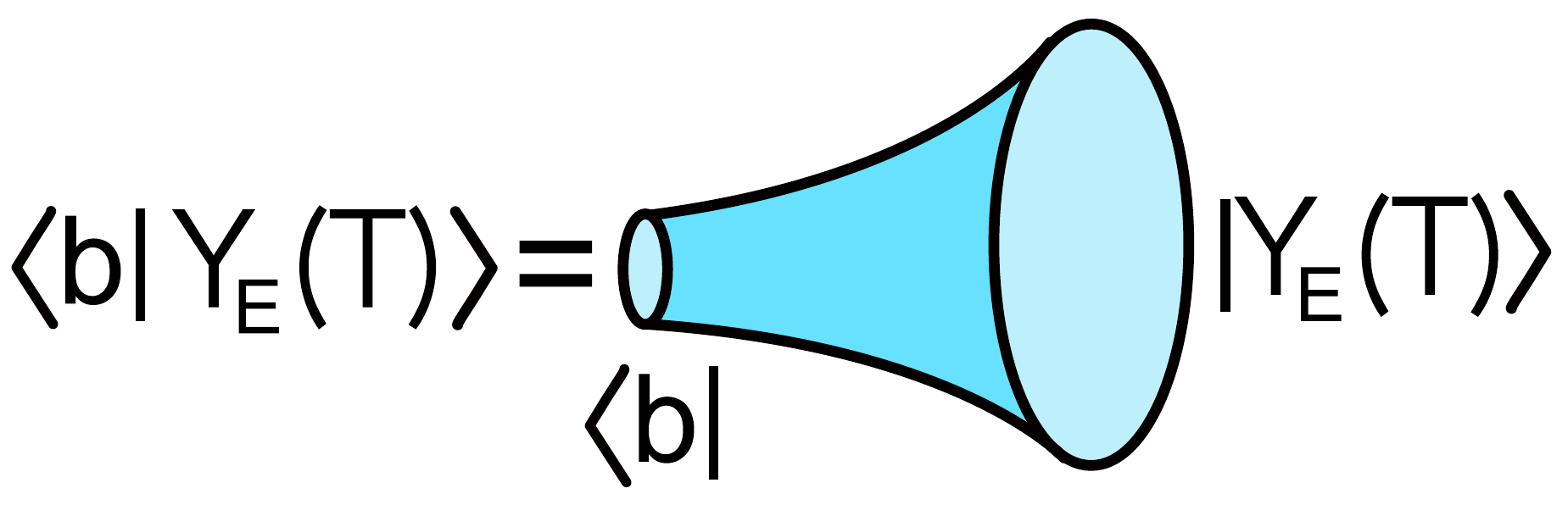}
\caption{\small The "trumpet" geometry gives the overlap between $|Y_E(T)\rangle$ and the one-universe state $|b\rangle$.}
\end{figure}

With fixed $\beta$, the path integral over cylindrical ``trumpet'' geometries is equal to the trumpet partition function $Z_{Tr}(\beta+ i T,b) = \frac{e^{-\frac{1}{2}\frac{b^2}{\beta+ i T}}}{\sqrt{2\pi}\sqrt{\beta+ i T}}$  \cite{Saad:2019lba}. Then, fixing the energy with the integral over $\beta$,
\be\label{trumpetdef}
\langle b| Y_E(T)\rangle = \frac{\Delta E}{2\pi i}\int_{\epsilon+ i \mathbb{R}} d\beta \;e^{E\beta+ \Delta E^2 \beta^2} Z_{Tr}(\beta+ i T,b) \equiv Y_{Tr}(T,E,b).
\ee
Now we use the fact that this cylinder can be expressed as two trumpets glued together along their geodesic boundaries, integrating over the size $b$ of this geodesic with the measure $b db$ \cite{Saad:2019lba},
\be\label{trumpetgluingwormhole}
\langle Y_E(T)| \hat{P}_1 |Y_{E'}(T')\rangle = \int_0^\infty b db\;  Y_{Tr}(T,E,b)^* Y_{Tr}(T',E',b).
\ee
The factor of $b$ in the measure $bdb$ accounts for  the relative twist angle allowed  when when gluing the $b$ boundaries together. 

Using \eqref{trumpetdef} and \eqref{trumpetgluingwormhole} to express both sides of \eqref{innerproductindirect} as integrals over a product of the $Y_{Tr}(T,E,b)$, we can identify the inverse inner product $(\langle b|b'\rangle)^{-1}$ as $b \delta(b-b')$. Then the resolution of the identity in the one-universe sector is $\hat{P}_1 = \int_0^\infty b db |b\rangle\langle b|$, and the inner product is
\be
\langle b|b'\rangle = \frac{1}{b} \delta(b-b').
\ee
The inner product between two $n$-universe states are symmetrized products of the single-universe inner product.

\subsubsection{$\alpha$ states in the $b$ basis}

With this basis in hand, we now apply the results of Section \ref{cgsspecies} to JT gravity with disks and cylinders.

The one-point function of $\hat{Y}_E(T)$ in a state $|\psi\rangle$ can be written as the overlap between the state $|Y_E(T)\rangle$ and $|\psi^2\rangle$, which picks out the zero and one-universe components of $|\psi^2\rangle$. We can express the one-universe contribution in the $b$ basis,\footnote{Following the conventions from Section \ref{SectionAlphaStates}, we express correlation functions as an overlap with the ket $|\psi^2\rangle$, so $\hat{Y}_E(T)$ then acts on the bra $\langle NB|$. In order to express this overlap in terms of the state $|Y_E(T)\rangle$, we take the complex conjugate, using the fact that the components of $|\psi^2\rangle$ is real, because $|\psi^2\rangle = |\psi(\hat{Y}_E(T))|^2|NB\rangle$.}
\begin{align}\label{JTonepointfunction}
\langle \hat{Y}_E(T) \rangle_{\psi} & = \langle Y_E(T) |\psi^2\rangle^*
\cr
& = \langle Y_E(T)|\hat{P}_0 |\psi^2\rangle^* + \langle Y_E(T)| \hat{P}_1 | \psi^2\rangle^*
\cr
& = \disk_E(T) + \int_0^\infty b db \; Y_{Tr}(T,E,b) \langle b|\psi^2\rangle.
\end{align}
In the second line, we introduced a factor of the identity as a sum of the projectors onto the zero-universe and one-universe states. In the third line, we recognized the zero-universe component of $|Y_E(T)\rangle$ as $\disk_E(T)$, and the zero-universe component of $|\psi^2\rangle$ as one, as required by the normalization of $|\psi\rangle$. We also expressed the projector $\hat{P}_1$ in the $|b\rangle$ basis and recognized the overlap $\langle Y_E(T)|b\rangle^* = Y_{Tr}(T,E,b)$.

We can represent this formula pictorially as 

\begin{figure}[H]
\centering
\includegraphics[scale=0.32]{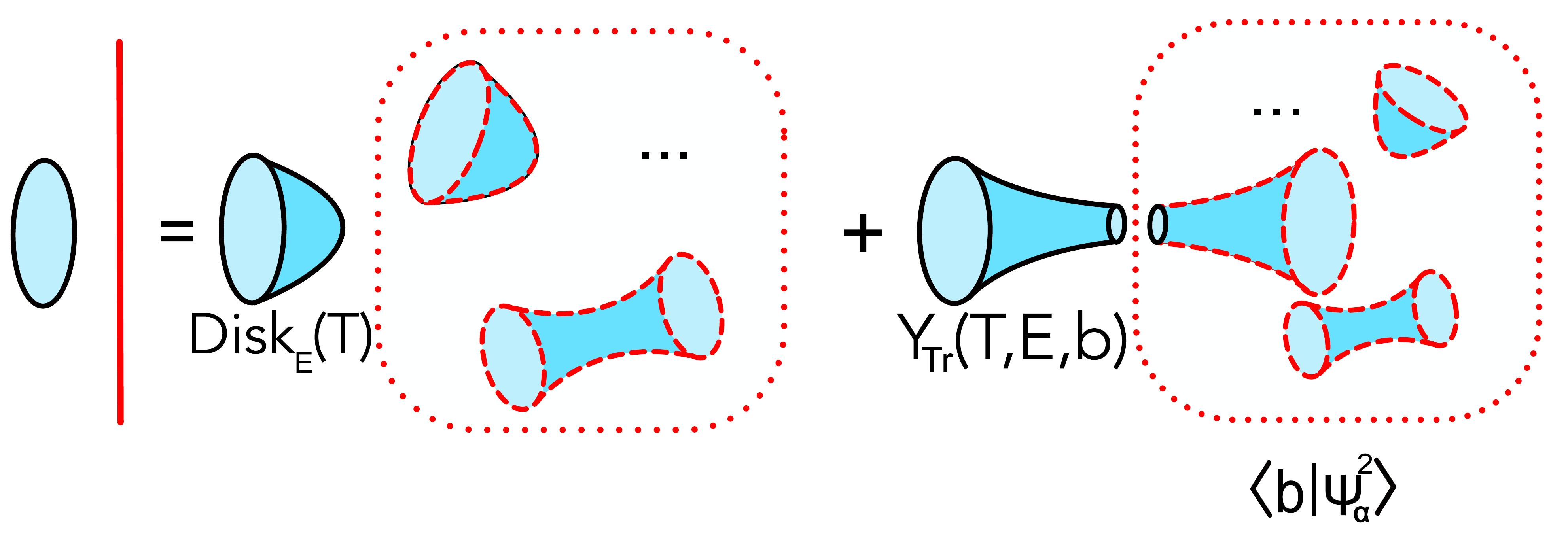}
\caption{\small Viewed as an overlap $\langle Y_E(T) |\psi^2\rangle^*$, the one-point function of $\hat{Y}_E(T)$ has a contribution in which the $\hat{Y}_E(T)$ boundary is capped off on a disk, pictured in the first term on the RHS, and a contribution in which the $\hat{Y}_E(T)$ boundary connects to the $|\psi^2\rangle$ boundaries with a cylinder, pictured in the second term on the RHS. In the first term on the RHS, the disconnected spacetimes factor our and are normalized away, while in the second term, they contribute to the one-universe wavefunction $\langle b|\psi^2\rangle$. This figure is the JT gravity generalization of Figure \ref{OnePointFunction}.}
\end{figure}

If we take $|\psi\rangle$ to be an $\alpha$ state $|\psi_\alpha\rangle$, we can think of the one-universe component $\langle b|\psi^2_\alpha\rangle$ as a random function of $b$ which parametrizes the ensemble of boundary theories. In the disk-and-cylinder approximation, each member $\alpha$ of the boundary ensemble corresponds to a wavefunction $\langle b|\psi^2_\alpha\rangle$, rather than a set of discrete energy levels,
\be
\{E_1, E_2,\dots E_L \} \hspace{20pt} \longrightarrow \hspace{20pt} \langle b|\psi^2_\alpha\rangle.
\ee
This wavefunction is not the only natural choice of variables to parametrize the ensemble; one could instead describe the ensemble in terms of the eigenvalues of boundary operators $\hat{Z}[\mathcal{J}]$, such as $\hat{Z}(\beta+i T)$ or $\hat{Y}_E(T)$, or in terms of the boundary density of states. The wavefunction $\langle b|\psi^2_\alpha\rangle$ is related to the eigenvalues $\hat{Y}_E(T)$ through \eqref{JTonepointfunction}, and then to (a continuous approximation to) the boundary energy levels by \eqref{JTYdef}. In this section we will focus on the discription in terms of $\langle b|\psi^2_\alpha\rangle$, but we will make use of the discription in terms of a random continuous density of states later in Section \ref{jtapproximatealpha}.

The wavefunction $\langle b|\psi^2_\alpha\rangle$ plays a similar role to the random wavefunction $\langle 1|\psi^2_\alpha\rangle$ in the MM model. In the MM model, the wavefunction $\langle 1|\psi^2_\alpha\rangle$ has simple statistics inherited from the No-Boundary state; in the disk-and-cylinder approximation, $\langle 1|\psi^2_\alpha\rangle$ is Gaussian distributed with mean zero and variance one.

In the present case, we also find that $\langle b|\psi^2_\alpha \rangle$ has simple Gaussian statistics. Again using the symbol $\mathbb{E}[\dots]_{NB}$ to denote averaging over the No-boundary distribution, we can infer from equations \eqref{JTonepointfunction} and \eqref{trumpetgluingwormhole},
\be
\mathbb{E}\big[ \langle b|\psi^2_\alpha\rangle \big]_{NB}=0,\hspace{20pt} \mathbb{E} \big[ \langle b|\psi^2_\alpha\rangle  \langle b'|\psi^2_\alpha\rangle^*\big]_{NB} = \frac{1}{b} \delta(b-b').
\ee
Factorization in an $\alpha$ state then implies that the $n$-universe components of $|\psi^2_\alpha\rangle$ are redundant, and simply fixed in terms of the one-universe component. We can illustrate this with the relationship for the two-universe component,
\be\label{factorizationidentityjt}
\langle b_1 b_2 |\psi^2_\alpha\rangle + \frac{1}{b_1} \delta(b_1-b_2) = \langle b_1|\psi^2_\alpha\rangle\langle b_2 |\psi^2_\alpha\rangle.
\ee
This equation is the analog of equation \eqref{eq:twouniversefactorizespecies} in the continuous $|b\rangle$ basis. Here the "diagonal" delta function represents the cylinder in the $|b\rangle$ basis. The "diagonal" part of the RHS of this equation is then related to the cylinder by a "diagonal = cylinder" identity, along the lines of equation \ref{approxdiagcylidentity}.\footnote{In this continuous basis there are some subtleties in defining the "diagonal" part.}

\subsubsection{Effective model with $\Psi$ boundaries}

We now turn to the effective description of JT gravity in an $\alpha$ state. The first step is to describe the random $\Psi$ boundary condition with which we replace the contributions from the $\alpha$ state. In contrast to the MM model, we now have a \textit{choice} of the type of boundary condition for the $\Psi$ boundaries. The $\Psi$ boundary condition replaces the one-universe state $\hat{P}_1|\psi^2_\alpha\rangle$; the choice of type of boundary condition for the $\Psi$ boundaries corresponds to a choice of basis of one-universe states to express $\hat{P}_1|\psi^2_\alpha\rangle$ in.

We focus on a particular choice which we view as natural, which corresponds to the $b$ basis. In the path integral, a one-universe state in the $b$ basis is specified on circular geodesic spatial slices. Then the corresponding $\Psi$ boundaries are circular geodesics, and the random boundary condition is described by a function of the size $b$ of that circular boundary. We call this function $\Psi(b)$. Then geometrically, a broken cylinder with a $\Psi$ boundary is a trumpet, with the size $b$ of the geodesic boundary integrated over with the weight $\Psi(b)$. For example, 
\be\label{JTeffectivecyl}
Y_E^{(\Psi)}(T) \supset \int_0^\infty b db\; Y_{Tr}(T,E,b) \Psi(b).
\ee
Comparing with equation \eqref{JTonepointfunction}, we can see directly that $\Psi(b)$ replaces the one-universe component of $|\psi^2_\alpha\rangle$ in the $b$ basis,
\be
\langle b|\psi^2_\alpha\rangle \hspace{10pt} \longrightarrow \hspace{10pt} \Psi(b).
\ee
Note that this implies that $\Psi(b)$ is real.

\begin{figure}[H]
\centering
\includegraphics[scale=0.4]{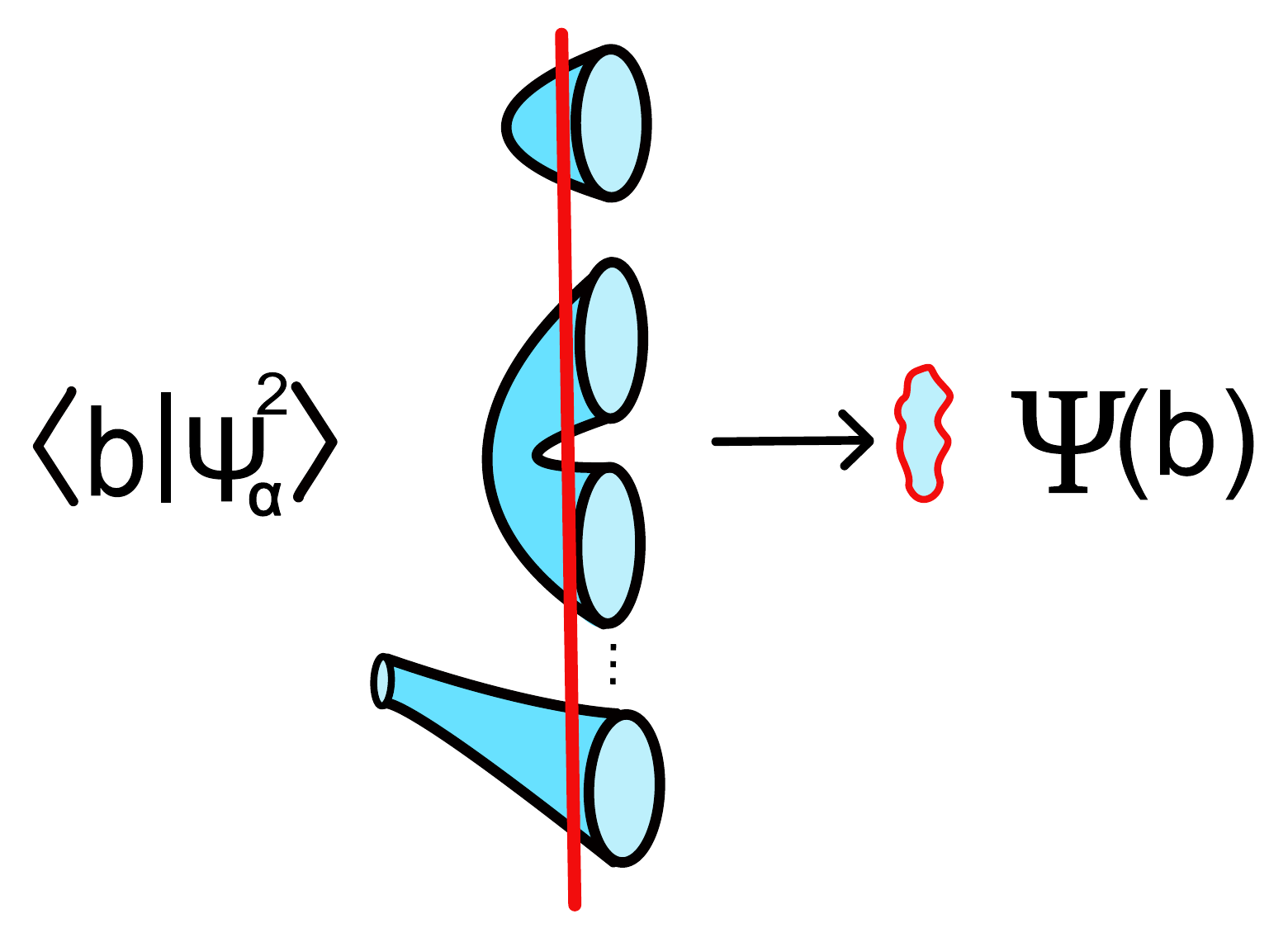}
\caption{\small The one-universe wavefunction $\langle b|\psi^2_\alpha\rangle$ is computed by a sum over spacetimes with a single $|b\rangle$ boundary and many $\alpha$ state boundaries. In the effective description, the boundary condition corresponding to this wavefunction is replaced with the random $\Psi(b)$ boundary condition, denoted by a wiggly red boundary.}
\end{figure}

We should also take $\Psi(b)$ to be drawn from the same distribution as $\langle b|\psi^2_\alpha\rangle$. Denoting averaging over the random $\Psi$ boundary conditions with the symbol $\mathbb{E}[ \dots ]_{\Psi}$
\be\label{psistatisticsjt}
\mathbb{E}\big[\Psi(b) \big]_\Psi=0,\hspace{20pt} \mathbb{E}\big[\Psi(b)\Psi(b')\big]_\Psi = \frac{1}{b}\delta(b-b').
\ee

As we will see momentarily, this implies that averaging glues the broken cylinders together along their $\Psi$ boundaries to make whole cylinders.
\footnote{Douglas Stanford actually suggested such a boundary condition some time ago as a model for how the double trumpet could ``break.''}

As discussed in Section \ref{effectivemodel}, if we compute a product of multiple observables using the effective description, we have two choices of effective description. We may include contributions from cylinders pairing the observable, at the cost of using "linked" $\Psi$ boundaries, or we can use unlinked $\Psi$ boundaries but disallow the cylinder contributions. 

In this case, the linked $\Psi$ boundary condition for a group of $n$ $\Psi$ boundaries is described by a function $\Psi(b_1\dots b_n)$, which replaces the contribution of the $n$-universe component of $|\psi^2_\alpha\rangle$. Then the linked boundary condition $\Psi(b_1\dots b_n)$ is related to the unlinked boundary condition $\Psi(b)$ through equations such as Equation \eqref{factorizationidentityjt}, with the components of $|\psi^2_\alpha\rangle$ replaced appropriately by the $\Psi(b_1\dots b_n)$. As in Section \ref{effectivemodel}, we can think of this relation as due to an ad-hoc "exclusion rule", instructing us to view the linked boundary condition as a product of the unlinked boundary conditions, with the "diagonal" cylinder excluded: $\Psi(b,b') = \Psi(b)\Psi(b')-\frac{1}{b}\delta(b-b')$.

\begin{figure}[H]
\centering
\includegraphics[scale=0.45]{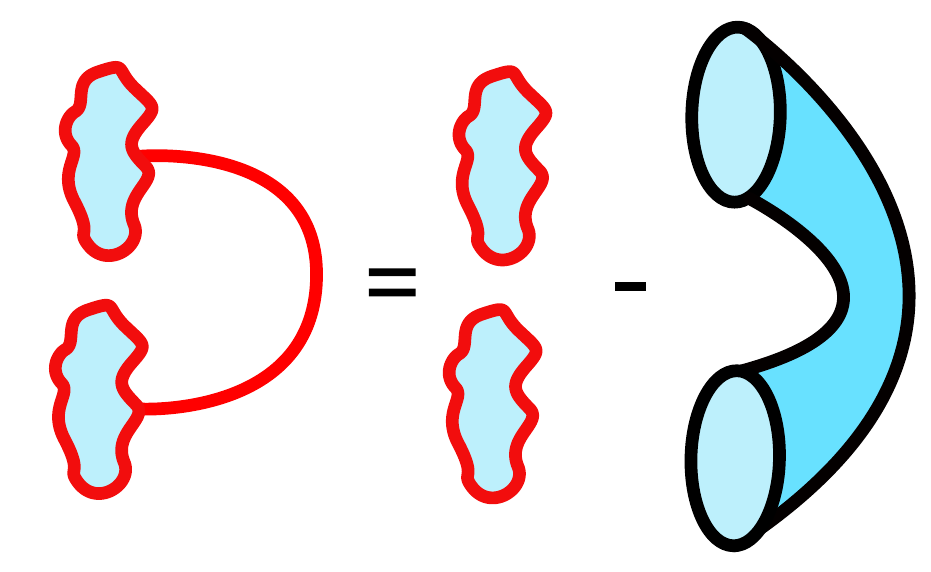}
\caption{\small The linked $\Psi$ boundary condition, represented on the LHS, is equal to a product of unlinked $\Psi$ boundary conditions minus the cylinder: $\Psi(b,b') = \Psi(b)\Psi(b')-\frac{1}{b}\delta(b-b')$. In appropriate applications of the effective model, we can use an approximate "diagonal=cylinder" identity to think of both sides of this equation as "off-diagonal".}
\end{figure}

Having introduced this effective description of JT gravity, we can now illustrate some of its uses. 

The first, and simplest, use is to reproduce the wormhole computations of moments upon averaging. To illustrate this, consider the contribution of a pair of broken cylinders, with unlinked $\Psi$ boundaries, to a product of observables $Y_E^{(\Psi)}(T) Y_{E'}^{(\Psi)}(T')$.
\be\label{JTsffeffectivemodel}
\int_0^\infty b db \; Y_{Tr}(T,E,b) \Psi(b) \int_0^\infty b' db' \; Y_{Tr}(T',E',b') \Psi(b') .
\ee
If we average this product using \eqref{psistatisticsjt}, we set $b=b'$ and integrate. This integral glues together the two fixed-energy trumpets along the $b$ circle, resulting in the cylinder contribution to $\langle \hat{Y}_E(T) \hat{Y}_{E'}(T')\rangle_{NB}$ in the full theory.

\begin{figure}[H]
\centering
\includegraphics[scale=0.4]{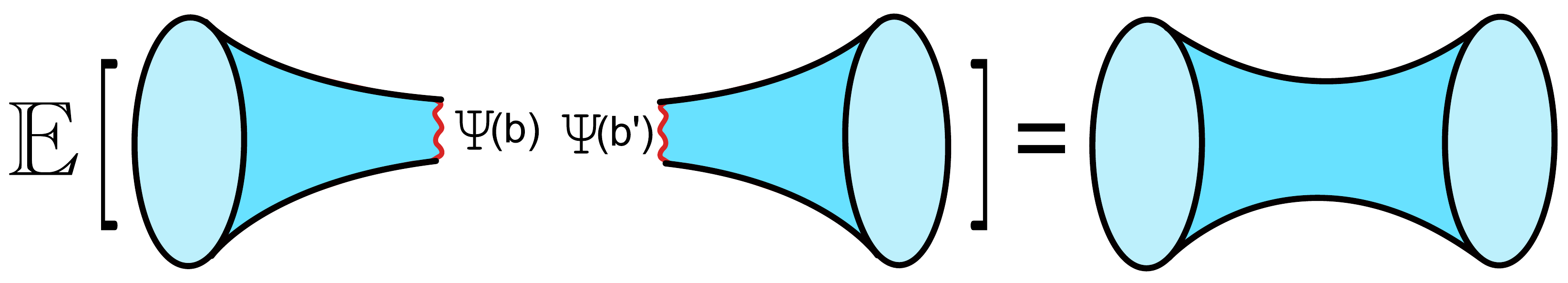}
\caption{\small Averaging a pair of broken cylinders using \eqref{psistatisticsjt} sets $b=b'$, so that the broken cylinders are glued together along their $\Psi$ boundaries to form a full cylinder}
\end{figure}

We can reproduce the wormhole computation of higher moments of $Y_E^{(\Psi)}(T)$ using the unlinked $\Psi$ boundaries in an analogous fashion.  Upon averaging, pairs of broken cylinders are glued together along their $\Psi$ boundaries by the mechanism we have just seen. Summing over all of the ways to glue together pairs of broken cylinders, we reproduce the wormhole computation in the full theory.

If we use the effective description with whole cylinders and linked $\Psi$ boundaries, we reproduce the averaged description more immediately. The contributions of broken cylinders with linked $\Psi$ boundaries average to zero, leaving behind just the whole cylinders.

We also expect that time-averaging, rather than ensemble averaging, may glue the broken cylinders together to make wormholes. We can use this effective description to confirm this expectation, and verify that time-averaging the spectral form factor over an appropriate time window has the same effect as ensemble-averaging.

Focusing on the contribution \eqref{JTsffeffectivemodel} to the fixed-energy spectral form factor, we now average in time with a Gaussian time window, with a width $\Delta_T$. We find
\begin{align}\label{JTtimeave1}
\int_{-\infty}^\infty dT'\; \frac{e^{-\frac{(T-T')^2}{2\Delta_T^2}}}{\sqrt{2\pi \Delta_T^2}} \;|Y_E(T')|^2 &\supset \int_0^\infty b db \int_0^\infty b' db' \; \Psi(b) \Psi(b')
\cr
&\hspace{20pt} \times \bigg[\int_{-\infty}^\infty dT'\frac{e^{-\frac{(T-T')^2}{2\Delta_T^2}}}{\sqrt{2\pi \Delta_T^2}} Y_{Tr}(T',E,b) Y_{Tr}(T',E, b')\bigg] .
\end{align}
Let's focus on the integral in the square brackets. We can study the fixed energy trumpets $Y_{Tr}(T',E,b)$ by doing the integral \eqref{trumpetdef} by saddle point. Each trumpet is peaked at $T' \sim b^2/E$. For $b\sim b'$, the product product $Y_{Tr}(T',E,b) Y_{Tr}(T',E, b')$ is then peaked near $T' \sim b^2/E$. Around this peak, this product oscillates in $T'$ with a frequency proportional to $b^2-b'^2$. Then the function in the square brakets, given by an integral over $T$, decays in $|b^2-b'^2|$. For $\Delta_T$ much bigger than the autocorrelation time $1/\Delta E$, this decay is sharp, and we can treat the function in the square brackets like an approximate delta function. This approximate delta function approximately glues together the $\Psi$ boundaries, so this contribution to the time-averaged spectral form factor is schematically
\be\label{JTtimeave2}
\int_0^\infty b db \int_0^\infty b' db' \; \Psi(b) \Psi(b') \; |Y_{Tr}(T,E,b)|^2 \;\delta_{Approx}(b-b') \equiv \int_0^\infty bdb \; \big(\Psi(b)^2\big)_{Reg.} \; |Y_{Tr}(T,E,b)|^2.
\ee
Here we have defined a regulated version of $\Psi(b)^2$, obtained by integrating the product $\Psi(b)\Psi(b')$ with the approximate delta function.

\begin{figure}[H]
\centering
\includegraphics[scale=0.3]{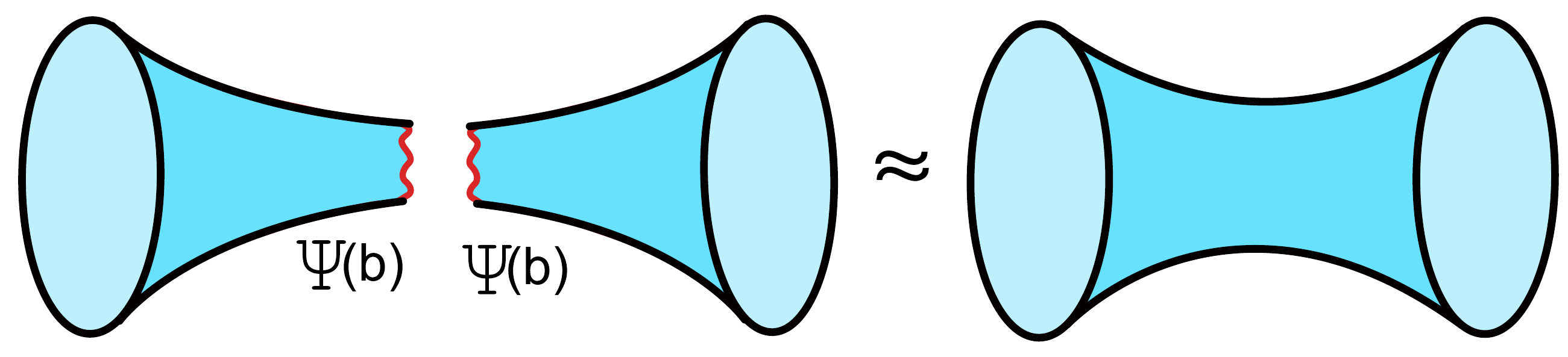}
\caption{\small Time-averaging sets $b\approx b'$ in the pair of $\Psi$ boundary conditions. This has the effect of approximately gluing together the broken cylinders to form a full cylinder.}
\end{figure}

This expression looks somewhat like the expression \eqref{trumpetgluingwormhole} for the cylinder as two trumpets glued together. However, this depends explicitly on the function $\Psi(b)$, so it is not exactly equal to the cylinder. Equation \eqref{trumpetgluingwormhole} would result from replacing the factor $\big(\Psi(b)^2\big)_{Reg.}$ by its averaged value of one.

Fortunately, we can show that for a typical draw of $\Psi(b)$, we can replace the factor $\big(\Psi(b)^2\big)_{Reg.}$ (or equivalently the product $\Psi(b)\Psi(b')$) with its average value up to a small error, so the result \eqref{JTtimeave2} from the time-averaging is \textit{approximately} equal to the cylinder. To do this, it is useful to take a somewhat indirect approach and show that the LHS of \eqref{JTtimeave1} is \textit{self-averaging}: it is equal to its ensemble-average up to a small error. Then \eqref{JTtimeave2} is also self-averaging. Intuitively, what is happening is that the sample-to-sample fluctuations in of $\big(\Psi(b)^2\big)_{Reg.}$ add incoherently in  \eqref{JTtimeave2}, so while $\big(\Psi(b)^2\big)_{Reg.}$ itself has large fluctuations for fixed $b$, these fluctuations cancel out when integrated over. \footnote{This is a similar phenomenon that we used to explain the diagonal=cylinder relation in Appendix \ref{diagselfaveraging}.}

To show that the LHS of \eqref{JTtimeave1} is self-averaging, we calculate its variance and showing that it is small if we take the size of the time window to be sufficiently large. The variance is computed by the the connected correlator of the LHS with itself,
\be\label{JTtimeaveselfaveraging}
\int dT'  \frac{e^{-\frac{(T-T')^2}{2\Delta_T^2}}}{\sqrt{2\pi \Delta_T^2}} \int dT'' \frac{e^{-\frac{(T-T'')^2}{2\Delta_T^2}}}{\sqrt{2\pi \Delta_T^2}} \; \mathbb{E}\big[ |Y_E^{(\Psi)}(T')|^2 |Y_E^{(\Psi)}(T'')|^2\big]_{\Psi,c}.
\ee
The connected contribution to the correlator is computed by wormholes connecting the $Y_E(T'), Y_E(T')^*$ to the $Y_E(T''),Y_E(T'')^*$. The wormholes connecting two $Y$s or two $Y^*$s give small contributions for large $T$, so we ignore them. We are left with the modulus squared of the wormhole contribution to $\mathbb{E}[Y_E(T') Y_E(T'')^*]$, which as we see in Appendix \ref{autoJTappend} decays as a Gaussian in $|T'-T''|$ with a width $\sim 1/\Delta E$,
\be\label{JTconnectedcorrelator}
\mathbb{E}\big[ |Y_E(T')|^2 |Y_E(T'')|^2\big]_{\Psi,c} \approx C (T'+T'')^2 \exp\big[- \frac{3}{2} \Delta E^2 (T'-T'')^2\big].
\ee
Here $C$ is a normalization factor.

We now consider two limiting cases:  $\Delta_T \ll 1/\Delta E$, and $\Delta_T \gg 1/\Delta E$. Inputting \eqref{JTconnectedcorrelator} into \eqref{JTtimeaveselfaveraging}, we see that for small $\Delta_T$, the integrals over $T'$ and $T''\approx T$ are sharply peaked around $T'\approx T''$, so that we can approximate \eqref{JTconnectedcorrelator} by its value at $T'=T''=T$. This gives the square of a linear ramp in $T$. Time averaging this over the small time window has little effect, so we can ignore the time averaging altogether. The result is approximately equal to the disconnected contributions to the correlator of two copies of the the LHS of \eqref{JTtimeave1}, so the fractional variance is large, and thus time-averaging over a small window fails to average out the noise in the spectral form factor, as we expect.

However, if we take the size of the time window large compared to $1/\Delta E$, the autocorrelation function still forces the integral to be sharply peaked around $T' = T''$, relative to the size of the time windows. If we perform the integrals over $T'$ and $T''$, only a small portion of the time windows contribute, with $T'\approx T''$. Since the time windows are chosen to be normalized, the result from just the center of the time windows is small, $\sim \frac{1}{\Delta E \Delta_T}$. 

We can understand this suppression by thinking of $|Y_E(T)|^2$ for times separated by an autocorrelation time $\sim 1/\Delta E$ as independent random variables. Then averaging over a time window of size $\Delta_T$, we find an average of $N_{windows}\sim \Delta_T \Delta E$ independent random numbers. The fractional variance of such a sum is of order $1/N_{windows}$  So for time windows $\Delta_T\gg 1/\Delta E$, the size of this connected contribution is smaller than the disconnected contribution by a factor $\sim \frac{1}{\Delta E \Delta_T}$. This means that for sufficiently large time time windows, the time-averaged spectral form factor has a small variance and thus is self-averaging, again as we expect.

\subsubsection{Comments on dynamical realizations of the $\Psi$ boundaries}\label{eowbranesection}

In the effective model, the $\Psi$ boundaries are assigned a random boundary condition, in correspondence with the random one-universe wavefunction $\langle b|\psi^2_\alpha\rangle$ in the full description. It is natural to consider a variant of this effective model, which may not match as precisely with JT gravity, but may serve as a model which is more appropriate for understanding conventional non-averaged examples of gauge/gravity duality. 

In this variant of the effective model we would introduce \textit{dynamical} end-of-the-world (EOW) branes. Even in amplitudes without external EOW branes, there can be dynamical boundaries with the topology of a circle, corresponding to "loops" of these EOW branes. For appropriately chosen EOW brane dynamics, the path integral with a $Y_E(T)$ boundary and a circular EOW brane boundary might produce a function like \eqref{JTeffectivecyl}, but with the random function $\Psi(b)$ replaced by a \textit{pseudorandom} function $\Psi_{EOW}(b)$, where the pseudorandomness would be the result of some chaotic dynamics of the brane. This pseudorandom function would behave like a typical draw of the random function $\Psi(b)$.  Such an EOW brane model might serve as a step towards a microscopic understanding of analogs of the broken cylinders with $\Psi$ boundaries in non-averaged theories.

Some examples of JT gravity with EOW branes have been studied in \cite{Kourkoulou:2017zaj,Goel:2020yxl,Gao:2021uro,Antonini:2021xar}.\footnote{A version of JT gravity coupled to a scalar field with a random boundary condition was recently described in \cite{Garcia-Garcia:2021squ}, and related to the SYK model. In this model, the random boundary conditions seems to play a similar role to the random $\Psi$ boundary, or EOW brane boundary.} We have not been able to construct any compelling models of EOW branes that reproduce the effective description of JT gravity in an $\alpha$ state, so rather than describing our attempts, we briefly highlight some of the challenges.
\begin{itemize}{}

\item There are a number of natural boundary conditions for EOW branes in JT gravity, outlined in \cite{Goel:2020yxl}. A natural choice for our purposes corresponds to EOW branes that live on geodesic boundaries, with a given tension. Branes of this type, possibly with random tensions or with some chaotic internal dynamics, are a natural first guess for reproducing the effective $\Psi(b)$ boundaries.

In the effective model, $\Psi(b)$ was a random Gaussian function with mean zero. A typical draw oscillates erratically around zero as a function of $b$. If we choose to include EOW branes with the geodesic boundary condition, reproducing this erratic oscillatory behavior seems to require some unusual features of the brane dynamics or random tensions. The basic problem is that these branes follow Euclidean trajectories and so naturally produce functions $\Psi_{EOW}(b)$ which behave more like thermal partition functions at inverse temperature $\propto b$. In particular they would normally be positive.\footnote{An example of an unusual feature would be to allow imaginary tensions to produce oscillations, an approach that Henry Lin has also explored.}

\item The erratic oscillations needed are perhaps more naturally produced by EOW branes which follow Lorentzian trajectories, or which are separated from the $b$ circle by some Lorentzian region of the spacetime. For example, the pseudorandom phases produced by the chaotic real-time evolution of a brane following a Lorentzian trajecory might result in an appropriate erratic function. The role of these Lorentzian EOW brane trajectories might then be somewhat subtle if we consider analtyically continuing these geometries.

\item These EOW branes may have additional difficulties in reproducing the early-time behavior of $Y_E(T)$. The early time behavior of the cylinder with $Y_E(T)$ boundaries is dominated by small $b$, suggesting that short EOW brane trajectories may be important at these timescales. In Section \ref{periodicorbits}, analogous "short orbits" give the leading density of states, rather than the small fluctuations; similarly, the early time contributions of EOW branes pose a danger of overcounting the disk contribution to $Y_E(T)$.\footnote{One possibility is that the EOW branes only serve as a good model for sufficiently late times/long trajectories. Another possibility is that the EOW brane loop contribution \textit{replaces} the disk. This possibility was discussed briefly in \cite{Gao:2021uro}.}

\item Perhaps the  most difficult task for the EOW brane models would be to reproduce the \textit{linked} $\Psi$ boundaries. To do this seems to require a nonlocal coupling between EOW branes on disconnected spacetimes, or some other principle to "exclude" the diagonal part of the contribution of two EOW branes. One possibility is that there are two versions of the theory, one version with the cylinders excluded, and another version with the cylinders included, but with a nonlocal coupling between different EOW branes. This possibility is reminiscent of what happens in the SYK model \cite{Saad:2021rcu}. Another, perhaps related, scenario is that there is some principle in the bulk theory which instructs us to identify the cylinder and the diagonal part of the pair of EOW brane boundaries as \textit{equivalent} in some sense; then instead of overcounting by summing over both contributions, we have a choice of including one or the other. Perhaps such a "quantum equivalence" can be viewed as some sort of gauge equivalence. We comment further on these  points in the \hyperref[discussion]{Discussion}.

\end{itemize}

In the \hyperref[discussion]{Discussion}, we briefly discuss another posible dynamical realization of the broken cylinders with random $\Psi$ boundaries, in which the role of the $\Psi$ boundary is played by the black hole singularity in ageometry related to the "double-cone" \cite{Saad:2018bqo}.

\subsection{Approximate $\alpha$ states in JT gravity}\label{jtapproximatealpha}

In the remainder of this section we study the validity of the disk-and-cylinder approximation for correlators in certain approximate $\alpha$ states in JT gravity. We begin by explaining the relationship between the disk-and-cylinder approximation and the approximation of the boundary theory in terms of Gaussian fluctuations of a coarse-grained, continuous density of states. This prompts us to consider approximate $\alpha$ states which roughly correspond to (approximately) fixing a given draw of this coarse-grained density. 

Correlation functions  of operators which are insensitive to the fine details of the spectrum computed in such states should be well-approximated with just disks and cylinders.\footnote{Though the D-brane effects described in \cite{Saad:2019lba} are given by many copies of the disk and cylinder spacetimes, we do not allow such effects in our meaning of "disk-and-cylinder approximation".} Intuitively, we expect that correlation functions of small numbers of operators $Y_E(T)$ provide an example of such correlation functions, as long as the times $T$ are much less than the plateau time.\footnote{At the plateau time, we expect that nonperturbative "D-brane" effects become important \cite{Saad:2019lba}.} After explaining these ideas we conclude this section with a brief overview of a rough estimate for a bound which we describe in more detail in  Appendix \ref{appendixbound}.  This argument in turn is based on  the conjectured behavior for certain nonperturbative contributions, which we describe in Appendix \ref{appendixmoments}.  Though we do not expect that this bound is optimal, it at least indicates that the disk-and-cylinder approximation can be accurate in "typical" approximate $\alpha$ states with an exponentially small variance for exponentially long times (but still exponentially shorter than the plateau time).

\subsubsection{Describing JT gravity in terms of a continuous density of states}

In Section \ref{SectionAlphaStates} we found it useful to focus on a choice of independent Gaussian observables. In order to connect more closely with the CGS model, we would then  like to study approximately independent Gaussian variables in JT gravity. We have already seen that the $Y_E(T)$ are an example of such variables, but in this section we will take a different route towards finding these variables which will be useful for understanding the disk-and-cylinder approximation.

The correlation functions of interest to us in JT gravity can be expressed as integral transforms of the density of states $\hat{\rho}(E)$. In the No-Boundary state, these correlators can be expressed using the boundary description as a double-scaled random matrix integral. For simplicity, we will instead work with an approximate description in terms of a finite-dimsensional matrix integral over an $L\times L$ matrix with eigenvalues $E_i$, which we think of as describing the theory within some energy window centered at energy $E_w$. The typical spacing between neighboring eigenvalues is of order $1/\rho_0(E_w)$, where $\rho_0(E_w)$ is the leading order density of states computed by the disk topology, so for an energy window of width $\Delta E$, we take $L \sim \Delta E \rho_0(E_w) = \Delta E e^{S(E_w)}$.  Since we are focusing on the case $\Delta E\sim S(E_w)^0$, we can take $L\sim e^{S(E_w)}$.

 Correlators of the density at external energies $E^{ext}_1\dots E^{ext}_n$ and the matrix integral can be expressed as in integral over just the eigenvalues of the matrix,
\be\label{MatrixInt}
\langle \hat{\rho}(E^{ext}_1)\dots\hat{\rho}(E^{ext}_n)\rangle_{NB} \propto \int d\{E_i \} \; P(\{ E_i\}) \; \sum_{i_1} \delta(E^{ext}_1- E_{i_1})\dots \sum_{i_n} \delta(E^{ext}_n- E_{i_n}).
\ee 
The probability density $P(\{ E_i\})= \Delta(\{E_i\})^2 \exp\big[- L \sum_i V(E_i) \big]$ is a product of the square of the Vandermonde determinant of the matrix and a factor involving the eigenvalue potential $V(E)$. It is useful to combine these two factors into an effective action for the eigenvalues,
\be\label{effectiveactioneigen}
I_{eff}(\{E_i\}) = -\frac{1}{L^2} \sum_{i\neq j} \log |E_i- E_j|^2 + \frac{1}{L} \sum_i V(E_i),
\ee
so that $P(\{E_i\}) = \exp[-L^2 I_{eff}(\{E_i\})]$. The log term describes a long-ranged repulsive force between the eigenvalues.

Rather than directly studying the form \eqref{MatrixInt} of the matrix integral, we change variables from the eigenvalues to the density of states. It is convenient to use a normalized density of states,
\be
\tilde{\rho}(E) \equiv \frac{1}{L} \sum_i \delta (E - E_i).
\ee
 We can then rewrite the action \eqref{effectiveactioneigen} in terms of $\tilde{\rho}(E)$,
\be\label{effectiveactionrho}
I_{eff}[\tilde{\rho}(E)] = -\int' dE dE' \; \tilde{\rho}(E) \tilde{\rho}(E')\; \log|E-E'|^2 + \int dE \;\tilde{\rho}(E) \;V(E).
\ee
Here the prime on the integral denotes removing the coincident energy singularity and corresponds to the omission of the $i=j$ term in \eqref{effectiveactioneigen}. In the equations of motion resulting from varying this action this can be understood as a principal part prescription for the integral.

In terms of $\tilde{\rho}(E)$, the classical effective action is \textit{quadratic}, and independent of $L$. Of course, we should also keep in mind corrections from the Jacobian for this change of variables. We will simply assert that to leading order, this Jacobian is one.\footnote{This can be confirmed by comparing computations in the $\tilde{\rho}(E)$ variables with computations using the original variables. At subleading order, one find corrections to this Jacobian which give non-Gaussian interactions for $\tilde{\rho}(E)$. Unlike the quadratic terms, perturbatively these higher order terms do not give contributions to the density correlators which are singular as energies approach each other. So they give contributions to correlation functions of $Y_E(T)$ which decay exponentially in time.} This is intuitive, because we expect that the distinction between a continuous density of states and a discrete density of states is unimportant to leading order for many observables.

 Then, to leading order the $\tilde{\rho}(E)$ varibles are Gaussian, with the fluctuations of these variables described in the bulk by cylinders. It is natural to then diagonalize the propagator for the $\tilde{\rho}(E)$ to find a set of (approximately) \textit{independent} Gaussian variables. Expanding around the saddle point by shifting $\tilde{\rho}(E) = \tilde{\rho}_0(E)+\delta \tilde{\rho}(E)$, with $\tilde{\rho}_0(E)$ determined by varying \eqref{effectiveactionrho}, we find that the propagator for $\delta \tilde{\rho}(E)$ is obtained by inverting the logarithm term in \eqref{effectiveactionrho}. As this depends only on the difference of energies, to diagonalize the propagator we introduce the Fourier transform variable\footnote{Because we are working in a finite energy window, it may be more appropriate to express $\delta\tilde{\rho}(E)$ as a sum of discrete Fourier modes. In the next subsection we will take this point of view.}
\be
\delta\tilde{\rho}(s) = \int dE\; e^{- i E s} \; \delta\tilde{\rho}(E).
\ee
If we are careful with regulating the double integral in \eqref{effectiveactionrho}, we find that the $\delta\tilde{\rho}(s)$ have a propagator
\be
\langle \delta \tilde{\rho}(s) \delta\tilde{\rho}(s')\rangle_{\text{Matrix Integral}} \propto s \;\delta(s-s').
\ee
One might have expected this result, as $\delta\tilde{\rho}(s)$ is essentially the fluctuating part of $Y_{E_w}(T)$ after identifying $s$ and $T$. The formula \eqref{JTYdef} expresses $Y_{E_w}(T)$ as a similar Fourier transform of the density of states, but restricted to an energy window. Because our effective matrix integral \eqref{MatrixInt} is meant to describe JT gravity within an energy window, we can roughly identify these two variables.
This means that we can roughly think of \eqref{effectiveactionrho} as an effective action for $Y_{E_w}(T)$ in the disk-and-cylinder approximation. As long as this effective action is valid, the disk-and-cylinder approximation is valid.

The description of JT gravity in terms of the continuous density of states is closely related to the effective description described earlier in the section. Writing $Y_E(T)$ for fixed density $\tilde{\rho}(E)= \tilde{\rho}_0(E)+ \delta \tilde{\rho}(E)$ as
\be
Y_E(T) = \int dE' \; e^{-\frac{(E-E')^2}{2\Delta E^2} + i E' T } \big(\tilde{\rho}_0(E')+ \delta \tilde{\rho}(E')\big),
\ee
we can identify the term involving $\tilde{\rho}_0(E)$ as coming from the disk topology, and the term involving $\delta \tilde{\rho}(E)$ as corresponding to a broken cylinder with a $\Psi(b)$ boundary. Using \ref{JTeffectivecyl}, we may express $\delta \tilde{\rho}(E)$ as an integral transform of $\Psi(b)$, and vice versa.

As was the case for the MM model described  in Section \ref{MMmodel}  there are two main failure modes for the disk-and-cylinder approximation: perturbative non-Gaussian effects, and effects from the discreteness of the boundary spectrum. We would  like to understand these issues from the point of view of this effective action. The perturbative non-Gaussian effects correspond to the contributions of surfaces with more than two boundaries, and we will be able to study these corrections with a straighforward generalization of the analysis in Section \ref{MMmodel}. Nonperturbative corrections related to the discreteness of the boundary spectrum are more difficult to study. To  get some intuition we will briefly explain some expectations for the limitations of the continuous density approximation due to the effects from this discreteness.

\subsubsection{Effects from discreteness and coarse-graining the density of states}

We certainly expect  the effective action \eqref{effectiveactionrho} to break down at sufficiently high frequencies $s$, corresponding to long times $T$. For $s\sim e^{S(E_w)}$, modes $\tilde{\rho}(s)$ describe structure in the spectrum at the scale of the typical spacing between the discrete eigenvalues. Instead, we should think of $\tilde{\rho}(s)$ as only defined for $s$ less than some cutoff frequency $s_{cutoff}\sim e^{S(E_w)}$. Using the rough equivalence between the frequency $s$ and the time $T$, we expect the effective action \eqref{effectiveactionrho}  to only be accurate for describing correlators on timescales less than the plateau time $T_{plateau}\sim  e^{S(E_w)}$.\footnote{One consequence of this is that we should not think of $Y_{E_w}(T)$ for $T>T_{plateau}$ as an independent degree of freedom, but one built out of a linear combination of $Y_{E_w}(T)$ for $T< T_{plateau}$.}

As a check, we can compare the number of independent modes $\tilde{\rho}(s)$ with the number $L \sim \Delta E e^{S(E_w)}$ of discrete eigenvalues within our energy window. As we are working within a finite energy window, we can describe $\tilde{\rho}(E)$ in terms of modes $\tilde{\rho}(s_n)$ with discrete frequencies $s_n \sim n/ \Delta E $. The $\tilde{\rho}(s_n)$ are roughly equivalent to $Y_{E_w}(T_n)$ where the $T_n$ are times evenly spaced by the autocorrelation time $\Delta T\sim 1/\Delta E$, so we are thinking of the $Y_{E_w}(T)$ for times spaced by this amount as the independent Gaussian variables.\footnote{See Appendix \ref{autoJTappend} for more details.} To count the total number of independent variables, we divide the maximum frequency $s_{cutoff} \sim \rho_0(E_{window})$ by the frequency spacing, or equivalently the plateau time by the autocorrelation time. This gives an estimate of $\sim \Delta E e^{S(E_w)}\sim L$ independent modes, roughly matching the number of eigenvalues in the window. 

This leads us to expect that the disk-and-cylinder approximation, or equivalently the approximation using the effective action \eqref{effectiveactionrho}, is valid for correlators of (order one numbers of) $Y_{E_w}(T)$ in the No-Boundary state, for $T \ll T_{plateau}$ but not for longer timescales. In examples such as the spectral form factor,this matches our expectations from random matrix theory; the ramp in the spectral form factor for $T<T_{plateau}$ is desribed by the cylinder, but the pleateau is described by a many-universe "D-brane" effect.

What do we expect for the validity of this continuous-density description for $\alpha$ states, or approximate $\alpha$ states? For exact $\alpha$ states, the discrete eigenvalues are exactly fixed, so to accurately describe a given $\alpha$ state we must describe the spectrum at down to scales below the average level spacing. The continuous density approximation with modes of frequency $s<s_{cutoff}$ is insufficient to describe these states, and so we expect the disk-and-cylinder approximation to fail.\footnote{Using eigenbranes \cite{Blommaert:2019wfy} to approximate and exact $\alpha$ state in an energy window involves inserting $\sim e^{S(E_w)}$ eigenbranes.  It appears that the disk-and-cylinder approximation is not controlled in this case, and nonperturbative "D-brane" effects are important,  in accordance with our expectations. See \cite{Blommaert:2021gha} for progress in controlling this highly nonperturbative regime.}
On the other hand, we may consider approximate $\alpha$ states which correspond to approximately fixing the modes $\tilde{\rho}(s_n)$ for $s_n$ less than some maximum $s$, which should be less than $s_{cutoff}$. In states where we only fix this \textit{coarse-grained} density, the density is allowed to fluctuate at small energy scales, but correlators of small numbers of $\hat{Y}_{E_w}(T)$ for times before the corresponding time cutoff should only depend weakly on these fluctuations. 

We now discuss such fixed coarse-grained density states in more detail, and then turn to  the validity of this continuous density approximation, and thus the disk-and-cylinder approximation, in some of these states.

\subsubsection{Random draws of the coarse-grained density and approximate $\alpha$ states}\label{JTaprroxalphastate}

To recap the previous section, correlators of operators\footnote{We simplify our notation in this section by denoting $\hat{Y}_{E_w}(T)$ as $\hat{Y}_{E}(T)$, without the explicit "w" subscript. This should not cause confusion, as this is the only appearance of an "$E$" variable in the remainder of this section.} $\hat{Y}_{E}(T)$ are sensitive only to features in the density of states at energy scales $\sim 1/T$, so they can be described by a coarse-grained, continuous density of states with Fourier modes of frequency up to $s_{max}\sim T$ fixed. It is natural then to consider approximate $\alpha$ states in which the density is approximately fixed to a given random draw of this coarse-grained density, with a small variance. We expect that as long as the variance is not too small, so that the coarse-grained density is not fixed too sharply, the continuous-density approximation (and correspondingly the disk-and-cylinder approximation) will continue to be valid.

In practice, to construct these states we use the fact that the operators $\hat{Y}_E(T)$ roughly correspond to Fourier modes of the density with frequency $\sim T$. We then define our approximate $\alpha$ states by approximately fixing $\hat{Y}_E(T)$ for $n_{max}$ evenly spaced times $T_n$, with the spacing between times $\Delta T \equiv T_{n+1}-T_n$.
\begin{align}\label{JTapproxalphastate}
|\{y_E(T_n)\}\rangle &\equiv \prod_{n=1}^{n_{max}} \exp\bigg[\frac{|\hat{Y}_E(T_n)- y_E(T_n) |^2}{ 4 \Delta^2}\bigg] |NB\rangle
\cr
& \propto \prod_{n=1}^{n_{max}} \int_{-\infty}^\infty dp_n dq_n \exp\bigg[-\Delta^2(p_n^2 + q_n^2) + i p_n \text{Re}\big(\hat{Y}_E(T_n)- y_E(T_n)\big) 
\cr
&\hspace{100pt}+ i q_n \text{Im}\big(\hat{Y}_E(T_n) - y_E(T_n)\big)\bigg] |NB\rangle.
\end{align}
The complex numbers $\{y_E(T_n)\}$ are the fixed values of $\hat{Y}_E(T_n)$. Here we will take the $\{y_E(T_n)\}$ to be a the values of $\hat{Y}_E(T)$ for a typical draw from the No-Boundary ensemble, so that $|\{y_E(T_n)\}\rangle$ approximates a typical $\alpha$ state. In these states, $\hat{Y}_E(T)$ is strongly correlated at nearby times.

\begin{figure}[H]
\centering
\includegraphics[scale=0.45]{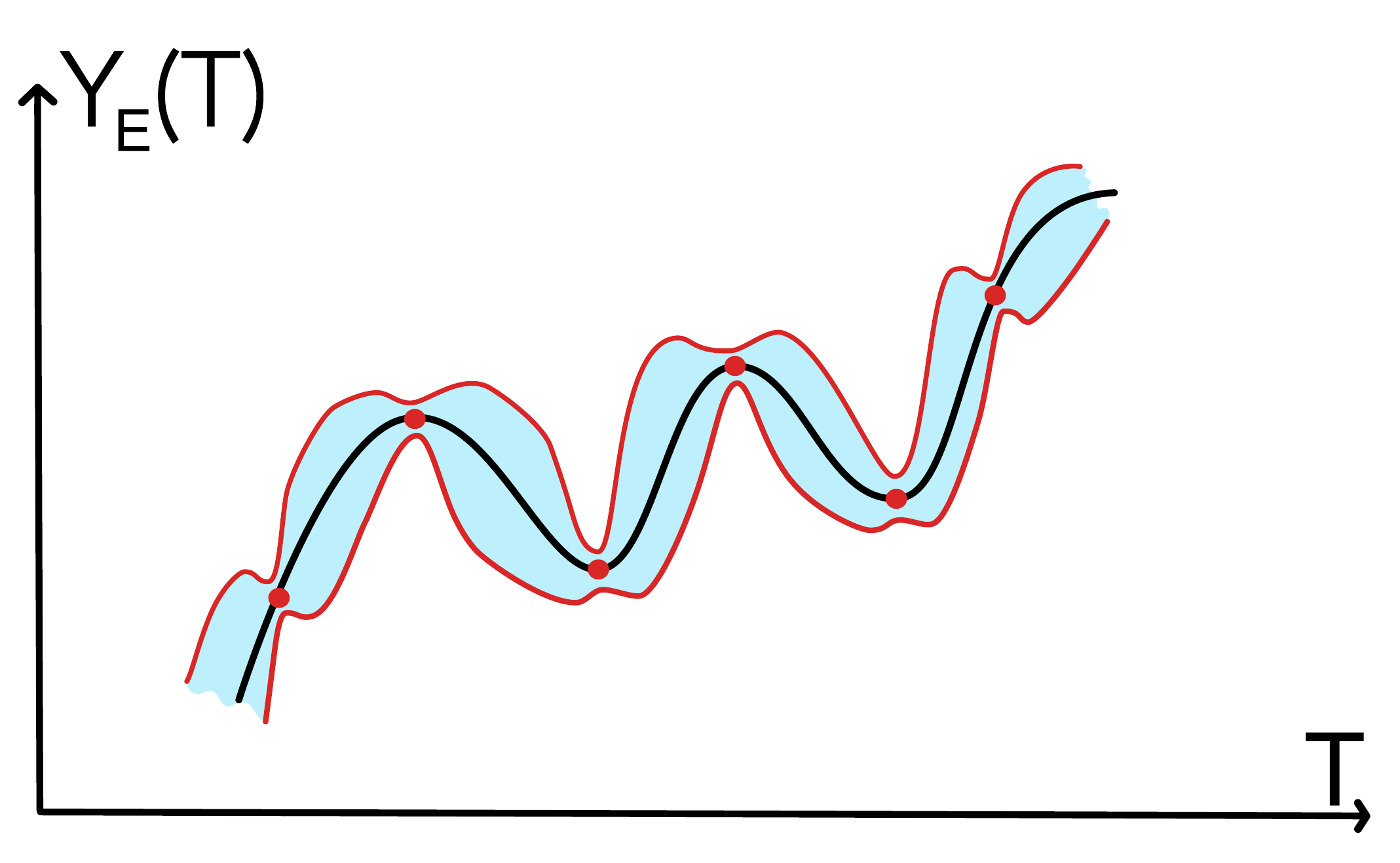}
\caption{\small Here we have schematically pictured the distribution of $\hat{Y}_E(T)$ in an approximate $\alpha$ state. The black curve is the value of $\hat{Y}_E(T)$ in the exact $\alpha$ state which we are approximating. The blue regions surrounded by the red curves denote the width of the wavefunction around the black curve. At the times $T_n$, denoted with the red dots, the wavefunction is fixed tightly, with a small width $\Delta$. Between the $T_n$, the width of the wavefunction (the variance of $\hat{Y}_E(T)$) grows as $\Delta E^2 |T-T_n|^2$, where $T_n$ is the nearest fixed time. (This image greatly exaggerates the scale of variation of the black curve relative to the width of the approximate $\alpha$ state. In a typical $\alpha$ state, $\hat{Y}_E(T)$ varies by order one on a timescale of order one, while the width of the approximate $\alpha$ state can be made exponentially small. So in the cases of interest, the blue region would have a width that is exponentially small compared to the scale of variation of the black curve.)}
\end{figure}

We can think of this state as an approximation to an exact $\alpha$ state with eigenvalues $\{y_E(T_n)\}$ of $\hat{Y}_E(T_n)$. If the spacing $\Delta T$ is sufficiently small (but not too small), the approximate $\alpha$ state $|\{y_E(T_n)\}\rangle$ is \textit{almost} an eigenstate of $\hat{Y}_E(T)$ for all times $T$ between $T_1$ and $T_{n_{max}}\equiv T_{max}$, in the sense that the (fractional) variance of $\hat{Y}_E(T)$ for these times is small. 

This is clear if we focus on $\hat{Y}_E(T)$ at the discrete times $T_n$. At these times the state \eqref{JTapproxalphastate} is a Gaussian with width $\sim \Delta$, which as we will argue in the next section we may take to be small. In between times $T_n$ and $T_{n+1}$, the variance of $\hat{Y}_E(T)$ is larger, but if we take the time spacing $\Delta T$ sufficiently small compared to the autocorrelation time $\sim 1/\Delta E$, then we expect that the variance will still be small. Working in the disk-and-cylinder approximation, and assuming that $\Delta T\ll 1/\Delta E$, we find that the variance at times in between $T_{n}$ and $T_{n+1}$ is roughly given by the sum of the variance at the fixed times, which is of order $\Delta^2$, and a term given by the size of the decay of the autocorrelation function $\langle \hat{Y}_E^\dagger(T) \hat{Y}_E(T_n)\rangle_{NB}$.\footnote{If we simply assume the disk-and-cylinder approximation, we find that computations in the state \eqref{JTapproxalphastate} give the expected answer; $\hat{Y}_E(T)$ is equal to $y_E(T_n)$ at times $T_n$ with a variance of order $\Delta^2$. But this does not mean that the disk-and-cylinder approximation is correct; for example, in the MM model we saw that if $\Delta$ is too small, corrections to the disk-and-cylinder approximation can be large order-by-order, though they add to give a numerically small correction. In the next section we will also find that we must take $\Delta$ to be sufficiently small if the corrections are to be controlled.} The autocorrelation function is a Gaussian of width $\sim 1/\Delta E$, so expanding for small time differences $\sim \Delta T$, this contribution is of order $\Delta T^2 \Delta E^2$. For simplicity, we consider the case in which $\Delta T^2 \Delta E^2\gg \Delta^2$. Then the fractional variance of $\hat{Y}_E(T)$ in between times $T_n$ and $T_{n+1}$ is at most of order $\Delta T^2\Delta E^2$.\footnote{More generally, the errors from our approximate fixing at times $T_n$ and from the fluctuations in between times $T_n$ add in quadrature, so the variance is roughly bounded above by $c_1\Delta ^2+ c_2\Delta T^2\Delta E^2$, for some constants $c_1$ and $c_2$ which are order one in the entropy.}

To summarize, in these approximate $\alpha$ states, assuming that $\Delta \ll \Delta T/\Delta E$ the variance $V(T)\equiv \langle|\hat{Y}_E(T)|^2 \rangle_{\{y_E(T_n)\}}  - |\langle  \hat{Y}_E(T) \rangle_{\{y_E(T_n)\} }|^2$ is roughly bounded by
\be\label{JTvariancebound}
\Delta^2 \lesssim V(T) \lesssim \Delta T^2\Delta E^2,\hspace{20pt} T_1 < T< T_{max}.
\ee

If $\Delta T \Delta E$ is of order one, as would be the case if we spaced the $T_n$ by an autocorrelation time, the variance in between the $T_n$ can become of order one. We would like to avoid this by taking $\Delta T \ll 1/\Delta E$.\footnote{In the previous subsection, we gave an intuitive picture in which we considered fixing the coarse-grained density by fixing its independent modes, which we roughly identified as the $\hat{Y}_E(T)$ at times $T_n$ spaced by an autocorrelation time. Here we see that we must space the $T_n$ more closely than an autocorrelation time if we want our approximate $\alpha$ state to have a small variance, because the $\hat{Y}_E(T)$ at times that are closer than the autocorrelation time are not completely correlated. Our intuitive picture was too rough to account for this. } As we have discussed in the previous subsection, we expect that we can only fix $\hat{Y}_E(T)$ at $n_{max} \sim e^{S(E)}\sim L$ times. To be conservative, we should take $n_{max}\ll e^{S(E)}$. Then if we choose $\Delta T\Delta E \sim e^{-c S(E)}$ with $c<1$, in order to make the maximum variance \eqref{JTvariancebound} exponentially small, we find that $T_{max}= n_{max}\Delta T$ may be exponentially long. 

For example, using the bound on $\Delta$ we argue for in the next section, taking $n_{max} \sim e^{(1-\epsilon) S(E)}$, and $c<\frac{2}{3}$ so that we can approximate $\Delta T \Delta E \gg \Delta$, the variance is roughly bounded by
\be
e^{-(2-c) S(E)} \lesssim V(T)\lesssim e^{-2 c S(E)}, \hspace{20pt} T \lesssim e^{(1-c-\epsilon) S(E)}.
\ee
 Here we can see that there is a competition between making $T_{max}$ longer and making the variance smaller. If we wanted to take $T_{max} \sim e^{S(E)}$ so it covers an appreciable fraction of the ramp, we would have to take $c\approx 0$ so that the maximum variance is large. However for $c$ of order one, but less than one (for example $c= 1/2$),  the variance is exponentially small for an exponentially long time (but still an exponentially small fraction of the plateau time). \footnote{It may be that if we approximately fix $\hat{Y}_E(T)$ at $L$ (or $\sim L$) times, then the variance will be small even between the $T_n$. This is because the $L$ variables $\hat{Y}_E(T_n)$ are a complete set of variables for our theory in an energy window with $L$ eigenvalues; exactly fixing these $\sim L$ values of $\hat{Y}_E(T_n)$ should pick out an exact $\alpha$ state with zero variance at all times, so perhaps approximately fixing these variables is enough to approximately fix $\hat{Y}_E(T)$ for all times on the ramp. However, it is not clear that the disk-and-cylinder approximation is valid in such a state.}

Our goal in the next section is to argue that for these values of $c$, the disk and cyinder approximation is still valid. We emphasize that our goal is not to provide an optimal bound on the range of validity of the disk-and-cylinder approximation, but instead to simply argue for its validity over  considerable timescale.

\subsubsection{Conjecture for a lower bound on the range of validity of the disk-and-cylinder approximation} \label{JTapproxbound}

In this section be briefly summarize an argument described in Appendix \ref{appendixbound} for a bound on the range of validity of the disk-and-cylinder approximation for computations of correlators in the approximate $\alpha$ states we have just described. This argument relies on some conjectured behavior about correlation functions of many copies of $\hat{Y}_E(T)$ in the No-Boundary state, which we discuss in Appendix \ref{appendixmoments}. Our strategy is to first fix the $n_{max}$ choices of $T_n$ at which $\hat{Y}_E(T)$ is approximately fixed with a small Gaussian width $\sim \Delta$, assuming $n_{max}\ll e^{S(E)}$, $T_{n_{max}}\ll T_{plateau}$, then estimate the smallest value $\Delta$ for which the disk-and-cylinder approximation is still valid for computations of correlators (of a small number of operators) in this state.

First, for simplicity, we consider the case in which we approximately fix $\hat{Y}_E(T)$ at only one moment of time, $T_1$, to a value $y_E(T_1)$, using a Gassian wavefunction of with a small width $\sim \Delta$. Our goal is to find a lower bound on the value of $\Delta$ for which the disk-and-cylinder approximation is still valid for correlators of a small number of $Y_E(T)$ in this state. If we make $\Delta$ too small, we expect that nonperturbative effects will become important. We also focus on the case that $y_E(T_1)$ is a \textit{typical} value of $\hat{Y}_E(T)$ in the No-Boundary ensemble. 

The logic of our argument follows the logic of the MM model discussion in Section \ref{MMmodel}. For simplicity, we focus on the norm of our approximate $\alpha$ state, which we can write as
\be\label{approxalphanormjtonetime}
\int_{-\infty}^\infty dp_1 dq_1\; e^{-\frac{\Delta^2}{2}(p_1^2+q_1^2)} \Big\langle e^{ i p_1 \text{Re}[\hat{Y}_E(T_1)- y_E(T_1)] + i q_1 \text{Im}[\hat{Y}_E(T_1)- y_E(T_1)]}\Big\rangle_{NB}.
\ee
We can then evaluate the expectation value as\footnote{The $k=1$ term in this sum is normalized to zero.}
\be\label{cumumlantjtforargumentsummary}
\Big\langle e^{ i p_1 \text{Re}[\hat{Y}_E(T_1)] + i q_1 \text{Im}[\hat{Y}_E(T_1)]}\Big\rangle_{NB} = \exp\bigg[\sum_{k=1}^\infty \frac{1}{k!} \bigg\langle \bigg( i p_1 \text{Re}[\hat{Y}_E(T_1)] + i q_1 \text{Im}[\hat{Y}_E(T_1)] \bigg)^k\bigg\rangle_{NB,c} \bigg].
\ee
The disk-and-cylinder approximation corresponds to keeping only the disk for $k=1$ and the cylinder for $k=2$, then dropping all terms for $k\geq 3$. Inserting this into \eqref{approxalphanormjtonetime}, the resulting integral over $p_1,q_1$ can be treated as a Gaussian integral around an order one value of $p_1$ and of $q_1$ (with the center of the Gaussian determined by $y_E(T_1)$).\footnote{We emphasize that the disk-and-cylinder approximation is only valid for the $k=1,2$ terms for $T_1<T_{plateau}$, which we are assuming is the case for all times under consideration.} 

To justify this approximation we must argue that both perturbative and nonperturbative corrections to this approximation are small. Perturbative corrections to disk-and-cylinder approximation, given to leading order by the $k=3$ terms in \eqref{cumumlantjtforargumentsummary}. The $k=3$ term is given to leading order by the three-holed sphere, which is exponentially suppressed $\sim e^{- S(E)}$, and also decaying in $T_1$. This term can be enhanced at large $y_E(T_1)$, but we find that for typical values of $y_E(T_1)$ in the No-Boundary ensemble, this term is still small even without taking the decay in $T_1$ into account.

The nonperturbative corrections to the disk-and-cylinder approximation come from regions of large $p_1, q_1$ in the integral \eqref{approxalphanormjtonetime}. At sufficiently large $p_1,q_1$, \eqref{cumumlantjtforargumentsummary} cannot be approximated using just the disk and cylinder. Instead, we expect that the terms for all $k$ will be important, and that the sum over $k$ will lead to a function in $p_1, q_1$ which, after inserting into \eqref{approxalphanormjtonetime}, will lead to additional saddle points at large $p_1,q_1$. The expansion around these saddle points would not be well-approximated by disks and cylinders, and would instead depend on terms for large $k$. For $k$ much less than $L$, these are approximately given by the $k$-holed sphere, but for sufficiently large $k$, we expect that nonperturbative effects in the JT gravity matrix integral dominate these terms. 

In the MM model, we could explicitly study the analogous terms for all $k$ (see equation \eqref{eq:nonperturbperiodic}) and sum the series to find additional saddle points. In the present case we do not have enough control to do this. Instead, we will simply estimate the region in the $p_1, q_1$ integration space for which the $k\geq 3$ terms in \eqref{approxalphanormjtonetime} are unimportant. Then we can choose $\Delta$ to be large enough that the Gaussian in \eqref{approxalphanormjtonetime} suppresses the potentially dangerous region of the integration space where these terms may be important. This estimate relies on the behavior of the large $k$ terms in \eqref{cumumlantjtforargumentsummary}, which should be dominated by nonperturbative effects. In Appendix \ref{appendixmoments}, we give a conjecture for this behavior. 

The end result is that we estimate  the $k\geq 3$ terms to be unimportant if $|p_1|, |q_1| \ll L\sim e^{S(E)}$.\footnote{As a reminder, $L$ and $e^{S(E)}$ only differ by a factor of $\Delta E$, which we take to be of order one, so we often use $L$ and $e^{S(E)}$ interchangeably.} Then we can choose $\Delta \gg 1/L$ to suppress these contributions.\footnote{One might worry that the additional saddle points have a large action, which overwhelms the suppression from the Gaussian term in $\Delta$. However, for real $p_1,q_1$, the expectation value \eqref{cumumlantjtforargumentsummary} is an average of phases with a normalized probability distribution, so it is at most equal to one. Typically, we expect that it is much smaller than one. Note that in the MM model, the analogous expectation value is equal to its minimum value of one at values of $p$ spaced by $2\pi e^{S_0}$; this was possible because the eigenvalues of $\hat{Z}$ are evenly spaced.} This matches the intuition that the $\hat{Y}_E(T)$ for a single time $T$ behave like a single copy of $\hat{Z}$ in the MM model.

It is simple to modify this argument to include the case in which we approximately fix $\hat{Y}_E(T)$ at $n_{max}$ times $T_n$, for $n_{max}\ll L$ but still exponentially large in $S(E)$. As discussed earlier, in order to make the variance \eqref{JTvariancebound} small for all times in between the $T_n$, we should choose the spacing $\Delta T$ between the $T_n$ to be small compared to the autocorrelation time $1/ \Delta E$. The main additional complication in the argument is that we are now approximately fixing variables $\hat{Y}_E(T_n)$ which are correlated with $n_{coupled}\sim \frac{1}{\Delta T \Delta E} \gg 1 $ other $\hat{Y}_E(T_n)$. Accordingly, we must be careful to take the fixed values $y_E(T_n)$ to correspond to a typical draw from the No-Boundary ensemble, so that they are appropriately correlated. 

In Appendix \ref{appendixbound} we argue that if $\Delta \gg \sqrt{n_{coupled}}/L$, then the disk-and-cylinder approximation is valid. This bound allows for the variance \eqref{JTvariancebound} to be made exponentially small for an exponentially long time while still remaining in the regime of validity of the disk-and-cylinder approximation. To illustate this bound, we consider an example in which we fix $n_{max} \sim e^{c_1 S(E)}$ modes with $c_1<1$, at times spaced by $\Delta T \sim e^{- c_2 S(E)}/\Delta E$ with $c_1 > c_2$ so that the maximum time is $T_{max} \sim e^{(c_1- c_2) S(E)} \Delta E$. Taking the energy window to have a size of order one, we can satisfy this bound by taking $\Delta \gg e^{- (1-c_2/2) S(E)}$. As an explicit example we can take  $\Delta \sim \Delta T \sim e^{-\frac{1}{2} S(E)}$, and $T_{max}\sim e^{\frac{1}{2} S(E)}$, while still being far from saturating this bound.

We end by emphasizing that in the case where we fix many modes our bound is quite rough, and may well be far from optimal. To estimate the contribution of the $k$-holed spheres, we ignored the decaying behavior in the times, and we were somewhat conservative in estimating the nonperturbative contributions to the large $k$ terms in \eqref{cumulantmanytimes}. It may well be that with a more careful analysis, one would find that nonperturbative effects in \eqref{JTfixmanytimes} come from even larger values of $p_n, q_n$ than we estimated. However, we do expect that nonperturbative effects must eventually become important, as perturbation theory should break down in an exact $\alpha$ state. Motivated by the behavior of the MM model, it seems reasonable to expect that $\Delta$ cannot be made as small as $e^{-S(E)}$ while remaining within the regime of validity of the disk-and-cylinder approximation.

\section{Discussion}\label{discussion}
The models discussed in this paper -- CGS, MM and JT -- are defined as ensemble averages, which we do not expect to be the case for standard holographic systems.   As a consequence these systems are  naturally formulated in a description involving arbitrarily many boundaries, a dramatic extension of the usual AdS/CFT framework.\footnote{We should note though that for finite dimensional Hilbert spaces the notion of number of boundaries, that is the number of traces in an observable, is not unique.  In particular $\text{Tr} [ e^{-iHT}]$ for large enough $T$ is not linearly independent of products of traces with shorter time evolutions. This is a null state effect in a many-boundary Hilbert space. In the ramp region this is not an important effect.}

One of the main goals of this paper has been to show how an effective description can be constructed that employs only a finite number of extra ($\Psi$) boundaries, equal to the number of probe partition functions in the correlator (as in the periodic orbit analogy, but with ``fundamental'' wormholes).  But this description comes with a price.   
We must posit an ad hoc
 ``exclusion rule'' that eliminates either the ``diagonal''
  contribution, or the spacetime wormhole to avoid overcounting (Furthermore, these two contributions must be adjusted to be numerically equal, by the tuning that picks out an $\alpha$ state.).   Where could such a rule come from in a ``few boundary'' description?

\subsection{SYK model}

One hint comes from the study of another ensemble system, the SYK model.   With Douglas Stanford we have recently studied this system with {\it fixed} fermion couplings, so it is described by a single boundary Hamiltonian \cite{Saad:2021rcu}.   We have been able to study aspects of a simple model of this system in some detail, arriving at a plausible description of the full SYK model with fixed couplings.

If we consider two decoupled SYK-like systems labelled L and R we can construct in the standard way a collective field representation using collective fields $G_{LL}, \Sigma_{LL}$ describing correlations within the L system, $G_{RR}, \Sigma_{RR}$ and fields $G_{LR}, \Sigma_{LR}$ describing correlations between L and R.   A saddle point with nonzero LR fields is the analog of a wormhole.   Although these collective fields are usually employed after averaging over fermion couplings, it is possible to construct such a description when the fermion couplings are {\it fixed} at the expense of having a complicated
 coupling-dependent weighting factor.\footnote{Note that the function $\Psi(\sigma)$ in \cite{Saad:2021rcu} does \textit{not} play an analogous role to the $\Psi$ boundary condition described in this paper. Rather, it is the function $\Phi(\sigma)$, which encodes the dependence on the random couplings, which is more closely analogous to the contribution of a random $\Psi$ boundary.}  One major finding of \cite{Saad:2021rcu} is that there are new saddle points when couplings are fixed, called ``half-wormholes'' in \cite{Saad:2021rcu}, which seem to be an analog of the broken cylinders discussed in this paper.

 But with fixed couplings we have a choice of collective field description.   We can include the LR fields or not, at our discretion.   Without the LR fields the collective field description manifestly factorizes, into unlinked broken cylinders.  But with it, there can be a wormhole saddle point and factorization is not manifest.  There are other saddle points, called ``linked half-wormholes'' in \cite{Saad:2021rcu}, that are the analog of the linked broken cylinders pictured in Figure \ref{figlinkedpsi}.  The equivalence of these two descriptions relies on an approximate identity relating the contributions of the linked and unlinked half-wormholes; a direct analog of our equation \ref{eq:twouniverserelation}. This relation encodes a semiclassical ``exclusion rule'' and ``diagonal $=$ cylinder'' identity.
 
 In the SYK model, the $G,\Sigma$ collective field integral serves as a proxy for a bulk description. The two formulations of the problem, with and without the LR fields, correspond to two different bulk descriptions, which must be equivalent. Explaining this equivalence is necessary to explain the relation between the linked and unlinked half-wormholes (the ``exclusion rule'' and ``diagonal $=$ cylinder'' identity). This equivalence may be established by integrating the auxiliary LR fields along their defining contour, undoing the procedute of introducing them in the first place. The integral over the defining contour does not have an apparent geometric interpretation, so in SYK the exclusion rule might be explained through a "nongeometric" mechanism.

\subsection{Some open questions}
There are many additional open questions. Here we enumerate a few.

\begin{itemize}
\item Are there other models described by a variant of this effective description, for which the exclusion rule follows from a different principle than the one seen in the many-universe description? Such a principle would give some sort of "quantum equivalence" between the wormhole and the "diagonal" contribution of broken cylinders. 

Examples of possibly related quantum equivalences have appeared in a number of recent contexts. For the tensionless string, Eberhardt \cite{Eberhardt:2021jvj} demonstrated that perturbative fluctuations around a fixed background \textit{include} the contributions of other semiclassical geometries. Jafferis and Schneider \cite{Jafferis:2021ywg} studied examples of worldsheet dualities between backgrounds with a two-sided black hole and backgrounds corresponding to entangled states of disconnected spatial geometries. Marolf and Maxfield \cite{Marolf:2020xie} described how semiclassically distinct states, describing geometries with possibly different numbers of spatial components, can be "gauge equivalent", differing only by a null state.\footnote{This seems to be an equivalence of a different type, however. Roughly, the gauge equivalence described by Marolf and Maxfield is an equivalence between states on (possibly disconnected) spatial slices, not an equivalence between the contributions of \textit{spacetimes} to the path integral. This gauge equivalence does not require us to "gauge fix" the gravitational path integral by identifying contributions from different spacetimes, while an equivalence between the cylinder and the diagonal broken cylinders would require such a "gauge fixing". We comment more on the relationship between the gauge equivalence identified by Marolf and Maxfield and the exclusion rule in Appendix \ref{appendixsharing}.} 

\item These issues about quantum equivalences may in fact show up before thinking about wormholes. It has been suggested for a long time that in a microscopic description of quantum gravity, the Euclidean black hole should be replaced by a sum over horizonless geometries, with the horizon at the tip of the cigar replaced by some sort of microstructure.\footnote{For further discussion of these issues see \cite{Harlow:2020bee}. } Fuzzballs \cite{Mathur:2012np,Mathur:2014nja,Mathur:2014dia} are an example. This would be related to the ER=EPR statement that the Hartle-Hawking state can be viewed as an entangled state of one-sided black holes. On the other hand, in the examples we have looked at in this paper (as well as in the SYK model), one does not replace the black hole, but adds a small noisy contribution. The difference is illustrated in Figure \ref{punctureddisk}

\begin{figure}[H]
\centering
\includegraphics[scale=0.35]{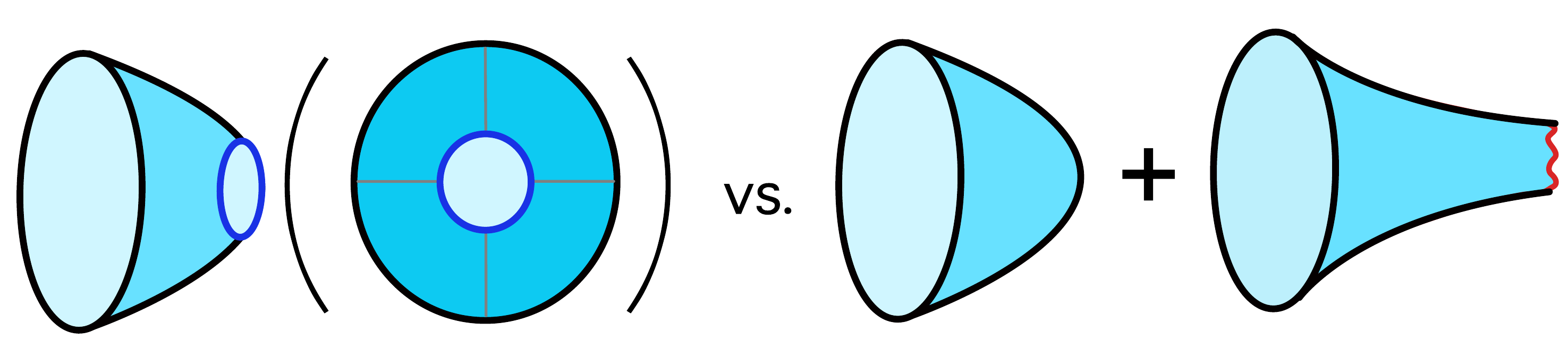}
\caption{\small Here we illustrate the two scenarios. On the left we picture a punctured disk (in a side and front view). We can view this contribution as a trace over microstates, where the grey lines indicate time slices. On the right we picture the scenario seen in the ensemble models studied in this paper, in which we add a noisy broken cylinder to the conventional black hole.}
\label{punctureddisk}
\end{figure}

In the periodic orbit examples, "short orbits" play the role of the black hole, giving the smooth density of states, while the long orbits contribute a small noisy correction. In the MM model and JT gravity, the "broken cylinder" was \textit{added} to the disk, giving a small noisy correction.

We don't know what the correct picture is in a more general theory of gravity, but perhaps there is some choice between these two pictures, along the lines of the exclusion rule and quantum equivalences discussed in the previous point. In this case there would be two ways of computing the partition function, one with the black hole, and one without, related by some sort of equivalence.\footnote{One can think of the disk in SYK as built up out of perturbation corrections around the half wormhole (and vice versa), somwhat unifying the two pictures \cite{Mukhametzhanov:2021nea}. This is reminsicent of Eberhardt's demonstration that perturbative fluctuations around one background may build up the contributions of other backgrounds \cite{Eberhardt:2021jvj,Eberhardt:2020bgq}}$^,$\footnote{Perhaps the dualities described in \cite{Jafferis:2021ywg} may be related to such a choice.}

\item Are there microscopic realizations of the "broken cylinders", or "half-wormholes", in conventional AdS/CFT? In Section \ref{eowbranesection} we briefly discussed models in which the random $\Psi$ boundaries are replaced by dynamical EOW brane boundaries. Another intriguing possibility is that the role of the $\Psi$ boundaries is played by the black hole singularity.\footnote{This was pointed out by Yiming Chen and Juan Maldacena.} Here we briefly discuss this possibility.

We begin by reviewing the "double cone" solution for the microcanonical spectral form factor, $Y_E^\dagger(T) Y_E(T)$ \cite{Saad:2018bqo}. The double cone solution can be constructed from the the eternal black hole geometry with energy $E$ by a periodic identification in Schwarzschild time, $t\sim t+ T$. The resulting geometry has a conical singularity at the location of the bifurcate horizon, $r=r_h$, but one can obtain a nonsingular geometry by taking the $r$ coordinate slightly into the complex plane to avoid the singularity at $r_h$. This complexifies the geometry.

A possible "broken cylinder", or "half-wormhole" solution which is related to the double cone can be constructed by a similar quotient of the eternal black hole spacetime. For simplicity, we focus on the computation of one copy of $Y_E(T)$. Beginning with the eternal black hole, we again periodically identify the geometry under $t\sim t+T$, but we now take the $r$ coordinate to run from infinity to zero, going around the singularity at $r=r_h$. For $r<r_h$, this describes the evolution of a closed universe (the periodically identified Einstein-Rosen bridge) in the timelike $r$ direction, with this evolution ending at the black hole singularity. This geometry has the required boundary conditions to contribute to $Y_E(T)$, and is a solution to Einstein's equations away from $r=0$. However, to compute the contribution of this "solution" to $Y_E(T)$, one would need to evaluate the spacetime action near the singularity, which is UV-sensitive.

One expects that in a UV-complete theory of quantum gravity, the black hole singularity is resolved. One might hope that this proposed half-wormhole is a genuine solution in such theories, and gives a noisy contribution to $Y_E(T)$, where the noise is a result from some UV physics near the singularity. 

This solution could make contact with the final state proposal \cite{Horowitz:2003he,Gottesman:2003up} for a resolution to the black hole information problem. In the proposed half-wormhole solution, one could view the (periodically identified) singularity as projecting onto a state $|\Psi\rangle$ of a single closed universe; the contribution to $Y_E(T)$ would be related to the wavefunction of this state.

In order to produce the correct noise in the spectral form factor, one expects that the state $|\Psi\rangle$ imposed at the singularity should look like a random state.\footnote{Douglas Stanford \cite{StanfordTalk}, as well as Marolf and Maxfield \cite{Marolf:2020rpm}, have suggested that averaging over random black hole final states produces wormholes in this way. Stanford has also emphasized the importance of corrections to the random Gaussian behavior of the state, required to ensure unitarity. The authors of \cite{Stanford:2021bhl} have identified effects which they point out may explain this.} For example, if we take $|\Psi\rangle$ to be a Gaussian random state, averaging over pairs of these half-wormholes should glue them together to make the double cone (after deforming the $r$ contour away from $r=0$ to wrap around the singularity at $r_h$)

\item In order to have a controlled disk-and-cylinder approximation in JT gravity and the MM model, we needed to work in approximate $\alpha$ states, rather than in exact $\alpha$ states. In models where there is no ensemble, we do not expect that there is a controlled toplogoical expansion. But are there other ways to regain control? Perhaps there are other kinds of averaging, such as time averaging, which are sufficient to recover a controlled disk and cylinder approximation.\footnote{For example, is time averaging the spectral form factor over a small energy window (small enough so that the order one fluctuations are not averaged out) sufficient to find a controlled approximation?}

\item Perhaps the most remarkable result from \cite{Marolf:2020xie}, which we have not emphasized here, is that contributions of higher topologies lead to the existence of null states. Removing these null states reduces the size of the physical Hilbert space, or equivalently the number of $\alpha$ states. Marolf and Maxfield, as well as Macnamara and Vafa \cite{McNamara:2020uza}, proposed that a way to reconcile wormholes and conventional-non-averaged AdS/CFT is that this phenomenon occurs in the extreme, rendering the Hilbert space one-dimensional. This implies, among other things, that the No-Boundary state is (the unique) $\alpha$ state.\footnote{Assuming that the No-Boundary state exists, which seems to be a reasonable assumption, then if the Hilbert space of closed universes is one-dimensional, then all states are proportional to the No-Boundary state. If one attempts to form other states by acting on the No-Boundary state with boundary operators to create a state $f(\hat{Z}[\mathcal{J}])|NB\rangle$, we can use the fact that correlation functions factorize in the No-Boundary state, so that it is an eigenvector of the $\hat{Z}[\mathcal{J}]$ with eigenvalues $Z[\mathcal{J}]$, so that this state is simply equal to $f(Z[\mathcal{J}])|NB\rangle\propto |NB\rangle$. Said another way, in states that are "gauge-equivalent" to the No-Boundary state, the boundaries can be normalized away.} Then partition functions would factorize in the No-Boundary state.

Computations in the No-Boundary state, in all existing formulations, correspond to a sum over geometries. In Appendix \ref{appendixsharing} we try to emphasize the importance of the extra $\alpha$ state boundaries for factorization, in particular the fact tht they are "shared" between the operators in the correlation function. This idea generalizes the geometric "exclusion effect". In the No-boundary state, there are no boundaries of this sort to give such an effect. This seems to suggest that factorization in the No-Boundary state requires some other, nongeometrical, mechanism.\footnote{We emphasize that an orthodox sum over geometries can be sufficient for a theory to have the null states described by Marolf and Maxfield. For example, the MM model was defined as an unrestricted sum over topologies, without an identification like the exclusion rule. So a theory for which removing null states reduces the dimension of the Hilbert space to one then must be in some way qualitatively different than theories for which the dimension of the Hilbert space is reduced by removing null states, but remains greater than one.}

\end{itemize}

\section*{Acknowledgements} 
We thank Adam Levine, Don Marolf, Henry Maxfield, Douglas Stanford, Zhenbin Yang, and Ying Zhao for valuable discussions.   This paper has taken a very long time to appear.  We thank the organizers of the conferences Black Microstructure 2020 and Island Hopping 2020 \cite{Saadtalk} for the opportunity to present preliminary versions of these ideas.   PS was supported by the Marvin L. Goldberger Membership and W. M. Keck Foundation Fund at the Institute for Advanced Study. SS and SY were supported in part by NSF grant PHY-1720397. 
\appendix
\section{The diagonal sum is self-averaging}\label{diagselfaveraging}
In this appendix, we derive (\ref{eq:diagequalscyl}). In the CGS model with species, $\langle I | \psi_{\{z_{I}\}}^{2}\rangle$ is a Gaussian random number which satisfies $\mathbb{E}[\langle I | \psi_{\{z_{I}\}}^{2}\rangle\langle J | \psi_{\{z_{J}\}}^{2}\rangle]_{NB}=\delta_{IJ}$, with $\mathbb{E}[\dots]_{NB}$ denoting the average over the No-Boundary ensemble. For notational simplicity, in this appendix we will denote $\langle I | \psi_{\{z_{I}\}}^{2}\rangle$ as $r_I$. To show
\begin{align}\label{approxdiagcylidentity}
\text{Diag}(\mathcal{J},\mathcal{J}')=\sum_{I=1}^{d_{1}}\langle Z[\mathcal{J}] | I\rangle\langle Z[\mathcal{J}^{\prime}] | I\rangle r_I^2 &=\sum_{I=1}^{d_{1}}\langle Z[\mathcal{J}] | I\rangle\langle I | Z[\mathcal{J}^{\prime}]\rangle\left(1+\mathcal{O}\left(\sqrt{\frac{1}{d_{1}}}\right)\right)
\cr
& = \cyl(\mathcal{J},\mathcal{J}')\left(1+\mathcal{O}\left(\sqrt{\frac{1}{d_{1}}}\right)\right),
\end{align}
we first demonstrate that the average of $\text{Diag}(\mathcal{J},\mathcal{J}')$ in the No-Boundary ensemble is equal to $\text{Cyl}(\mathcal{J},\mathcal{J}')$, then show that the variance of $\text{Diag}(\mathcal{J},\mathcal{J}')$ variance is of order $1/d_1$ subject to some assumptions. Using $\mathbb{E}[r_i^2]_{NB}=1$, we find the average value
\be
\mathbb{E}\left[ \text{Diag}(\mathcal{J},\mathcal{J}')\right]=\sum_{I=1}^{d_{1}}\langle Z[\mathcal{J}] | I\rangle\langle I | Z[\mathcal{J}^{\prime}]\rangle=\text{Cyl}(\mathcal{J},\mathcal{J}').
\ee
Here we recognized the projector onto one-universe states $\hat{P}_1 = \sum_{I}|I\rangle\langle I|$, as well as the definition of the cylinder, $\cyl(\mathcal{J},\mathcal{J}') = \langle Z[\mathcal{J}] |\hat{P}_1| Z[\mathcal{J}^{\prime}]\rangle$

Now we compute the variance,
\be
\mathbb{E}[\text{Diag}^2]_{NB}-\mathbb{E}[\text{Diag}]_{NB}^2=\sum_{I,J} \langle Z[\mathcal{J}] | I\rangle\langle I | Z[\mathcal{J}^{\prime}]\rangle \langle Z[\mathcal{J}] | J\rangle\langle J | Z[\mathcal{J}^{\prime}]\rangle \mathbb{E}[r_I^2 r_J^2]_{NB}-\text{Cyl}(\mathcal{J},\mathcal{J}')^2 .
\ee
Since $r_I$ is a gaussian variable, $\mathbb{E}[r_I^2r_J^2]_{NB}=1+2\delta_{IJ}$. The normalized variance is:
\be
\frac{\mathbb{E}[\text{Diag}^2]-\mathbb{E}[\text{Diag}]^2}{\mathbb{E}[\text{Diag}]^2}=2\frac{\sum_I \langle Z[\mathcal{J}] | I\rangle^2\langle I | Z[\mathcal{J}^{\prime}]\rangle^2}{\text{Cyl}(\mathcal{J},\mathcal{J}')^2}
\ee
In the models that we study, the cylinder is of order $d_1^0$, so $\text{Cyl}(\mathcal{J},\mathcal{J}')=\sum_{I=1}^{d_{1}}\langle Z[\mathcal{J}] | I\rangle\langle I | Z[\mathcal{J}^{\prime}]\rangle\sim O(1)$. Assuming the wavefunctions $\langle I | Z[\mathcal{J}]\rangle$ vary "smoothly" with $I$ (they are not sharply peaked around particular values of $I$), then typical components have a size of order $1/\sqrt{d_1}$, so that $\langle Z[\mathcal{J}] | I\rangle\langle I | Z[\mathcal{J}^{\prime}]\rangle \sim O(1/d_1)$. Thus, the normalized variance scales as $1/d_1$.\footnote{In JT gravity, the one-universe Hilbert space is infinite-dimensional. However, one may be able to define a finite effective Hilbert space dimension which generalizes $d_1$. In the cylinder contribution to $\langle \hat{Y}_E(T)^\dagger \hat{Y}_E(T)\rangle_{NB}$ \ref{trumpetgluingwormhole}, the integral over $b$ is dominated by a saddle point in $b$. The width of the Gaussian expansion around this saddle point may be an appropriate generalization of $d_1$ in this case.}

\section{Explicit calculations in the CGS model with species}\label{appendixexplicitcalculationspecies}
In this appendix we will discuss the computations of the one-universe and two-universe components of an $\alpha$ state in the CGS model with species in more detail. Our goal will be to understand the ``factorization'' property of the off-diagonal two-universe components into the product of one-universe components,
\be
\langle \{I_1 I_2\}|\psi_{\{z_I\}}^2\rangle = \langle I_1|\psi_{\{z_I\}}^2\rangle \langle I_2|\psi_{\{z_I\}}^2\rangle,\hspace{20pt} I_1\neq I_2.
\ee
This property may be puzzling from the point of view of the geometrical picture described at the end of the Section \ref{cgsspeciesalphastates}. The computation of the two-universe component $\langle \{I_1 I_2\}|\psi_{\{z_I\}}^2\rangle$ only involves one copy of the $|\psi^2{\{z_I\}}\rangle$ boundaries, but the computation of the product $ \langle I_1|\psi_{\{z_I\}}^2\rangle \langle I_2|\psi_{\{z_I\}}^2\rangle$ involves two.

We start with the expression (\ref{psi2spec}) for the one-universe component,
\be\label{oneunivamp}
\langle I_1 | \psi_{\{z_I\}}^2 \rangle=\frac{1}{(\sqrt{2\pi})^{d_1}} \int_{-\infty}^\infty \prod_I dp_{+I}\; e^{-\sum_I \frac{p_{+I}^2}{2} + i p_{+I}(z_I-\disk_I)} (i p_{+I_1}).
\ee
We can give this state a sort of geometric picture by viewing $|\psi_{\{z_I\}}^2 \rangle$ as a superposition of "spacetime D-brane" states labeled by $p_{+I}$. A spacetime D-brane with can emit any number of identical closed universes, each in a state proportional to $\sum_I p_{+I}|I\rangle$. The amplitude for emitting one universe is given by (\ref{oneunivamp}). 

If we want to study factorization, we look at the two-universe components:
\be
\langle \{I_1I_2\}| \psi_{\{z_I\}}^2\rangle =\frac{1}{(\sqrt{2\pi})^{d_1}} \int_{-\infty}^\infty \prod_I dp_{+I} \; e^{-\sum_I \frac{p_{+I}^2}{2} + i p_{+I}(z_I-\disk_I)} (i p_{+I_1})(i p_{+I_2}).
\ee
Since $p_{+I_1}$ and $p_{+I_2}$ are independent Gaussian variables, we are free to multiply the above expression by
\be
1=\frac{1}{(\sqrt{2\pi})^{d_1}}\int_{-\infty}^\infty \prod_I dp'_{+I}\; e^{-\sum_I \frac{(p'_{+I})^2}{2} + i p'_{+I}(z_I-\disk_I)},
\ee
which corresponds to introducing a second, decoupled, spacetime D-brane, and relabel $p_{I_2}$ as $p'_{I_2}$ to find
\be
\langle \{I_1I_2\}| \psi^2\rangle=\langle I_1 |\psi^2\rangle \langle I'_2| \psi^2 \rangle=\langle I_1 |\psi^2\rangle \langle I_2| \psi^2 \rangle ,\hspace{20pt} I_1\neq I_2.
\ee
It is straightforward to generalize this computation to derive analogous relationships between components with more than two universes and the one-universe component, such as \eqref{threeuniversefactorizebasisindep}.

\section{The autocorrelation function in JT gravity} \label{autoJTappend}
In this appendix we will discuss the autocorrelation function in JT gravity. The autocorrelation function for the microcanonical partition function is
\be
\langle \hat{Y}_{E}(T) \hat{Y}_{E'}(T')^\dagger\rangle_{NB} = \frac{(\Delta E)^2}{(2\pi i)^2} \int d\beta d\beta' \; e^{\beta E +\frac{1}{2}\beta^2 \Delta E^2}e^{\beta' E' +\frac{1}{2}\beta'^2 \Delta E^2} \langle \hat{Z}(\beta+i T) \hat{Z}(\beta' - i T') \rangle_{NB},
\ee
in which $\langle \hat{Z}(\beta+i T) \hat{Z}(\beta' - i T') \rangle_{NB}$ is the autocorrelation function for the analytically continued partition function \cite{Saad:2019lba}
\be
\langle \hat{Z}(\beta+i T) \hat{Z}(\beta' - i T') \rangle_{NB} = \int_0^\infty b db\; \frac{e^{-\frac{1}{2}\frac{b^2}{\beta+ i T}}}{\sqrt{2\pi}\sqrt{\beta+i T}}\frac{e^{-\frac{1}{2}\frac{b^2}{\beta'- i T'}}}{\sqrt{2\pi}\sqrt{\beta'-i T'}}.
\ee
The autocorrelation function decays when $T'-T$ is large. For $\delta T \equiv T'-T \ll T$, we approximate
\be
\frac{1}{\beta' - i T'} \approx \frac{1}{\beta' - i T} - \frac{i \delta T}{(\beta' - i T)^2}.
\ee
For the time scales we are interested in, in which $T$ is on the ramp, $\beta,\beta'\ll T$. Thus,
\be
\frac{1}{\beta+i T}+ \frac{1}{\beta'-i T} \approx \frac{\beta+\beta'}{T^2}.
\ee
In this approximation, the autocorrelation function for the analytically continued partition function is
\be
\langle \hat{Z}(\beta+i T) \hat{Z}(\beta' - i T') \rangle_{NB} \approx \frac{1}{2\pi \sqrt{T T'}} \int_0^\infty b db \; e^{- \frac{\beta+\beta'+i \delta T}{2} \frac{b^2}{T^2} }.
\ee
Let $x=b/T$. Then we denote $\beta+\beta' = \beta_+$ and $\beta-\beta' = \beta_-$. Doing the integral over $\beta_-$,
\be
\langle \hat{Y}_{E}(T) \hat{Y}_{E'}(T')^\dagger\rangle_{NB} \approx \frac{\sqrt{2\pi}}{i\sqrt{2}}\frac{T^2\Delta E}{(2\pi)^3 \sqrt{T T'}} e^{-\frac{(E-E')^2}{4\Delta E^2}} \int d\beta_+ \int x dx \; e^{\frac{1}{2}\beta_+ (E+E') + \frac{1}{4}\beta_+^2 \Delta E^2} e^{-  \frac{\beta_+ +i \delta T}{2} x^2}
\ee
We then perform the Gaussian integral over $\beta_+$ to find a gaussian integral over $y=x^2$, with $y$ positive.
\be
\frac{1}{2}\frac{1}{\Delta E}\frac{ T^2 \Delta E}{(2\pi)^2\sqrt{TT'}} e^{-\frac{(E-E')^2}{4\Delta E^2}}\int_0^\infty dy \; e^{-\frac{(y-(E+E'))^2}{4\Delta E^2}+ i\frac{1}{2} \delta T y}
\ee
Notice that there is a saddle point at $y$ near $E+E'$
\be
y=(E+E')+i\delta T \Delta E^2,
\ee
which means $b$ near $T \sqrt{(E+E')}$. For $\delta T$ nonzero $y$ is slightly imaginary, which means $b$ is slightly imaginary.

For simplicity, we focus on energy windows which have small overlap with the edge of the spectrum, with $E/\Delta E, E'/\Delta E \gg 0$. Then we can ignore the endpoint at $y=0$ and approximate the integral by the contribution of this saddle point to find
\be
\langle \hat{Y}_{E}(T) \hat{Y}_{E'}(T')^\dagger\rangle_{NB}\approx \frac{\Delta E}{\sqrt{2}(\sqrt{2\pi})^3} T e^{-\frac{(E-E')^2}{4\Delta E^2}} e^{- \frac{3}{4}\Delta E^2 \delta T^2} e^{\frac{1}{2}i (E+E') \delta T}.
\ee
This result contains some important features. In particular, the Gaussian decay in $\delta T$ and $|E-E'|$ tells us that the $\hat{Y}_E(T)$ for times separated by much more than the autocorrelation time $1/\Delta E$, and energies separated by much more than the size of the energy window $\Delta E$, are approximately uncorrelated and so behave as approximately independent variables.

\section{Bound on the disk-and-cylinder approximation in JT gravity}\label{appendixbound}

In this appendix we explain the argument described in section \ref{JTapproxbound} in more detail. We begin with the computation of the norm of the approximate $\alpha$ state (\ref{approxalphanormjtonetime}), in which $\hat{Y}_E(T)$ is approximately fixed at a single time $T_1$. Our discussion of the computation of the norm of the approximate $\alpha$ state in JT gravity is essentially generalization of the analysis for the MM model in Section \ref{MMmodel} in the case that we fix $\hat{Y}_E(T)$ at a single time (with small additional complications in the more general case). For simplicity, we will not consider more general correlation functions.\footnote{Following our procedure in Section \ref{MMmodel}, it is straighforward to generalize our arguments to correlation functions of a small number of $\hat{Y}_E(T)$ in an approximate $\alpha$ state.}

First, we condsider the perturbative corrections to the disk-and-cylinder approximation. For $k\ll L$, the $k$'th term in the cumulant expansion \eqref{cumumlantjtforargumentsummary} can be approximated with the $k$-holed sphere, which is suppressed by a factor of $e^{-(k-2)S(E)}$.\footnote{In an $L\times L$ random matrix integral, connected $k$-point functions are typicaly suppressed by powers of $1/L$. Here $L \sim \Delta E e^{S(E)}$, but we will always take $\Delta E$ to be of order one so we do not need to distinguish between $L$ and $e^{S(E)}$ in our estimates.} We expect that this contribution decays in $T_1$; for sufficiently large $T_1$ other contributions may dominate, but to be conservative in our estimates we ignore the decay. Additionally, the contribution of the $k$-holed sphere grows as $k!$, due to the growth of the moduli space of Riemann surfaces with $k$ boundaries \cite{Manin:1999wj,zograf2008large}. Altogether, for $k\ll L$ the $k$'th term in \eqref{cumumlantjtforargumentsummary} is a homogeneous polynomial in $p_1$ and $q_1$ with coefficients that are at most of order $e^{-(k-2) S(E)}$ at large $S(E)$.\footnote{Additional combinatorical factors from the many terms in this polynomial can be ignored.}

To study the corrections to the disk-and-cylinder approximation, we view the $k\geq 3$ terms as perturbations to a shifted Gaussian integral with a variance described by the cylinder contribution to the $k=2$ term and center determined by $y_E(T_1)$.\footnote{Here for simplicity we approximate the disk as small compared to $y_E(T_1)$.} We then shift $p_1$ and $q_1$ so that the Gaussian part of the integral is centered at $p_1,q_1=0$. $p_1$ and $q_1$ are shifted by an amount of order $|y_E(T_1)|/\cyl_E(T_1,T_1)$. For typical choices of $y_E(T_1)$ in the No-Boundary distribution, this is of order $1/\sqrt{T_1}$. Shifting the variables in the $k$'th term in \eqref{cumumlantjtforargumentsummary} results in lower order polynomials with coefficients that go as powers of $|y_E(T_1)|/\cyl_E(T_1,T_1)$ times the original coefficients in the polynomial, times combinatorical factors. If $p_1,q_1$ are shifted by an exponentially large amount, as would be the case for highly atypical choices of $y_E(T_1)$, these lower order polynomials could have large coefficients, and compete with the contributions of the disk and cylinder; for example, the $k=3$ term would lead to a quadratic term in the shifted variables, which could compete with the quadratic contribution of the cylinder. However, for typical choices of $y_E(T_1)$ the shift is small enough that the $k\geq 3$ terms lead to small corrections to the Gaussian disk-and-cylinder approximation to the integral.

So far, this just establishes that the $k\geq 3$ terms in the integral lead to small corrections in the expansion around $p_1, q_1$ near zero. However, at large $p_1, q_1$, the $k\geq 3$ terms in \eqref{cumumlantjtforargumentsummary} can be large and lead to large contributions in \eqref{approxalphanormjtonetime}. For example, in the analogous computation in the MM model, there were additional saddle points at large $p$, which involved all of the terms in the cumulant expansion (\ref{eq:ksumexpression}). These effects are nonperturbative, as they involve terms for all $k$, which in turn may be dominated by nonperturbative effects in the JT gravity matrix integral for sufficiently large $k$. In order to ensure that the disk-and-cylinder approximation is valid, we may choose $\Delta$ such that the region of large $p_1,q_1$ is suppressed, via the Gaussian term in \eqref{approxalphanormjtonetime}. Then any contributions which lead to large corrections to the disk-and-cylinder approximation to \eqref{cumumlantjtforargumentsummary} will be suppressed in the computation of the norm of the approximate $\alpha$ state.

We now argue that as long as $|p_1|,|q_1|\ll L$, then the additional terms in \ref{cumumlantjtforargumentsummary} are unimportant. For sufficiently small $k$ compared to $L$, we expect that these terms are dominated by the $k$-holed sphere. However, for $k\sim L^2$, we expect that nonperturbative effects take over. In appendix \ref{appendixmoments}, we explain our expectations in more detail, giving a conjecture for the behavior of these terms. For now, we simply assume this behavior and estimate the dependence of the size of these terms on $|p_1|, |q_1|$.

As we have already discussed, the terms for $k\ll L^2$, which are dominated by the $k$-holed sphere, are homogeneous polynomials of degree $k$ in $p_1$ and $q_1$ with coefficents that go as $ e^{-(k-2)S(E)}$ times a decaying function in $T_1$. To estimate the effect of these terms, we should factor out an overall power of $e^{2 S(E)}$, so that the exponent in \eqref{cumumlantjtforargumentsummary} is equal to $e^{2 S(E)}$ times a series in $p_1, q_1$, where for the terms for $k\ll L^2$ in this series each power of $p_1$ and $q_1$ comes with a power of $e^{-S(E)}$. Then we would like to understand when this series differs appreciably from the the disk-and-cylinder result. Including only contributions from the $k$-holed sphere, and ignoring the decay in time to be conservative, we would find that the corrections to the disk-and-cylinder result become important when $p_1, q_1 \sim e^{S(E)}$.

For $k \gtrsim L^2$, we expect that nonperturbative effects dominate. As we explain in appendix \ref{appendixmoments}, we expect that in this regime, the coefficients of the polyomial in $p_1, q_1$ begin to decrease quickly with $k$, as the correlators in \eqref{cumumlantjtforargumentsummary} do not have a $k!$ growth to cancel the $1/k!$ in the cumulant expansion. Instead, expect that the coefficients of the polynomial go at most as $L^k/k!$. With this behavior, we also expect that the large $k$ terms in \eqref{cumumlantjtforargumentsummary} only become important for $|p_1|, |q_1|$ at least of order $e^{S(E)}$. 

Altogether, we expect that if we suppress the region of integration space with $|p_1|, |q_1|\sim e^{S(E)}$, then nonperturbative effects will be suppressed. We may do this by choosing $\Delta \gg e^{-S(E)}$, so that the Gaussian weight $e^{-\frac{\Delta^2}{2}(p_1^2+q_1^2)}$ in \eqref{approxalphanormjtonetime} is small in that region. One might worry that the remaining part of the integrand in \eqref{approxalphanormjtonetime} could be large enough in this region to compensate for this suppression; however, this remaining piece of the integrand is an average over unit-norm phases. Then the average can be at most of order one.\footnote{In the MM model, there are regions in the $p$ integration space where the analogous expression to \ref{cumumlantjtforargumentsummary}, $\langle e^{i p \hat{Z}}\rangle_{NB}$, reached its maximum value of one. These regions contribute the nonperturbative effects we discussed. This is possible due to the fact that the eigenvalues of $\hat{Z}$ are evenly spaced, allowing the phases in $\langle e^{i p \hat{Z}}\rangle_{NB}$ to align. In JT gravity, there does not seem to be a clear reason for \ref{cumumlantjtforargumentsummary} to be of order one for $|p_1|,|q_1|\sim L$. Instead, we typically expect it to be highly suppressed away from $p_1,q_1=0$. As a result, we expect that our strategy gives a rather conservative bound on $\Delta$. Our strategy is based on choosing $\Delta$ to suppress the region of the $p_1,q_1$ integration space where the factor \ref{cumumlantjtforargumentsummary} in the integrand is dominated by nonperturbative effects. However, it may well be that \ref{cumumlantjtforargumentsummary} is small in this region.}

The connection between these nonperturbative contributions to the discreteness of the spectrum of the boundary Hamiltonian is not as clear as the analogous connection between nonperturbative effects in the MM model and the discreteness of the spectrum of $\hat{Z}$.\footnote{One rather indirect way to see this connection is to note that the large $k$ behavior of the terms in \eqref{cumumlantjtforargumentsummary} is dominated by effects which rely on the discretness of the boundary energy levels, as described in appendix \ref{appendixmoments}.} However, we can heuristically understand this result in a way which makes closer contact with the discreteness of the boundary spectrum by thinking of the integral \eqref{approxalphanormjtonetime}, for fixed $p_1,q_1$, as given by a computation in a version of JT gravity with a perturbed eigenvalue potential. The perturbation to the eigenvalue potential is a function of $p_1$ and $q_1$, $\delta V(E_i, p_1,q_1) \equiv i p_1  e^{-\frac{(E-E_i)^2}{2 \Delta E^2 }}\cos(E_i T_1) + i q_1 e^{-\frac{(E-E_i)^2}{2 \Delta E^2}} \sin(E_i T_1) $. Integrating over $p_1, q_1$ with the Gaussian weight averages the results of these computations with the perturbed potential. If the typical size of $p_1,q_1$ is sufficiently large, this perturbation to the eigenvalue potential may lead to large nonperturbative effects. In their Gaussian distrbution, $p_1, q_1$ are typically of order $1/\Delta$. This tells us that the typical size of the perturbation to the eigenvalue potential is of order $1/\Delta$, while the unperturbed potential is of order $e^{S(E)}$. Choosing $\Delta \gg e^{-S(E)}$, we ensure that this perturbation to the potential is small. However, if $\Delta$ is much smaller than $e^{S(E)}$, this potential can have a large effect and the saddle point for the eigenvalues can change dramatically and effects from the discretness of the spectrum can be important.

Now we turn to the case in which we approximately fix $\hat{Y}_E(T)$ at $1\ll n_{max}\ll e^{S(E)}$ times $T_n$, spaced by a time interval $\Delta T$. The analysis of this case is simplest if we choose $\Delta T$ to be at least one autocorrelation time, $\Delta T \sim 1/\Delta E$. In that case, the modes $\hat{Y}_E(T_n)$ are approximately independent, and the analysis essentially reduces to our previous analysis in the case that we fix only one mode. If we choose the spacings $\Delta T$ to be an exponentially small fraction of the autocorrelation time, $\Delta T \sim  e^{- c S(E)}/\Delta E$, with an exponent $c<1$, then the analysis is slightly more complicated but the maximum variance in the approximate $\alpha$ state in between the times $T_n$ is still exponentially small. 

Again, for simplicity we focus on computing the norm of the approximate $\alpha$ state. The norm of the approximate $\alpha$ state can be expressed as an integral over variables $p_n, q_n$, with $n= 1\dots n_{max}$, which is proportional to
\be\label{JTfixmanytimes}
\int_{-\infty}^\infty \prod_{n=1}^{n_{max}}\bigg[ dp_n dq_n e^{-\frac{\Delta^2}{2}(p_n^2+q_n^2)} \bigg] \; \bigg\langle e^{\sum_{n=1}^{n_{max}}\big(i p_n \text{Re}[\hat{Y}_E(T_n)- y_E(T_n)] + i q_n \text{Im} [ \hat{Y}_E(T_n)- y_E(T_n)]\big)}\bigg\rangle_{NB}.
\ee
We then express the expectation value of the exponential as an exponential of the sum over cumulants,
\begin{align}\label{cumulantmanytimes}
&\bigg \langle e^{\sum_{n=1}^{n_{max}}\big(i p_n \text{Re}[\hat{Y}_E(T_n)] + i q_n \text{Im} [ \hat{Y}_E(T_n)]\big)}\bigg\rangle_{NB} 
\cr
&= \exp\bigg[\sum_{k=0}^\infty \frac{1}{k!} \bigg\langle \sum_{n=1}^{n_{max}} \big( i p_n \text{Re}[\hat{Y}_E(T_n)] + i q_n \text{Im} [ \hat{Y}_E(T_n)]\big)^k \bigg\rangle_{NB,c}\bigg].
\end{align}
The $k$'th term in the expansion has $k$ sums over the times $n$, involving the connected correlators between $\hat{Y}_E(T_n)$ at different times $T_n$. If the times $T_n$ are chosen to be spaced by much more than an autocorrelation time, then these connected correlators should be small unless all the $T_n$ in the correlator are chosen to be equal. This would allow us to express the exponent in \eqref{cumulantmanytimes} as a single sum over $n$. Then the integral \eqref{JTfixmanytimes} is a product over $n_{max}$ integrals, each of the form \eqref{approxalphanormjtonetime}.

On the other hand, if we take the $T_n$ to be exponentially close, for example if we take $\Delta T \sim  e^{- c S(E)}/\Delta E$ $c<1$ so that the variance is exponentially small, then the $k$'th term in the expansion \eqref{cumulantmanytimes} couples many of the $p_n$ and $q_n$. Roughly, the number $n_{coupled}$ of other variables $p_m, q_m$ that a given variable $p_n$ or $q_n$ is coupled to goes as the ratio of the time spacing with the autocorrelation time, $n_{coupled}\sim \frac{1}{\Delta E \Delta T} \sim e^{c S(E)}$. As our main interest is in the situation where we space the $T_n$ closely, we will focus on this case.

As explained in Section \ref{JTaprroxalphastate}, if we take the times to be spaced more closely than the autocorrelation time, we must be careful to choose the $y_E(T_n)$ to be appropriately correlated, for example by taking the $y_E(T_n)$ to be the values of $\hat{Y}_E(T)$ at times $T_n$ for a typical draw from the No-Boundary ensemble. In appendix \ref{jtsphereerrors}, we show that for typical choices of $y_E(T_n)$, the perturbative error in the integral \eqref{JTfixmanytimes} due to the $3$-holed sphere is small if $n_{coupled} \ll e^{\frac{1}{3} S(E)}$. This estimate is derived without taking into account the decay in time of the three-holed sphere; for times well within the ramp region, it seems reasonable to expect that this decay essentially allows us to ignore the three-holed sphere (as well as other perturbative corrections) altogether.

Now we address the nonperturbative corrections in \eqref{JTfixmanytimes}. As in our analysis of the case for $n_{max}=1$, our strategy is to argue that nonperturbative corrections to the disk-and-cylinder saddle point can only come from large $p_n, q_n$. Then if we take $\Delta$ to be sufficiently large, contributions from this region of the integral should be suppressed.

We expect that the terms in \eqref{cumulantmanytimes} with $k \ll L^2$ should be given to leading order by the $k$-holed sphere if the times $T_n$ are sufficiently close, and negligibly small otherwise. After pulling out an overall factor of $e^{2 S(E)}$ in the exponent of \eqref{cumulantmanytimes}, these terms are homogeneous polynomials of degree $k$ in the $p_n, q_n$ with coefficients of order $e^{-k S(E)}$ times a decaying function of the times.\footnote{Again, the $k$-holed sphere comes with a factor of $k!$ from the Weil-Petersson volume of surfaces with $k$ holes, but this is cancelled by the explicit factor of $1/k!$ in \eqref{cumulantmanytimes}.} There are of order $ n_{coupled} \choose k$ terms in each polynomial coupling together $n_{coupled}$ $p_n$ and $q_n$, which grows with $k$ at most as $n_{coupled}^k$. Then if the $p_n$ and $q_n$ are much smaller than $\sim e^{S(E)}/n_{coupled}$, the contributions of the $k$-holed spheres are small. For this estimate, we again did not use the decaying behavior in the times $T_n$; this would only serve to strengthen our bound.

In appendix \ref{appendixmoments} we briefly discuss the nonperturbative contributions to the correlators in \eqref{cumulantmanytimes} for large $k$. If the times $T_n$ are not chosen to be equal, it is difficult to estimate the behavior of these correlators. However, we expect that these contributions to the correlators should decay rapidly if the times are separated. Then, assuming the conjectured behavior of the correlation functions described in Appendix \ref{appendixmoments}, the couplings between many $p_n$ and $q_n$ should be unimportant, and if $p_n, q_n$ are much smaller than $\sim e^{S(E)}/n_{coupled}$, the large $k$ terms in \eqref{cumulantmanytimes} should be small.

In order to suppress the region of the integral where the $p_n$ and $q_n$ are of order $e^{S(E)}/n_{coupled}$, we must take\footnote{To see this, we set $|p_n|=|q_n| = e^{S(E)}/n_{coupled}$ for a group of $n_{coupled}$ variables $p_n,q_n$ in the Gaussian weight $\exp[-\frac{\Delta^2}{2}\sum_{n=1}^{n_{max}} p_n^2+q_n^2]$, with the remaining variables allowed to be small. This gives a suppression $\exp[- \frac{\Delta^2}{2n_{coupled}} e^{2 S(E)}]$. In order for this to be small, we must take $\Delta \gg \sqrt{n_{coupled}} e^{-S(E)}$.} 
\be
\Delta \gg \sqrt{n_{coupled}} e^{-S(E)}.
\ee

We end by noting that in this appendix we have been somewhat conservative with our estimates. For instance, we have ignored the time dependence of the correlators appearing in \ref{cumumlantjtforargumentsummary} and \ref{cumulantmanytimes} into account. We expect these correlators to decay in time, which leads us to expect that the corrections to the disk-and-cylinder approximation to \ref{cumumlantjtforargumentsummary} and \ref{cumulantmanytimes} are smaller than we have assumed for our argument. Furthermore, we have not made any attempt to estimate the size of \ref{cumumlantjtforargumentsummary} and \ref{cumulantmanytimes}. Instead, we used the fact that these expressions are averages of unit-norm phases, and so are bounded by one. Then, choosing $\Delta$ to obey our bound, the contributions of large $p_n, q_n$ are small even if nonperturbative effects cause \ref{cumumlantjtforargumentsummary} and \ref{cumulantmanytimes} to reach their maximum values in this region. However, it may well be that \ref{cumumlantjtforargumentsummary} and \ref{cumulantmanytimes} are small in this potentially dangerous region, which would strenghten the bound on $\Delta$.

\section{Correlators of many copies of $\hat{Y}_E(T)$}\label{appendixmoments}

In this appendix we study the behavior of correlators $\langle | \hat{Y}_E(T)|^{2 k'}\rangle_{NB}$ for large $k'$. The connected parts of these correlators are related to the correlators appearing in \ref{cumumlantjtforargumentsummary} and \ref{cumulantmanytimes}, identifying $k = 2k'$

To get some intuition we start by studying a related problem, the computation of $\langle |\Tr [U]|^{2k}\rangle_{Haar}$, where $U$ is an $L\times L$ unitary matrix and we average over the unitary group with the Haar measure. Here we are viewing the random unitary matrix $U$ as analogous to the unitary matrix $e^{- i  H T}$, with $H$ drawn from the JT gravity ensemble. These unitary matrices have different statistics, but we will draw some lessons from the analysis of the simpler problem $\langle |\Tr [U]|^{2k}\rangle_{Haar}$.\footnote{For some background see \cite{Haake:1315494,mehta2004random}.}

Writing $\Tr[U]= \sum_{i=1}^L e^{i \theta_i}$, we can express the Haar average as an integral over the phases $\theta_i$ with a measure $\propto |\Delta(\{e^{i\theta_i}\})|^2 \prod_{i=1}^L d\theta_i$, with $\Delta(\{e^{i\theta_i}\})$ the Vandermonde determinant. We can combine the Vandermonde determinant and the insertions of $\Tr[U]$ into a potential for the phases $\theta_i$,
\begin{align}\label{unitarypotential}
\langle |\Tr [U]|^{2k'}\rangle_{Haar} & \propto \bigg[\prod_{i=1}^L\int_0^{2\pi}  d\theta_i \bigg]\; e^{- V(\{\theta_i\})},
\cr
-V(\{\theta_i\}) = \sum_{i<j} \log\bigg[ \sin^2 & \bigg(\frac{\theta_i-\theta_j}{2}\bigg)\bigg]+ k' \log\bigg[\sum_{i,j} \cos(\theta_i-\theta_j)\bigg].
\end{align}
It is useful to expand the first term in the potential as\footnote{The formula $\sum_{n=1}^{\infty} \frac{2\cos(n\theta)}{n}=-\log(2-2\cos(\theta))$ is helpful here.}
\be
\sum_{i<j}\log\bigg[ \sin^2 \bigg(\frac{\theta_i-\theta_j}{2}\bigg)\bigg] = - \sum_{i<j} \sum_{n=1}^\infty \frac{2\cos(n(\theta_i-\theta_j))}{n}+ const.
\ee
We can simplify the problem by introducing the normalized density $\rho(\theta)\equiv \frac{1}{L}\sum_{i=1}^L\delta(\theta-\theta_i)$, as well as its Fourier modes $\rho_n \equiv \int_0^{2\pi} d\theta e^{i n \theta} \rho(\theta)$, and expressing \eqref{unitarypotential} as an integral over the density with a potential
\begin{align}
\frac{V(\rho)}{L^2} \approx const. + \int_0^{2\pi} d\theta d\theta' & \rho(\theta) \rho(\theta') \sum_{n=1}^\infty \frac{\cos(n(\theta-\theta'))}{n}  
\cr
& - \frac{k'}{L^2} \log\bigg[\frac{1}{2}\int_0^{2\pi} d\theta d\theta' \rho(\theta) \rho(\theta') \cos(\theta-\theta') \bigg]
\cr
\approx const. + \sum_{n=1}^\infty \frac{1}{n} |\rho_n|^2 &- \frac{k'}{L^2} \log\bigg[\frac{1}{2}|\rho_1|^2\bigg].
\end{align}
By the normalization of the density, the Fourier mode $\rho_0$ is constrained to be equal to one.

So far we have ignored any Jacobian from the change of variables to the continuous density. We can check that to leading order it is equal to one, by comparing with exact calculations for low $k'$, as well as fix the constant contributions to the potential.

Let's now study this for $k'$ scaling with $L$ in various ways. Varying the $\rho_n$, the equations of motion set $\rho_n=0$ for $n\geq 2$, but are solved by a nonzero value of $\rho_1$. This nonzero value of $\rho_1$ makes the density slightly non-uniform, with a maximum at some angle $\theta_1$, and a minumum at the antipodal angle. The magnitude of $\rho_1$ is of order $\sqrt{k'/L^2}$, and the phase of $\rho_0$ is $\theta_1$, which is a zero mode. At $k'\sim L^2$, $|\rho_1|$ becomes of order one and the density can become negative in some places. At this point we expect that the $n\geq 2$ modes are turned on so that the density is everywhere nonnegative.

For $k'\ll L^2$ one finds through an exact computation that $\langle |\Tr [U]|^{2k'}\rangle_{Haar} \approx k'!$. The action of the saddle point reproduces Stirling's approximation to this factorial. This result can be heuristically understood as counting the number of "wormhole" pairings between pairs of $\Tr[U]$ and $\Tr[U^\dagger]$.\footnote{Haar integrals over matrix elements of products of an order one number of unitary matrices are given to leading order by Wick contractions between pairs of $U$ and $U^\dagger$. These Wick contractions are analogous to wormhole pairings.} This Gaussian result dominates over any fully connected contribution to the correlator $\langle |\Tr [U]|^{2k'}\rangle_{Haar}$. Note that this $k'!$ behavior is \textit{not} related to the $k!$ growth of the $k$-holed sphere contribution to the correlator of $k$ copies of $\hat{Y}_E(T)$.

For $k'\sim L^2$, the density approaches $\rho(\theta) \approx \frac{1}{2\pi}\big(1+ \cos(\theta- \theta_1)\big)$. This indicates that the phases begin to bunch up at $\theta= \theta_1$. This gives the behavior $\langle |\Tr [U]|^{2k}\rangle_{Haar} \sim L^{2k'} e^{-c L^2}$ for some order one number $c$. As we will explain in the analogous problem of $\langle |\hat{Y}_E(T)|^{2k'}\rangle_{NB}$, we expect that this effect gives a contribution to the fully connected correlator as well.

The $k'!$ behavior for $k'\ll L^2$ matches with this result if we take $k' = L^2$, so that the rapid factorial growth slows down to the exponential $L^{2k'}$ growth.

We can understand this result as follows: for $k'$ large, the expectation value $\langle |\Tr [U]|^{2k'}\rangle_{Haar}$ should be dominated by configurations of the eigenvalues which maximize $|\Tr[U]|$. The largest value attainable by $|\Tr[U]|$ is $L$, in which case all of the eigenvalues align. Such configurations do not contribute because of the Vandermonde measure factor, but we expect that for $k\gg L^2$, similar configurations, in which the eigenvalues are closely bunched around some angle $\theta_1$, dominate. For these configurations, the Vandermonde contributes an action $\sim L^2$, so a configuration in which the eigenvalues are bunched up contributes an amount $\sim L^{2k'} e^{- c L^2}$ for some $c$.\footnote{To be precise, the action $\sim c L^2$ is the difference between the action of this bunched configuration and action of the the unperturbed, constant density.} The vandermonde punishes such closely bunched configurations with a large action, while the factor of $|\Tr[U]|^{2k'}$ gives the $\sim L^{2k'}$ scaling with $k'$. Additionally, we expect a suppression at large $k'$ from the one-loop determinant around these saddle points.

Using this intuition, we can also study the behavior for $k'\gg L^2$. In the limit where the eigenvalues are closely spaced, is useful to rescale $\theta_i = \frac{L}{\sqrt{k'}} \omega_i$. Then for large $k'/L^2$, we approximate the measure as
\be
\prod_{i=1}^L d\theta_i |\Delta( \{e^{i \theta_i}\})|^2 \approx \bigg(\frac{L^2}{k'}\bigg)^{L^2/2} \prod_{i=1}^L d\omega_i \Delta(\{\omega_i\})^2 ,
\ee
and the insertion of $|\Tr[U]|^{2k'}$ as
\begin{align}
|\Tr[U]|^{2k'}=\exp\bigg[ k' \log\Big[ \sum_{i,j} e^{i\frac{L}{\sqrt{k'}}(\omega_i-\omega_j)} \Big] \bigg] &\approx \exp\bigg[k' \log\Big[ L^2- \frac{1}{2}\frac{L^2}{k'} \sum_{i,j} (\omega_i-\omega_j)^2\Big]\bigg]
\cr
&\approx L^{2k'} \exp\bigg[-\frac{1}{2} \sum_{i,j} (\omega_i-\omega_j)^2\bigg].
\end{align}
At large $k'/L^2$, we then find
\be
\langle |\Tr[U]|^{2k'} \rangle_{Haar} \propto \bigg(\frac{L^2}{k'}\bigg)^{L^2/2} L^{2k'} \times \int\prod_{i=1}^L d\omega_i \Delta(\{\omega_i\})^2 \exp\bigg[-\frac{1}{2} \sum_{i,j} (\omega_i-\omega_j)^2\bigg].
\ee
The integral over the $\omega_i$ is independent of $k'$. The action for this integral is of order $L^2$, and is an approximation to the action in the Haar integral without the insertion of $|\Tr[U]|^{2k'}$, in the limit that the eigenvalues are closely bunched up, so this integral goes as $e^{-cL^2}$ for some $c>0$. The factor of $\big(\frac{L^2}{k'}\big)^{L^2/2}$ resulted from rescaling the $\theta_i$, so we associate it with the one-loop determinant. Altogether, this result matches our expectations.

In this simple example it was not difficult to find these formulae for the behavior of $\langle |\Tr[U]|^{2k'} \rangle_{Haar}$. However, even without doing any work, we could have at least known that $\langle |\Tr[U]|^{2k'} \rangle_{Haar}$ is not any larger than $L^{2k'}$. This is because $L$ is the maximum value of $|\Tr[U]|$, and we are averaging over the Haar ensemble with a normalized measure. Part of our reason for studying the behavior in more detail is that our interest is in the behavior of the \textit{connected} correlator (or the analogous connected correlator in JT gravity). It is more difficult to compute the connected correlators directly. However, the saddle point computation of the full correlator informs our expectations for the behavior of the full correlator; for example, the fact that the full correlator is dominated by a saddle point induced by the insertion of $|\Tr[U]|^{2k'}$ leads us to expect that the connected correlator is similarly dominated by such configurations.

This intuition should continue to hold in our problem of interest, $\langle |\hat{Y}_E(T)|^{2k'} \rangle_{NB}$. For $k'\ll L^2$, perturbative wormhole pairings dominate, giving $\langle |\hat{Y}_E(T)|^{2k'} \rangle_{NB} \sim k'!\; \cyl_E(T,T)^{k'}$, with a small connected part described by the $2k'$-holed sphere. However for $k'\sim L^2$, we expect that perturbation theory breaks down, and nonperturbative contributions in which $|\hat{Y}_E(T)|$ is large dominate. The scale $k'\sim L^2$ follows from the fact that the eigenvalue action is of order $L^2$, so the insertion of $|\hat{Y}_E(T)|^{2k'}$ may compete with this action for $k'\sim L^2$.

The maximum value of $|\hat{Y}_E(T)|$ is of order $L$, which is attained by configurations of the eigenvalues which are spaced by integer multiples of $2 \pi/T$. In these configurations, the phases in $\hat{Y}_E(T)$ align. As the No-Boundary distribution is normalized, the maximum value of $\langle |\hat{Y}_E(T)|^{2k'} \rangle_{NB}$ is then of order $L^{2k'}$. So this factorial growth at small $k'$ cannot continue for $k' \sim L^2$. As in the problem we have just analyzed, we expect that for $k'\sim L^2$, the integral over the eigenvalues is dominated by these configurations which maximize the size of $\hat{Y}_E(T)$. 

This problem is somewhat more complicated than the problem for $\langle |\Tr[U]|^{2k'} \rangle_{Haar}$. In the saddle point analysis of the integral for $\langle |\hat{Y}_E(T)|^{2k'} \rangle_{NB}$, we can think of the insertion of $|\hat{Y}_E(T)|^{2k'}$ as contributing a term to the eigenvalue potential. This term is proportional to $k'$, and oscillates with a spacing $\sim 1/T$ between wells. For $T< T_{plateau}$, the spacing between these wells is larger than $e^{-S(E)}$, the unperturbed spacing between neighboring eigenvalues. We expect that for $T$ on the ramp, and for $k'\sim L^2$, there is a complicated interplay between the spacing of the wells, the Vandermonde determinant, and the unperturbed eigenvalue potential. However, for this regime of large $k'$, the integral should be dominated by these configurations in which the eigenvalues are approximately spaced by multiples of $2\pi/T$, and the correlator grows no faster than $L^{2k'}$. 

These expectations lead us to conjecture that the connected correlators appearing in \ref{cumumlantjtforargumentsummary} behave similarly for $k\gg L^2$, growing at most as $L^k$. Furthermore, we conjecture that for intermediate $k\sim L^2$, the correlators do not grow faster than $k!$. With such behavior of these correlators, the terms in \ref{cumumlantjtforargumentsummary} for $k >2$ can be ignored for $|p_1|,\; |q_1| \ll L$. As explained in Section \ref{JTapproxbound}, this translates into a bound on $\Delta$ which lines up with our intuition that $\hat{Y}_E(T)$ behaves like a single copy of $\hat{Z}$ in the MM model.

In the case that we fix $\hat{Y}_E(T)$ at many times $T_n$, it is more difficult to describe the behavior of the correlators in \ref{cumulantmanytimes} for large $k$. We now must consider correlators of many copies of $\hat{Y}_E(T)$ at possibly different times. Here we will simply point out that the role of the saddle points we have just described in the one-time case suggests that these correlators should be small if the times are separated. The saddle point configuration was understood as corresponding to the configuration which maximizes (or almost maximizes) $\hat{Y}_E(T)$ for a given $T$, by aligning the phases $e^{i E_n T}$. However, a configuration which maximizes $\hat{Y}_E(T_n)$ does \text{not} maximize $\hat{Y}_E(T_m)$ for $T_m$ not an integer multiple of $T_n$. For generic $T_m$, $\hat{Y}_E(T_m)$ will be small for such a configuration. In our argument for our bound on $\Delta$, we assume that the correlators appearing in \ref{cumulantmanytimes} for $k \sim L^2$ and $k \gg L^2$ decay as the times are separated, and are small for times separated by more than the autocorrelation time. This seems to be a reasonable assumption based on the effect we just described.

\section{Errors from the $3$-holed sphere in JT gravity}\label{jtsphereerrors}

In this section we will describe the leading perturbative correction to the disk and cylinder approximation in our approximate $\alpha$ states in JT gravity, given by the three-holed sphere. Similiar to the discussion in Section \ref{sec:MM3holesphere}, we look at the three-holed sphere correction to equation \ref{JTfixmanytimes}.
\be\label{JTappendfix}
\int_{-\infty}^\infty \prod_{n=1}^{n_{max}}\bigg[ dp_n dq_n e^{-\frac{\Delta^2}{2}(p_n^2+q_n^2)} \bigg] \; \bigg\langle e^{\sum_{n=1}^{n_{max}}\big(i p_n \text{Re}[\hat{Y}_E(T_n)- y_E(T_n)] + i q_n \text{Im} [ \hat{Y}_E(T_n)- y_E(T_n)]\big)}\bigg\rangle_{NB}.
\ee
One of the main differences between this intergral and the one in Sec.\ref{sec:MM3holesphere} is the correlation of $Y_E(T_n)$ between different times. To make our discussion simple, we will denote $p_i$ as both $p_n$ and $q_n$ in Eq.\ref{JTappendfix}

We can then express the contribution of the three-holed sphere to the action of the integral \ref{JTappendfix} as
\be
V_3 = e^{-S} \sum_{ijk} V_{ijk}(T_i , T_j, T_k) p_i p_j p_k .
\ee
Here we separate out the expected suppression in the entropy, and pack all the other details about three holed sphere in to $V_{ijk}(T_i,T_j,T_k)$.

When we only keep the disk and cylinder contribution in Eq.\ref{JTappendfix}, it is a Gaussian integral over $p_i$ variables with non-zero mean, determined by the $y_i$ (with $i$ labeling the real and imaginary parts of the $y_E(T_n)$). Shifting the $p_i$, we can express the result as
\be
V_3 = e^{-S} \sum_{ijk} V_{ijk}(T_i , T_j, T_k) (p_i - i \cyl^{-1}_{ia} y_a) (p_j - i \cyl^{-1}_{jb} y_b) (p_k - i \cyl^{-1}_{kc} y_c).
\ee
Expanding this out makes terms which are quadratic or linear in the $p_i$, as well as a constant term. The first two can be viewed as shifts to the existing linear and quadratic terms $D_i, \cyl_{ij}$,
\be
\Delta D_i \sim e^{-S} \sum_{jk} V_{ijk} \cyl^{-1}_{jb}\cyl^{-1}_{kc} ,y_b y_c
\ee
\be
\Delta \cyl_{ij} \sim e^{-S} \sum_k V_{ijk}\cyl^{-1}_{kc} y_c.
\ee
We focus on fixing to a typical member in the ensemble here, which means $y_i\sim \sqrt{T}$. The $y_i$ are also correlated for a typical member of the ensemble; this correlation is key to introducing cancellations in the size of these corrections.

Let's look at $\Delta \cyl_{ij}$ first. It averages to zero, and its variance is
\be
\langle \Delta \cyl_{ij} \Delta \cyl_{i'j'} \rangle \sim e^{-2S}\sum_{kk'} V_{ijk} V_{i'j'k'} \cyl^{-1}_{kk'}.
\ee
$\cyl^{-1}_{kk'}$ couples $\sim n_{coupled}$ pairs of times together, and is of order $1/T$, so
\be
\langle \Delta \cyl_{ij} \Delta \cyl_{i'j'} \rangle \lesssim \frac{1}{T} n_c^2 e^{-2S } \times \text{Typical size of $V_{ijk}^2$}.
\ee
For late times the factor of $1/T$ helps, but we ignore it to be conservative. The RMS size of this is $\sim n_{coupled} e^{-S} \times \text{Typical size of $V_{ijk}$}$. Again we ignore the time dependence in $V_{ijk}$ to be conservative. Note that taking the $y_i$ to be averaged over the No-Boundary ensemble introduced factors of $\cyl_{ij}$, which canceled factors of $\cyl^{-1}_{ij}$. These cancellations tell us that the size of these corrections are smaller than one might naively expect for \textit{typical} choices of $y_i$.

Then we also have $\Delta D_i$. It has a nonzero average
\be
\langle \Delta D_i\rangle \sim e^{-S} \sum_{jk} V_{ijk} \cyl^{-1}_{jk} \sim e^{-S} n_{coupled}^2 \times \text{Typical size of $V_{ijk}$}.
\ee
Similarly, it has a variance of the same order. 

Finally we address the constant shift. We can analysis its size by looking at its variance. This is of order $n_{coupled}^3 e^{-S} T^{-3/2} \times \text{Typical size of $V_{ijk}$}$.

We can see that the strictest bound comes from the constant shift. We need this to be small in absolute terms, so a strict requirement is $n_{coupled}^3 e^{-S} \ll 1$ (without taking the decay into account). If this is obeyed, the other terms are also small. However, as we stated in Section \ref{JTapproxbound}, it seems reasonable to expect that if we take the (likely exponential) decay in time of the correlators into account, we can ignore this restriction for most of the ramp region.

\section{Formal discussion about factorization in the No-Boundary state}\label{appendixsharing}

In the models we have studied, the extra boundaries introduced by the $\alpha$ state are crucial for factorization. In particular, the fact that the $\alpha$ state boundaries are "shared" by different by the observables in a correlation function plays a key role. For example, in the disk-and-cylinder approximation, this sharing allows the for the exclusion effect, illustrated in Figure \ref{exclusioneffect}. In the No-Boundary state, there are no $\alpha$ state boundaries, so it does not appear that we can take advantage of the mechanism for factorization we have seen in this paper. Disconnected spacetimes with no $\alpha$ state boundaries are "shared" by observables, but this is also not enough to allow for factorization. We can attempt to sharpen this idea somewhat, generalizing the "exclusion effect". Even though the point of this argument is somewhat self-evident, we try to explain this point somewhat formally, in the hope that this formal discussion may be useful for narrowing down possible exceptions.

First we formally describe the computation of the one-point function and two-point function of operators $\hat{Z}[\mathcal{J}]$ in an $\alpha$ state. We will write formulas which should apply quite generally, assumming that these correlation functions are computed by an "orthodox" sum over geometries. We can describe the required properties of such an orthodox sum over geometries as follows:
\begin{itemize}
\item We use the same bulk theory to compute all correlation functions.

\item Correlation functions can be computed by a well-defined sum over spacetimes.\footnote{Perhaps the arguments described in this section may be circumvented if the sum over geometries gives an asymptotic expansion, though it seems unlikely. For example, the sum over geometries is asymptotic in JT gravity, but the essential ideas described in this section still hold. } These spacetimes have boundary conditions described by the operator insertions and state for the correlation function.\footnote{This assumption allows for additional boundaries as well, such as the EOW brane boundaries discussed in Section \ref{JTgravity}.} We also assume that we sum over all distinct spacetimes with the given boundary conditions.\footnote{For example, we do allow for an "exclusion rule", which restricts the sum over geometries.}

\item The action of a given spacetime, with any number of connected components, should be a \textit{sum} over actions for each connected component. That is, we allow for the bulk theory to be nonlocal, but these nonlocal effects must only be nonlocal over a given connected component of spacetime.\footnote{The "linked $\Psi$ boundaries" in the effective model may be viewed as violating this requirement.}
\end{itemize}

Consider the one-point function $\langle \hat{Z}[\mathcal{J}]\rangle_\alpha$ in an $\alpha$ state. This is computed by a sum over spacetimes with the boundary conditions corresponding to the $\alpha$ state, with a single connected component that has a $\hat{Z}[\mathcal{J}]$ boundary. Label the set of these connected components $\{m_i^{(1)}\}$. A given spacetime $m_i^{(1)}$ may have any number of boundaries introduced by the $\alpha$ state. A spacetime that contributes to $\langle \hat{Z}[\mathcal{J}]\rangle_\alpha$ that includes $m_i^{(1)}$ as a component also has components with boundary conditions described by the $\alpha$ state. However, $m_i^{(1)}$ may have already "used up" some of the $\alpha$ state boundaries, so the remaining components of the spacetime do not include all of the $\alpha$ state boundaries.\footnote{For concreteness, we can expand the $\alpha$ state as a superposition of states created by a fixed number of boundary operators, for example by writing the state in the "Z basis". Then the path integral for the correlator can be written as a sum over the number of boundaries in both the bra and ket. For fixed numbers $n$ and $n'$ of boundaries in the bra and ket, and where $m_i^{(1)}$ has $k$ boundaries from the $\alpha$ state, with $k\leq n+n'$, then the remaining components of a given spacetime have $n+n'-k$ boundaries. For each choice of $n$, $n'$, and $k$, the component $m_i^{(1)}$ has "used up" $k$ boundaries.} Denote the sum over spacetimes contributing to $\langle \hat{Z}[\mathcal{J}]\rangle_\alpha$, which include $m_i^{(1)}$ as a component, as $Z_1(m_i^{(1)}) D_i\langle 1\rangle_{NB}$, where $Z_1(m_i^{(1)})$ is the weight in the path integral of $m_i^{(1)}$, and $D_i$ is the path integral over the remaining components, with the factor of $\langle 1\rangle_{NB}$, the sum over spacetimes with no $\alpha$ state boundaries, factored out. As we have just explained, $D_i$ depends on the choice of $m_i^{(1)}$. For now, we will normalize $\langle 1\rangle_{NB}=1$, though we will reexamine this assumption of the normalization of the No-Boundary state later. Summing over the choice of $m_i^{(1)}$, 
\be\label{generalonepointalpha}
\langle \hat{Z}[\mathcal{J}]\rangle_\alpha = \sum_i Z_1(m_i^{(1)}) D_i
\ee
Now we consider the computation of the two-point function $\langle \hat{Z}[\mathcal{J}]\hat{Z}[\mathcal{J}]\rangle_\alpha$. For notational simplicity, we choose the operators to have the same boundary conditions $\mathcal{J}$. The path integral includes a sum over spacetimes where the $\hat{Z}[\mathcal{J}]$ boundaries are on separate connected components, or on the same connected component (a wormhole). Denote the set of connected spacetimes with two $\hat{Z}[\mathcal{J}]$ boundaries, as well as any number of $\alpha$ state boundaries, as $m_i^{(2)}$.

Now consider the spacetimes contribution to this two-point function which include two connected components $m_i^{(1)}$ and $m_j^{(1)}$, each with a $\hat{Z}[\mathcal{J}]$ boundary. Both $m_i^{(1)}$ and $m_j^{(1)}$ can "use up" some of the $\alpha$ state boundaries, so the remaining components of the spacetime have boundaries from the $\alpha$ state, but these boundaries do not include the ones already belonging to $m_i^{(1)}$ and $m_j^{(1)}$. Denote the sum over these remaining components of spacetime, for fixed $m_i^{(1)}$ and $m_j^{(1)}$, by $D_{ij}$.\footnote{This includes a factor of $\langle 1\rangle_{NB}$ which we may separate out. For now we do not write it explicitly, as we have assumed it is normalized to one.} Note that $D_{ij}$ is not a product of the sums $D_i D_j$ which would appear in the product of one-point functions \eqref{generalonepointalpha}; the $\alpha$ state boundaries are "shared" by the two operators in the two-point function. We think of this fact as generalizing the exclusion effect.

The computation of the two-point function also includes a sum over spacetimes with wormhole components $m_i^{(2)}$. Denoting the path integral weight of the connected component $m_i^{(2)}$ as $Z_2(m_i^{(2)})$, we can write the expression for the full sum over spacetimes for the two-point function as
\be
\langle \hat{Z}[\mathcal{J}]\hat{Z}[\mathcal{J}]\rangle_\alpha = \sum_{i,j} Z_1(m_i^{(1)})Z_1(m_i^{(1)}) D_{ij} + \sum_i Z_2(m_i^{(2)}) D_i
\ee
This should be equal to the square of \eqref{generalonepointalpha}. Of course, this is possible if the $D_{ij}$ and $Z_2(m_i^{(2)})$ are tuned, as they must be in an $\alpha$ state.\footnote{The $D_i$ and $D_{ij}$ would then be related through a generalization of equations \ref{eq:twouniverserelation} and \ref{eq:twouniversefactorizespecies}. It would be interesting to see if our effective description can be generalized to incorporate corrections from more complicated topologies.} However, we can see that if the wormhole contributions are nonzero, then in an $\alpha$ state we cannot have $D_{ij}= D_i D_j$; the "shared" property of the $\alpha$ states in the computation of $D_{ij}$ is crucial.

It has been suggested that for some theories, the Hilbert space of closed universes may be one-dimensional, and this fact may be visible through careful computations including the full sum over geometries, and any stringy or highly quantum effects. Naively, this may be plausible, as \cite{Marolf:2020xie} demonstrates that including contributions from many topologies may drastically reduce the number of independent states of closed universes. However, a theory with a trivial Hilbert space of closed universes is \textit{qualitatively} different than a theory with many null states but a nontrivial closed universe Hilbert space. It seems difficult to reconcile a geometric picture, including an orthodox sum over spacetimes, with a theory that has a trivial Hilbert space of closed universes but still includes wormholes.\footnote{We emphasize that the MM model, which has many null states, is defined by an orthdox sum over geometries. So the "gauge equivalence" between semiclassicaly distinct states resulting from these null states does \textit{not} require us to violate the assumptions laid out at the beginning of this appendix. For example, we are not required to "gauge fix" the gravitational path integral in the MM model by restricting our sum over geometries, beyond the usual constraints of diffeomorphism invariance.}

To demonstrate this tension, we examine the possibility that the Hilbert space of closed universes is one-dimensional. This implies that correlation functions must factorize in the No-Boundary state, which is then the unique $\alpha$ state.\footnote{This point has been emphasized in \cite{McNamara:2020uza}.} All other states of closed universes are gauge-equivalent to the No-Boundary state. Consider the computation of the one-point function and two-point function in the No-Boundary state. In this case, the $D_i$ and $D_{ij}$ are both independent of $i$ and $j$. This is because the dependence on $i$ and $j$ followed from the fact that the spacetimes connected to the $\hat{Z}[\mathcal{J}]$ could use up $\alpha$ state boundaries, changing the boundary conditions required for the sum over the remaining components of the spacetime. In this case, there are no $\alpha$ state boundaries to be used up; the disconnected spacetimes are only those that contribute to $\langle 1\rangle_{NB}$, which factors out of both the one-point function and two-point function. 

To be careful, we can momentarily relax our assumption that $\langle 1\rangle_{NB}=1$. Then we find the expressions
\begin{align}
\langle \hat{Z}[\mathcal{J}]\rangle_{NB}^2 &= \bigg( \sum_i Z_1(m_i^{(1)}) \langle 1\rangle_{NB} \bigg)^2\equiv Z_1^2\; \langle 1\rangle_{NB}^2,
\cr
 \langle \hat{Z}[\mathcal{J}]^2\rangle_{NB} &= \bigg(\sum_{i,j} Z_1(m_i^{(1)})Z_1(m_i^{(1)}) + \sum_i Z_2(m_i^{(2)}) \bigg)\langle 1\rangle_{NB} \equiv (Z_1^2+ Z_2) \langle 1\rangle_{NB}
\end{align}
In the product of one-point functions, we find two factors of $\langle 1\rangle_{NB}$, while in the two-point function we only find one factor; the closed universes contributing to $\langle 1\rangle_{NB}$ are \textit{shared} in the two-point function. However, this sharing is not enough to ensure that the two-point function factorizes; by equating the two expressions and solving for $\langle 1\rangle_{NB}$, one finds that $\langle 1\rangle_{NB}$ should be equal to $1+\frac{Z_2}{Z_1^2}$. However, this depends on our choice of operator $\hat{Z}[\mathcal{J}]$, so if $Z_2$ is nonzero, we cannot choose the normalization of the sum over disconnected spacetimes $\langle 1\rangle_{NB}$ so that correlation functions factorize for all choices of $\mathcal{J}$.

To summarize, in an $\alpha$ state, if the wormhole contributions are nonzero, we require that $D_{ij} \neq D_i D_j$. This relies crucially on the "shared" property of the $\alpha$ state boudaries, generalizing the "exclusion effect", and does not hold in the No-Boundary state. In order for correlation functions to factorize in the No-Boundary state, the bulk theory should violate at least one of the assumptions laid out at the beginning of this section.

The effective model of factorization described in this paper is \textit{not} described by an orthodox sum over geometries, as it includes an exclusion rule which one may think of as violating any of these three assumptions. The exclusion rule can be thought of as a choice between bulk descriptions, as describing a "gauge choice" which restricts the sum over geometries, or as giving an action for broken cylinders which is not a sum over independent actions for each broken cylinder (so that the linked $\Psi$ boundaries are "shared").

\bibliography{references}

\bibliographystyle{utphys}

\end{document}